\documentclass[preprints,article,accept,moreauthors]{Definitions/mdpi}

\firstpage{1} 
\pubvolume{xx}
\issuenum{1}
\articlenumber{5}
\pubyear{2020}
\copyrightyear{2020}
\history{}

\usepackage{Style_Ali}
\usepackage{Style_Avi}

\Title{\LARGE The Broadcast Approach in Communication Networks
}

\Author{Ali Tajer$^{1}$ , Avi Steiner$^{2}$, Shlomo Shamai (Shitz)$^{3}$}

\AuthorNames{Ali Tajer, Avi Steiner, Shlomo Shamai (Shitz)}

\address{%
$^{1}$ \quad Rensselaer Polytechnic Institute, Troy, NY, USA; tajer@ecse.rpi.edu\\
$^{2}$ \quad Technion---Israel Institute of Technology, Haifa, Israel; steiner.avi@gmail.com\\
$^{3}$ \quad Technion---Israel Institute of Technology, Haifa, Israel; sshlomo@ee.technion.ac.il}

\abstract{This paper reviews the theoretical and practical principles of the broadcast approach to communication over state-dependent channels and networks in which the transmitters have access to only the probabilistic description of the time-varying states while remaining oblivious to their instantaneous realizations. When the temporal variations are frequent enough, an effective long-term strategy is adapting the transmission strategies to the system's ergodic behavior. However, when the variations are infrequent, their temporal average can deviate significantly from the channel's ergodic mode, rendering a lack of instantaneous performance guarantees. To circumvent a lack of short-term guarantees, the {\em broadcast approach} provides principles for designing transmission schemes that benefit from both short- and long-term performance guarantees. This paper provides an overview of how to apply the broadcast approach to various channels and network models under various operational constraints. }


\allowdisplaybreaks
\begin{document}

\clearpage
\nointerlineskip
\tableofcontents



\section{Motivation and Overview}

\subsection{What is the Broadcast Approach?}

The information- and communication-theoretic models of a communication channel are generally specified by the probabilistic description of the channel's input and output relationship. The output, subsequently, depends on the channel input and the state process of the channel. The channel's probabilistic description changes over time in various domains, rendering a time-varying channel state process. These, for instance, include mobile wireless communications, storage systems, and digital fingerprinting, where all have time-varying communication mediums. Reliable communication generally necessitates transmitting an encoded message over multiple channel uses. Therefore, temporal fluctuations in channel states can cause a significant impediment to sustaining reliable communications. When channel states are known to the transmitters, the encoders can be guided to adjust the transmission rates in response to the changes in the channel's actual states. When a transmitter is informed of the channel state (e.g., via side information or feedback), it can adopt variable-length channel coding, the fundamental performance limits of which are well-investigated~\cite{Burnashev,Tchamkerten,Shayevitz,PPV:ISIT2010,Tyagi}.

While desirable, informing the transmitters of the time-varying state process can be practically prohibitive in a wide range of existing or emerging communications technologies. In such circumstances, while the encoders cannot adapt their transmissions to channel states, there is still the possibility of adapting the decoders to the channel states. The information-theoretic limits of communication over such state-dependent channels when the transmitters have only access to the statistical description of the channel state process is studied broadly under the notion of {\em variable-rate} channel coding~\cite{Verdu10variable-ratechannel}. When the temporal variations are frequent enough, an effective long-term strategy is adapting the transmission strategies to the system's ergodic behavior. However, when these variations are infrequent, their temporal average can deviate significantly from the channel's ergodic mode, rendering the ergodic metrics (e.g., ergodic capacity) unreliable performance targets.

State-dependent channels appear in various forms in communication systems. A prevalent example is mobile wireless channels, which undergo fading processes. Fading induces time-varying states for the channel, resulting in uncertainty about the network's state at all transmitter and receiver sites~\cite{SH98}. Other examples include opportunistic scheduling, in which the transmitter adjusts encoding and transmission based on a quality-of-service metric that depends on the state of the channel~\cite{TelatarOpp,Sharif,asadi}, e.g., signal-to-noise ratio, latency, and throughput; opportunistic spectrum access (across time, space, and frequency); and cognitive radio communication, in which the quality of communication relies on the access to the spectrum resources~\cite{Zhao,Tanab}. This survey paper focuses primarily on the fading process in different network models and the mechanisms for circumventing transmitters' lack of information about random fading processes. Nevertheless, most techniques that we will review can be adjusted to cater to other forms of state-dependent channels as well.

When wireless channels undergo fading, a useful convention to circumvent uncertainties about the fading process is establishing training sessions to estimate channel states. Such sessions should repeat periodically commensurate to how frequently the states vary. Depending on the multiplexing mode in a communication channel, the training sessions are either bidirectional (e.g., in frequency-division multiplexing systems) or they are unidirectional and ensued by feedback sessions (e.g., in time-division multiplexing systems). While effective in delivering the channel state to the receiver sites, both mechanisms face various challenges for delivering the same information to the transmitters. For instance, establishing channels in both directions is not always feasible, and even when it is, feedback communication incurs additional costs and imposes additional latency. Such impediments are further exacerbated as the size of a network grows.

When the probabilistic model of the process is known, an alternative approach to channel training and estimation is hedging against the random fluctuations. When the fluctuations are rapid enough, an effective long-term strategy is adapting the transmission strategies to the system's ergodic behavior. A widely-used instance of this is the ergodic capacity as a reliable transmission rate for a channel that undergoes a fast-fading process. On the other hand, when the fluctuations occur in time blocks, which is often the case, an effective strategy is the outage strategy, aiming to meet target reliability with a pre-specific probabilistic guarantee. An example of an outage strategy is adopting the notion of outage capacity, which evaluates the likelihood of reliable communication at a fixed transmission rate~\cite{OZ98}. When the actual channel realization can sustain the rate, the transmission is carried out successfully; and otherwise, it fails, and no message is decoded~\cite{SH98,OZ98}. The notions of outage and delay-limited capacities are studied extensively for various networks, including the multiple access channel (c.f.~\cite{HanlyTse,LiJindalGoldsmith,narasimhan,Haghi,DasNarayan,jafar} and references therein).

While the ergodic and outage approaches provide long-term probabilistic performance guarantees, they lack instantaneous guarantees. That is, each communication session faces a chance of complete failure. For instance, when the channel's instantaneous realization does not sustain a rate equal to the ergodic or outage capacity, the entire communication session over that channel will be lost. To circumvent a lack of short-term guarantees, the {\em broadcast approach} provides principles for designing transmission schemes that benefit from both short- and long-term performance guarantees. In information-theoretic terms, the broadcast approach is called {\em variable-to-fixed} channel coding~\cite{Verdu10variable-ratechannel}.

\subsection{Degradedness \& Superposition Coding}

The broadcast approach ensures a minimum level of successful communication, even when the channels are in their weakest states.  In this approach, any channel realization is viewed as a broadcast receiver, rendering an equivalent network consisting of several receivers. Each receiver is designated to a specific channel realization, and it is degraded with respect to a subset of other channels. Designing a broadcast approach for a channel model has the following two pivotal elements.

\vspace{.05 in}

\noindent {\bf 1- Degradedness in channel realizations:} The first step in specifying a broadcast approach for a given channel pertains to designating a notion of degradedness that facilitates rank-ordering different realizations of a channel based on their relative strengths. The premise for assigning such degradedness is that if communication is successful in a specific realization, it will also be successful in all realizations considered {\em stronger}.  For instance, in a single-user single-antenna wireless channel that undergoes a flat-fading process, the fading gain can be a natural degradedness metric. In this channel, as the channel gain increases, the channel becomes stronger.  Adopting a proper degradedness metric hinges on the channel model. While it can emerge naturally for some channels (e.g., single-user flat-fading), in general, selecting a degradedness metric is rather heuristic, if possible at all. For instance, in the multiple access channel, the sum-rate capacity can be used as a metric to designate degradedness, while in the interference channel, comparing different network realizations, in general, is not well-defined. 

\vspace{.05 in}

\noindent {\bf 2- Degradedness in message sets:}  Parallel to degradedness in channel realization, in some systems, we might have a natural notion of degradedness in the message sets as well. Specifically, in some communication scenarios (e.g., video communication), the messages can be naturally divided into multiple {\em ordered} layers that incrementally specify the entire message. In such systems, the first layer conveys the baseline information (e.g., the lowest quality version of a video); the second layer provides additional information that incrementally refines the baseline information (e.g., refining video quality), and so on. Such a message structure specifies a natural way of ordering the information layers, which should also be used by the receiver to retrieve the messages successfully. Specifically, the receiver starts by decoding the baseline (lowest-ranked) layer, followed by the second layer, and so on. While some messages have inherent degradedness structures (e.g., audio/video signals), that is not the case in general. When facing messages without an inherent degradedness structure, a transmitter can still split a message into multiple, independently generated information layers. The decoders, which are not constrained by decoding the layers in any specific order, will decode as many layers as they afford based on the actual channel realization.

In a communication system, in general, the states of degradedness in channel realizations and degradedness in message sets can vary independently. Subsequently, designing a broadcast approach for a communication system hinges on its channel and message degradedness status. By leveraging the intuitions from the known theories on the broadcast channel, we briefly comment on different combinations of the degradedness states.
\begin{itemize}
\item {\bf Degraded message sets.} A message set with an inherent degradedness structure enforces a prescribed decoding order for the receiver.
\begin{itemize}
\item {\it Degraded channels.} When there is a natural notion of degradedness among channel realizations (e.g., in the single-user single-antenna flat-fading channel), we can designate one message to each channel realization such that the messages are rank-ordered in the same way that their associated channels are ordered.  At the receiver side, based on the actual realization of the channel, the receiver decodes the messages designated to the weaker channels, e.g., in the weakest channel realization, the receiver decodes only the lowest-ranked message, and in the second weakest realization, it decodes the two lowest-ranked messages, and so on. Communication over a parallel Gaussian channel is an example in which one might face degradedness both in the channel and the message~\cite{Kfir:IZS2020}.
\vspace{.1 in}
\item {\it General channels.} When lacking a natural notion of channel degradedness (e.g., in the single-user multi-antenna channel or the interference channel), we generally adopt an effective (even though imperfect) approach to rank order channel realizations. These orders will be used to prescribe an order according to which the messages will be decoded. The broadcast approach in such settings mimics the K\"orner-Marton coding approach for broadcast transmission with degraded message sets~\cite{KornaerMarton77}. This approach is known to be optimal for a two-user broadcast channel with a degraded set of messages, while the optimal strategy for the general broadcast approach is an open problem despite the significant recent advances, e.g.,~\cite{NairGamal09}.
 \vspace{.05 in}
\end{itemize}

\item {\bf General message sets.} Without an inherent degradedness structure in the message, we have more freedom to generate the message set and associate the messages to different channel realizations. In general, each receiver has the freedom to decode any desired set of messages in any desired order. The single-user multi-antenna channel is an important example in which such an approach works effectively~\cite{ShitzSteiner03}. In this setting, while the channel is not degraded in general, different channel realizations are ordered based on the singular values of the channel matrix's norm, which implies an order in channel capacities. In this setting, it is noteworthy that the specific choice of ordering the channels and assigning the set of messages decoded in each realization induces degradedness in the message set.

\end{itemize}

Built based on these two principles, and following the broadcast approach to compound channels~\cite{CO72}, the notion of broadcast strategy for slowly fading single-user channel was initially introduced for effective single-user communication~\cite{Shitz97broadcast}.

\subsection{Application to Multimedia Communication}

The broadcast approach has a wide range of applications that involve successive and incremental retrieval of information sources. Representative examples include image compression and video coding systems, which can be naturally integrated with the successive refinement techniques~\cite{bergergibson,WWZ}. Specifically, the broadcast approach's underlying premise is to allow the receivers to decode the messages only partially, as much as the channels' actual instantaneous realizations allow. This is especially relevant in audio/video broadcast systems, in which even partially decoding the messages still renders signals that are aurally or visually interpretable or recognizable. In these systems, a transmitter is often oblivious to the instantaneous realization of the channels, and the quality of its channel shapes the quality of the audio or video signal recovered. This is also the principle widely used in communication with successive refinement, in which a message is split into multiple layers. A baseline layer carries the minimal content that allows decoding an acceptable message. The subsequent layers successively and progressively add more details to the message, refining its content and quality. This approach enables  digitally achieving a key feature of  analog
audio and video transmission: the quality of communication is a direct function of the channel
quality, while there is no channel state information at the transmitter.

In this review paper, we start by reviewing the core ideas in designing a broadcast approach in the single-user wireless channel in Section~\ref{sec:BCApproach}. In this section, we address both single-antenna and multi-antenna systems under various transmission constraints. Next, we provide an overview of the applications to the multiple access channel in Section~\ref{sec:MAC}. This section discusses settings in which transmitters are either entirely or partially oblivious to the channel states. Sections~\ref{sec:IC} and \ref{sec:relay} will be focused on the interference channel and the relay channel, respectively. A wide range of network settings will be discussed in Section~\ref{sec:networks}, and finally, Section~\ref{sec:outlook} provides a perspective on the possible directions for extending the theory and applications of the broadcast approach.

\section{Variable-to-fixed Channel Coding}
\label{sec:BCApproach}

As pointed out earlier, the broadcast approach is, in essence, a variable-to-fixed channel coding \cite{Verdu10variable-ratechannel}  for a state-dependent channel, where the state realization is known only at the receiver. While being oblivious to the channel realizations, the transmitter has access to the probabilistic description of the channel. The key idea underpinning the broadcast approach is splitting the transmitted message into multiple independent layers and providing the receiver with the flexibility to decode as many layers as it affords, depending on the channel's actual state. While the concept is general and can be applied to a wide range of state-dependent channels, in this paper we focus on wireless channels.

\subsection{Broadcast Approach in Wireless Channels}

In wireless communications, the channels undergo random fading processes. In these systems, the channel state corresponds to a fading gain, and the channel state statistical description is characterized by the probability model of the fading process~\cite{ShitzSteiner03, Shitz97broadcast,AsSh08_1,SH98}.  The relative duration of the channel's coherence time to the system's latency requirement specifies the channel's fading condition. Specifically, slow (fast) fading arises when the channel's coherence time is large (small) relative to the system's latency requirement. In particular, slowly fading channels are commonly when a mobile front-end moves slowly relative to the data transmission rate. Such a model is especially apt in modern communication systems with high spectral efficiency and data rates.

In systems with slowly-fading channels, a receiver can estimate the channel fading coefficients with high accuracy. This motivates considering the instantaneous and perfect availability of the channel state information (CSI) at the receiver sites. On the other hand, acquiring such CSI at the transmitter sites (CSIT) can be either impossible, due to the lack of a backward channel from a receiver to its respective transmitter; prohibitive, due to the extensive costs associated with backward communication; or unhelpful, due to a mismatch between the stringent latency constraints and the frequency of backward communication. Hence, in these circumstances, properly circumventing the lack of perfect CSIT plays a pivotal role in designing effective communication schemes.

Capitalizing on the system's ergodic behavior (e.g., setting the transmission rate to the ergodic capacity of a channel) effectively addresses the lack of CSIT~\cite{SH98}. However, this is viable only when the transmission is not facing any delay constraints, and the system is allowed to have sufficiently long transmission blocks (relative to the fading dynamics). In particular, in a highly dynamic channel environment, stringent delay constraints imply that a transmission block, in spite of still being large enough for having reliable communication \cite{OZ98}), is considerably shorter than the dynamics of the slow fading process. To quantify the quality of communication in such circumstances, the notion of capacity versus outage was introduced and discussed in~\cite{OZ98} and \cite{SH98} (and references therein). A fundamental assumption in these systems is that the fading process variations throughout the transmission block are negligible. In an outage strategy, the transmission rate is fixed, and the information is reliably retrieved by the receiver when the instantaneous channel realizations allow. Otherwise, communication fails (an outage event). In such systems, the term outage capacity refers to the maximal achievable average rate. It can also be cast as the capacity of an appropriately defined compound channel \cite{SH98}. The main shortcoming of the outage approach to designing transmission is the possibility of outage events, which translates to possibly a significant loss in spectral efficiency.

The broadcast approach aims to avoid outage events while the transmitters remain oblivious to the state of their channels. In this approach, reliable transmission rates are adapted to the actual channel conditions without providing feedback from the receiver to the transmitter. This approach's origins are discussed in Cover's original paper \cite{CO72}, which suggests using a broadcast approach for the compound channel. Since the slowly-fading channel
can be viewed as a compound channel with the channel realization
as the parameter of the compound channel, transmission over these channels can be naturally viewed and analyzed from the perspective of the broadcast approach. This strategy is useful in various
applications, and in particular, it is in line with the successive refinement source coding approach of~\cite{successiveCover1991} and the subsequent studies in~\cite{RI99,NgTian07,Tian08,Ng09,NgTian12}. Specifically, the underlying premise is that the more the provided information rate, the less average distortion evident in the reconstructed source. 

An example of successive refinement of source coding is
image compression, in which a gross description exists at first,
and gradually with successive improvements of the description, the image quality is further refined. An application example is
progressive JPEG encoding, where additional coded layers serve to
refine the image quality. In the broadcast approach, the
transmitter sends layered coded information, and in view of the
receiver as a continuum of ordered users, the maximum number of
layers successively decoded is dictated by the fading channel
realization. Thus, the channel realization influences the received
quality of the data. The broadcast approach has a practical appeal
in  voice communication cellular systems, where a layered voice
coding is possible. Service quality, subsequently, depends on the channel realization. This facilitates using coding to achieve the basic features of analog communications, that is, the better the channel, the better the performance, e.g., the measured signal-to-noise ratio (SNR) or the received minimum mean-squared error (MMSE). All this is viable, while the transmitters are unaware of channel realizations. Other applications can be found in~\cite{DuhamelKieffer09}. The problem of layered coding suggests unequal error protection
on the transmitted data, which was studied in \cite[and references
therein]{TR96}. A related subject is the priority encoding
transmission (PET). The study in \cite{BO99} shows
that sending hierarchically-organized messages over lossy
packet-based networks can be analyzed using the
broadcast erasure channel with degraded message set, using the
information spectrum approach~\cite{AL96}. Finally, we remark~\cite{Woyach} extends the notion to settings in which the probabilistic model is unknown to the transmitter.

\subsection{Relevance to the Broadcast Channel}

Since the broadcast approach's foundations hinge on those of the broadcast channel, we provide a brief overview of the pertinent literature on the broadcast channel, which was first
explored by Cover \cite{CO72,Cover}. In a broadcast channel,
a single transmission is directed to a number of receivers, each
enjoying possibly different channel conditions, reflected in their
received SNRs. The Gaussian broadcast channel with a single transmit antenna coincides with the classical physically degraded Gaussian broadcast channel, whose capacity region is well known (see \cite{Cover} for the deterministic case and \cite{TS02,li98_1,li98_2} for the composite or ergodic cases). For multiple transmit antennas, the Gaussian broadcast channel is, in general, a non-degraded broadcast
channel, for which the capacity region with a general message set is not fully known
\cite{VI02,SV02,KR02,CA01_1,YU01}, and it cannot be reduced to an equivalent set of parallel degraded broadcast channels, as studied in \cite{GA80,TS02,li98_1,li98_2}. In the special case of individual messages without common broadcasting, the capacity region in the multi-antenna setting was characterized in \cite{Weingarten06}.

Broadcasting a single user essentially means broadcasting common
information. Information-theoretic results and challenges for
broadcasting a common source are discussed in \cite{SE03}, and in
light of endless information, data transmission is termed streaming in
\cite{FED01}. The interpretation of single-user broadcasting is the hierarchical broadcasting using
multi-level coding (MLC) \cite{SC97,SC98, SC99}.
The study in~\cite{SC98} demonstrates the spectral efficiency of
MLC with hierarchical demodulation in an additive white Gaussian noise (AWGN) channel and a fading channel. The study in~\cite{SA96} examines the fading interleaved channel with one bit of side
information about the fading process. The broadcast approach is
adapted to decode different rates for channels
taking these two distinct states (determined by whether the SNR is
above or below a threshold value). Since the channel is memoryless,
the average rate, given by the mutual information we have \textcolor{black}{$I(y,\hat{s} ; x)$} (where $x$ is the channel input, $y$ is the channel output, and $\hat{s}$ is the partial state information), is achievable. This is not the case with the broadcast approach, which seems to be unfit here, where
channel states are assumed to be independent and identically distributed (i.i.d.).

Finally, the study in~\cite{Takesh01} considers a superposition
coding scheme to achieve higher transmission rates in the slowly-fading channel. This study adopts the broadcast approach for the
single-input single-output (SISO) channel with a finite number of receivers. The number of receivers is the number of coded layers. It is evident from \cite{Takesh01} that for the SISO channel, a few
levels of coded layering closely approximates the optimal strategy
employing transmission of infinite code layers.

\subsection{The SISO Broadcast Approach - Preliminaries}

In this section, we elaborate on the original broadcast approach, first presented in \cite{Shitz97broadcast}, and we provide the derivation of the expressions related to the broadcast approach concept, an optimal power distribution, and the associated average achievable rates under different system constraints. We start by providing a canonical channel model for the single-user single-antenna system. The fading parameter realization can be interpreted as an index (possibly continuous), which designates the SNR at the receiver of interest. This model also serves as the basis for other channel models discussed in the rest of the paper. Specifically, consider the channel model:
\begin{align}\label{SISO1}
y \; = \;  h x \; + \; n ~ ,
\end{align}
where $x$ is the transmitted complex symbol, $y$ is the received symbol, and $n$ accounts for the AWGN with zero mean and
unit variance denoted by ${\mathcal{CN}}(0,1)$. Constant $h$ represents the fading coefficient. For each realization of $h$, there is an achievable rate. We are interested in the
average achievable rate for various independent transmission blocks. Thus we present the results in terms of average performance, averaged over the distribution of $h$.

Information-theoretic considerations for this simple model were
discussed in \cite[and references therein]{OZ98}, as a special
case of the multi-path setting. With the $h$ value known to
the transmitter, and with a short-term power constrain (excluding
power optimization in different blocks), the reliable rate
averaged over many block realizations is given by
\begin{align}\label{SISO2}
C_{\rm erg} = \bbe_s[ \log(1+s P)]\ ,
\end{align}
where $s\triangleq |h|^2$ is the random fading power. The
normalized SNR, following the channel
model definition (\ref{SISO1}), is denoted by $P=\bbe[|x|^2]$, where $\bbe$ stands for the expectation operator (when a subscript is added, it specifies the random variable with respect to which the expectation is taken).
\begin{figure}[tb]
	\unitlength=1in
	\begin{center}
		\psfragscanon
		\includegraphics[width=4in, height=4.5in]{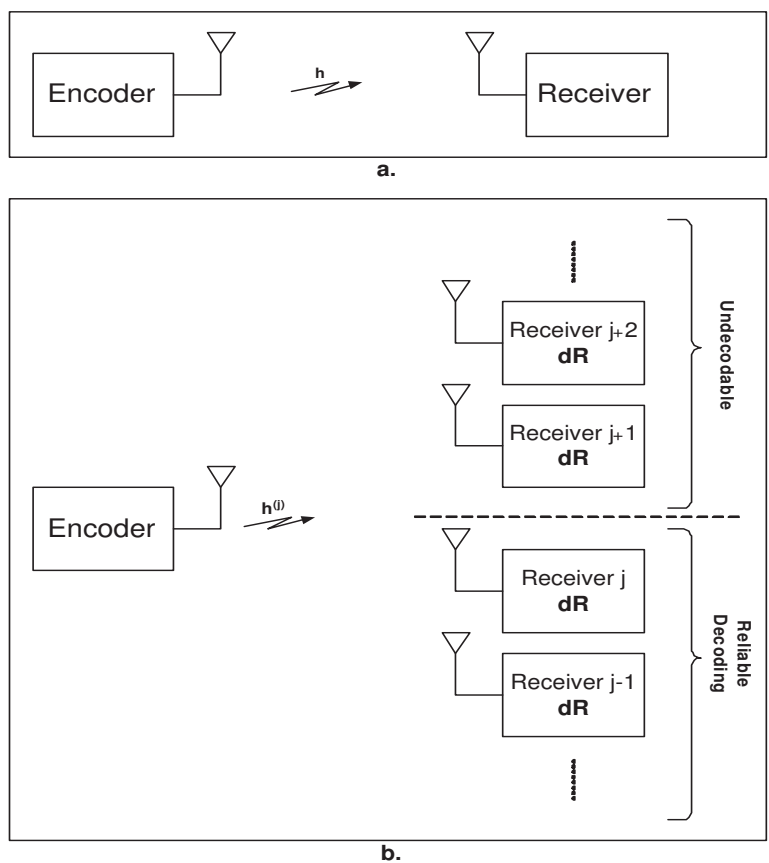}
	\end{center}
	\caption{ (a) A SISO channel with a fading parameter $h$. (b) The equivalent SISO broadcast channel model. For a channel realization $h^{(j)}$, only receivers indexed up to $j$ can decode their fractional rate $\d R$.}\label{fig:siso_broadcast}
\end{figure}

The SISO channel defined in (\ref{SISO1}) is illustrated in
Fig.~\ref{fig:siso_broadcast}(a), and its associated broadcast channel
is depicted in Fig.~\ref{fig:siso_broadcast}b. This figure
also illustrates the broadcast approach, according to which the transmitter
sends an infinite number of coded information layers. The receiver is
equivalent to a continuum of ordered users, each decoding a coded
layer if channel realization allows. In general, the number of coded
layers (and respectively, receivers) depends on the cardinality of
the fading power random variable (RV). Specifically, in a Gaussian fading channel, a continuum of coded layers is required. Predetermined ordering is achieved due to the degraded nature of the Gaussian SISO channel \cite{Cover}. Each of the users has to decode a fractional rate, denoted by $\d R$ in Fig.~\ref{fig:siso_broadcast}(b). The fractional rates $\d R$ of the different users are not equal but depend on their receiver index. For some fading
realization $h^{(j)}$, only the continuum of receivers up to
receiver $j$ can decode their fractional rates $\d R$. The first
receiver decodes only its own $\d R$, the second decode initially
the interference $\d R$ (information intended to the first user) and
then decodes its own $\d R$. Finally, receiver $j$ decodes all
fractional interferences up to layer $j-1$, and then decodes its
information layer $\d R$. Hence the total achievable rate for a
realization $h^{(j)}$ is the integral of $\d R$ over all
receivers up to $j$. This model is the general case of coded
layering. The broadcast approach in \cite{Shitz97broadcast} with a finite number of code layers, also termed superposition coding, is presented in \cite{Takesh01}. In finite level code layering, only a finite set of ordered receivers is required. This approach has
a lower decoding complexity. However, it is a broadcast sub-optimal
approach.

Next, assume that the fading power RV $S$ is continuous. Then for some channel realization $h^{(j)}$ of Fig.~\ref{fig:siso_broadcast}(b), with a fading power $s^{(j)}$, the
designated reliably conveyed information rate is denoted by
$R(s^{(j)})$. We now drop the superscript $j$, and refer to $s$ as
the realization of the fading power RV $S$. As illustrated, the
transmitter views the fading channel as a degraded Gaussian
broadcast channel \cite{Cover} with a continuum of receivers, each
experiencing a different effective receive SNR specified by
$s\cdot P$. The total transmitted power $P$ is also the SNR as
the fading and additive noise are normalized according to~\eqref{SISO1}.
The term $s$ is, therefore, interpreted as a continuous index. \textcolor{black}{By noting that for small enough $x>0$ $\log(1+x)\approx x$,} the
incremental differential rate is given by
\begin{align}\label{SISO3}
\d R(s)=\log
\myround{1+\frac{s\rho(s)\d s}{1+sI(s)}} = \frac{s\rho(s)\d s}{1+sI(s)}\ ,
\end{align}
where $\rho(s)\d s$ is the transmit power associated with a layer parameterized by $s$, intended for receiver $s$, which also designates the transmit power distribution. The right-hand-side equality is justified in \cite{V90}. Information streams intended for
receivers indexed by $u>s$ are undetectable and are treated as
additional interfering noise, denoted by $I(s)$. The interference
for a fading power $s$ is
\begin{align}\label{SISO4}
I(s)=\int\limits_s^\infty\rho(u)\d u\ ,
\end{align}
which is also a monotonically decreasing function of $s$. The total transmitted power is the overall collected power assigned to all layers, i.e., 
\begin{align}\label{SISO5}
P = \int\limits_0^\infty\rho(u)\d u = I(0)\ .
\end{align}
As mentioned earlier, the total achievable rate for a fading
realization $s$ is an integration of the fractional rates over all
receivers with successful layer decoding capability, rendering
\begin{align}\label{SISO6}
R(s) = \int_0^s\frac{u\rho(u)\d u}{1+uI(u)}\ .
\end{align}
The average rate is achieved with sufficiently many transmission
blocks, each viewing an independent fading realization. Therefore,
the total rate averaged over all fading realizations is
\begin{align}\label{SISO7}
R_{\rm bs} ~=~ \int\limits_0^\infty \d u~f(u)R(u) ~=~\int_0^\infty \d u (1-F(u))\frac{u\rho(u)}{1+uI(u)}\ ,
\end{align}
where $f(u)$ is the probability distribution function (PDF) of the
fading power, and 
\begin{align}
F(u)=\int\limits_0^u \d a f(a)\ ,
\end{align}
is the corresponding cumulative distribution function (CDF).

Optimizing $R_{\rm bs}$ with respect to the power
distribution $\rho(s)$ (or equivalently with respect to $I(u)$, where
$u\geq 0$) under the power constraint $P$ (\ref{SISO5}) is of
interest and can in certain cases be found by solving the
associated constrained E\"{u}ler equation \cite{GF91}. We turn
back to the expression in (\ref{SISO7}), corresponding to
$s_{\rm th}=0$, and explicitly write the optimization problem posed
\begin{align}\label{SISO10}
R_{\rm bs,max} ~=~ \max\limits_{I(u)}\int_0^\infty \d u
(1-F(u))\frac{u\rho(u)}{1+uI(u)}\ ,
\end{align}
where we maximize $R_{\rm bs}$ (\ref{SISO7}) over the residual
interference function $I(u)$. For an extremum function $I(x)$, the
variation of the functional (\ref{SISO10}) is zero \cite{GF91},
corresponding to a proper E\"{u}ler equation, which yields the
extremal solution for $I(x)$. Let us first present the functional
of (\ref{SISO10}) subject to maximization
\begin{align}\label{SISO11}
S(x,I(x),I'(x)) = (1-F(x))\frac{-xI'(x)}{1+xI(x)}\ .
\end{align}
The necessary condition for a maximum of the integral of
$S(x,I(x),I'(x))$ over $x$ is a zero variation of the functional
\cite[(Theorem 2, Section 3.2)]{GF91}. Correspondingly, the E\"{u}ler Equation is given by
\begin{align}\label{SISO12}
S_I - \frac{\d}{\d x}S_{I'} = 0\ ,
\end{align}
where
\begin{align}\label{SISO13}
S_I & = (1-F(x))\frac{x^2I'(x)}{(1+xI(x))^2}\ ,\\
S_{I'} & = (1-F(x))\frac{-x}{1+xI(x)}\ ,\\
\frac{\d}{\d x}S_{I'} & = \frac{xf(x)}{1+xI(x)} +
(1-F(x))\frac{x^2I'(x)-1}{(1+xI(x))^2}\ .
\end{align}
These relationships simplify from a differential equation
(\ref{SISO12}) to a linear equation by $I(x)$, providing the
following closed-form solution
\begin{align}\label{SISO14}
I(x) = \mycase{
	\begin{array}{cl}
	\frac{1-F(x)-x\cdot f(x)}{x^2f(x)} &~ x_0\leq x\leq x_1 \\
	0 &~ else
	\end{array}}~,
\end{align}
where $x_0$ is determined by $I(x_0)=P$, and $x_1$ by $I(x_1)=0$. All the analyses are also valid for the single-input multiple-output (SIMO) and multiple-input single-output (MISO) channels as long as the channels are degraded regardless of the
number of receive antennas in SIMO or transmit antennas in MISO. The number of transmit or receive antennas only affects the fading power distribution CDF. As an example, consider a SISO Rayleigh flat fading channel for which  the fading power $S$ has an exponential distribution 
with pdf
\begin{align}\label{SISO15}
f(u)=e^{-u}\ , \qquad \mbox{and} \qquad F(u)=1-e^{-u},~~~u\geq 0\ .
\end{align}
The optimal transmitter power distribution that maximizes
$R_{\rm bs}$ in (\ref{SISO10}) is specified by substituting  $f(u)$ and
$F(u)$ from (\ref{SISO15}) into (\ref{SISO14}), resulting in
\begin{align}\label{SISO16}
\rho(s) = -\frac{\d}{\d s}I(s) = \mycase{
	\begin{array}{cl}
	\frac{2}{s^{3}}-\frac{1}{s^{2}}\ , & s_0\leq s\leq s_1\\
	&\\
	0\ , & \mbox{else}
	\end{array}}\ .
\end{align}
Constant $s_0$ is determined by solving $I(s_0)=P$, and it is given by
\begin{align}
s_0 = \frac{2}{1+\sqrt{1+4P}}\ .
\end{align}
Similarly, $s_1$ can be found by solving $I(s_1)=0$, which indicates $s_1 = 1$. The corresponding
rate $R(s)$ using (\ref{SISO6}) is
\begin{align}\label{SISO17}
R(s) = \mycase{
	\begin{array}{cl}
	0 &,~ 0\leq s \leq s_0\\
	&\\
	2\ln(\frac{s}{s_0}) - (s-s_0) &,~ s_0\leq s\leq 1\\
	&\\
	-2\ln(s_0) - (1-s_0) &,~  s\geq 1
	\end{array}}\ ,
\end{align}
and following (\ref{SISO7}), the associated total average rate  is
\begin{align}\label{SISO18}
R_{\rm bs}=2E_i(s_0)-2E_i(1)-(e^{-s_0}-e^{-1})\ ,
\end{align}
where
\begin{align}\label{SISO19}
E_i(x)=\int\limits_x^\infty\frac{e^{-t}}{t}\;\d t,~~~x\geq 0
\end{align}
is the exponential integral function. The limiting behavior of
$R_{\rm bs}$ is found to be
\begin{align}\label{SISO20}
R_{\rm bs} \thickapprox \mycase{
	\begin{array}{cl}
	\ln\frac{P}{9.256} &,~ P\rightarrow\infty \\
	&\\
	\frac{1}{e}P &,~ P\rightarrow 0
	\end{array}}\ .
\end{align}
The ergodic capacity in this case is given by \cite{OZ98},
\begin{align}\label{SISO21}
C_{\rm erg} = e^{1/P}\cdot E_i(\frac{1}{P})\thickapprox \mycase{
	\begin{array}{cl}
	\ln\frac{P}{1.78} &,~ P\rightarrow\infty \\
	&\\
	P &,~ P\rightarrow 0
	\end{array}}\ .
\end{align}
The average achievable rate of the standard outage
approach, depends on the outage probability $P_{\rm out} = \PP\{s\leq
s_{\rm th}\}=1-e^{-s_{\rm th}}$. Thus, the achievable outage rate is given
by
\begin{align}\label{SISO22}
R_{\rm o}(s_{\rm th})=e^{-s_{\rm th}}\log(1+s_{\rm th}P)\ ,
\end{align}
where $R_{\rm o}(s_{\rm th})$ is the average achievable rate of a single
layered code for a parameter $s_{\rm th}$. That is, a rate of
$\log(1+s_{\rm th}P)$ is achieved when the fading power realization is
greater than $s_{\rm th}$, with probability
$e^{-s_{\rm th}}$.
The outage capacity is the product of maximizing the achievable
outage average rate (\ref{SISO22}) with respect to the outage
probability (or the fading power threshold $s_{\rm th}$). This yields
an outage capacity
\begin{align}\label{SISO23}
R_{\rm o,max}=e^{-s_{\rm th,opt}}\log(1+s_{\rm th,opt}P)\ ,
\end{align}
where $s_{\rm th,opt}$ solves the equation
\begin{align}\label{SISO24}
\log(1+s_{\rm th,opt}P)=\frac{P}{1+s_{\rm th,opt}P}\ ,
\end{align}
and it can be expressed in closed-form as
\begin{align}\label{SISO24_1}
s_{\rm th,opt} = \frac{P-W_L(P)}{W_L(P)\cdot P}\ ,
\end{align}
where $W_L(P)$ is the Lambert-W function, also known as the Omega function, which is the inverse of the function $f(W)=We^W$.
Subsequently, the outage capacity is given by \cite{AvestimehrTse07}
\begin{align}\label{SISO24_2}
R_{\rm o,max} = e^{-(P-W_L(P))/W_L(P)/P}\cdot\log\left(P/W_L(P) \right)\thickapprox \mycase{
	\begin{array}{cl}
	\ln\frac{P}{W_L(P)} &,~ P\rightarrow\infty \\
	&\\
	\frac{1}{e}P &,~ P\rightarrow 0
	\end{array}}\ .
\end{align}
The study in \cite{BustinSh15} provides an interesting interpretation for the basics of the broadcast approach \cite{Shitz97broadcast} from the I-MMSE perspective.

When a transmitter has full CSI and transmits at a fixed power $P$, the transmission rate can be adapted to channel state, and single-layer transmission can achieve the ergodic capacity. When variability in transmission power is allowed, and we face an average power constraint, a water-filling approach can be used. This facilitates adapting the transmission power and rate to the fading state, which is advantageous in terms of the expected rate. However, when lacking the perfect CSIT, the SISO broadcast approach can be optimized as studied in \cite{AsSh08}. In this approach, the CSI is quantized by the receiver and fed back to the transmitter. This allows for short latency, and the optimized achievable expected rate can be characterized as a function of the CSI accuracy.

The studies in \cite{AsSh08_2,ShenLiuFitz08} investigate various multi-layer encoding hybrid automatic repeat request (HARQ) schemes \cite{AsSh08_4}. The
motivation for extending the conventional HARQ schemes to
multi-layer coding is to achieve high throughput efficiency with
low latency. The study in \cite{AsSh08_2} focuses on finite-level coding with incremental redundancy HARQ, where
every coded layer supports incremental redundancy coding. The multi-layer bounds were investigated through continuous broadcasting by defining different broadcasting protocols that coherently combine HARQ and broadcasting incremental redundancy HARQ. Optimal power distribution cannot be obtained for continuous broadcasting. However, it was observed that even with a sub-optimal broadcasting power
distribution, significantly high gains of $\sim 3$ dB over an outage
approach could be achieved for low and moderate SNRs in the long-term static channel model, with latency as short as two blocks. In the long-term static channel model, the channel is assumed to remain in the same fading state within the HARQ session. This is especially
interesting as the conventional broadcast approach (without HARQ),
has only marginal gains over the outage approach for low SNRs.
The retransmission protocol of~\cite{AsSh08_2} is also an interesting approach, which uses
retransmissions for sending new information at a rate matched to
the broadcasting feedback from the first transmission. The optimal
broadcasting power distribution for outage approach retransmission was fully characterized in \cite{AsSh08_2}, and numerical results showed that it is the most efficient scheme for high SNRs, and at the same time, it closely approximates the broadcasting incremental redundancy-HARQ for low SNRs. However, in broadcasting incremental redundancy HARQ, only sub-optimal power distributions were used and finding the broadcasting optimal power distribution is still an open problem. It may also turn out that the broadcasting incremental redundancy HARQ with an optimal power distribution has more gains over the outage approach retransmission scheme.

\begin{figure}[tb]
	\unitlength=1in
	\begin{center}
		\psfragscanon
		\includegraphics[width=4.5in]{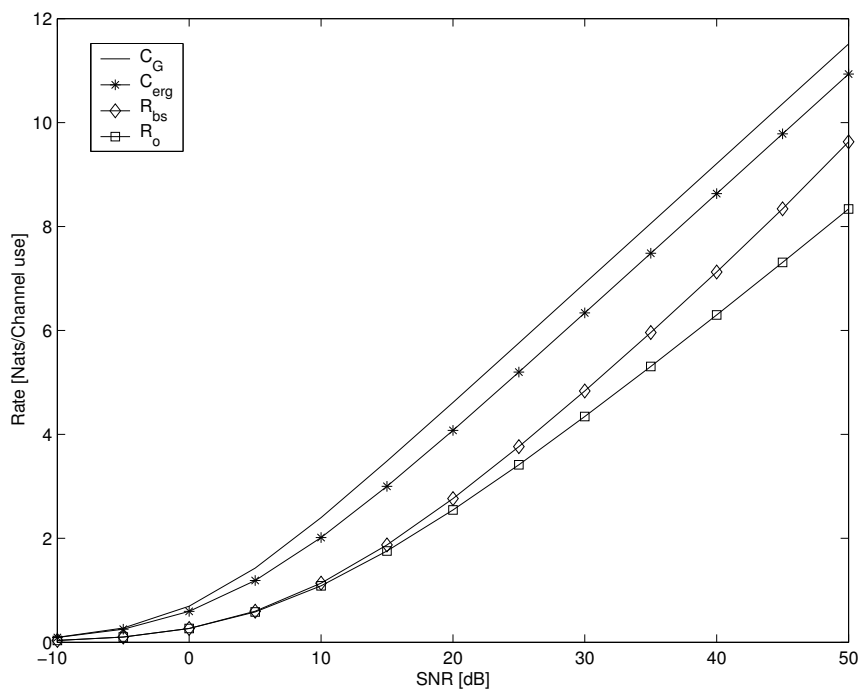}
	\end{center}
	\caption{SISO broadcast achievable average rate $R_{\rm bs}$, outage capacity $R_{\rm o}$, ergodic capacity $C_{\rm erg}$ and Gaussian channel upper bound $C_G$ versus SNR.}\label{fig:all_siso}
\end{figure}

Next, we  present the results on the achievable rates for the single-user
SISO Rayleigh flat fading channel under the broadcast
approach. Figure~\ref{fig:all_siso} demonstrates the SISO broadcast
achievable average rate $R_{\rm bs}$ (\ref{SISO18}), outage capacity
$R_{\rm o}$ (\ref{SISO23}), the ergodic capacity $C_{\rm erg}$
(\ref{SISO21})  upper bound, and the Gaussian capacity $C_G =
\log(1+P)$ as a reference. Clearly, $R_{\rm bs}>R_{\rm o}$ as the latter is
achieved by substituting $\rho(s)$ with $P\delta(s-s_{\rm th,opt})$ in
lieu of the optimized $\rho(s)$ in (\ref{SISO6}). Outage capacity
is equivalent to optimized single-layer coding rather than the
optimized continuum of code layers in the broadcast
approach. This difference is more pronounced in the high SNRs. Such a
comparison of the single- level code layer and two-level
achievable rates is presented in \cite{Takesh01}. This comparison shows
that two-level code layering is already very close to the optimum
$R_{\rm bs}$. The ergodic capacity in the general SIMO case, with $N$ receive antennas, is given by \cite[(9)]{T99}:
\begin{align}\label{eq:C_erg_1xM}
C_{\rm erg} = \frac{1}{\Gamma(N)}\int_0^\infty \d x \log(1+P\cdot
x)x^{N-1}e^{-x}\ ,
\end{align}
\textcolor{black}{where $\Gamma$ denotes the Gamma function.} The probability density of the total fading power for $N$ receive
antennas, is given by \cite{T99}
\begin{align}
f(\lambda)={\rm const}(N)\cdot \lambda^{N-1}e^{-\lambda}\ ,
\end{align}
where ${\rm const}(N)$ is a normalization constant.

\subsection{The MIMO Broadcast Approach}

Next, we review the multiple-input multiple-output (MIMO) channel. MIMO channels, in general,  are non-degraded broadcast channels. 
The MIMO capacity region is known for multiple users with private messages~ \cite{ Weingarten06}, and for two users with a common message \cite{GengNair14}. A complete characterization of the broadcast approach requires the full solution of the most general MIMO broadcast channel with a general degraded message set, which is not yet available. Hence, suboptimal ranking procedures are studied. Broadcasting with degraded message sets is not only unknown in general channels, but also, it is unknown for MIMO channels \cite{Chong14,Chong18}. Various approaches to transmitting degraded message set with sub-optimal ranking at the receiver are studied in \cite{ShitzSteiner03,AsSh04,BustinPaySh13}. 
The ranking of channel matrices (as opposed to a vector in a SIMO
case) can be achieved via supermajorization ranking of the singular
values of $HH^{\sf H}$. The variational problem for
deriving the optimal power distribution for the MIMO broadcast
strategy is characterized in \cite{ShitzSteiner03}, but seems not to lend itself to closed-form expressions. Thus a sub-optimal solution using majorization is considered and demonstrated for the Rayleigh fading channel.

We adopt the broadcast approach described earlier for the SISO and SIMO channels, in which the receivers opt to detect the highest possible rate based on the
actual realization of the propagation matrix $H$ not available to
the transmitter. In short, as $H$ improves, it sustains higher reliable rates. This is because the MIMO setting is equivalent to the general
broadcast channel (from the perspective of infinite layer coding), rather than a degraded broadcast channel as in
the single-input case. In the sequel, we demonstrate a broadcast
approach suited for this MIMO scenario. The approach suggests
an ordering of the receivers based on supermajorization of singular values of the channel norm matrix. Consider the following flat fading MIMO channel with $M$ transmit antennas and $N$ receive antennas:
\begin{equation}\label{1}
\boldy \;= \; H\boldx \; + \; \boldn \, ,
\end{equation}
where $\boldx$ is the input $(M\times 1)$ vector, $\boldn$
is the $ (N\times 1) $ noise vector with complex Gaussian i.i.d.
${\mathcal{CN}}(0,1)$  elements. The propagation matrix
$ (N\times M) $ is designated by $H$ and also possesses  complex
Gaussian i.i.d. ${\mathcal{CN}}(0,1)$ elements. The
received $ (N\times 1) $ vector is denoted by $ \boldy $. We
adhere to the non-ergodic case, where $H$ is fixed throughout the
code word transmission. We assume that the receiver is aware of
$H$ while the transmitter is not. The total transmit power constraint is $P$, i.e., $\bbe[\textrm{tr}\{\boldx \boldx^{\sf H}\}]\leq P$.

\subsubsection{Weak Supermajorization}\label{subsubWS}

First, we introduce some partial ordering relations based on
classical theory of majorization \cite{MO79}. Let $ \boldal =
\{\al_i\} , \, \boldbeta = \{\beta_i\} $ be two sequences of
length $K$. Let $ \{ \al_{(i)}\} \, , \{ \beta_{(i)}\} $ be the
increasing ordered permutations of the sequences, i.e., 
\begin{align}
\label{2}
\al_{(1)} & \le \al_{(2)} \, \dotsb \le \al_{(K)}\ , \\
\beta_{(1)} & \le \beta_{(2)} \, \dotsb \le \beta_{(K)} \, .
\end{align}
Let $ \boldal $ be weakly supermajorized by $ \boldbeta $, $
\boldal \prec^w\boldbeta $, that is
\begin{equation}
\label{3} \sum\limits_{i=1}^k\,\al_{(i)} \ge \sum\limits_{i=1}^k\,
\beta_{(i)} \; , \quad  k=1 \, \dotsc \, , K \, .
\end{equation}
Then, the relation $ \boldal \prec^w \boldbeta $
implies that \cite{MO79}
\begin{equation}
\label{4} \sum\limits_{i=1}^K\,\phi (\al_i) \le \dsum_{i=1}^K\,
\phi (\beta_i) \, ,
\end{equation}
for all continuous decreasing convex functions $ \phi (\cdot) $.

\subsubsection{Relation to Capacity}

Next, consider the received signal in (\ref{1}), where the
undetectable code layers are explicitly stated as
\begin{align}
\boldy = H (\boldx_S+\boldx_I) +\boldn \, ,
\end{align}
where $ \boldx_S $ and $ \boldx_I $ are decodable information and
residual interference Gaussian vectors, respectively. Their average norms are denoted by $P_S$ and $P_I$, respectively, and the total transmit power  $P=P_I+P_S$. $\boldn$ is an i.i.d. Gaussian complex vector with unit
variance per component. The mutual information between $\boldx_S$ and $\boldy$ is given by
\begin{align}\label{6}
I (\boldy ; \boldx_S) & = I(\boldy;\boldx_S,\boldx_I) - I(\boldy;\boldx_I|\boldx_S)\\
& = \log\det\, \left(I+\dfr{P_S+P_I}{M} HH^{{\sf H}}\right) -
\log\det \, \left(I+\dfr{P_I}{M}\, H H^{{\sf H}}\right)\\
& = \sum\limits_{k=1}^{J} \,\log \,\left(1+ \dfr{P_S \la_k}{1+P_I \la_k}\right)\\
& \triangleq C(\boldla ; P_S,P_I) \, .
\end{align}
Parameters $ \{\la_k\}$ for $k=1 \,\dots \, J$, where $J \triangleq \min (N,M) $,
designate the singular values (or eigenvalues) of the matrix $
\frac{1}{M}\, H^{{\sf H}} H $ for $ M\le N$, or $ \frac{1}{M}\, H
H^{{\sf H}} $ for $ N\le M $ \cite{T99}. 
Finally, if $ \boldla \prec^w \boldelta $, we have
\begin{equation}
\label{7} C(\boldla ; P_S,P_I) \ge C(\boldelta ; P_S,P_I) \, .
\end{equation}

\subsubsection{The MIMO Broadcast Approach Derivation}
We discuss the MIMO channel broadcast approach via supermajorization layering for the simple case of $M=N=2$. The signal $\boldx$ is
composed of a layered double indexed data stream with indices
denoted by $u$ and $v$. We refer to layer ordering by columns bottom-up, where $u$ and $v$
are described as a pair of indices taking integer values within
the prescribed region. This is only for demonstration purposes, as
indices $u$ and $v$ are continuous singular values of
$\frac{1}{2}HH^{\sf H}$. Say $u$ and $v$ are associated with the
\textbf{minimal} eigenvalue $ \la_2 $ and the sum of eigenvalues $
\la_2 + \la_1 $, respectively. Evidently, $ u\ge 0 , \, v\ge 2u $.
Say that $ \la_2 , \, \la_1 $ take on the set of integer values $
\{0,1,2,3,4\} $, then the layered system is described by $ (u,v) $
in the order: $ (0,0), \, (0,1) , \, (0,2) , \, (0,3) , \, (0,4) ,
\, (1,2) , \, (1,3) $, $ (1,4) , \, (2,4)$. The actual ordering of
the layers is in fact immaterial, as will be shown decoding is not
done successively as in the SISO case \cite{Shitz97broadcast}, but rather according to what is decodable adhering to partial ordering.

We envisage all possible realizations of $H$ and order them by $u
= \la_2$, $v = \la_2 + \la_1$ where $ \la_2 $ and $ \la_1 $ are,
respectively, the minimal and maximal eigenvalues of $ \frac{1}{2}
\, HH^{{\sf H}} $ (a $ 2\times 2 $ matrix in our case).
Supermajorization ordering dictates that all streams decodable for
realization $H$ will be decodable for realization $ H' $ as long
as
\begin{equation}\label{8}
\la_2^{'} \ge \la_2 ,  \,~~ \la_2^{'} + \la_1^{'} \gt \la_2 +
\la_1 \, .
\end{equation}
Thus, we visualize all possible realizations of $H$ as channels
referring to different users in a broadcast setting, and we
investigate the associated rates of the users, which we have ranked
as in Section \ref{subsubWS}, via a degraded ordering. It is
evident that the current approach specifies an achievable rate
region, but by no means is it claimed to be optimal. In fact, it even has some inherent limitations.

Let $ u = \la_2 $ and $ v=\la_1 $ be the eigenvalues of
$\frac{1}{2}HH^{\sf H}$ for some channel realization such that $ v\ge
u \ge 0 $. Let $ \rho (u,v) \, \d u \d v $ be the power associated
with the information stream indexed by $(u,v)$ where $v\ge u$, and
featuring the incremental rate $\d^2R(u,v)$. Again, for a given $u$
and $v$, all rates associated with the indices $ (a,b)\, , \; a
\le u $, $ b \le v $ can be decoded, as $( \la_2 , \, \la_1 )$ is
supermajorized by $( \la_2 = a , \, \la_1 = b )$. A natural optimization problem, in parallel to that posed and
solved for the single dimensional case, is to optimize the power
density $\rho(u,v)$, or the related interference pattern $I(u,v)$
maximizing the average rate, under the power constraint $I(0,0) =
P$. Let $I(u,v)$ designate the residual interference at $(u,v)$. Hence,
\begin{equation}\label{9}
I (u,v) = P - \int\limits_0^u \d a \int\limits_{a}^v\, \d b\, \rho
(a,b)\ .
\end{equation}
The associated incremental rate $ \d^2R(u,v) $, based on
(\ref{SISO3}) and (\ref{6}), is then given by
\begin{align}\label{10}
\d^2R(u,v)&=\log \,\left(1+ \frac{u\rho(u,v) \, \d u \d v}{1+uI
	(u,v)}\right) + \log \,\left(1+ \frac{v\rho(u,v) \, \d u \d v}{1+vI
	(u,v)}\right)\\
& =\,\frac{u\rho(u,v) \, \d u \d v}{1+uI (u,v)} + \frac{v\rho(u,v) \,
	\d u \d v}{1+vI (u,v)}\ .
\end{align}
The power density is the second order derivative of the residual
interference function (\ref{9}), i.e., 
\begin{equation}\label{10.5}
\rho(u,v) = -\frac{\partial^2}{\partial u\partial v} I(u,v) \triangleq
I_{uv}\ ,
\end{equation}
and the incremental rate may be expressed as
\begin{equation}\label{10.6}
\d^2R(u,v,I,I_{uv}) = -\dfr{uI_{uv}(u,v)\d u\d v}{1+uI(u,v)}  -
\dfr{vI_{uv}(u,v)\d u \d v}{1+vI(u,v)}\ .
\end{equation}
The accumulated reliable rate decoded at $(u,v)$ is
\begin{equation}\label{11}
R(u,v) = \int\limits_0^u\int\limits_{a}^v\, \d^2
R(a,b)\ .
\end{equation}
The expected rate, averaged over various channel realizations, is then given by
\begin{equation}\label{12}
R_{\rm ave} = \int\limits_0^{\infty}\,\int\limits_0^{\infty}\, f(u,v)
\, R(u,v) \d u\d v\ ,
\end{equation}
where $ f(u,v) $ designates the joint PDF of the ordered eigenvalues of $\frac{1}{2}HH^{\sf H}$,
random variables $u$ and $v$. For a Gaussian $H$ with i.i.d. components, 
the joint density function of $ \la_2 , \, \la_1 $ is given by
\cite{T99}
\begin{equation}\label{13}
f_{\la_2,\la_1}(u,v) = 16\, e^{-2v-2u} (v-u)^2 , v\ge u\ge 0\ .
\end{equation}
The optimal expected rate is a product of an optimal selection of the
power distribution $\rho(u,v)$. Specifying the power distribution uniquely
specifies the residual interference function $I(u,v)$ (\ref{9}) and
(\ref{10.5}). Hence, optimizing $R_{\rm ave}$ can instead be carried out with respect to the $I(u,v)$, i.e., 
\begin{align}\label{eq:R_MAX}
R_{\rm ave}^{\max}=\max\limits_{I(u,v)} \int\limits_0^\infty
\d a\int\limits_0^\infty \d b f(a,b)\int\limits_0^a\d u
\int\limits_{u}^b\d v R_F(u,v,I,I_{uv})\ ,
\end{align}
where $f(a,b)$ is defined in (\ref{13}), and we have set $R_F(u,v,I,I_{uv})
\triangleq \frac{\d^2R(u,v,I,I_{uv})}{\d u \d v}$ from (\ref{10.6}), which depends on
the interference function $I(u,v)$ and the power density
$I_{uv}(u,v)$ from (\ref{9}) and (\ref{10.5}), respectively. Maximizing $R_{\rm ave}$ with respect to the functional
$I(u,v)$ is a variational problem~\cite[Appendix A]{ShitzSteiner03}. Consequently, the optimization problem may be stated in the form of a partial differential equation (PDE),
\begin{align}\label{eq:pardif_1}
S_I + \frac{\partial^2}{\partial uv}S_{I_{uv}} = 0\ ,
\end{align}
where
\begin{align}
S(a,b,I,I_{ab})\triangleq\myround{1+F(a,b)-F(a)-F(b)}\cdot
R_F(a,b,I,I_{ab})\ ,
\end{align}
and $S_I$ is the partial derivative with respect to the function
$I(u,v)$, $S_{I_{uv}}$ is the partial derivative with respect to
the function $I_{uv}$, and $I_{uv}$ is the second-order partial
derivative of $I(u,v)$ with respect to $u$ and $v$. The necessary condition for the extremum is given in \cite[Appendix A]{ShitzSteiner03} in terms of a non-linear second order PDE and does not appear to have a straightforward analytical solution.
Therefore, we demonstrate a single-dimension approximation to the optimal solution. This approximation approach is called the 1-D approximation, and it is developed for the $2\times 2$ channel, i.e., two transmit and two receive antennas. It suggests breaking the mutual dependency of the optimal power distribution $\rho(a,b)$ by requiring $\rho(a,b) = \rho(a)\rho(b)$. Such a representation bears two independent solutions, obtained from solving the optimal SISO broadcast strategy. Another sub-optimal solution could be obtained based on a finite-level code layering, as suggested in \cite{Takesh01} for
the SISO scheme. Accordingly, a single layer (outage)
coding with and without employing majorization ranking at the
receiver is suggested by~\cite{ShitzSteiner03}. A two-layer coded scheme for the $2\times 2$ channel is also studied and compared with the outage approach in \cite{ShitzSteiner03}.
Another sub-optimal approach to the MIMO channel involves modeling the MIMO channel as a multiple-access channel (MAC), where each antenna transmits an independent stream~\cite{ShitzSteiner03}. In a MAC approach for the MIMO channel, instead of performing joint encoding for all transmit antennas, each antenna has an independent encoder. Thus the receiver views a MAC. When each encoder performs layered coding, we essentially get a MAC-broadcast strategy. This approach was first presented in \cite{SH00} for the multiple-access channel, employing the broadcast approach at the receiver. The advantage of this approach is that each transmitter views an equivalent degraded broadcast channel, and the results of the SISO broadcast strategy may be directly used.

\subsubsection{Degraded Message Sets}

Next, we briefly outline the formulation of the general MIMO broadcasting with degraded message sets. The key step for addressing the continuous broadcast approach for MIMO channels with degraded message sets involves decoupling the layering index and the channel state. In many previous studies on the continuous broadcast approach (e.g., \cite{ShitzSteiner03,Tian08, SteinerShamai2007}) the layering index is associated with the channel fading gain. However, for the MIMO case with degraded message set, it is proposed that the continuous layering indices are associated with only the power allocation and layer rates.

Consider the MIMO channel model in~\eqref{1}. The source transmits layered messages with a power density distribution function $\rho(s)$, where $s\in [0, \infty)$. The first transmitted message is associated with $s=0$, and can be considered as a common message for all receivers. The next layer indexed by $\d s$, cannot be decoded by the first user, but it is a common message for all other users.
The capacity of the channel in~\eqref{1} for a given channel state is the mutual information given by
\begin{eqnarray}\label{eq5}
	I(\boldy ; \boldx) = \log\det\myround{I+\frac{P}{M}HH^{\sf H}}\ ,
\end{eqnarray}
which can also be expressed using the eigenvalues of $\frac{1}{M}HH^{\sf H}$ \cite{T99},
\begin{eqnarray}\label{eq6}
	I(\boldy ; \boldx) = \sum\limits_{k=1}^K \log\myround{1 + P\lambda_k}\ ,
\end{eqnarray}
where $K = \min(M,N)$ is the degree of freedom of the MIMO channel, and $\{\lambda_k\}_{k=1}^K$ are the eigenvalues of $\frac{1}{M}HH^{\sf H}$. The singular value decomposition (SVD) of $\frac{1}{M}HH^{\sf H} = U\Lambda V^{\sf H}$ where $U$ and $V$ are unitary matrices and $\Lambda$ is a $[K x K]$ diagonal matrix of singular values of $\frac{1}{M}HH^{\sf H}$. The equivalent receive signal of \eqref{1} multiplied by $H$ is $\boldy'=U\Lambda V^{\sf H}\boldx + \boldn'$, and multiplying the received signal by $U^{\sf H}$ creates a parallel channel $U^{\sf H}\boldy'=\Lambda \boldx' + \boldn''$, where $\boldx'=V\boldx$. This makes the channel of \eqref{1} an effective parallel channel when $\boldx'$ is transmitted. However $V$ is known at the receiver, and therefore the transmitter does not have to perform any precoding, and layering can be performed with respect to singular values distribution of $\frac{1}{M}HH^{\sf H}$. The fractional achievable rate for a power allocation $\rho(s)\d s$, and under successive decoding, is given by
\begin{eqnarray}\label{eq7}
	\sum\limits_{k=1}^K \log\myround{1 + \frac{\lambda_k \rho(s)\d s}{1+\lambda_k I(s)}} = \sum\limits_{k=1}^K \frac{\lambda_k \rho(s)\d s}{1+\lambda_k I(s)}\ ,
\end{eqnarray}
where $I(s)$ is the residual layering power. $I(s)$ serves as interference for decoding layer $s$. The relationship between power density distribution and the residual interference is $\rho(s) = -\frac{\d I(s)}{\d s}$.  It is achievable for the set of eigenvalues $\{\lambda_k\}_{k=1}^K$ such that
\begin{eqnarray}\label{eq8}
	\d R(s) \leq \sum\limits_{k=1}^K \frac{\lambda_k\rho(s)\d s }{1+\lambda_k I(s)} \triangleq \d I_K(\lambda_1,...,\lambda_K,s)\ .
\end{eqnarray}
Feasibility of successive decoding here results from the fact that the function $\d I_K(\lambda_1,...,\lambda_K,s)$ is an increasing function of $\lambda_k$, $\forall ~k \in\{1,..,K\}$. Define a fractional rate allocation function $r(s)$, such that $r(s)\rho(s)=\d R(s)$. The cumulative rate achievable for a layer index $s$ is simply
\begin{eqnarray}\label{eq9}
	R(s) = \int\limits_{0}^{s} r(u)\rho(u)\d u\ .
\end{eqnarray}
The probability of achieving $R(s)$ is given by
\begin{eqnarray}\label{eq10}
	F^c(s)= \PP\myround{ r(s) \leq \sum\limits_{k=1}^K\frac{\lambda_k }{1+\lambda_k I(s)} }\ ,
\end{eqnarray}
where $F^c(s)$ is the complementary CDF of the layering index $s$, i.e., $F^c(s)=1-F(s)$. The expected broadcasting rate is then
\begin{align}\label{eq11}
	R_{\rm bs} = \int\limits_0^\infty \d s (1-F(s))r(s)\rho(s) = \int\limits_0^\infty \d s   \PP\myround{ r(s) \leq \sum\limits_{k=1}^K\frac{\lambda_k }{1+\lambda_k I(s)} } r(s)\rho(s)\ .
\end{align}
We focus now on the case of $K=2$, i.e., $\min(M,N)=2$. In this case the fractional rate $r(s)$ is decipherable if
\begin{eqnarray}\label{eq12}
	r(s) \leq \frac{\lambda_1 }{1+\lambda_1 I(s)} + \frac{\lambda_2 }{1+\lambda_2 I(s)}\ .
\end{eqnarray}
An alternative formulation is for a given $\lambda_1$, the eigenvalues $\lambda_2$ for which $r(s)$ can be reliably decoded are given by
\begin{eqnarray}\label{eq13}
	\lambda_2 \geq \frac{r(s) + r(s)\lambda_1 I(s) - \lambda_1 }{1 + (2\lambda_1-r(s))I(s) - r(s)\lambda_1 I^2(s) } \triangleq G(\lambda_1, s, I, r)\ ,
\end{eqnarray}
where the inequality holds only for $G(\lambda_1, s, I, r) \geq 0$. An alternative representation of the decoding probability of layer $s$ is thus
\begin{align}
 	F^c(s)& = \PP\myround{ \lambda_2 \geq G(\lambda_1, s, I, r) } \\
	& =  \int\limits_0^\infty \d u \int\limits_{ G(u, s, I, r) }^\infty \d v f_{\lambda_1,\lambda_2}(u,v) \cdot \textbf{1}\mat{G(u, s, I, r)\geq 0}\\
\label{eq14}	& = \int\limits_0^\infty \d u \myround{f_{\lambda_1}(u) -  Q_{\lambda_1,\lambda_2}(u,G(u, s, I, r) ) } \cdot \textbf{1}\mat{G(u, s, I, r)\geq 0}\ ,
\end{align}
where $\textbf{1}(x)$ is the indicator function, and 
\begin{align}
f_{\lambda_1,\lambda_2}(u,v)=\frac{\partial ^2 F_{\lambda_1,\lambda_2}(u,v)}{\partial u \partial v}\ ,
\end{align}
is the joint PDF of $(\lambda_1,\lambda_2)$, and 
\begin{align}
Q_{\lambda_1,\lambda_2}(u,v) \triangleq \frac{\partial F_{\lambda_1,\lambda_2}(u,v)}{\partial u}\ .
\end{align}
The expected rate for a general layering function $r(s)$ and a layering power allocation function $I(s)$ is given by
\begin{align}\label{eq15}
	R_{\rm bs} = \int\limits_0^\infty \d s    r(s)\rho(s) \cdot \int\limits_0^\infty \d u  \mat{f_{\lambda_1}(u) -  Q_{\lambda_1,\lambda_2}(u,G(u, s, I, r) ) } \cdot \textbf{1}\mat{G(u, s, I, r)\geq 0}\ .
\end{align}
Clearly, the optimization problem for expected broadcasting rate maximization is given by
\begin{align}\label{eq16}
	R_{\rm bs,opt} = \max\limits_{r(s)\geq 0,~I(s),~\textrm{s.t.}~ I(0)=P,~ \rho(s)\geq 0}\int\limits_0^\infty \d s \;  J(s,I,I',r)\ ,
\end{align}
where the integrand functional $J(s,I,I',r)$ is given by
\begin{align}\label{eq17}
	J(s,I,I',r) = r(s)\rho(s)  \int\limits_0^\infty \d u  \mat{f_{\lambda_1}(u) -  Q_{\lambda_1,\lambda_2}(u,G(u, s, I, r) ) } \cdot \textbf{1}\mat{G(u, s, I, r)\geq 0}\ .
\end{align}
The necessary conditions for extremum are given by the Euler equations \cite{GF91}
\begin{eqnarray}\label{eq17_1}
	J_r = 0\label{OutlookEq17_2}\ ,\\
	J_I - \frac{\partial}{\partial s}J_{I'} = 0\label{OutlookEq17_3}\ ,
\end{eqnarray}
where $J_r$ is the partial derivative of $J$ with respect to $r(s)$. The extremum condition for $r(s)$ in \eqref{OutlookEq17_2} can be expressed as follows:
\begin{align}\label{eq17_2}
	\int\limits_0^\infty \d u  \mytwist{\mat{f_{\lambda_1}(u) -  Q_{\lambda_1,\lambda_2}(u,G) } \cdot \myround{\frac{1}{r(s)}\textbf{1}\mat{G\geq 0} + \delta(G) }  -f_{\lambda_1,\lambda_2}(u,G)\frac{\partial}{\partial s}G\cdot \textbf{1}\mat{G\geq 0}  }= 0\ ,
\end{align}
where for brevity, $G(u, s, I, r)$ is replaced by $G$, and $\delta(x)$ is the Dirac delta function.
The extremum conditions as stated in \eqref{OutlookEq17_2} and \eqref{OutlookEq17_3} do not lend themselves into closed-form analytical solutions even though $K=2$, and characterizing them remains an open problem for future research.

\subsection{On Queuing and Multilayer Coding}

Classical information theory generally assumes an infinitely long queue of data ready for transmission, which is motivated by maximizing communication throughput (Shannon capacity). In network theory, on the other hand, the input data is usually a random process that controls writing to a buffer (serving as a queue), and the readout from this buffer is another random process. In these settings, the design goal of transmission concentrates on
minimizing the queue delay for the input data. However, designing the data queue and transmission algorithm
cannot be decoupled in the presence of stringent delay constraints on input data transmission. This is because the objective is no longer only maximizing the throughput. 
This conceptual difference between network theory and information theory can be overcome by posing a common optimization problem and jointly minimizing the delay of a random input process under a power (rate) control constraint. This becomes a cross-layer optimization problem involving the joint optimization of two layers of the seven-layer open systems interconnection (OSI) model. Other
fundamental gaps between network theory and information theory are
covered in detail in~\cite{TG95,HAJEK98,GALLAGER85,YooLiuShamai2012}.

Queuing and channel coding for a block fading channel
with transmit CSI only, for a single user, is discussed in
\cite{IDO01}. In this section, we first consider optimizing rate and power
allocation for a single layer code transmission. For this scheme, the outage capacity \cite{OZ98} maximizes the achievable throughput. Rate and power are optimized jointly to minimize the overall delay. The delay is measured from the arrival of a packet at the queue until successfully decoded, including, if needed, retransmission due to outage events.

\begin{figure}[tbh] 
	\hspace*{-.4cm}\scalebox{0.4}{\includegraphics{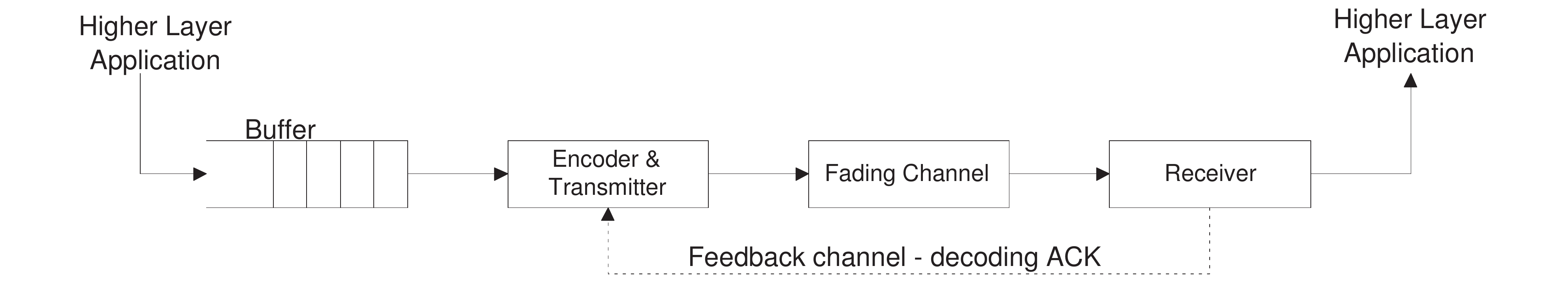}}
	\caption{A schematic communication system with a queue buffer followed by a wireless transmitter.}\label{fig:queue}
\end{figure}

The study in \cite{AsSh10} considers a cross-layer system optimization approach for a single-server queue followed by a multi-layer wireless channel encoder, as depicted in Fig.~\ref{fig:queue}. The main focus is on minimizing the average delay
of a packet measured from entering the queue until successful service completion.

\subsubsection{Queue Model  --  Zero-padding Queue}\label{S:queue_model}

Next we consider the zero-padding queue model described in \cite{AsSh10}.   It is assumed that the transmission is performed every time the queue is not empty. If
the available queue data is less than a packet size, a frame can be
generated with zero-padding to have a valid frame for the channel
encoder. We define the {\em queuing time} as the time from arrival to
completion of service, and the {\bf waiting time} as the time measured from
arrival until initially being served. The queue's waiting time analysis can be done at embedded points: the beginning of every time slot. The random process of packet arrival random at each slot is a deterministic process denoted by
$\lambda$ (bits/channel use).

The queue waiting time can be  measured directly based on the queue size, as stated on Little's theorem \cite{WOLFF89}, by
normalizing the queue size by the inverse of the input rate
$\lambda$. Notice that Little's theorem does not consider the instantaneous quantities to the average waiting time and average
queue size. The following equation defines the queue size:
		\begin{align}\label{n1}
		Q_{n+1}=\mycase{
			\begin{array}{ll}
			N\lambda_{n+1} + Q_n -NR_n & N\lambda_{n+1} + Q_n -NR_n\geq 0\\
			&\\
			0 & {\rm otherwise}
			\end{array}\ ,
		}
		\end{align}
where $N$ is the number of channel uses between
slots, which is also the block length, and $\lambda_{n+1}$ is a deterministic queue input rate
$\lambda$. It is noteworthy that in a single-layer coding, $R_n$ is a fixed $R$
with probability $p$, and it is 0 with probability $1-p$. This waiting
time equation is also analyzed in \cite[chapter 5]{IDO02} for a single-layer coding approach and a deterministic arrival process, where tight bounds
on the expected waiting time are obtained. For simplicity, by normalization of the queue size by the block-length $N$, the Lindley equation is obtained~
\cite{kleirock_v2}:
\begin{align}\label{n2}
\widetilde{q}_{n+1}=\mycase{
	\begin{array}{ll}
	\widetilde{q}_n + \lambda_{n+1}-R_n & ~\widetilde{q}_n + \lambda_{n+1}-R_n\geq 0\\
	&\\
	0 & ~\widetilde{q}_n + \lambda_{n+1}-R_n < 0
	\end{array}\ ,
}
\end{align}
where $\widetilde{q}_n$ is now the queue size in units of blocks of
data corresponding to $N$ arrivals to the queue. In an outage
approach, we have $R_n=R$ with probability $p$, and $R_n=0$ with a complementary probability $1-p$, which is also the outage probability. For the rest of the analysis, the queue equations will be normalized following
(\ref{n2}). We specify the queuing
time equation for completeness of the definitions, which is the overall system delay for the zero-padding queue. The overall delay must always take into account the additional delay of service time beyond the queue's waiting time. The normalized queue size is the waiting time equivalent, i.e., 
\begin{align}\label{n4}
q_{n+1}=\mycase{
	\begin{array}{ll}
	q_n + \frac{\lambda_{n+1}}{\lambda}-\frac{R_n}{\lambda} \ ,& ~q_n -\frac{R_n}{\lambda}\geq 0\\
	&\\
	\frac{\lambda_{n+1}}{\lambda} \ ,& otherwise
	\end{array}\ ,
}
\end{align}
where $q_n$ is a normalized queue size at a renewal slot $n$. In a single-layer coding approach, it is possible to analyze the queue
delay by adopting the standard M/G/1 queue model. The input random process of an M/G/1 model follows a Poisson process, and its service distribution
is another general random process. In an outage approach, a geometrically distributed random variable characterizes the time between services. For using the M/G/1 model, an important assumption on the system model is made: input arrives in blocks that have the same length as the coded transmission blocks. That is, the queue equation is
normalized to the data block size of its corresponding transmission. The number of arrivals is measured in block units, and the input process has a rate of $\lambda_{\rm norm}$.

Having the arrival blocks are equal in size to transmitted blocks is a limiting constraint since a change of transmission rate means a change in input block size. Therefore, the M/G/1 queue model is not adopted in \cite{AsSh10}, and in the following, we use the zero-padding queue model as described earlier.

\subsubsection{Delay Bounds for a Finite Level Code	 Layering}\label{S:k_level_queue}

We consider here $K$ multi-layer coding, and describe the Lindley equation \cite{WOLFF89}. The queue update equation is given by
\begin{align}\label{l1}
w_{n+1}=\mycase{
	\begin{array}{ll}
	w_n + x_n & ~w_n + x_n\geq 0\\
	0 & ~w_n + x_n < 0
	\end{array}\ ,
}
\end{align}
where  $x_n$ is the update random variable, which depends on the number
of code layers. Its value represents the difference
between the queue
input $\lambda$ and the number of layers successfully decoded, i.e., 
\begin{align}\label{l2}
x_n\triangleq \lambda-\sum\limits_{i=1}^{K}\nu_{i,n}R_i\ .
\end{align}
Random variables $\{\nu_{i,n}\}_{i=1}^{K}$ are associated with the outage probability as function of layer index. The corresponding fading power thresholds are denoted by
$\{s_{{\rm th},i}\}_{i=1}^{K}$. Random variables
$\{\nu_{i,n}\}_{i=1}^{K}$ are related to the fading thresholds as
follows
\begin{align}\label{l3}
\nu_{i,n}=\mycase{
	\begin{array}{ll}
	1 & s_{{\rm th},i}\leq s_n\leq s_{{\rm th},i+1}\\
	&\\
	0 & \textrm{otherwise}
	\end{array}}\ ,
\end{align}
where $s_n$ is the fading power realization at the $n^{\rm th}$
time-slot, and $s_{{\rm th},K+1}=\infty$. Every random variable $\nu_{i,n}$ has a probability of being 1, denoted by $p_{K-i+1}$. Note that outage
probability is 
\begin{align}
\overline{p}=1-\sum\limits_{i=1}^Kp_{i}\ ,
\end{align} 
in which
$\overline{p}$ represents the probability that all layers cannot be decoded.
The CDF of the queue size at these embedding points requires computing the CDF at every time instant. In this setting,
the probability density $\d F_X(\tau)$ of $X$ (\ref{l2}) is given by
\begin{align}\label{l5}
\d F_X(x) = \sum\limits_{i=1}^K p_i \delta\left(x -
(\lambda-\sum_{j=1}^{K-i+1}R_j)\right) + \overline{p}\delta(x-\lambda)\ ,
\end{align}
where $p_i = \Pr\{s_{{\rm th},i}\leq s_n\leq s_{{\rm th},i+1}\}$ for
$i\in\{1,\dots, K\}$ and $s_{{\rm th},K+1}=\infty$. The next theorem discussed in details in \cite[Appendix B]{AsSh10} establishes upper and lower bounds on $\bbe[W_K]$.
\begin{Theorem}[\cite{AsSh10}]
For a $K$-layer coding, the expected queue size is upper and lower bounded by 
		\begin{align}\label{l6}
		\bbe[w_{K}]\geq \frac{(\Re_{K}-\lambda)
			(\sum\limits_{i=1}^Kp_i\Re_{K-i+1}-\lambda) - (\Re_K-\lambda)^2
			+\sum\limits_{i=1}^Kp_i(\Re_K-\Re_{K-i+1})^2 +
			\overline{p}\Re_K^2}{2(\sum\limits_{i=1}^Kp_i\Re_{K-i+1}-\lambda)}\ ,
		\end{align}
		and
		\begin{align}\label{l6.1}
		\bbe[w_{K}]\leq \frac{2(\Re_{K}-\lambda)
			(\sum\limits_{i=1}^Kp_i\Re_{K-i+1}-\lambda) - (\Re_K-\lambda)^2
			+\sum\limits_{i=1}^Kp_i(\Re_K-\Re_{K-i+1})^2 +
			\overline{p}\Re_K^2}{2(\sum\limits_{i=1}^Kp_i\Re_{K-i+1}-\lambda)}\ ,
		\end{align}
	where $\Re_V\triangleq \sum_{j=1}^{V}R_j$.
\end{Theorem}\label{theorem3}
The variance of the achievable rate random variable $\sigma^2_{R_{\rm KL}}$ is given by
\begin{align}\label{l100}
\begin{array}{lll}
\sigma^2_{R_{\rm KL}} &\triangleq& \sum\limits_{i=1}^Kp_i\Re_{K-i+1}^2
- (R_{\rm KL,av})^2\ ,
\end{array}
\end{align}
where
\begin{align}\label{l101}
R_{\rm KL,av}\triangleq \sum\limits_{i=1}^Kp_i\Re_{K-i+1}\ .
\end{align}

\begin{Corollary}
	Queue expected size and expected delay for $K$-layer coding are upper bounded by
	\begin{align}\label{l102}
	\bbe[w_{\rm KL}]\leq \frac{\sigma^2_{R_{KL}}}{2(R_{\rm KL,av}-\lambda)} -
	(1-\frac{\lambda}{R_{\rm KL,av}})\frac{\sigma^2_{R_{KL}}}{2R_{\rm KL,av}}\ ,
	\end{align}
	and the expected delay is upper bounded by
	\begin{align}\label{l103}
	\bbe[w_{\lambda, \rm KL}]\leq \frac{\sigma^2_{R_{KL}}}{2\lambda
		(R_{\rm KL,av}-\lambda)} -
	(1-\frac{\lambda}{R_{\rm KL,av}})\frac{\sigma^2_{R_{KL}}}{2R_{\rm KL,av}\lambda}\ ,
	\end{align}
	where $\sigma^2_{R_{\rm KL}}$ and $R_{\rm KL,av}$ are given by
	(\ref{l100}) and (\ref{l101}) respectively.
\end{Corollary}

\subsubsection{Delay bounds for Continuum Broadcasting}\label{S:broadcasting_queue}

A continuous broadcasting approach is considered in this section. In this approach, the transmitter also sends multi-layer coded data. Unlike $K$-layer coding, the layering is a continuous function of the channel fading gain parameter. The
number of layers is not limited, and an incremental
rate with a differential power allocation is associated with every
layer. The differential per layer rate is 
$\d R(s)=\frac{s\rho(s)\d s}{1+sI(s)}$
and $\rho(s)\d s$ is the transmit power of a layer $s$. This also determines the
transmission power distribution per layer \cite{V90}. The residual interference for a fading power $s$ is $I(s)=\int_s^\infty\rho(u)\d u$ \eqref{SISO4}. The total achievable rate for a fading gain realization $s$ is $R(s) = \int_0^s\frac{u\rho(u)\d u}{1+uI(u)}$ \eqref{SISO6}. It is possible to extend the $K$-layer coding bounds shown above to this continuous broadcast setting. The bounds in
(\ref{l6}) and (\ref{l6.1}) could be used for broadcasting after
performing the following modifications:
\begin{enumerate}
	\item The number of layers is unlimited, that is
	$K\rightarrow\infty$.
	\item Since the layering is continuous, every layer $i$ is associated with a fading gain parameter $s$. Every Rate $R_i$ is
	associated  with a differential rate $\d R(s)$ specified in (\ref{SISO3}).
	\item The cumulative rate $\Re_K$ should be replaced by
	\begin{align}\label{b1}
	R_T = \int\limits_0^\infty  \d R(s)\ .
	\end{align}
	\item The sum $\sum\limits_{i=1}^Kp_i\Re_{K-i+1}$ is actually
	the average rate and it turns to be $R_{\rm bs}$ (\ref{SISO7}) for
	the continuum case.
	\item Finally, in finite-level coding the expression $\sum\limits_{i=1}^Kp_i(\Re_K-\Re_{K-i+1})^2 + \overline{p}\Re_K^2$
	turns out to be
	\begin{align}\label{b8} 
	R^2_{\rm d,bs} &\triangleq \int\limits_0^\infty \d u f(u)\mat{R_T
		-\int\limits_0^u \d R(s)}^2\\
	&=\int\limits_0^\infty \d u f(u)\mat{\int\limits_u^\infty
		\d R(s)}^2\\
	&=2\int\limits_0^\infty \d u F(u)\d R(u)\int\limits_u^\infty
	\d R(s)\ ,
	\end{align}
	in the continuous case, where $\d R(u)$ and $R(u)$ are specified in
	(\ref{SISO3}) and (\ref{SISO6}), respectively.
	
\end{enumerate}

\begin{Corollary}
The queue average size for a continuous
	code layering is upper and lower bounded by
	\begin{align}\label{b6}
	\bbe[w_{\rm bs}]\geq \frac{R_T-\lambda}{2}+\frac{R^2_{\rm d,bs} -(R_T-\lambda)^2
	}{2(R_{\rm bs}-\lambda)}\ ,
	\end{align}
	\begin{align}\label{b6.1}
	\bbe[w_{\rm bs}]\leq
	(R_T-\lambda)+\frac{R^2_{\rm d,bs} -(R_T-\lambda)^2}{2(R_{\rm bs}-\lambda)}\ ,
	\end{align}
	and the average delay is lower and upper bounded by
	\begin{align}\label{b7}
	\bbe[w_{\lambda, \rm bs}]\geq
	\frac{R_T-\lambda}{2\lambda}+\frac{R^2_{\rm d,bs} -(R_T-\lambda)^2
	}{2\lambda (R_{\rm bs}-\lambda)}\ ,
	\end{align}
	\begin{align}\label{b7.1}
	\bbe[w_{\lambda, \rm bs}]\leq
	\frac{R_T-\lambda}{\lambda}+\frac{R^2_{\rm d,bs} -(R_T-\lambda)^2
	}{2\lambda (R_{\rm bs}-\lambda)}\ ,
	\end{align}
	where $R_{\rm bs}$, $R_T$, and $R^2_{\rm d,bs} $ are specified in
	(\ref{SISO7}), (\ref{b1}), and (\ref{b8}) respectively.
\end{Corollary}
The variance of the achievable rate random variable $\sigma^2_{R_{\rm bs}}$ is given by
\begin{align}\label{b9}
\sigma^2_{R_{\rm bs}} &\triangleq \int\limits_0^\infty \d u
f(u)\mat{R(u)}^2 - R_{\rm bs}^2\\
&=\int\limits_0^\infty \d u f(u)\mat{\int\limits_0^u \d R(s)}^2 -
R_{\rm bs}^2\\
&=2\int\limits_0^\infty \d u (1-F(u))\d R(u)\int\limits_0^u \d R(s) -
R_{\rm bs}^2\\
&= 2\int\limits_0^\infty \d u (1-F(u))\d R(u)R(u) - R_{\rm bs}^2\ .
\end{align}

\begin{Corollary}
The queue average size for a continuous
	code layering is upper bounded by
	\begin{align}\label{b10}
	\bbe[w_{\rm bs}]\leq \frac{\sigma^2_{R_{\rm bs}}}{2(R_{\rm bs}-\lambda)} -
	(1-\frac{\lambda}{R_{\rm bs}})\frac{\sigma^2_{R_{\rm bs}}}{2R_{\rm bs}}\ ,
	\end{align}
	and the average delay is upper bounded by
	\begin{align}\label{b11}
	\bbe[w_{\lambda, \rm bs}] \leq \frac{\sigma^2_{R_{\rm bs}}}{2\lambda
		(R_{\rm bs}-\lambda)} -
	(1-\frac{\lambda}{R_{\rm bs}})\frac{\sigma^2_{R_{\rm bs}}}{2R_{\rm bs}\lambda}\ ,
	\end{align}
	where $R_{\rm bs}$ and $\sigma^2_{R_{\rm bs}}$ are given by (\ref{SISO7})
	and (\ref{b9}), respectively.
\end{Corollary}

For minimizing the expected delay in the continuous layering case, it is required
to obtain the optimal $\rho(s)$ (\ref{SISO4}) which minimizes the average queue size upper bound. As in multi-layer coding, an analytic solution is not available and remains an open problem for further research. However, numerical optimization is impossible here. The constraint of optimization is a continuous function. The target functional in the optimization problem for continuous layering does not have a localization
property \cite{GF91}. A functional with localization property can
be written as an integral of some target function. Our
functional contains a ratio of integrals and further
multiplication of integrals, which cannot be converted to an
integral over a single target function. Such functional
is also denoted as a nonlocal functional in \cite{GF91}. In such cases, it is preferable to look for an approximate representation of the nonlocal functional, which has
the localization property. Alternatively, approximate target
functions with reduced degrees of freedom may be optimized.

An interesting observation from the numerical results of \cite{AsSh10} is that when considering delay as a performance measure, code layering could give noticeable performance gains in terms of delay, which are more impressive than those associated with throughput. This makes
layering more attractive when communicating under stringent delay
constraints.

Analytic resource allocation optimization for delay minimization, under the simple queue model in \cite{AsSh10}, remains an open problem for further research. In general, when layering is adopted at the transmitter, in conjunction with successive decoding at the receiver, the first layer is decoded
earlier than other layers, and it has the shortest service time. Accounting for a different service delay per layer, the basic queue size update equation (the Lindley equation) should be modified accordingly. The analysis of the broadcast approach with a per layer queue is a subject for further research.
The queue model which was used in \cite{AsSh10} is a zero-padding
queue. In this model, the frame size is kept fixed every
transmission, and if the queue is nearly empty, the transmission includes
zero-padded bits on top of queue data. Optimizing the transmission
strategy as a function of the queue size, such that no zero-padding is
required, can further increase layering efficiency and minimize the expected delay. This is a possible direction for further research.

\subsection{Delay Constraints}

There are various aspects in which delay constraints in communications may impact the system design.
Stringent delay constraints might not allow to capture the channel ergodic distribution, and may benefit from a broadcast approach. This is while relaxed delay constraints may allow transmission of long codewords that capture the channel ergodicity. When there is a mixture of delay requirements on data using the same physical transmission resources, interesting coded transmission schemes can be considered. This is studied in \cite{CohenSteinerShamai12} as discussed in next subsections and also widely covered in \cite{Nikbakht2019,Nikbakht_2020}. Another aspect is decoding multiple independent blocks, as considered in \cite{YEH01}, and studied by its equivalent channel setting, which is the MIMO Parallel channel \cite{Kfir:IZS2020} and discussed in detail in the next subsections.

\subsubsection{Mixed Delay Constraints}

The work in \cite{CohenSteinerShamai12} considers the problem of transmission with delay-constrained (DC) and non-delay-constrained (NDC) streams are transmitted over a SISO channel, with no CSIT adhering to the broadcast approach for the DC stream. The DC stream comprises layers that have to be decoded within a short period of a single transmission block. The NDC stream comprises layers that may be encoded over multiple blocks and decoded after the complete codeword is received, potentially observing the channel ergodicity. Three overall approaches are suggested in \cite{CohenSteinerShamai12}, trying to maximize the expected sum rate.
Their achievable rate regions over DC and NDC are examined. A DC stream is always decoded in the presence of an NDC stream, which is treated as interference. However, before decoding an NDC stream, the decodable DC layers can be removed, allowing NDC decoding at the highest signal-to-interference-plus-noise ratio (SINR).
A closed-form solution of the sum-rate maximization problem can be derived for the outage and broadcast DC stream in parallel to a single NDC layer. When NDC transmission is also composed of multi-layers, the optimization problem of the expected sum-rate becomes much more complicated.

The joint strategy of accessing both DC and NDC parts on a single channel uses a two-level block nesting.
Every $L$ samples define a block for the DC stream, while the NDC
stream is encoded over $K$ such blocks, consisting of $L \cdot K$ samples. The NDC
block is called a super block.
$L$ is large enough for reliable communication for the DC part, but it is
much shorter than the dynamics of the slow fading process. $K$ is
large enough to enable the empirical distribution of the fading
coefficient to be similar to the real one. Two independent streams of
information are encoded.
The \emph{DC stream} is decoded at the
completion of each block at the decoder, at a rate dependent upon
the realization of the channel fading coefficient for that block.
The \emph{NDC stream} is decoded only at the completion of
the super block. All of the following proposed schemes assume
superposition coding, equivalent to symbol-wise additivity of the
DC and NDC code letters. Denote by $w^L$
the $L$-length codeword for the DC code for each block, and
$z^{KL}$ the $KL$-length codeword for the NDC code for each super
block. Define one super block as
\begin{align}
	y_{k,i}&=\sqrt{s_k} \cdot (w_{k,i} + z_{k,i}) + n_{k,i} \quad,   \qquad \mbox{for}\;\;\; i=1, 2,\dotsc, L\ , \quad \quad k=1, 2,\dotsc, K	\ , \label{eq: model double index}
\end{align}
where the double sub-index $\{k,i\}$ is equivalent to the time
index $(k-1) \cdot L+i$. Note that slow fading channel nature was
used by defining $s_{k,i}=s_k$. This scheme reflects a power
constraint of the form $\bbe[\abs{w_{k,i} + z_{k,i} }^2] \leq P$.
Define $R_{\rm DC}(s)$ as the \emph{achievable rate for a fading
	power realization $s$} per block. The \emph{total expected DC rate over
	all fading power realizations} is given by
\begin{align}
	R_{\rm DC}&=\int_0^\infty{f_S(u)R_{\rm DC}(u) \d u}\ .	\label{eq: R_DC = int R_DC (s) ds}
\end{align}
Let $R_{\rm NDC}$ designates the \emph{rate of
	the NDC part}, which experiences enough such realizations throughout
communication. When relaxing the stringent delay constraint, coding over sufficient large blocks achieves
\emph{ergodic capacity}, denoted by $C_{\rm erg} = \bbe_{S}[ {\log \left(1+S P\right)}]$. Clearly, for any coding scheme $R_{\rm DC} + R_{\rm NDC}\leq C_{\rm erg}$.

\subsubsection{Broadcasting with Mixed Delay Constraints}

The superposition of DC and NDC is employed by allocating a fixed amount of power per
stream. Define the \emph{DC relative power portion} as
$\beta \in [0,1]$, that is $\beta \cdot P$ is the power allocated for
the DC stream and the rest $(1-\beta) \cdot P$ for the NDC stream.
The DC part uses the broadcast approach. During decoding of the DC part, the NDC is treated as additional interference since during the decoding of each DC block the
NDC codeword cannot be completely received, and thus cannot be decoded nor reconstructed to assist the DC decoding. The NDC decoder
is informed of all DC decoded layers per DC codeword, and it cancels out the decoded part
from the corresponding NDC block, maximizing its SINR for NDC decoding. By designing the two encoders like described earlier, we can justify that
both DC and NDC parts communicate over a flat fading channel with additive Gaussian noise.
The imposed noise for each part consists of the white channel noise along with undecodable codewords
of those that are undecoded yet from both parts.

The DC encoder uses superposition of an infinite number of layers,
ordered using channel fading realization $s$
in a manner that forms a degraded broadcast channel. Per DC message,
the transmitted codeword of length $L$ is given by
\begin{align}
	w^L(m_1, m_2, \dotsc, m_\infty)&= \sum_{j=1}^\infty w_j^L(m_j)\ .		 \label{eq: encoding DC as sum of infinite single message codes}
\end{align}
Designate $\rho(s)$ to be the \emph{DC layering power distribution},
which will be optimized later on, and each layer communication scheme will try to overcome a Gaussian channel where the fading is known to
both sides. The NDC encoder sends a single message through a block of length $L\cdot K$. By random coding over a Gaussian channel, the codewords can be generated. A total of $e^{L \cdot K \cdot R_{\rm NDC}}$ codewords can be used, where the channel rate $ R_{\rm NDC}$
relies on the optimized channel power $\rho(s)$ as well.

The decoders are activated by order. First, the DC decoder
works on every $L$-block and by \emph{successive decoding}
can reveal as many layers as the channel permits.
It is similar to the classic broadcast approach, except
all layers suffer from an undecodable (at this stage) interference.
 All DC decoders' outputs are fed to the NDC decoder, which works after $K$
such blocks. After removal of the decodable DC codewords of all blocks,
the NDC part is decoded with a minimal residual interference, where the interference includes only the undecoded DC layers. Calculating the DC rate in the presence of NDC is a direct extension of \cite{ShitzSteiner03},
which is a special case for $\beta=1$.
Define the \emph{DC interference for a fading power $s$} as $I(s)$, implying
\begin{align}
	I(s) &= \int_s^\infty {\rho(u) \d u}
	\ ,\qquad \mbox{and} \qquad 
	\rho(s) = - {\frac{\d}{\d s} I(s)}\ . \label{eq: I(s) = int(rho(s)) quad rho(s) = - d/ds I(s)}
\end{align}
It associates the undecodable layers upon a channel fading
realization $s$ as noise to the transmission. It is restricted to
the total DC allocated power
$$I(0) = \int_0^\infty {\rho(u) \d u} = \beta P\ ,$$ with $0\leq\beta\leq 1$.

\begin{Lemma}[Achievable Expected DC Rate  \cite{CohenSteinerShamai12}]\label{lem: Achievable Expected DC Rate}
	Any  total expected DC rate $R_{\rm DC}$, which is averaged over all fading realizations, that satisfies
\begin{align}
		R_{\rm DC}  &\leq \int_{u=0}^\infty {(1-F_S(u)) \frac{u \rho(u)}{1+u I(u) +(1-\beta)P u} \d u} \ ,\label{eq: R_DC definition}
\end{align}
	is achievable.
\end{Lemma}


\begin{Lemma}[Achievable Expected NDC Rate  \cite{CohenSteinerShamai12}]
	Any {total expected NDC rate $R_{\rm NDC}$, which is averaged over all fading realizations}, that satisfies 
	\begin{align}
		R_{\rm NDC} &\leq \int_0^\infty {f_S (u)  \log \left( 1+ \frac{(1-\beta) P u}{1+u I(u)} \right) \d u}\ ,	\label{eq: R_NDC definition}
	\end{align}
	is achievable.
\end{Lemma}
It is possible to derive the optimal power allocation for DC layering that maximizes the sum rate $(R_{\rm DC}+R_{\rm NDC}$ as stated in \eqref{eq: R_DC definition} and
\eqref{eq: R_NDC definition}, respectively. It is a function that depends on $I(s)$ according to
\eqref{eq: I(s) = int(rho(s)) quad rho(s) = - d/ds I(s)}. Specifically, the optimization problem is
\begin{align}
	I^* (s) &= \argmax_{I(s)} \left\{ R_{\rm DC} + R_{\rm NDC} \right\}	\qquad \mbox{s.t. } \quad  I(0)=\beta P \ ,\quad \mbox{and} \quad I(\infty)=0 \ . \label{eq: optimal problem: R_DC + mu R_NDC}
\end{align}
The outage approach is a simple special case of layering, where a single DC coded layer is used. In this case, the power distribution $I(s)$ is explicitly given by
\begin{align}
	I(s) &=
	\begin{cases}
		\beta P & \text{if } 0 \leq s \leq s_{\rm th} \\
		0 		& \text{if } s > s_{\rm th}			
	\end{cases}\ ,	\label{eq: optimal I for outage}\\
	\rho(s) &= \beta P \cdot \delta(s-s_{\rm th})\ ,	\label{eq: optimal rho for outage}
\end{align}
where $\delta$ is the Dirac function and $s_{\rm th}$ is a parameter
set prior to the communication. Constant $s_{\rm th}$ may be interpreted as the fading gain threshold for single layer coding.
The advantages of this approach are low implementation complexity and ease of analysis.
The disadvantage is its sub-optimality.
The outage approach is designed for a channel with fixed fading of $s_{\rm th}$.
On the one hand, if $s \geq s_{\rm th}$, the message can be transmitted error-free at a
rate adjusted for $s_{\rm th}$. On the other hand, if $s < s_{\rm th}$,
the specific transmission is useless.

\begin{Proposition}[Joint Optimality by Outage DC \cite{CohenSteinerShamai12}]
	The maximizer $I_o(s)$ of \eqref{eq: optimal problem: R_DC + mu R_NDC} subject
	that satisfies the form in \eqref{eq: optimal I for outage} is specified by $s^*_{\rm th}$, which can be found as the solution to
	\begin{align}
		f_S(s_{\rm th}^*) \log \left(1+\beta P s_{\rm th}^*\right) =	(1-F_S(s_{\rm th}^*)) \frac{\beta P}{(1 + P s_{\rm th}^*)(1 + (1-\beta)P s_{\rm th}^*)}\ .	\label{eq: s_th * necessary general condition}
	\end{align}
	The {optimal expected DC outage rate} and the \emph{optimal expected NDC outage rate}, which together maximize the sum rate are 
	\begin{align}
		R_{\rm DC,o} & =  \left(1-F_S(s_{\rm th}^*)\right) \log \left( 1+\frac{\beta P s_{\rm th}^*}{1 + (1-\beta)P s_{\rm th}^*} \right)\ , \label{eq: R_DC,o}\\
		R_{\rm NDC,o} & =  \int_0^{s_{\rm th}^*} {f_S (u)  \log \left( 1+ \frac{(1-\beta) P u}{1+ \beta P u} \right) \d u}  +\: \int_{s_{\rm th}^*}^\infty {f_S (u)  \log \left( 1+ (1-\beta) P u\right) \d u}\ .	\label{eq: R_NDC,o}
	\end{align}
\end{Proposition}
Maximizing \eqref{eq: optimal problem: R_DC + mu R_NDC} can be
derived analytically by developing an E\"{u}ler Equation in a
similar way to \cite{ShitzSteiner03}. This is done by enlarging
the class of admissible functions $I(s)$ (as opposed to the outage approach) to be continuously
differentiable and to satisfy the boundary conditions $I(0)=\beta P$ and $I(\infty)=0$.

\begin{figure}
	\centering
	\includegraphics[width=5in]{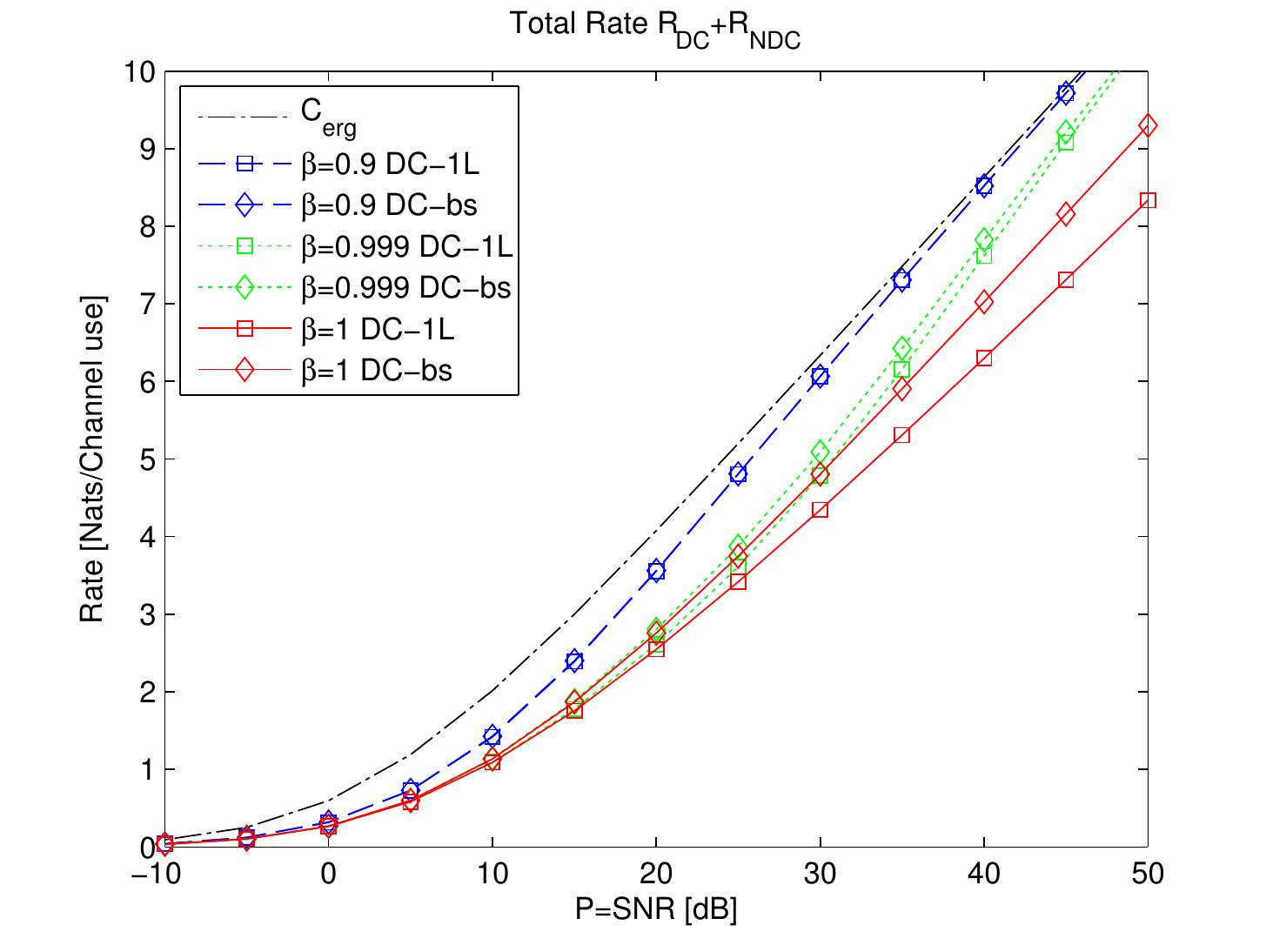}
	\caption{Total rate for several $\beta$ values and the ergodic capacity vs. the SNR $P$, for the flat Rayleigh channel.}
	\label{fig: total rate vs. P}
\end{figure}

\begin{Proposition}[Joint Optimality by Broadcast DC ]\label{prps: Joint Optimality by Broadcast DC}
	The maximizer $I_{bs}(s)$ of \eqref{eq: optimal problem: R_DC + mu R_NDC} when considering all
	continuously differentiable boundary conditioned functions is
	$I_{\rm bs}(s)={[ \tilde{I}(s)]}_0^{\beta P}$, where
	\begin{align}
		\tilde{I}(x)&=	\frac{1}{x} \left( \frac{-b(x) + \sqrt{b^2(x)-4 a(x) c(x)}}{2 a(x)}-1 \right)\ , \label{eq: optimal Itilde(x)}\\
		a(x)		&=	x f_S (x) \ ,\\
		b(x)		&=	2(1-\beta) P f_S(x) x^2 -(1-F_S(x))\ ,\\
		c(x)		& =	 (1-\beta)^2 P^2 f_S(x) x^3\ .
	\end{align}
	The associated rates $R_{\rm DC,bs}$ and $R_{\rm NDC,bs}$ can be achieved by substituting it in
	\eqref{eq: R_DC definition} and \eqref{eq: R_NDC definition}.
\end{Proposition}
The square root in \eqref{eq: optimal Itilde(x)} can impose a finite-length domain for $\tilde{I}(s)$,
that can result in discontinuity at $I(s)$. This situation is addressed by
assigning a Dirac function at $\rho(s)$, which can be interpreted as a superposition of single-layer coding and continuous layering. Figure~\ref{fig: total rate vs. P} shows the relation of
$R_{\rm DC}+R_{\rm NDC}$ for the joint outage approach and the joint broadcast
approach, for selected values of $\beta$. The total expected sum-rate is the sum of the
DC rate and the NDC rate. As may be observed, if $\beta\leq 0.9$, then the ergodic capacity can be nearly achieved in high SNRs.

\subsubsection{Parallel MIMO Two-state Fading Channel}

Broadcasting over MIMO channels is still an open problem, and only sub-optimal achievable schemes are known \cite{ShitzSteiner03}. The work in \cite{Kfir:IZS2020} considers a two-state parallel MIMO channel, which is equivalent to a SISO two-state channel, where decoding can be done over multiple consecutive blocks, as studied in \cite{YEH01}. The work in \cite{Kfir:IZS2020} considers the slow (block) fading parallel MIMO channel~\cite{T99}, where channel state is known at the receiver only. Under this channel model, the transmitter may adopt a broadcast approach~\cite{ShitzSteiner03}, which can optimize the expected transmission rate under no transmission CSI, which is essentially characterized by the \emph{variable-to-fixed coding}~\cite{Verdu10variable-ratechannel}.

The study in~\cite{GA80} composes two degraded broadcast
channels~\cite{CO72,Cover1998CommentsBroadcast} into a three-user setup: an encoder with two
outputs, each driving a dual-output broadcast channel; two decoders, where each is fed by one less-noisy broadcast channel output and one more-noisy output of the other channel (called
{\em unmatched}). This channel is referred to as \emph{degraded broadcast product channel}. 
For the AWGN case, the capacity region (private and common rates) of this channel was derived \cite{GA80}. In \cite{Kfir:IZS2020}, the MIMO setting for the broadcast approach is revisited, with new tools that differ from those in~\cite{ ShitzSteiner03,SteinerShamai2007}. This is by analyzing the finite-state parallel MIMO channel, where the capacity region in~\cite{GA80} is used to address the multi-layering optimization problem for maximizing the expected rate of a two-state fading~\cite{zohdy2019broadcast,YEH01,Tajer18} parallel MIMO channel.

\subsubsection{Capacity of Degraded Gaussian Broadcast Product Channels}

\begin{figure*}[tb] 
	\begin{center}
		\scalebox{0.8}{
			\begin{tikzpicture}[>=latex]
			\node (f_AA)              [rectangle, draw, minimum height = 1.2cm, minimum width
			= 1.5cm] at (0, 1.7) {$f_{\rm AA}(\cdot)$};
			\node (f_cr)              [thick,red, rectangle, draw, minimum height = 1.2cm, minimum width = 1.5cm] at (0, 0) {$f_{\rm cr}(\cdot)$};
			\node (f_BB)              [rectangle, draw, minimum height = 1.2cm, minimum width
			= 1.5cm] at (0,-1.7) {$f_{\rm BB}(\cdot)$};
			\node (p1left)            [thick,red, circle, draw, above right= 0.25 and 4.5 of f_cr] {$+$};  
			\node (p2left)            [thick,red, circle, draw, below right= 0.25 and 4.5 of f_cr] {$+$}; 
			\draw  	  		  [thick,red, <-] (p1left.-90) -- node[below=0.1] {${\scriptstyle \mathcal{CN}(0,\nu_b^{-1})}$} ($(p1left.-90) + (0,-0.3)$) ;
			\draw                [thick,red, <-] (p2left.+90) -- node[above=0.1] {${\scriptstyle \mathcal{CN}(0,\nu_b^{-1})}$} ($(p2left.+90) + (0,+0.3)$) ;
			\draw                     [<-] (f_AA.west) -- node[above]{$w_{\rm AA}$              } ($(f_AA.west)+(-2.5,0)$); 
			\draw                     [thick,red, <-] (f_cr.west) -- node[above]{$w_{0},w_{BA },w_{\rm AB}$} ($(f_cr.west)+(-2.5,0)$); 
			\draw                     [<-] (f_BB.west) -- node[above]{$w_{\rm BB}$              } ($(f_BB.west)+(-2.5,0)$); 
			\node (p1right)           [thick,red, circle, draw, right= 2   of p1left] {$+$};
			\node (p2right)           [thick,red, circle, draw, right= 2   of p2left] {$+$};
			\node (topplus)           [circle, draw, left = 1.5 of p1left] {$+$};
			\node (botplus)           [circle, draw, left = 1.5 of p2left] {$+$};
			\draw                     [->] (f_AA. 20) -- ($(f_AA. 20) + (0.5,0)$) --  ($(topplus.160)+(-0.5,0)$) --  (topplus.160);
			\draw                     [thick,red,->] (f_cr. 20) -- ($(f_cr. 20) + (0.5,0)$) --  ($(topplus.180)+(-0.5,0)$) --  (topplus.180);
			\draw                     [->] (f_BB. 20) -- ($(f_BB. 20) + (0.5,0)$) --  ($(topplus.200)+(-0.5,0)$) --  (topplus.200);
			\draw                     [->] (f_AA.-20) -- ($(f_AA.-20) + (0.5,0)$) --  ($(botplus.160)+(-0.5,0)$) --  (botplus.160);
			\draw                     [thick,red,->] (f_cr.-20) -- ($(f_cr.-20) + (0.5,0)$) --  ($(botplus.180)+(-0.5,0)$) --  (botplus.180);
			\draw                     [->] (f_BB.-20) -- ($(f_BB.-20) + (0.5,0)$) --  ($(botplus.200)+(-0.5,0)$) --  (botplus.200);
			\draw                     [thick,red,<-] (p1right.-90) -- node[below=0.1] {${\scriptstyle \mathcal{CN}(0,\nu_a^{-1}-\nu_b^{-1})}$} ($(p1right.-90) + (0,-0.3)$) ;
			\draw                     [thick,red,<-] (p2right.+90) -- node[above=0.1] {${\scriptstyle \mathcal{CN}(0,\nu_a^{-1}-\nu_b^{-1})}$} ($(p2right.+90) + (0,+0.3)$) ;
			\draw                     [thick,red,->] (p1left) -- (p1right);
			\draw                     [thick,red,->] (p2left) -- (p2right); 
			\draw                     [thick,red,->] (topplus) -- node[above=0.1]{$x_1$} (p1left);
			\draw                     [thick,red,->] (botplus) -- node[below=0.1]{$x_2$} (p2left); 
			\coordinate (p1leftright) at ($(p1left.east)!0.5!(p1right.west)$); 
			\coordinate (p2leftright) at ($(p2left.east)!0.5!(p2right.west)$); 
			\coordinate (p1lefttemp) at ($(p1left)-(0,0.35)$); 
			\coordinate (p2lefttemp) at ($(p2left)+(0,0.35)$); 
			\node  (gBA)              [thick,red, rectangle, draw, right= 6 of p1lefttemp, minimum height = 1.2cm] {$g_{\rm BA}(\cdot)$};
			\node  (gAB)              [thick,red, rectangle, draw, right= 6 of p2lefttemp, minimum height = 1.2cm] {$g_{\rm AB}(\cdot)$};
			\node  (gAA)              [rectangle, draw, above= 0.35 of gBA,    minimum height = 1.2cm] {$g_{\rm AA}(\cdot)$};
			\node  (gBB)              [rectangle, draw, below= 0.35 of gAB,    minimum height = 1.2cm] {$g_{\rm BB}(\cdot)$};
			\draw                     [thick,red,->] (p1leftright)  -- ++(0, 0.85) -- node[above,pos=1.0](Y1){$y_1$} ++(3.25,0) -- ($(gBA.150)-(0.5,0)$) -- (gBA.150);
			\draw                     [->] (p1leftright)  -- ++(0, 0.85) --                                ++(3.25,0) -- ($(gBB.150)-(0.5,0)$) -- (gBB.150);
			\draw                     [thick,red,->] (p1right.east) 				 -- node[above,pos=0.9](Z1){$z_1$} ++(1.5 ,0) -- ($(gAB.150)-(0.5,0)$) -- (gAB.150);
			\draw                     [->] (p1right.east) 				 --                                ++(1.5 ,0) -- ($(gAA.150)-(0.5,0)$) -- (gAA.150);
			\draw                     [thick,red,->] (p2right.east) 			     -- node[below,pos=0.9](Y2){$y_2$} ++(1.5 ,0) -- ($(gBA.210)-(0.5,0)$) -- (gBA.210);
			\draw                     [->] (p2right.east) 			     --                                ++(1.5 ,0) -- ($(gAA.210)-(0.5,0)$) -- (gAA.210);
			\draw                     [thick,red,->] (p2leftright)  -- ++(0,-0.85) -- node[below,pos=1.0](Z2){$z_2$} ++(3.25,0) -- ($(gAB.210)-(0.5,0)$) -- (gAB.210);
			\draw                     [->] (p2leftright)  -- ++(0,-0.85) --                                ++(3.25,0) -- ($(gBB.210)-(0.5,0)$) -- (gBB.210);
			\draw                     [->] (gAA) -- node[above]{$\hat{w}_{\rm AA}^{(AA)}                                         											$} ($(gAA)+(4,0)$);
			\draw                     [thick,red, ->] (gBA) --
			node[above]{$\textcolor{black}{\hat{w}_{\rm AA}^{(BA)},}\hat{w}_{0 }^{(BA)},\hat{w}_{\rm BA}^{(BA)}											$} ($(gBA)+(4,0)$); 
			\draw                     [thick,red, ->] (gAB) --
			node[above]{$\textcolor{black}{\hat{w}_{\rm AA}^{(AB)},}\hat{w}_{0 }^{(AB)},\hat{w}_{\rm AB}^{(AB)}											$} ($(gAB)+(4,0)$); 
			\draw                     [->] (gBB) -- node[above]{$\hat{w}_{\rm AA}^{(BB)},\hat{w}_{0 }^{(BB)},																$} ($(gBB)+(4,0)$);
			\draw                     [->] (gBB) -- node[below]{$                                        \hat{w}_{\rm BA}^{(BB)},\hat{w}_{\rm AB}^{(BB)},\hat{w}_{\rm BB}^{(BB)}	$} ($(gBB)+(4,0)$); 
			\coordinate (p1righttop)  at ($(p1right)+(1, 1)$);
			\coordinate (p2righttop)  at ($(p2right)+(1,-1)$);
			\node (topbox)            [draw, rounded corners, thick, fit = (p1left) (p1righttop) , dashed, inner sep = 2pt, label=above:{\footnotesize Gaussian BC channel 1}] {};
			\node (botbox)            [draw, rounded corners, thick, fit = (p2left) (p2righttop) , dashed, inner sep = 2pt, label=below:{\footnotesize Gaussian BC channel 2}] {};
			\end{tikzpicture}
		}
	\end{center}
	\caption{Encoding-decoding scheme of the 2 receiver Gaussian degraded product broadcast channel with users: AA, AB, BA, BB}
	\label{fig_GaussianDegradedBroadcastAllStates}
\end{figure*}
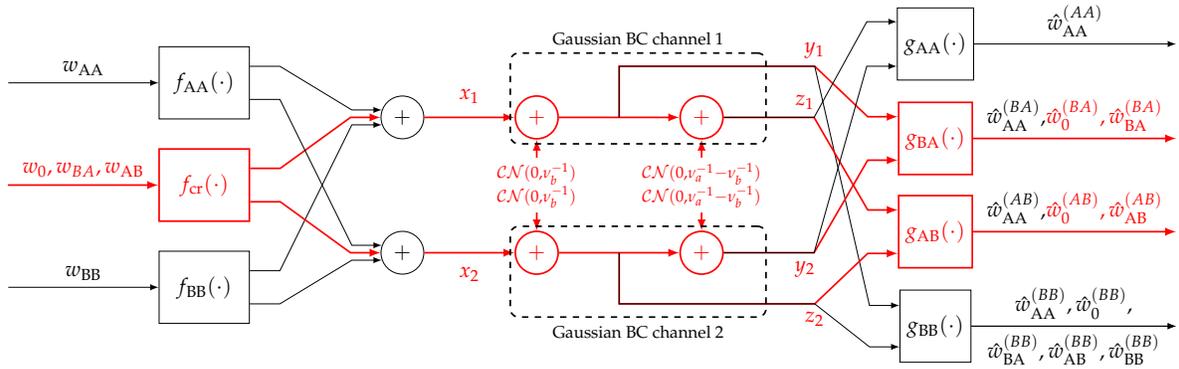

Consider the model introduced in~\cite{GA80}, which is a two-receiver discrete memoryless degraded product
broadcast channel. The Gaussian case was addressed as a special case.
A single transmitter encodes two $n$-length codewords consisting of a common message
$w_{0}\in\{1,...,2^{n R_0}\}$ to be decoded by both users, and two private messages $w_{\rm BA}\in\{1,...,2^{n R_{\rm BA}}\}$ and $w_{\rm AB}\in\{1,...,2^{n R_{\rm AB}}\}$, one for each of the two decoding users. A single function encodes these three messages into two codewords, where each undergoes parallel degraded broadcast sub-channels
\begin{equation}
\mycase{
	\begin{aligned}
	y_1 &= x_1 + n_{11} \\
	z_1 &= y_1 + n_{12}
	\end{aligned}\ ,
}
\quad\quad \mbox{and} \qquad 
\mycase{
	\begin{aligned}
	z_2&= x_2+ n_{21}\\
	y_2&= z_2+ n_{22}
	\end{aligned}\ ,
}
\end{equation}
where $n_{11},n_{21}\sim\mathcal{CN}(0,\nu_b^{-1})$ , $n_{21},n_{22}\sim\mathcal{CN}(0,\nu_a^{-1}-\nu_b^{-1})$. As depicted in the bold and red parts of Fig.~\ref{fig_GaussianDegradedBroadcastAllStates}, 
two users (namely $AB$ and $BA$) receive 
both common and private messages from the transmitter and independently decode the messages.
This is an unmatched setting, as $y_1$ is less noisy than $z_1$, and $z_2$ is less noisy than $y_2$. Hence, each of the users has one less-noisy channel output alongside another, which is the noisier output of the other sub-channel. 
Following \cite[Theorem 2]{GA80}, which shows this case, and exploiting symmetry for equal power allocation to both sub-channels, optimal allocation is expected to be achieved by equal common rate allocation to every user (state). Denoting $\bar{\alpha}=1-\alpha$, the capacity region $(R_0,R_{\rm BA},R_{\rm AB})$ is
\begin{align}
&R_0 					\leq \log \left( 1+\tfrac{\nu_a 	 \alpha P}{1+\nu_a
	\bar{\alpha}P}\right)  	+ \log \left( 1+\tfrac{\nu_b 	 \alpha
	P}{1+\nu_b\bar{\alpha}P}\right) \ , \\
&R_0+R_{\rm BA}=R_0+R_{\rm AB}	\leq \log\! \left(\ns 1+\tfrac{\nu_a\alpha P}{1+\nu_a
	\bar{\alpha}P}\ns\right)\!  \! +\! \log( 1+ \nu_b  P) \ ,  \\
&R_0+R_{\rm BA}+R_{\rm AB} 		\leq \log \left( 1+      \nu_b
P                      	\right) 	+ \log \left( 1+\tfrac{\nu_a 	 \alpha
	P}{1+\nu_a\bar{\alpha}P}\right) + \log \left( 1+\nu_b \bar{\alpha}P\right).  \label{eq_capacitySymmetricBcChannel}
\end{align}

\subsubsection{Extended Degraded Gaussian Broadcast Product Channels}

The classical product channel is extended by introducing two dual-input receivers in addition to the original two. The first gets the two more noisy channel outputs $(z_1,y_2)$, whereas the second receives the two less noisy outputs $(z_2,y_1)$.
To support this, two messages $w_{\rm AA}$ and $w_{\rm BB}$ are added. The total two $n$-length codewords
are the superposition of three codewords by independent encoders as follows
$(\textbf{X}_1,\textbf{X}_2)=f_{\rm AA}(w_{\rm AA})+f_{\rm cr}(w_0,w_{\rm BA},w_{\rm AB})+f_{\rm BB}(w_{\rm BB})$, where
subscript ${\rm cr}$ stands for {\em crossed} states ($(A,B)$ or $(B,A)$).
See Fig.~\ref{fig_GaussianDegradedBroadcastAllStates} for an illustration.

Stream AA is decoded first, regardless of whether the others can be decoded (this is done by
treating all the other streams as interference). Then, both streams AB and BA, including their common stream subscripted $0$ can be decoded after removing the AA impact from their decoder inputs (treating the BB stream as interference). Finally, removing all the above  decoded streams allows decoding stream BB. From \eqref{eq_capacitySymmetricBcChannel}, we have
\begin{align}
&        R_{\rm AA}                            \leq    2 \log \left( 1+\tfrac{             \alpha_{\rm AA}
	P}{\nu_a^{-1}+ \bar{\alpha}_{\rm AA}                        P}\right) \ ,  \\
&        R_{\rm AA}+R_0                        \leq    2 \log \left( 1+\tfrac{             \alpha_{\rm AA}
	P}{\nu_a^{-1}+ \bar{\alpha}_{\rm AA}                        P}\right) 
+  \log \!\left( 1+\tfrac{       \alpha\alpha_{\rm cr} P}{\nu_b^{-1}+(\bar{\alpha}\alpha_{\rm cr}+
	\alpha_{\rm BB})P}\!\right)   
+  \log\! \left( 1+\tfrac{       \alpha\alpha_{\rm cr}
	P}{\nu_a^{-1}+(\bar{\alpha}\alpha_{\rm cr}+
	\alpha_{\rm BB})P}\!\right)  \ ,  \\
&        R_{\rm AA}+R_0+R_{\rm BA}                 =       R_{\rm AA}+R_0+R_{\rm AB}\ 
\nonumber\\
&                                    \qquad\qquad\qquad        \leq    2 \log \left( 1+\tfrac{             \alpha_{\rm AA} P}{\nu_a^{-1}+ \bar{\alpha}_{\rm AA}                        P}\right) 
+  \log \left( 1+\tfrac{      \alpha \alpha_{\rm cr} P}{\nu_a^{-1}+(\bar{\alpha}\alpha_{\rm cr}+
	\alpha_{\rm BB})P}\right)    +  \log \left( 1+\tfrac{             \alpha_{\rm cr}
	P}{\nu_b^{-1}+                              \alpha_{\rm BB} P}\right)  \ ,  \\
&        R_{\rm AA}+R_0+R_{\rm BA}+R_{\rm AB} \nonumber \\ &          \qquad \qquad \qquad \leq    2 \log \left( 1+\tfrac{             \alpha_{\rm AA} P}{\nu_a^{-1}+ \bar{\alpha}_{\rm AA}                        P}\right)   
+  \log \left( 1+\tfrac{             \alpha_{\rm cr} P}{\nu_b^{-1}+
	\alpha_{\rm BB} P}\right) \nonumber  \\
&          \qquad \qquad \qquad                                       +  \log \left( 1+\tfrac{      \alpha \alpha_{\rm cr} P}{\nu_a^{-1}+(\bar{\alpha}\alpha_{\rm cr}+     \alpha_{\rm BB})P}\right)
+  \log \left( 1+\tfrac{ \bar{\alpha}\alpha_{\rm cr} P}{\nu_b^{-1}+
	\alpha_{\rm BB} P}\right)  \ ,  \\
&        R_{\rm AA}+R_0+R_{\rm BA}+R_{\rm AB}+R_{\rm BB}\nonumber \\ &   \qquad \qquad \qquad\leq    2 \log \left( 1+\tfrac{             \alpha_{\rm AA} P}{\nu_a^{-1}+ \bar{\alpha}_{\rm AA}                        P}\right)   
+  \log \left( 1+\tfrac{             \alpha_{\rm cr} P}{\nu_b^{-1}+
	\alpha_{\rm BB} P}\right)  \nonumber \\
&                       \qquad \qquad \qquad                          +  \log \left( 1+\tfrac{      \alpha \alpha_{\rm cr} P}{\nu_a^{-1}+(\bar{\alpha}\alpha_{\rm cr}+     \alpha_{\rm BB})P}\right)
+  \log \left( 1+\tfrac{ \bar{\alpha}\alpha_{\rm cr} P}{\nu_b^{-1}+
	\alpha_{\rm BB} P}\right)                            +2 \log \left( 1+\tfrac{             \alpha_{\rm BB}
	P}{\nu_b^{-1}                                            }\right)\ ,\label{eq_capacitySymmetricBcChannelAllStates}
\end{align}
where $\alpha_{\rm AA},\alpha_{\rm cr},\alpha_{\rm BB}\in[0,1]$ are the relative power allocations for the subscripted letters $\alpha_{\rm AA}+\alpha_{\rm cr}+\alpha_{\rm BB}=1$, and $\alpha\in[0,1]$ is the single user private power allocation within the unmatched channel.

\subsubsection{Broadcast Encoding Scheme}
\begin{figure*}[tb] 
	\begin{center}
		\scalebox{0.75}{
			\begin{tikzpicture}[>=latex]
			\node (splt)              [rectangle, draw, minimum height = 5.0cm, minimum width = 1.0cm] at (-3.5cm, 0) {splitter};
			\node (f_AA)              [rectangle, draw, minimum height = 1.2cm, minimum width = 1.5cm] at (-1cm, 2) {$f_{\rm AA}(\cdot)$};
			\node (f_cr)              [rectangle, draw, minimum height = 1.8cm, minimum width = 1.5cm] at (-1cm, 0) {$f_{\rm cr}(\cdot)$};
			\node (f_BB)              [rectangle, draw, minimum height = 1.2cm, minimum width = 1.5cm] at (-1cm,-2) {$f_{\rm BB}(\cdot)$};
			\node (p1left)            [isosceles triangle, draw, above right= 0.5 and 3.5 of f_cr] {$H_1$}; 
			\node (p2left)            [isosceles triangle, draw, below right= 0.5 and 3.5 of f_cr] {$H_2$};
			\draw                     [<-] ($(splt.west)+( 0.0, 0.0)$)   -- node[above]{$w$     } ($(splt.west)+(-1.0, 0.0)$); 
			\draw                     [<-] ($(f_AA.west)+( 0.0, 0.0)$)   -- node[above]{$w_{\rm AA}$} ($(f_AA.west)+(-1.1, 0.0)$); 
			\draw                     [<-] ($(f_cr.west)+( 0.0,+0.5)$)   -- node[above]{$w_{0}$ } ($(splt.east)+( 0.0,+0.5)$); 
			\draw                     [<-] ($(f_cr.west)+( 0.0,-0.5)$)   -- node[above]{$w_{\rm BA}$} ($(splt.east)+( 0.0,-0.5)$); 
			\draw                     [<-] ($(f_cr.west)+( 0.0, 0.0)$)   -- node[above]{$w_{\rm AB}$} ($(splt.east)+( 0.0, 0.0)$); 
			\draw                     [<-] ($(f_BB.west)+( 0.0, 0.0)$)   -- node[above]{$w_{\rm BB}$} ($(f_BB.west)+(-1.1, 0.0)$);
			\node (p1right)           [circle, draw, right= 0.5   of p1left] {$+$};
			\node (p2right)           [circle, draw, right= 0.5   of p2left] {$+$};
			\node (h1inv)             [isosceles triangle, draw, right = 1.0 of p1right, scale = 0.75] {$H_1^{-1}$};
			\node (h2inv)	          [isosceles triangle, draw, right = 1.0 of p2right, scale = 0.75] {$H_2^{-1}$};
			\node (topplus)           [circle, draw, left = 0.7 of p1left] {$+$};
			\node (botplus)           [circle, draw, left = 0.7 of p2left] {$+$};
			\draw                     [->] (f_AA. 20) -- ($(f_AA. 20) + (0.5,0)$) --  ($(topplus.160)+(-0.5,0)$) --  (topplus.160);
			\draw                     [->] (f_cr. 20) -- ($(f_cr. 20) + (0.5,0)$) --  ($(topplus.180)+(-0.5,0)$) --  (topplus.180);
			\draw                     [->] (f_BB. 20) -- ($(f_BB. 20) + (0.5,0)$) --  ($(topplus.200)+(-0.5,0)$) --  (topplus.200);
			\draw                     [->] (f_AA.-20) -- ($(f_AA.-20) + (0.5,0)$) --  ($(botplus.160)+(-0.5,0)$) --  (botplus.160);
			\draw                     [->] (f_cr.-20) -- ($(f_cr.-20) + (0.5,0)$) --  ($(botplus.180)+(-0.5,0)$) --  (botplus.180);
			\draw                     [->] (f_BB.-20) -- ($(f_BB.-20) + (0.5,0)$) --  ($(botplus.200)+(-0.5,0)$) --  (botplus.200);
			\draw                     [<-] (p1right.-90) -- node[below=0.1] {${\scriptstyle \mathcal{CN}(0,1)}$} ($(p1right.-90) + (0,-0.3)$) ;
			\draw                     [<-] (p2right.+90) -- node[above=0.1] {${\scriptstyle \mathcal{CN}(0,1)}$} ($(p2right.+90) + (0,+0.3)$) ;
			\draw                     [->] (p1left)  -- (p1right);
			\draw                     [->] (p2left)  -- (p2right); 
			\draw                     [->] (p1right) -- node[above,pos=0.6](Y1){$\textbf{Y}_1$} (h1inv);
			\draw                     [->] (p2right) -- node[below,pos=0.6](Y2){$\textbf{Y}_2$} (h2inv); 
			\draw                     [->] (topplus) -- node[above=0.1,pos=0.5]{$\textbf{X}_1$} (p1left);
			\draw                     [->] (botplus) -- node[below=0.1,pos=0.5]{$\textbf{X}_2$} (p2left); 
			\node  (gBA)              [rectangle, draw, below right = 0.2 and 6.0 of p1left,    minimum height = 1.0cm] {$g_{\rm BA}(\cdot)$};
			\node  (gAB)              [rectangle, draw, above right = 0.2 and 6.0 of p2left,    minimum height = 1.0cm] {$g_{\rm AB}(\cdot)$};
			\node  (gAA)              [rectangle, draw, above       = 0.5         of gBA,       minimum height = 1.0cm] {$g_{\rm AA}(\cdot)$};
			\node  (gBB)              [rectangle, draw, below       = 0.5         of gAB,       minimum height = 1.0cm] {$g_{\rm BB}(\cdot)$};
			\draw                     [->] (h1inv.east) -- ++(0.5 ,0) -- ($(gBA.150)-(0.5,0)$) -- (gBA.150);
			\draw                     [->] (h1inv.east) -- ++(0.5 ,0) -- ($(gBB.150)-(0.5,0)$) -- (gBB.150);
			\draw                     [->] (h1inv.east) -- ++(0.5 ,0) -- ($(gAB.150)-(0.5,0)$) -- (gAB.150);
			\draw                     [->] (h1inv.east) -- ++(0.5 ,0) -- ($(gAA.150)-(0.5,0)$) -- (gAA.150);
			\draw                     [->] (h2inv.east) -- ++(0.5 ,0) -- ($(gBA.210)-(0.5,0)$) -- (gBA.210);
			\draw                     [->] (h2inv.east) -- ++(0.5 ,0) -- ($(gAA.210)-(0.5,0)$) -- (gAA.210);
			\draw                     [->] (h2inv.east) -- ++(0.5 ,0) -- ($(gAB.210)-(0.5,0)$) -- (gAB.210);
			\draw                     [->] (h2inv.east) -- ++(0.5 ,0) -- ($(gBB.210)-(0.5,0)$) -- (gBB.210);
			\draw                     [->] (gAA) -- node[above]{$\hat{w}_{\rm AA}^{(AA)}                                         											$} ($(gAA)+(4,0)$) node[right]{$AA$};
			\draw                     [->] (gBA) -- node[above]{$\hat{w}_{\rm AA}^{(BA)},\hat{w}_{0 }^{(BA)},\hat{w}_{\rm BA}^{(BA)}											$} ($(gBA)+(4,0)$) node[right]{$BA$}; 
			\draw                     [->] (gAB) -- node[above]{$\hat{w}_{\rm AA}^{(AB)},\hat{w}_{0 }^{(AB)},\hat{w}_{\rm AB}^{(AB)}											$} ($(gAB)+(4,0)$) node[right]{$AB$}; 
			\draw                     [->] (gBB) -- node[above]{$\hat{w}_{\rm AA}^{(BB)},\hat{w}_{0 }^{(BB)},																$} ($(gBB)+(4,0)$);
			\draw                     [->] (gBB) -- node[below]{$                                        \hat{w}_{\rm BA}^{(BB)},\hat{w}_{\rm AB}^{(BB)},\hat{w}_{\rm BB}^{(BB)}	$} ($(gBB)+(4,0)$)  node[right]{$BB$};
			\coordinate (mux_nw)      at (14.6,+3.0);
			\coordinate (mux_sw)      at (14.6,-3.0);
			\coordinate (mux_ne)      at (15.6,+2.5);
			\coordinate (mux_se)      at (15.6,-2.5);
			\draw                     [-] (mux_nw) -- node[right]{mux}    (mux_sw) -- (mux_se) -- (mux_ne) -- (mux_nw);
			\draw                     [->] (15.6,0)   -- node[above]{$\hat{w}$     } (16.6,0); 
			\coordinate (p1righttop)  at ($(p1right)+(1,+0.9)$);
			\coordinate (p2rightbot)  at ($(p2right)+(1,-0.9)$);
			\coordinate (encoder_nw)  at (-4.4,+3.2);
			\coordinate (encoder_se)  at ( 2.6,-3.2);
			\coordinate (channel_nw)  at ( 3.1,+3.2);
			\coordinate (channel_se)  at ( 6.1,-3.2);
			\coordinate (decoder_nw)  at ( 6.7,+3.2);
			\coordinate (decoder_se)  at (15.8,-3.2);
			\node (encoderbox)        [draw, rounded corners, thick, fit = (encoder_nw) (encoder_se)    , dashed, inner sep = 0pt, label=above:{\footnotesize Encoder}] {};
			\node (channelbox)        [draw, rounded corners, thick, fit = (channel_nw) (channel_se)    , dashed, inner sep = 0pt, label=above:{\footnotesize Parallel Channel}] {};
			\node (decoderbox)        [draw, rounded corners, thick, fit = (decoder_nw) (decoder_se)    , dashed, inner sep = 0pt, label=above:{\footnotesize Decoder}] {};
			\end{tikzpicture}
		}
	\end{center}
	\caption{Encoding and decoding scheme of the two receiver Gaussian degraded product broadcast channel broadcast approach}
	\label{fig_BroadcastAllStatesBlockFadingChannels}
\end{figure*}
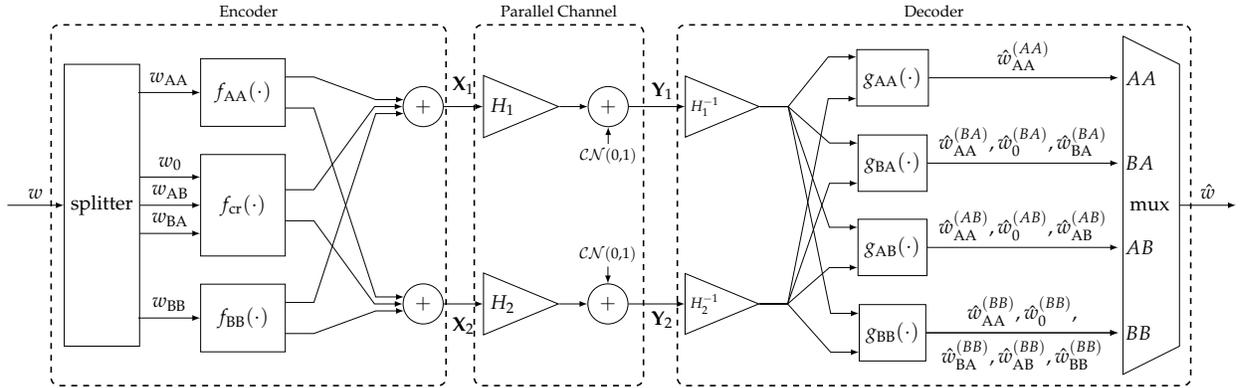

Adding a message splitter at the transmitter and
channel state-dependent message multiplexer at the receiver enriches the domain. Figure
\ref{fig_BroadcastAllStatesBlockFadingChannels} illustrates the encoding and decoding
schemes. During decoding, the four possible channel states $\mathbf{S}=(S_1,S_2)$ impose different decoding capabilities.
If $\mathbf{S}=\rm (A,A)$,  then $g_{\rm AA}(\cdot)$ can reconstruct $w_{\rm AA}$ to achieve a total rate of $R_{\rm AA}$.
For $\mathbf{S}=\rm (B,A)$, $g_{\rm BA}(\cdot)$ is capable of reconstructing three messages $(w_{\rm AA},w_0,w_{\rm BA})$ with sum rate of $R_{\rm AA}+R_0+R_{\rm BA}$.
Similarly for $\mathbf{S}=\rm (A,B)$, $g_{\rm AB}(\cdot)$ reconstructs $(w_{\rm AA},w_0,w_{\rm AB})$ with sum rate $R_{\rm AA}+R_0+R_{\rm AB}$.
When both channels are permissive $\mathbf{S}=\rm (B,B)$, all five messages $(w_{\rm AA},w_0,w_{\rm BA},w_{\rm AB},w_{\rm BB})$ are reconstructed at $g_{\rm BB}(\cdot)$ under the rate $R_{\rm AA}+R_0+R_{\rm BA}+R_{\rm AB}+R_{\rm BB}$. Recall that a single user transmission is of interest here, thus the expected rate of the parallel channel underhand can be expressed by
\begin{align} \label{eq R_ave definition two state channel}
R_{\text{ave}}	&=               P_A^2             R_{\rm AA}                                 
+       P_A      P_B     (R_{\rm AA}+R_0+R_{\rm AB}
)   \nonumber   \\         
&  \quad +       P_B      P_A     (R_{\rm AA}+R_0+R_{\rm BA}
)      \nonumber\\
&  \quad +       P_B^2            (R_{\rm AA}+R_0+R_{\rm BA}+R_{\rm AB}+R_{\rm BB} )\ . 
\end{align}
Using \eqref{eq_capacitySymmetricBcChannelAllStates}, and since both channels
have identical statistics leading to $R_{\rm AB}=R_{\rm BA}$, the achievable average rate is
\begin{align}
R_{\text{ave}}  &=  2 (P_A+P_B)^2 \log \left( 1+ \nu_a P \right) + R_0 ( 1-\alpha_{\rm AA}
) \nonumber \\ &\quad+ R_1 ( 1-\alpha_{\rm AA}-\alpha\alpha_{\rm cr} ) + R_2 ( 1-\alpha_{\rm AA}-\alpha_{\rm cr} )  \ ,
\end{align}
where the new notations are
\begin{align}\label{eq R0(alpha0)}
R_0(\alpha_0) &= [(P_A+P_B)^2-P_A^2]   \log ( 1+ \nu_b \alpha_0 P )-
[(P_A+P_B)^2+P_A^2] \log ( 1+ \nu_a \alpha_0 P )\ , \\
R_1(\alpha_1) &= P_B^2 \log ( 1+ \nu_b \alpha_1 P ) - [(P_A+P_B)^2-P_A^2]
\log ( 1+ \nu_a \alpha_1 P )\ , \\
R_2(\alpha_2) &= -2 P_A P_B  	        \log ( 1+ \nu_b \alpha_2 P ) \   .
\end{align}
and the arguments satisfy $\alpha_0 = 1-\alpha_{\rm AA}$, $\alpha_1 = 1-\alpha_{\rm AA}-\alpha\alpha_{\rm cr}$, and $\alpha_2 = 1-\alpha_{\rm AA}-\alpha_{\rm cr} = \alpha_{\rm BB}$.
Note that $R_0(\alpha_0)$ and $R_1(\alpha_1)$ are not obliged to be positive, as they can be negative for some scenarios, and $R_2(\alpha_2)$ is non-positive by definition. Denoting the domain $D'$ of valid power allocations vector $\mathbold{\alpha}'=[\alpha,\alpha_{\rm AA},\alpha_{\rm cr},\alpha_{\rm BB}]^T\in[0,1]^4$ and the operator  $[x]_+ = \max\{0,x\}$ yield the following proposition.

\begin{Proposition}\label{prop max over alpha1}
	The maximal sum rate of the symmetric two parallel two state channel over all power allocations is
	\begin{align}
	\max_{\mathbold{\alpha}' \in D'} R_{\textup{ave}}(\mathbold{\alpha}') &= 2(P_A+P_B)^2\log(1+
	\nu_a P) +\! \max_{0 \leq \alpha_{\rm AA} \leq 1}\!\! \left\{  R_0 ( 1-\alpha_{\rm AA} ) + R_1
	(\alpha_1^\textup{opt}(\alpha_{\rm AA})) \right\}\ ,
	\end{align}
	where
	\begin{align}\label{eq alpha_1 opt}
	\alpha_1^\textup{opt} 	(\alpha_{\rm AA})	&= \max \{ 0 , \min\{1-\alpha_{\rm AA} , \alpha_1^* \}
	\}\ , \\
	\alpha_1^* 		                        &= \frac{ P_B^2 \nu_b - [(P_A+P_B)^2-P_A^2] \nu_a}{ [(P_A+P_B)^2-P_A^2-P_B^2] \nu_a\nu_b P } \ ,
	\end{align}
	and the latter solves $\frac{\partial  }{\partial \alpha_1   } R_1(\alpha_1^*) =0$.
\end{Proposition}

\begin{Corollary}\label{crl alpha_BB^opt=0}
	The optimal power allocation for the state $(B,B)$ is $\alpha_{\rm BB}^{\textup{opt}}=0$.
\end{Corollary}
This is true for any set of parameters $\nu_a,\nu_b,P_A,P_B$, even if $P_B\rightarrow1$ and $\nu_b\gg\nu_a$. Inherently, a penalty occurs when trying to exploit the double permissive state.
\begin{Corollary}
	Under the optimal power allocation, $\alpha^\textup{opt}(\alpha_{\rm AA})=1-\alpha_1^\textup{opt}(\alpha_{\rm AA}) / (1-\alpha_{\rm AA}) $. 
\end{Corollary}
This removes a degree of freedom in the optimization problem.

Using these corollaries, and the notation $ \mathbold{\alpha}'=[\alpha, \alpha_{\rm AA}, \alpha_{\rm cr},
\alpha_{\rm BB}]^T$ instead of $ \mathbold{\alpha}=[\alpha_0, \alpha_1, \alpha_2]^T $,  we have the following theorem.
\begin{Theorem}\label{thm max sum rate 2 parallel channel}
	The maximal sum rate of the symmetric two-parallel two-state channel over all  allocations $\mathbold{\alpha}' \in D'$ is
	\begin{align}
	R_{\textup{ave}}^{\textup{opt}} &= 2(P_A+P_B)^2\log(1+\nu_a P) + \sol[l]{\max_{0 \leq
			\alpha_{\rm AA} \leq 1}}\!\! \left\{\ns  R_0 ( 1\!-\!\alpha_{\rm AA} ) \!+\! R_1
	((1\!-\!\alpha_{\rm AA})\!\cdot\!(1\!-\!\alpha^\textup{opt}(\alpha_{\rm AA})))\! \right\}\ ,
	\end{align}
	where
	\begin{align}{rCl}
		\alpha^\textup{opt}	(\alpha_{\rm AA})  &=  \left[\min\left\{1 , 1- \tfrac{ P_B^2 \nu_b
			- [(P_A+P_B)^2-P_A^2] \nu_a}{ 2P_A\cdot P_B\cdot \nu_a\nu_b P (1-\alpha_{\rm AA})}  \right\}
		\right]_+ \ .
	\end{align}
	Denoting the argument of the maximization as $\alpha_{\rm AA}^\textup{opt}$, the optimal power allocation vector is
	\begin{equation*}
	\mathbold{\alpha}'^{\textup{opt}}=[\alpha^\textup{opt}(\alpha_{\rm AA}), \alpha_{\rm AA}^\textup{opt}, 1-\alpha_{\rm AA}^\textup{opt}, 0]^\top         .
	\end{equation*}
\end{Theorem}
From Proposition~\ref{prop max over alpha1} and by setting $\alpha_1=1-\alpha_{\rm AA}-\alpha\alpha_{\rm cr}=(1-\alpha_{\rm AA})(1-\alpha)$, it can be observed that the optimal allocation for state BB is $\alpha_{\rm BB}=0$. For evaluation of the advantage of the joint $\alpha_{\rm AA}$ and $\alpha$, the following sub-optimal schemes are compared: a) independent broadcasting; b) privately broadcasting; and c) only common broadcasting. A scheme for which the encoder disjointly encodes different messages into each single channel of the parallel channel using the broadcast approach over the fading channel is denoted by \textbf{independent broadcasting}. The broadcast approach for fading SISO channel relies on two main operations: superposition coding by layering at the transmitter; and successive interference cancellation at the receiver. 
The maximal expected sum rate of the symmetric two parallel two state channel under independent broadcasting is
\begin{align*}
R^{\textup{ind-bc,opt}}_\textup{ave} 
&=    2(P_A+P_B) \log\left(\tfrac{ 1+ \nu_a P }{1+ \nu_a (1
	- \alpha^{\textup{ind-bc,opt}})  P} \right) + 2 P_B      \log\left( 1+ \nu_b (1 - \alpha^{\textup{ind-bc,opt}} )  P \right),
\end{align*}
\begin{equation}
\alpha^\textup{bc,opt} = \left[ \min\left\{1 ,  1-\tfrac{ P_B \nu_b - (P_A+P_B) \nu_a}{ P_A
	\nu_a\nu_b P }  \right\} \right]_+   .
\label{eq_alpha_bc_opt}
\end{equation}  
A scheme for which no power is allocated for the common stream in the $\rm (B,A)$ and $\rm (A,B)$ states (message $w_0$) is called \textbf{privately broadcasting}.
This scheme is equivalent to setting $\alpha=0$ in Theorem \ref{thm max sum rate 2 parallel channel}, thus allocating encoding power from the common stream ($R_0=0$) to the other streams $R_{\rm AA},R_{\rm AB},R_{\rm BA}$ and $R_{\rm BB}$, 
which achieves optimality for
\begin{equation}
\alpha_{\rm AA}^\textup{prv-bc,opt} = \left[  \min\left\{1 ,  1-\tfrac{ [P_B-P_A] \nu_b
	- [P_B+P_A]\nu_a}{ 2 P_A \nu_a\nu_b P }  \right\} \right]_+. 
\end{equation}
A scheme for which all of the crossed state power is allocated only to the common stream (message $w_0$) and no power is allocated to the private messages (no allocation for messages $w_{\rm AB}$ and $w_{\rm BA}$) is called \emph{only common broadcasting}.
This scheme is equivalent to setting $\alpha=1$ in Theorem \ref{thm max sum rate 2 parallel channel}, thus allocating encoding power from the private streams ($R_{\rm AB}=R_{\rm BA}=0$) to the other streams $R_{\rm AA},R_0$ and $R_{\rm BB}$, 
which achieves optimality for
\begin{equation}
\alpha_{\rm AA}^\textup{cmn-bc,opt}=  \left[\min\!\left\{\!1,\!1\!-\!\tfrac{ [(P_A\!+\!P_B)^2\!-\!P_A^2] \nu_b
	\!-\! [(P_A\!+\!P_B)^2\!+\!P_A^2]\nu_a}{ 2 P_A^2 \nu_a\nu_b P }\!  \right\} \right]_+. 
\end{equation}
The result in Theorem \ref{thm max sum rate 2 parallel channel} differs from the one presented
in \cite{YEH01} for the two-parallel two state channel. In \cite{YEH01} it is chosen to transmit
only common information to the pairs $\rm (A,B)$ and $\rm (B,A)$. \cite[equation (39)]{YEH01} clearly states that
for the crossed states (A,B) and (B,A) only common rate is used without justification. It is further claimed that this is an expected rate upper bound for some power allocation. The result in \eqref{eq_alpha_bc_opt} proves that broadcasting common information only, i.e., $\alpha=1$ is sub-optimal, and does not yield the maximal average rate. 

\subsection{Broadcast Approach via Dirty Paper Coding}

We conclude this section by noting the relevance of dirty paper coding (DPC) to the broadcast approaches discussed. Even though the central focus of the broadcast approaches discussed is superposition coding, all these approaches can be revisited by instead adopting dirty paper coding. Information layers generated by a broadcast approach interfere with one another with the key property that the interference is known to the transmitter. DPC enables effective transmission when the transmitted signal is corrupted by interference (and noise in general)  terms that are known to the transmitter. This is facilitated via precoding the transmitted signal by accounting for and canceling the interference.

DPC plays a pivotal role in the broadcast channel. It is an optimal (capacity-achieving) scheme for the multi-antenna Gaussian broadcast channel~\cite{Weingarten06,CA01_1} with general message sets and effective, in the form of binning, for the general broadcast channel with degraded message sets~\cite{NairGamal09}. To discuss the application of the DPC in the broadcast approach, consider the single-user channel with a two-state fading process, that is for the model in~\eqref{SISO1} we have $h\in\{h_w, h_s\}$ where $|h_w|<|h_s|$, rendering the following two models in these two states:
\begin{align}
y_w \; & = \; h_w x\; + \; n_w \ , \\
y_s \; & = \; h_s x\; + \; n_s \ ,
\end{align}
which can be also considered a broadcast channel with two receivers with channels $h_w$ and $h_s$. The broadcast region for this channel can be achieved by both superposition coding and DPC. When the noise terms have standard Gaussian distribution, the capacity region is characterized by all pairs
\begin{align}
R_w \; & \leq  \; \frac{1}{2}\log\left(1+\frac{\alpha P|h_w|^2}{1+(1-\alpha)P|
h_w|^2}\right)\ , \\
R_s \;& \textcolor{black}{leq} \; \frac{1}{2}\log\left(1+(1-\alpha)P|h_s|^2\right)\ .
\end{align}
over all  $\alpha\in[0,1]$. This capacity region is achievable by superposition coding of two information layers $x_w$ and $x_s$ with rates $R_w$ and $R_s$ to transmit $x=x_w+x_s$. The same region can be achieved by DPC, where $x_w$ is generated and decoded as done in superposition coding, and $x_s$ is designed by treating $x_w$ as the interference known to the transmitter non-causally. It is noteworthy that the original design of DPC in~\cite{Costa1983} the non-causally known interference term is modeled as additive Gaussian noise. However, as shown in~\cite{CohenISIT2012}, the interference term can be any sequence, like a Gaussian codeword, and still achieve the same capacity region.

The operational difference of superposition coding and DPC at the receiver side is that when using superposition coding, at the stronger receiver, the layers $x_w$ and $x_s$ have to be decoded sequentially, while when using DPC, the two layers can be decoded in parallel. This observation alludes to an operational advantage of DPC over superposition coding: while both achieving the capacity region, DPC imposes shorter decoding latency.

\section{The Multiple Access Channel}
\label{sec:MAC}

\subsection{Overview}

As discussed in detail in Section~\ref{sec:BCApproach}, CSI uncertainties result in degradation in communication reliability. Such degradations can be further exacerbated as we transition to multiuser networks consisting of a larger number of simultaneously communicating users.  Irrespective of multiuser channel models, a common realistic assumption is that slowly varying channels can be estimated by the receivers with high fidelity, providing the receivers with the CSI. Acquiring the CSI by the transmitters can be further facilitated via feedback from the receivers. However, feedback communication is often infeasible or incurs additional communication and delay costs, which increase significantly as the number of transmitters and receivers grows in the network.

This section focuses on the multi-access channel, consisting of multiple users with independent messages communicating with a common receiver. 
The channels undergo slow fading processes. Similar to the setting considered in Section~\ref{sec:BCApproach}, it is assumed that the receivers can acquire the CSI with high fidelity (e.g., through training sessions). While the receiver has perfect and instantaneous access to the states of all channels, the transmitters are either entirely or partially oblivious to the CSI, rendering settings in which the transmitters face CSI uncertainty. The information-theoretic limits of the MAC when all the transmitters and receivers have {\sl complete} CSI are well-investigated~\cite{ahlsewde,KimElGamal2011,SH98}. Furthermore, there is a rich literature on the MAC's information-theoretic limits under varying degrees of availability of {\sl instantaneous} CSIT. Representative studies on the capacity region include the impact of degraded CSIT~\cite{CemalSteiberg}, quantized and asymmetric CSIT~\cite{SenComoYuksel}, asymmetric delayed CSIT~\cite{BasherShiraziPermuter}, non-causal asymmetric partial CSIT~\cite{SenAlajajiYuksel}, and symmetric noisy CSIT~\cite{SenAlajajiYiikselComo:2013}. Bounds on the capacity region of the memoryless MAC in which the CSIT is made available to a different encoder in a causal manner are characterized in~\cite{LapidothSteinberg:2013}. Counterpart results are characterized for the case of common CSI at all transmitters in~\cite{LapidothSteinberg:2013com}, which are also extended in~\cite{LiSimeoneYener:2013} to address the case in which the encoder compresses previously transmitted symbols and the previous states. The study in~\cite{kotagiri2008multiaccess} provides an inner bound on the capacity region of the discrete and Gaussian memoryless two-user MAC in which the CSI is made available to one of the encoders non-causally. An inner bound on the capacity of the Gaussian MAC is derived in~\cite{lapidoth2010multiple} when both encoders are aware of the CSI in a strictly causal manner. The capacity region of a cooperative MAC with partial CSIT is characterized in \cite{PermuterShamaiSomekh:2011}. The capacity region of the multi-user Gaussian MAC in which each interference state is known to only one transmitter is characterized within a constant gap in \cite{Wang:2012}. A two-user generalized MAC with correlated states and non-causally known CSIT is studied in \cite{EmadiKhormujiSkoglundAref:2014}. In \cite{MonemizadehBahmaniHodtaniSeyedin:2014}, a two-user Gaussian double-dirty compound MAC with partial CSIT is studied. The capacity regions of a MAC with full and distributed CSIT are analyzed in \cite{sreekumar2015distributed}. A two-user cooperative MAC with correlated states and partial CSIT is analyzed in \cite{EmadiZamanighomiAref:2012}. The study in \cite{PermuterWeissmanChen:2009} characterizes inner and upper bounds on the capacity region of a finite-state MAC with feedback.

Despite the rich literature on the MAC with full CSIT, when the transmitters can acquire only the probability distribution of the fading channel state, without any instantaneous CSIT, the performance limits are not fully known. The broadcast approach is investigated for the two-user MAC with {\sl no} CSIT in~\cite{SH00,Minero:ISIT07,Tajer18,KazemiTajer2017, Zou13, zohdy2019broadcast}. The multiple access channel is primarily studied in~\cite{SH00,Minero:ISIT07,Tajer18,KazemiTajer2017, Zou13,zohdy2019broadcast}. Specifically, the effectiveness of a broadcast strategy for multiuser channels is investigated in~\cite{SH00,Minero:ISIT07,Tajer18} for the settings in which the transmitters are oblivious to all channels, and in~\cite{Zou13,zohdy2019broadcast} for the settings in which each transmitter is oblivious to only channels linking other users to the receiver. Specifically, when the transmitters are oblivious to all channels, the approaches in~\cite{SH00} and \cite{Minero:ISIT07} adopt the broadcast strategy designed for single-user channels and directly apply it to the MAC. As a result, each transmitter generates a number of information streams, each adapted to a specific realization of the direct channel linking the transmitter to the receiver. The study in~\cite{Tajer18} takes a different approach based on the premise that the contribution of each user to the overall performance of the multiple access channel not only depends on the direct channel linking this user to the receiver, but it is also influenced by the {\em relative} qualities of the other users' channels. Hence, it proposes a strategy in which the information streams are generated and adapted to the channel's combined state resulting from incorporating all individual channel states. The setting in which the transmitters have only local CSIT, that is, each transmitter has the CSI of its direct channel to the receiver while being unaware of the states of other users' channels, is studied in~\cite{Zou13,zohdy2019broadcast}. Medium access without transmitter coordination is studied in~\cite{Cao2007}.

The remainder of this section is organized as follows. This section focuses primarily on the two-user MAC, for which we provide a model in Section~\ref{sec:MACmodel}. We start by discussing the settings in which the transmitters have access to only the statistical model of the channel, and they are oblivious to the channel model in Section~\ref{sec:NCSI} with an emphasis on continuous channel models. Next, we focus on the setting in which the receiver has full CSI, and the transmitters have only the statistical model of the CSI and review two broadcast approaches in Sections~\ref{sec:CSIR_SU}~and~\ref{sec:CSIR_MU}. The focus of these two subsections is two-state discrete channel models. Their generalization to multi-state channels will be discussed in Section~\ref{sec:CSIR_MU_multi}. Finally, we will review two broadcast approach solutions for settings with local (partial) CSIT in Sections~\ref{sec:LCSIT_fixed}~ and~\ref{sec:LCSIT_MU}. The focus of these two subsections are on the two-state discrete channel models, and their generalization to the multi-state models is discussed in Section~\ref{sec:LCSIT_MU_multi}.

\begin{figure}[t]
\centering
\includegraphics[height=2.2in]{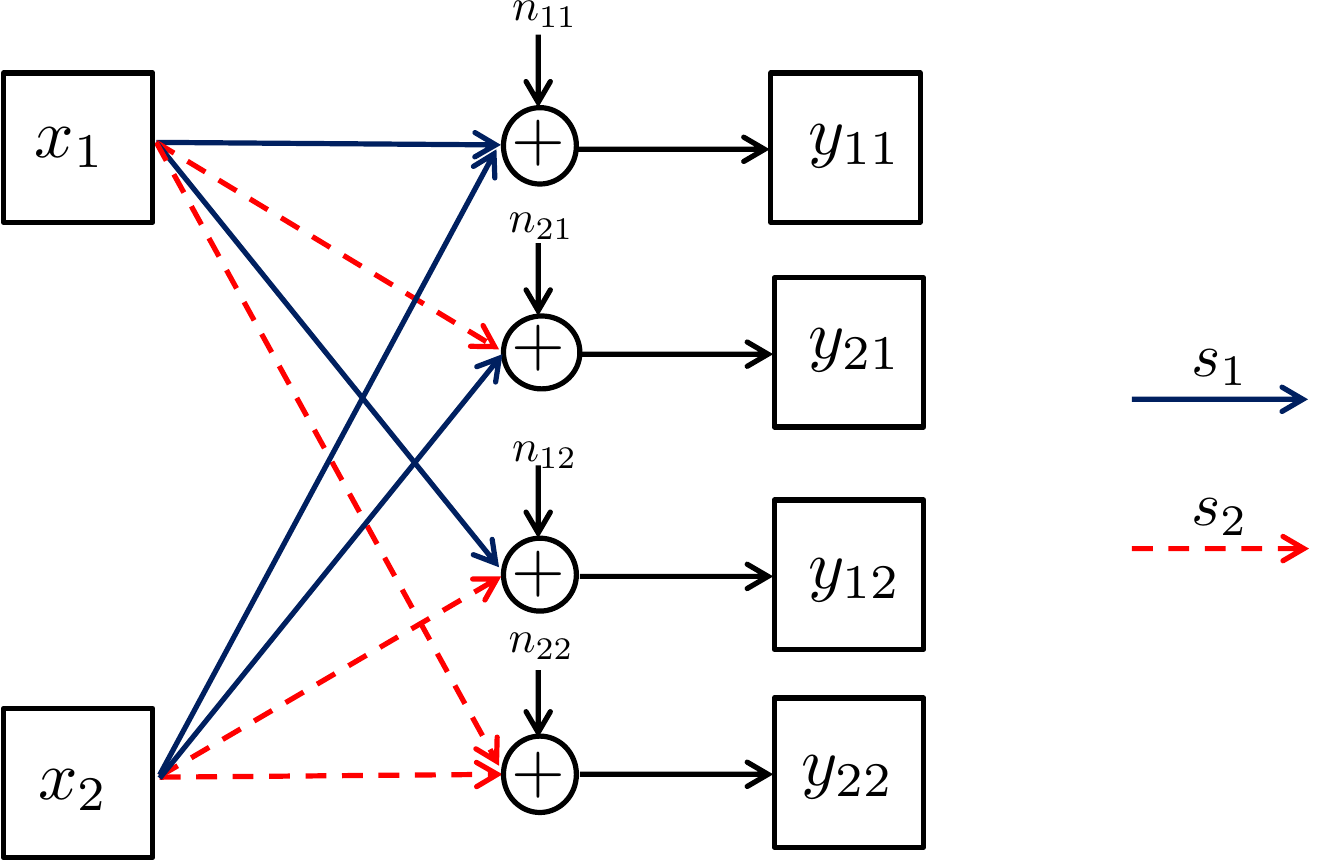}
\caption{Equivalent degraded broadcast channel corresponding to a two user four state multiple access channel with channel gains ${s}_1$ and ${s}_2$.}
\label{fig:network}
\end{figure}

\subsection{Network Model}
\label{sec:MACmodel}

Consider a two-user multiple access channel, in which two independent users transmit independent messages to a common receiver via a discrete-time Gaussian multiple-access fading channel.  All the users are equipped with one antenna, and the random channel coefficients are statistically independent.  The fading process is assumed to remain unchanged during each transmission cycle and can change to independent states afterward. The users are subject to an average transmission power constraint $P$. By defining $x_i$ as the signal of transmitter $i\in\{1,2\}$ and $h_i$ as the coefficient of the channel linking transmitter $i\in\{1,2\}$ to the receiver, the received signal is
\begin{equation} \label{eq:model}
y = h_1x_1 + h_2x_2 + n\ ,
\end{equation}
where $n$ accounts for the AWGN with mean zero and variance 1. We consider both continuous and discrete models for the channel.\vspace{.05 in}

\subsubsection{Discrete Channel Model} 
Each of the channels, independently of the other one, can be in one of the finite distinct states. We denote the number of states by $\ell\in\mathbb{N}$ and denote the distinct values that $h_1$ and $h_2$ can take by  $\{\sqrt{s}_m: m\in\{1,\dots,\ell\}\}$. Hence the multiple access channel can be in one of the combined $\ell^2$ states. By leveraging the broadcast approach (c.f. \cite{Shitz97broadcast, ShitzSteiner03}, and \cite{Minero:ISIT07}), the communication model in \eqref{eq:model} can be equivalently presented by a broadcast network that has two inputs $x_1$ and $x_2$ and $\ell^2$ outputs, each corresponding to one channel state combination. Each output corresponds to one possible combinations of channels $h_1$ and $h_2$. We denote the output corresponding to the combination $h_1=\sqrt{{s}_m}$ and $h_2=\sqrt{{s}_n}$ by
\begin{equation} \label{eq:model2}
y_{mn} = \sqrt{{s}_m} x_1 + \sqrt{{s}_n}x_2 + n_{mn}\ ,
\end{equation}
where $n_{mn}$ is a standard Gaussian random variable for all $m,n\in \{1,\dots,\ell\}$. Figure~\ref{fig:network} depicts this network for the case of the two-state channels ($\ell=2$). Without loss of generality and for the convenience in notations, we assume that channel gains $\{{s}_m: m\in\{1,\dots,\ell\}\}$ take real positive values and are ordered in the ascending order, i.e., $0<{s}_1< {s}_2< \dots < {s}_\ell$. We define $p_{mn}$ as the probability of the state $(h_1,h_2)=({{s}_m},{{s}_n})$. Accordingly, we also define $q_m =  \sum_{n=1}^\ell p_{mn}$ and $p_n =  \sum_{m=1}^\ell p_{mn}$. We will focus throughout the section on the case of symmetric average transmission power constraints, i.e., $P_1 = P_2 = P$, whereas the generalization to the case of asymmetric power constraints is straightforward.  \vspace{.05 in}

\subsubsection{Continuous Channel Model} 
In the continuous channel model, the fading coefficients $h_1$ and $h_2$ take a continuous of values that follow known statistical models. These statistical models are known to the transmitter and receiver. We denote the fading powers by $s_1= |h_1|^2$ and $s_2= |h_2|^2$. Depending on channel realizations, denote the channel output when the channel gains are $s_1$ and $s_2$ by 
\begin{equation} \label{eq:model2}
y_{s_1s_2} = \sqrt{{s}_1} x_1 + \sqrt{{s}_2} x_2 +n_{s_1s_2}\ .
\end{equation}
Throughout this section, we use the notation $C(x,y)= \frac{1}{2}\log_2\big(1+\frac{x}{y+\frac{1}{P}}\big)$.\vspace{.05 in}

We review settings in which the transmitters are either fully oblivious to all channels or have local CSIT. That is, each transmitter 1 (2) knows channel $h_1$ ($h_2$) while being unaware of channel $h_2$ ($h_1$). We refer to this model by L-CSIT, and similarly to the N-CSIT setting, we characterize an achievable rate region for it.

\subsection{Degradedness and Optimal Rate Spitting}

The broadcast approach's hallmark is a designating an order of degradedness among different network realizations based on their qualities. Designating degradedness in the single-user single-antenna channel arises naturally, as discussed in Section~\ref{sec:BCApproach}. However, for the multiuser networks, there is no natural notion of degradedness, and any ordering approach will at least bear some level of heuristics.  In the broadcast approaches that we discuss in this section for the MAC, we use the capacity region of the multiple access channels under different network states. Based on this notion of degradedness, once one of the channels improves, the associated capacity region expands, alluding the to the possibility of sustaining higher reliable rates.

\subsection{MAC without CSIT -- Continuous Channels}
\label{sec:NCSI}

We start by discussing the canonical Gaussian multiple-access channel in which the channels undergo a continuous fading model in~\eqref{eq:model2}. This is the setting that is primarily investigated in~\cite{SH00}. To formalize this approach, we define $R_i(s)$ as the reliability communicated information rate of transmitter $i$ at fading level $s$. Similarly to the single-user channel, we define $\rho_i(s)$ as the power assigned to the infinitesimal information layer of transmitter $i$ corresponding to fading power $s$. Accordingly, we define the interference terms
\begin{align}
I_i(s) = \int_s^\infty \rho_i(u)\; \d u\ .
\end{align}
When the channels fading powers are $s_1$ and $s_2$, we define ${\sf SNR}_i(s_1,s_2)$ as the effective SNR of transmitter $i$. These SNR terms satisfy
\begin{align}
{\sf SNR}_1(s_1,s_2)=\frac{s_1}{1+s_2I_2({\sf SNR}_2(s_1,s_2))}\ , \qquad \mbox{and} \qquad 
{\sf SNR}_2(s_1,s_2)=\frac{s_1}{1+s_2I_1({\sf SNR}_1(s_1,s_2))}\ .
\end{align}
Hence, corresponding to this channel combination, the rate that transmitter $i$ can sustain reliability is
\begin{align}
R_i(s_1,s_2)=\int_0^{{\sf SNR}_i(s_1,s_2)}  \frac{-u\d I_i(u)}{1+uI_i(u)}\ ,
\end{align}
and subsequently, the expected rate of transmitter $i$ is
\begin{align}
\bar R_i=\mathbb{E}[R_i(s_1,s_2)] = \int_0^\infty \left(1-F_i(u)\right)\frac{-u\d I_i(u)}{1+uI_i(u)}\ ,
\end{align}
where \textcolor{black}{$F_i$} denotes the CDF of ${\sf SNR}_i(s_1,s_2)$. Any resource allocation or optimization problem over the average rates $\bar R_1$ and $\bar R_2$ consists in determining the power allocation functions $\rho_i(s)$. For instance, finding the transmission policy that yields the maximum average rate $\bar R_1+\bar R_2$ boils down to designing functions $\rho_1$ and $\rho_2$. The same formulation can be readily generalized to the $K$-user MAC, in which we designate a power allocation function to each transmitter, accordingly define the interference functions, the achievable rates for each specific channel realization, and the average rate of each user.

\subsection{MAC without CSIT -- Two-state Channels: Adapting Streams to the Single-user Channels}

\label{sec:CSIR_SU}

We continue by reviewing finite-state multi-access channels.
 This setting is first investigated in~\cite{Minero:ISIT07} for the two-state discrete channel model. As suggested in ~\cite{Minero:ISIT07}, one can readily adopt the single-user strategy of~\cite{Shitz97broadcast} and split the information stream of a transmitter into two streams, each corresponding to one fading state, and encodes them independently. Recalling the canonical model in~\eqref{eq:model2}, let us refer to the channel with the fading gains $s_1$ and $s_2$ as  {\em weak} and {\em strong} channels, respectively\footnote{We will use this strong versus weak dichotomous model throughout Section~\ref{sec:MAC}}. The two encoded information streams are subsequently superimposed and transmitted over the channel. One of the streams,  denoted by $W_1$, is always decoded by the receiver, while the second stream, denoted by $W_2$, is decoded only when the channel is {\em strong}.

This strategy is adopted and directly applied to the multiple access channel in~\cite{Minero:ISIT07}. Specifically, it generates two coded information streams per transmitter, where the streams of user $i\in\{1,2\}$ are denoted by $\{W^i_1,W^i_2\}$. Based on the channels' actual realizations, a combination of these streams is successively decoded by the receiver.  In the first stage, the baseline streams $W^1_1$ and $W^2_1$, which constitute the minimum amount of guaranteed information, are decoded. Additionally, when the channel between transmitter $i$ and the receiver, i.e., $h_i$ is strong, in the second stage information stream $W^i_2$ is also decoded. Table~\ref{table:tse} depicts the decoding sequence corresponding to each of the four possible channel combinations.
{
\begin{table}[!h]
\renewcommand{\arraystretch}{1.2}
\caption{Successive decoding order when adapting the layers to the single-user channels,}
\label{table:tse}
\centering
\begin{tabular}{ |c||c|c|c|c| } 
 \hline
 $(h_1^2,h_2^2)$ & Decoding stage 1 & Decoding stage 2\\ 
 \hline\hline
 $({s}_1,{s}_1)$ & $W^1_1,W^2_1$ &\\ 
 \hline
 $({s}_2,{s}_1)$ & $W^1_1,W^2_1$ & $W^1_2$\\ 
 \hline
 $({s}_1,{s}_2)$ & $W^1_1,W^2_1$ & $W^2_2$\\ 
 \hline
 $({s}_2,{s}_2)$ & $W^1_1,W^2_1$ & $W^1_2,W^2_2$\\ 
 \hline
\end{tabular}
\end{table}
}

\begin{figure}[t]
\centering
\includegraphics[height=2.2in]{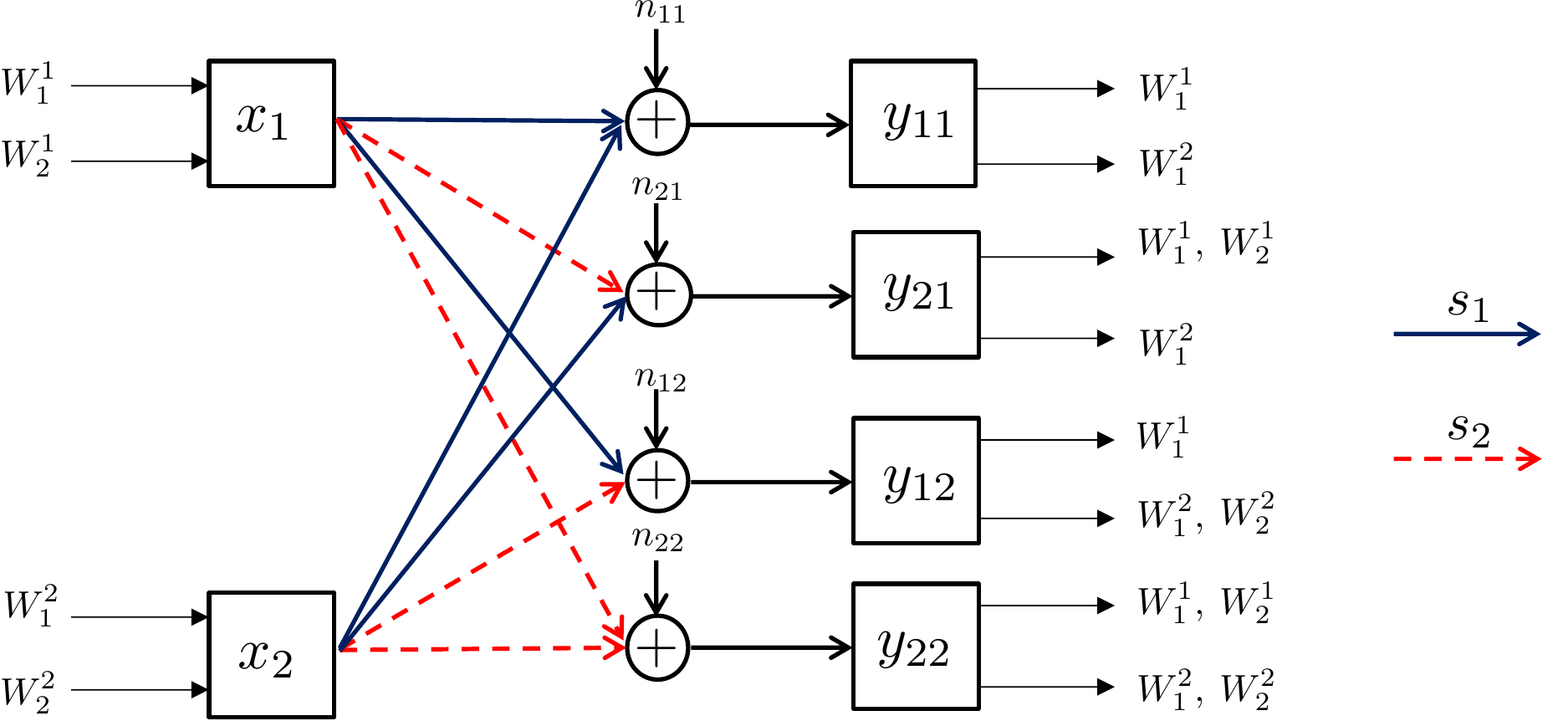}
\caption{Equivalent network when adapting the layers to the single-user channels (no CSIT).}
\label{chapter3_network2}
\end{figure}

Based on the codebook assignment and decoding specified in Table~\ref{table:tse}, the equivalent multiuser network is depicted in~Fig.~\ref{chapter3_network2}. The performance limits on the rates are characterized by delineating the interplay among the rates of the four codebooks $\{W_1^1, W_1^2, W_2^1,W_2^2\}$. We denote the rate of codebook $W_i^j$ by $R(W_i^j)$. There are two ways of grouping these rates and assessing the interplay among different groups. One approach would be analyzing the interplay between the rate of the codebooks adapted to the weak channels and the codebooks' rate adapted to the strong channels. The second approach would be analyzing the interplay between the rates of the two users. In a symmetric case and in the face of CSIT uncertainty, a natural choice will be the former approach. For this purpose, define $R_w=R_1^1+R_1^2$ and $R_s=R_2^1+R_2^2$ as the rate of the codebooks adapted to the weak and strong channels, respectively. The study in~\cite{Minero:ISIT07} characterizes the capacity region of the pair $(R_w,R_s)$ achievable in the Gaussian channel, where it is shown that superposition coding is the optimal coding strategy.  The capacity region of this channel is specified in the following theorem.

\begin{Theorem}[\cite{Minero:ISIT07}] \label{theorem:tse}
The $(R_w,R_s)$ capacity region for the channel depicted in Fig.~\ref{chapter3_network2} is given by the set of all rates satisfying
\begin{align}
R_w & \leq C(2s_1(1-\beta) \; , \; 2s_1\beta)\ ,\\
R_s & \leq C(2s_2\; , \; 0)\ ,
\end{align}
for all $\beta\in[0,1]$. 
\end{Theorem}

\subsection{MAC without CSIT -- Two-state Channels: State-dependent Layering}
\label{sec:CSIR_MU}

In the approach of Section~\ref{sec:CSIR_SU}, each transmitter adapts its transmission to its direct link to the receiver without regards for the channel linking the other transmission to the receiver. However, the contribution of user $i\in\{1,2\}$ to a network-wide performance metric (e.g., sum-rate capacity) depends not only on the quality of the channel $h_i$, but also on the quality of the channel of the other user. This motivates adapting the transmission scheme of each transmitter to the MAC's combined state instead of the individual channels. As investigated in~\cite{KazemiTajer2017,Tajer18} adapting to the network state can be facilitated by assigning more information streams to each transmitter and adapting them to the {\em combined} effect of {\em both} channels. Designing and assigning more than two information streams to each transmitter allows for a finer resolution in successive decoding, which in turn expands the capacity region characterized in~\cite{Minero:ISIT07}.

To review the encoding and decoding scheme as well as the attendant rate regions, we start by focusing on the two-state discrete channel model. This setting furnishes the context to highlight the differences between streaming and successive decoding strategy in this section and those investigated in Section~\ref{sec:CSIR_SU}. By leveraging the intuition gained, the general multi-state discrete channel model will be discussed in Section~\ref{sec:CSIR_MU}.

\begin{figure}[h]
\centering
\includegraphics[width=4in]{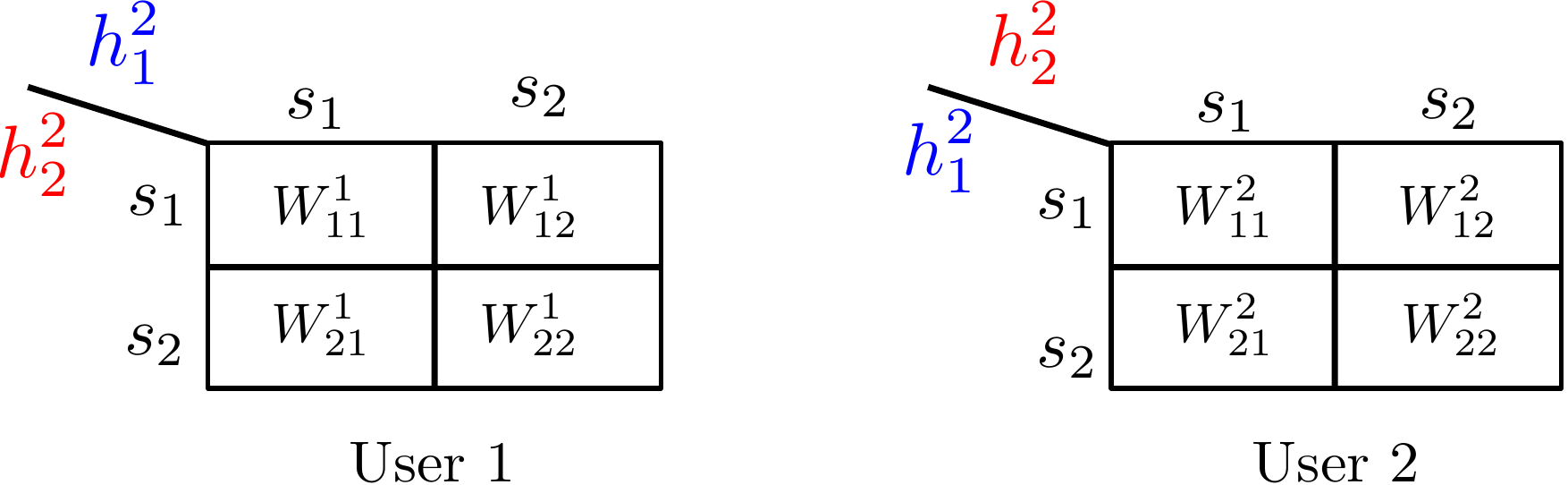}
\caption{Streaming and codebook assignments by user 1 and user 2.}
\label{fig:code} 
\end{figure}

In the approach that adapts the transmissions to the combined network states, each transmitter splits its message into {\em four} streams corresponding to the four possible combinations of the two channels. These codebooks for transmitter $i\in\{1,2\}$ are denoted by $\{W^i_{11},W^i_{12},W^i_{21},W^i_{22}\}$, where the information stream $W^i_{uv}$ is associated with the channel realization in which the channel gain of user $i$ is ${s}_v$, and the channel gain of the other user is ${s}_u$. These stream assignments are demonstrated in Fig.~\ref{fig:code}. The initial streams $\{W^1_{11},W^2_{11}\}$ account for the minimum amount of guaranteed information, which are adapted to the channel combination $(h^2_1,h^2_2)=({s}_1,{s}_1)$ and they should be decoded by all four possible channel combinations. When at least one of the channels is strong, the remaining codebooks are grouped and adapted to different channel realizations according to the assignments described in Fig.~\ref{fig:code}. Specifically: 
\begin{itemize}
\item The second group of streams $\{W^1_{12},W^2_{21}\}$ are reserved to be decoded in addition to  $\{W^1_{11},W^2_{11}\}$ when $h_1$ is strong, while $h_2$ is still weak. 
\item Alternatively, when $h_1$ is weak and $h_2$ is strong, instead the third group of streams, i.e., $\{W^1_{21},W^2_{12}\}$, are decoded.
\item Finally, when both channels are strong, in addition to all the previous streams, the fourth group $\{W^1_{22},W^2_{22}\}$ is also decoded. 
\end{itemize}

\begin{figure}[!t]
\centering
\captionsetup{justification=centering}
\includegraphics[height=2.2in]{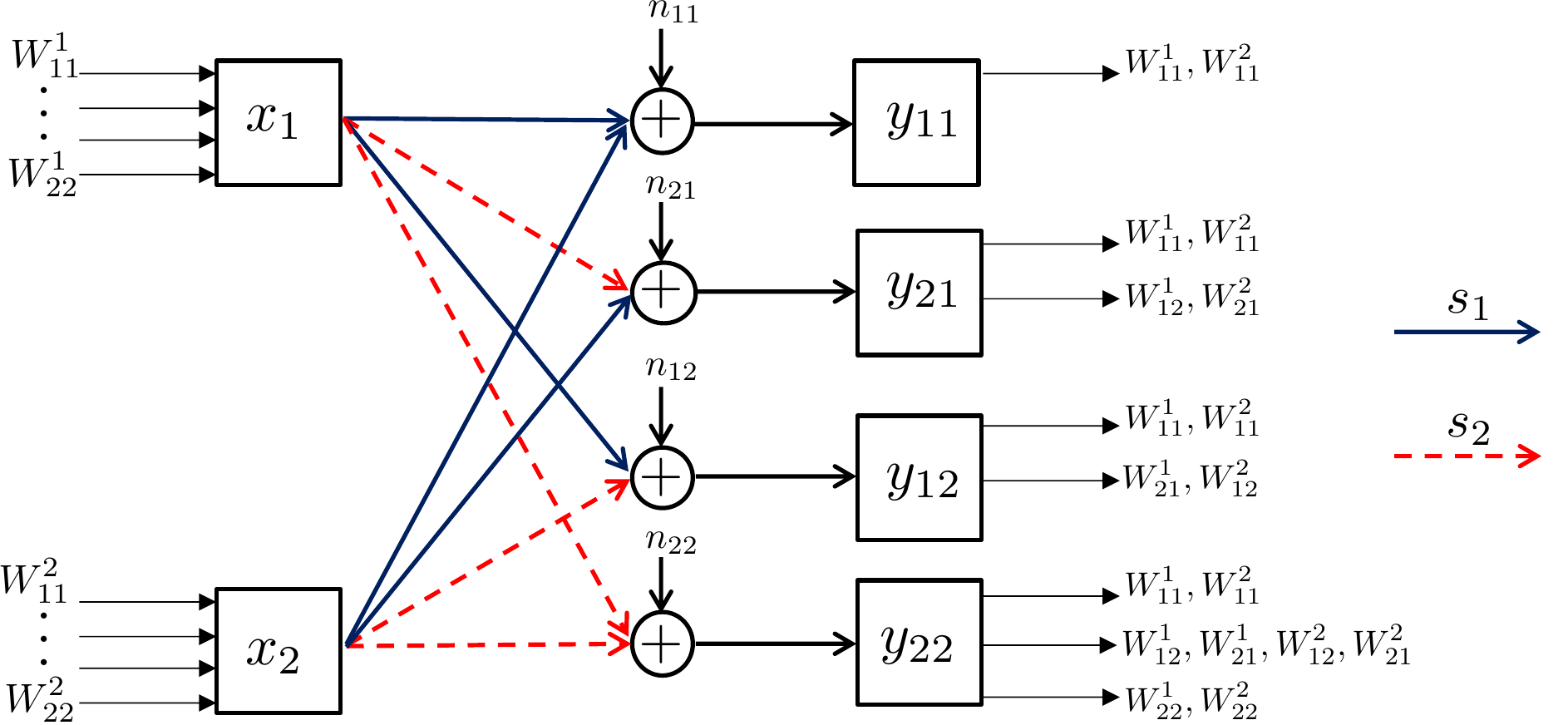}
\caption{Equivalent network when adapting the layers to the MAC (no CSIT).}
\label{fig:chapter3_network3}
\end{figure}

The order in which the codebooks are successively decoded in different network states is presented in Table~\ref{table:MAC}.  Based on this successive decoding order, channel gain state $({s}_1, {s}_1)$ is degraded with respect to all other states (i.e., the capacity region of the MAC corresponding to receiver $y_{11}$ is strictly smaller than those of the other three receivers), while $({s}_1, {s}_2)$ and $({s}_2,{s}_1)$ are degraded with respect to $({s}_2,{s}_2)$.  Clearly, the codebook assignment and successive decoding approach presented in Table~\ref{table:MAC} subsumes the one proposed in~\cite{Minero:ISIT07} presented in Table~\ref{table:tse}. In particular, Table~\ref{table:tse} can be recovered as a special case of Table~\ref{table:MAC} by setting the rates of the streams $\{W^1_{21},W^2_{21},W^1_{22},W^2_{22}\}$ to zero. The codebook assignment and decoding order discussed leads to the equivalent multiuser network with two inputs $\{x_1,x_2\}$ and four outputs $\{y_{11},y_{12},y_{21},y_{22}\}$, as depicted in Fig.~\ref{fig:chapter3_network3}.  Each receiver is designated to decode a pre-specified set of codebooks. 

\begin{table}[!h]
\renewcommand{\arraystretch}{1.2}
\caption{Successive decoding order of the streams adapted to the MAC}
\label{table:MAC}
\centering
\begin{tabular}{ |c||c|c|c|c| } 
 \hline
 $(h^2_1,h^2_2)$ & stage 1 & stage 2 & stage 3\\ 
 \hline\hline
 $({s}_1,{s}_1)$ & $W^1_{11},W^2_{11}$ & & \\ 
 \hline
 $({s}_2,{s}_1)$ & $W^1_{11},W^2_{11}$ & $W^1_{12}, W^2_{21}$ & \\ 
 \hline
 $({s}_1,{s}_2)$ & $W^1_{11},W^2_{11}$ & $W^1_{21}, W^2_{12}$ & \\ 
 \hline
 $({s}_2,{s}_2)$ & $W^1_{11},W^2_{11}$ & $W^1_{12}, W^2_{12},W^1_{21}, W^2_{21}$ & $W^1_{22},W^2_{22}$\\ 
 \hline
\end{tabular}
\end{table}

Next, we delineate the region of all achievable rates $R^i_{uv}$ for  $i,u,v\in\{1,2\}$, where $R^i_{uv}$ accounts for the rate of codebook $W^i_{uv}$. Define $\beta^i_{uv}\in[0,1]$ as the fraction of the power that transmitter $i$ allocates to stream $W^i_{uv}$ for $u\in\{1,2\}$ and $v\in\{1,2\}$, where we clearly have $\sum_{u=1}^2\sum_{v=1}^2\beta^i_{uv}=1$. For the simplicity in notations, and in order to place the emphasis on the interplay among the rates of different information streams, we focus on a symmetric setting in which relevant streams in different users have identical rates, i.e., rates of information streams $W^1_{uv}$ and $W^2_{uv}$, denoted by $R^1_{uv}$ and $R^2_{uv}$ respectively, are the same, and it is denoted by $R_{uv}$, i.e., $R_{uv}= R^1_{uv}=R^2_{uv}$.  

\begin{Theorem}[\cite{Tajer18}]
\label{theorem:achievable_rate2}
The achievable rate region of the rates $(R_{11}, R_{12}, R_{21}, R_{22})$ for the channel depicted in Fig.~\ref{fig:chapter3_network3} is the set of all rates satisfying:
\begin{align}
\label{R:11} R_{11} \; & \; \leq \; r_{11}\\
\label{R:12} R_{12} \; & \; \leq \; r_{12} \\
\label{R:21} R_{21} \; & \; \leq \;  r_{21}\\
\label{R:1221} R_{12}+R_{21} \; &\leq  \;  r_1 \\
\label{R:21221} 2R_{12}+R_{21} \; & \leq \;  r_{12}'\\
\label{R:12221} R_{12}+2R_{21} \; & \;  \leq \; r_{21}'\\
\label{R:22} R_{22} \; & \; \leq r_{22}\ ,
 \end{align}
over all possible power allocation factors  $\beta^i_{uv}\in[0,1]$ such that $\Sigma_{u=1}^2\Sigma_{v=1}^2\beta^i_{uv}=1$,  where by setting $\bar\beta_{uv}= 1-\beta_{uv}$ we have defined
\begin{align}
  r_{11} &  = \min \Big \{\frac{1}{2}\; C\big(2s_1 \beta_{11}, 2s_1\bar\beta_{11}\big) ,  C\big(s_1 \beta_{11},(s_1+s_2)\bar \beta_{11}\big)\;\Big\}\ , C\big(s_2 \beta_{12},s_1(\beta_{12}+\beta_{22})+s_2(\beta_{21}+\beta_{22}))\big) \Big\}\ ,\\
 r_{12}& = \min \Big \{\frac{1}{2}\; C\big(2\alpha_2 \beta_{12}, 2\alpha_2\beta_{22}\big)\; ,\; C\big(\alpha_2 \beta_{12},\alpha_1(\beta_{12}+\beta_{22})+\alpha_2(\beta_{21}+\beta_{22}))\big) \Big\}\ ,\\
 r_{21}& = \min \Big \{\frac{1}{2}\; C\big(2s_2 \beta_{21}, 2s_2\beta_{22}\big) \; , C\big(s_1 \beta_{21},s_1(\beta_{12}+\beta_{22})+s_2(\beta_{21}+\beta_{22})\big)\; \Big\}\ ,\\
\nonumber r_{1} & = \min \Big \{ \frac{1}{2}\; C\big(2s_2 (\beta_{12}+\beta_{21}), 2s_2\beta_{22}\big)
,   C\big(s_1\beta_{21}+s_2 \beta_{12},s_1(\beta_{12}+\beta_{22})+s_2(\beta_{21}+\beta_{22})\big)\Big\}\ ,\\
\label{r12_p} r_{12}' & = \; C\big ( s_2(2 \beta_{12}+\beta_{21}) \; ,\; 2s_2\beta_{22} \big)\ ,\\
\label{r21_p} r_{21}' & =  \; C\big (s_2(\beta_{12}+2\beta_{21})\; ,\; 2s_2\beta_{22}\big) \ ,\\ 
\label{r22} r_{22} & = \;  \frac{1}{2} C\big (2s_2\beta_{22}\; , \; 0\big)\ .
\end{align}
\end{Theorem}
\begin{proof}
The proof follows from the structure of the rate-splitting approach presented in Fig.~\ref{fig:code} and the decoding strategy presented in Table~\ref{table:MAC}. The detailed proof is provided in \cite[Appendix~B]{Tajer18}.
\end{proof}

In order to compare the achievable rate region in Theorem~\ref{theorem:achievable_rate2} and the capacity region presented in~Theorem~\ref{theorem:tse}, we group the information streams in the way that they are ordered and decoded in~\cite{Minero:ISIT07}. Specifically, the streams $\{W^1_{21},W^2_{21},W^1_{22},W^2_{22}\}$ are allocated zero power. Information streams $W_{11}^1$ and $W_{11}^2$ are adapted to the weak channels, and the information streams $W_{12}^2$ and $W_{12}^2$ are reserved to be decoded when one or both channels are strong. Information streams adapted to the {strong} channels are grouped, and their rates are aggregated, and those adapted to the weak channels are also groups, and their rates are aggregated. Based on this, the region presented in Theorem~\ref{theorem:achievable_rate2} can be used to form the sum-rates $R_w= (R^1_{11}+R^2_{11})$ and $R_s= (R^1_{12}+R^2_{12})$.

\begin{Theorem}[\cite{Tajer18}]\label{corollary:1}
By setting the power allocated to streams $\{W^1_{21},W^2_{21},W^1_{22},W^2_{22}\}$ to zero, the achievable rate region characterized by Theorem~\ref{theorem:achievable_rate2} reduces to the following region, which coincides with the capacity region characterized in \cite{Minero:ISIT07}.
\begin{align}
\label{eq:Rw} R_w\;&\; \leq \min\{a_3,a_6,a_9,a_4+a_8\}\ ,\\
\label{eq:Rs} \mbox{and}\quad R_s \;&\; \leq  C\left(  {{s}_2 \beta^1_{12} +{s}_2 \beta^2_{12} \; , \; 0} \right)\ ,
\end{align}
where we have defined
\begin{align}
a_3= & C\left(s_1(\beta^1_{11}+\beta^2_{11}) ,  s_1 (\bar\beta^1_{11}+\bar\beta^2_{11})\right)\ , \\
a_4 =&  C\left(s_1 \beta^1_{11}\;,\;s_1\bar\beta^1_{11}+s_2\bar\beta^2_{11}\right)\ ,\\
a_6= &
C\left(s_1\beta^1_{11}+s_2\beta^2_{11}\; , \; s_1 \bar\beta^1_{11}+s_2\bar\beta^2_{11}\right)\ , \\
a_8 = & C\left(s_1 \beta^2_{11}\; ,\; s_2 \bar\beta^1_{11}+s_1\bar\beta^2_{11}\right)\ , \\
a_9= & C\left(s_2\beta^1_{11}+s_1\beta^2_{11}\; , \; s_2 \bar\beta^1_{11}+s_1\bar\beta^2_{11}\right) \ .
\end{align}
\end{Theorem}
\begin{proof}
See \cite[Appendix D]{Tajer18}.  
\end{proof}

 \begin{figure}[t]
\centering
\includegraphics[width=5in]{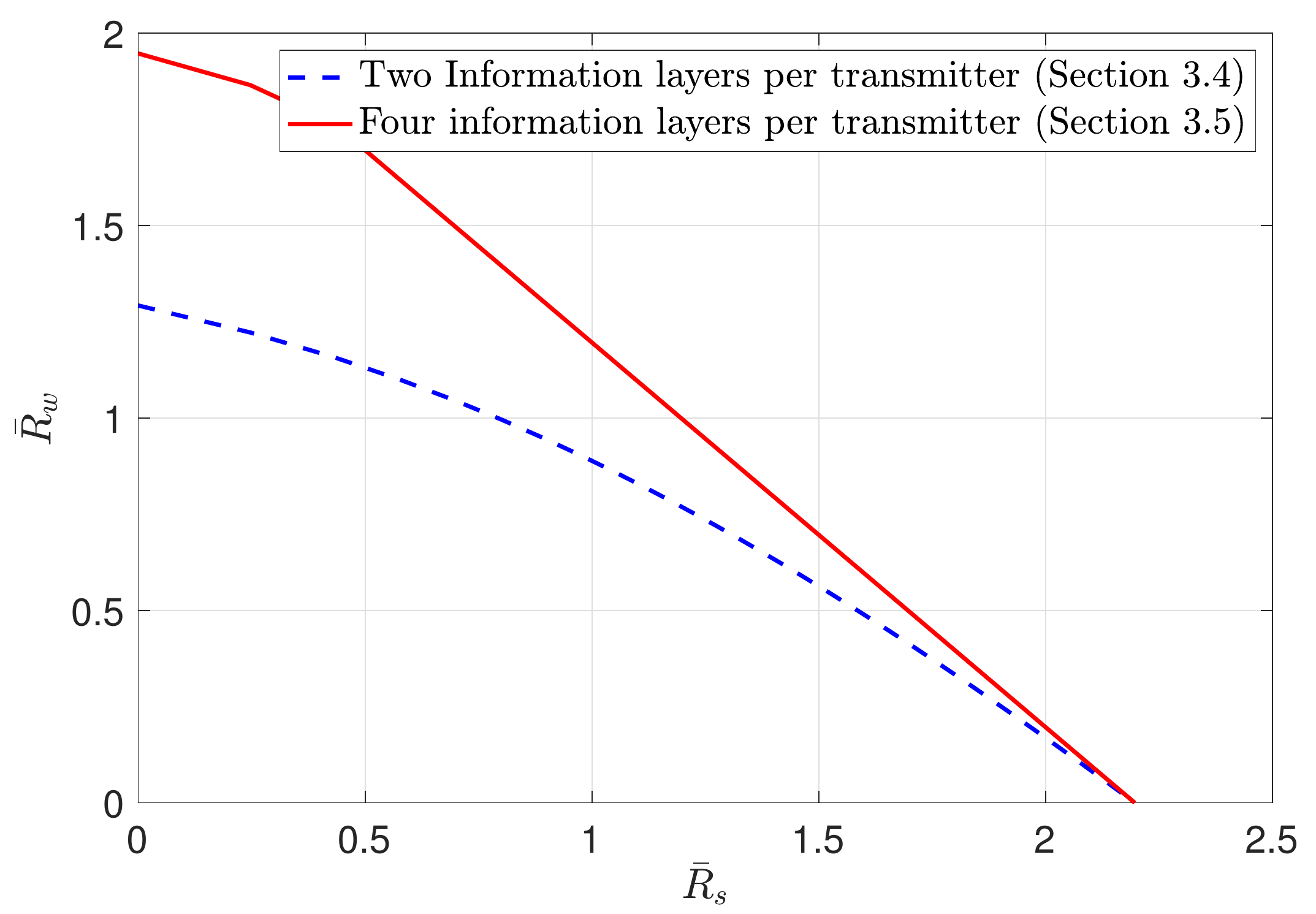}
\caption{Comparison of the capacity region in Section~\ref{sec:CSIR_SU} and achievable rate region in Section~\ref{sec:CSIR_MU} demonstrating the trade-off between $R_s$ and $R_w$, and $\bar R_s$ and $\bar R_w$. Transmission SNR is 10, and the channel gains are $(\sqrt{{s}_1}, \sqrt{{s}_2}) = (0.5, 1)$.}
\label{fig:region_snr_10}
\end{figure}

Figure~\ref{fig:region_snr_10} quantifies and compares the achievable rate region characterized in Theorem~\ref{theorem:achievable_rate2} and Theorem~\ref{corollary:1} with the capacity region characterized in Theorem~\ref{theorem:tse}.  The regions presented in theorems~\ref{theorem:achievable_rate2} and  \ref{corollary:1}  capture the interplay among the rates of the individual codebooks and the capacity region of Theorem~\ref{theorem:tse} characterize the trade-off between the sum-rates of the information streams adapted to the weak and strong channels. To have a common ground for comparisons, the result of theorems~\ref{theorem:achievable_rate2} and  \ref{corollary:1} can be presented to signify the codebooks of the weak and strong channel states. Recall that earlier we defined the sum-rates
\begin{align}
\label{eq:Rsw2}
R_w = R^1_{11}+R^2_{11} \ ,\quad  \mbox{and}\quad R_s  = R^1_{12}+R^2_{12}\ .
\end{align}
Accordingly, for the coding scheme (Table~\ref{table:MAC}) we define
\begin{align}
\label{eq:Rsw4}
\bar R_w & = R^1_{11}+R^2_{11}+R_{21}^1+R_{21}^2+R_{12}^1+R_{12}^2 \ , \\  \quad  \mbox{and}\quad \bar R_s &  = R^1_{22}+R^2_{22}\ .
\end{align}
Based on these definitions, Fig.~\ref{fig:region_snr_10} demonstrates the regions described by $(R_w,R_s)$ and $(\bar R_w,\bar R_s)$, in which the transmission SNR is 10, the channel coefficients are $(\sqrt{{s}_1},\sqrt{{s}_2})=(0.5,1)$, and the regions are optimized over all possible power allocation ratios.  The  numerical evaluation in Fig.~\ref{fig:region_snr_10} depict that the achievable rate region in Theorem~\ref{theorem:achievable_rate2} subsumes that of Theorem~\ref{corollary:1} (and subsequently, that of~\ref{theorem:tse}), and the gap between the two regions diminishes as the rates of the information layers adapted to the strong channels increases, i.e., $R_s$ and $\bar R_s$ increase. Next, in order to assess the tightness of the achievable rate regions, we present an outer bound on the capacity region of the network in Fig.~\ref{fig:chapter3_network3}.

\begin{figure}[t]
\centering
\includegraphics[width=5in]{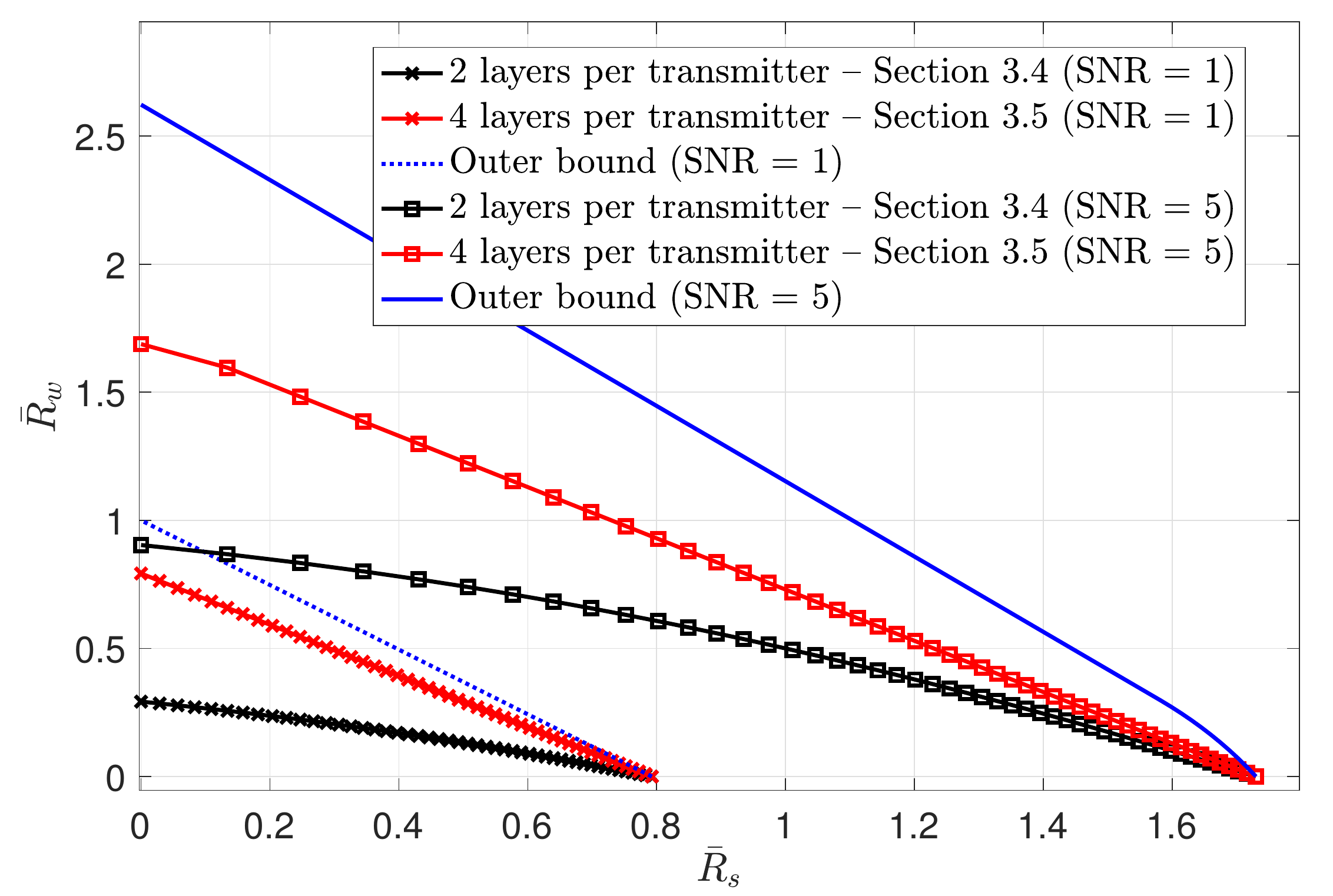}
\caption{Comparison of the capacity region in Section~\ref{sec:CSIR_SU} and achievable rate region and outer bounds in Section~\ref{sec:CSIR_MU}.}
\label{fig:outer_bound2}
\end{figure}

\begin{Theorem}[\cite{Tajer18}]
\label{theorem:outer_bound2}
An outer bound for the capacity region of the rates $(R_{11}, R_{12}, R_{21}, R_{22})$ for the channel depicted in Fig.~\ref{fig:chapter3_network3} is the set of all rates satisfying:
\begin{align*}
R_{11} \leq \frac{1}{2}a_{3} \; , \; R_{12}  \leq  \frac{1}{2}a_{24} \; , \; R_{21}  \leq \frac{1}{2}a_{27} \; , \; R_{22} \leq r_{22}\ ,
\end{align*}
where we have defined
\begin{align}
\label{R'_1_12_2_12}  a_{24} &  =
C\left(  {s_2 \beta^1_{12} +s_2 \beta^2_{12} \; , \;s_2 \beta^1_{22} +s_2 \beta^2_{22}  } \right)\ ,\\
\label{R'_1_21_2_21} a_{27}  & = C\left(  {s_2 \beta^1_{21} +s_2 \beta^2_{21} \; , \;s_2 \beta^1_{22} +s_2 \beta^2_{22}  } \right)\ ,\\
r_{22} & =\;  \frac{1}{2} C\big (2s_2\beta_{22}\; , \; 0\big)\ .
\end{align}
\end{Theorem}

Figure~\ref{fig:outer_bound2} compares the outer bound specified in Theorem~\ref{theorem:outer_bound2} and the achievable rate region presented in Theorem~\ref{theorem:achievable_rate2} for SNR values 1 and 5, and the choice of $(\sqrt{s_1},\sqrt{s_2}) = (0.5,1)$. Corresponding to each SNR, this figure illustrates the capacity region obtained in Theorem~\ref{theorem:tse}, as well as the achievable rate region and the outer bound reviewed in this section.  

\begin{figure}[t]
  \centering
  \includegraphics[width=5in]{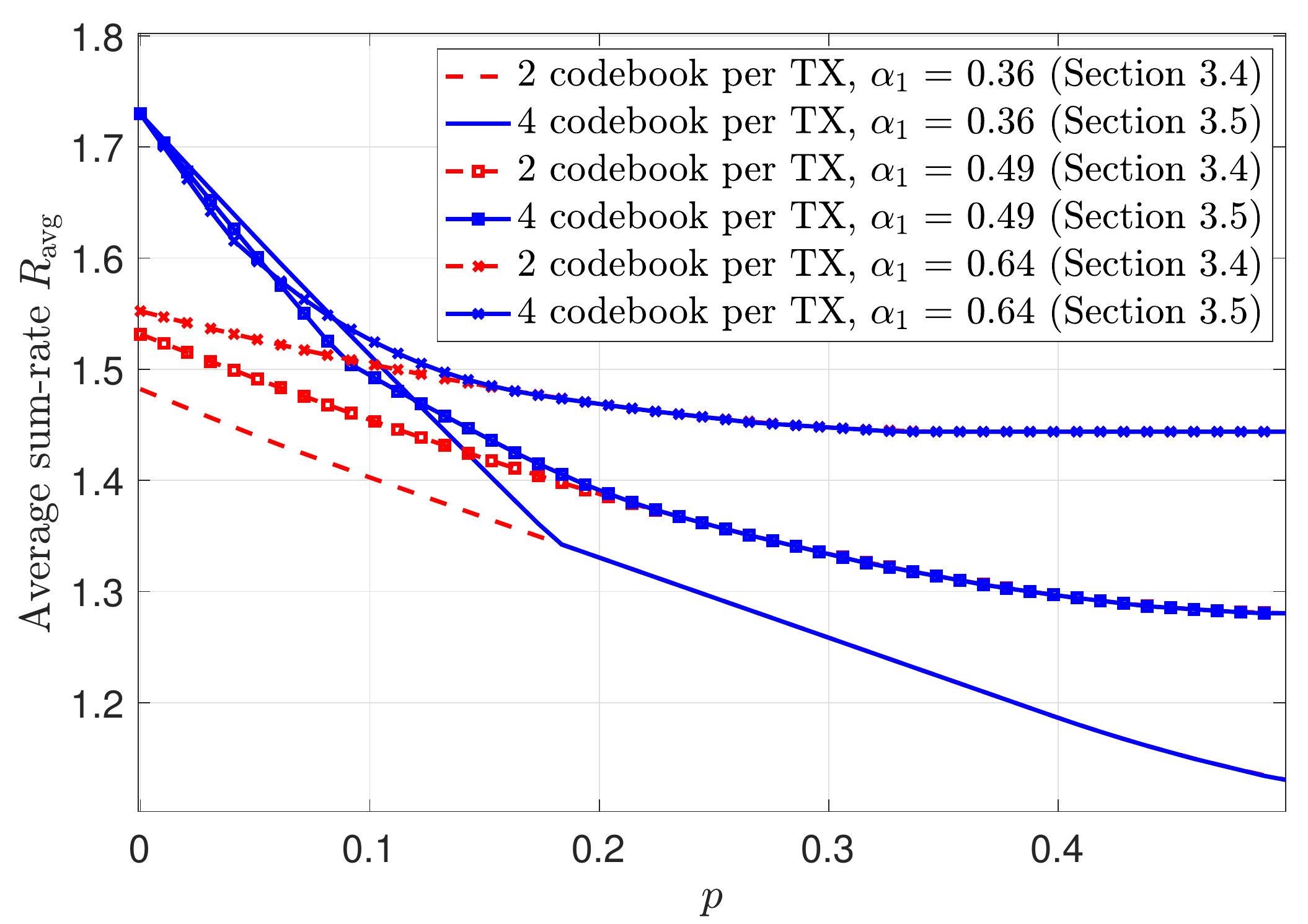}
  \caption{Average sum-rate versus $p$ for different values of ${s}_1$ (${s}_2=1$ and SNR=5).}
  \label{fig:rate_prob}
\end{figure}

To evaluate the average rate as a long-term relevant proper measure capturing the expected rate over a large number of transmission cycles, where each cycle undergoes an independent fading realization. Consider a symmetric channel, in which the corresponding information streams are allocated identical power and have the same rate, and set $R_{uv} =  R^1_{uv}=R^2_{uv}$ for $u,v\in\{1,2\}$. Also, consider a symmetric distribution for $h_1$ and $h_2$ such that  $\mathbb{P}(h_1^2={s}_i)=\mathbb{P}(h_2^2={s}_i)$ for $i\in\{1,2\}$, and define $p= \mathbb{P}(h_1^2={s}_1)=\mathbb{P}(h_2^2={s}_1)$. By leveraging the stochastic model of the fading process, the average rate is
\begin{align}\label{eq:Rave}
R_{\rm ave} = 2[R_{11} + (1-p)(R_{12}+R_{21})+(1-p)^2R_{22}]\ .
\end{align}
Figure~\ref{fig:rate_prob} depicts the variations of the average sum-rate versus $p$ for different values of ${s}_1$. The observations from this figure also confirm that higher gain levels are exhibited as $p$ decreases. It is noteworthy that the results from Fig.~\ref{fig:region_snr_10} validates the observations from  Fig.~\ref{fig:rate_prob} that improvement in average rate is significant when the probability of encountering a weak channel state is low since the rate distribution considered in the achievable rate region comparison will correspond to average rate if the probability of observing ${s}_1$ is zero.

\subsection{MAC without CSIT -- Multi-state Channels: State-dependent Layering}
\label{sec:CSIR_MU_multi}

The idea of adapting the transmission to the combined network states can be extended to devise codebook assignment and decoding strategy schemes for the general multiple-state channel. Similarly to the two-state channel, in the $\ell$-state channel model, $\ell^2$ codebooks are assigned to each transmitter. Hence,  corresponding to the combined channel state $(h_1^2,h_2^2)=({s}_q,{s}_p)$  codebook $W^1_{pq}$ is assigned to transmitter~1 and codebook $W^2_{qp}$ is assigned to transmitter~2. By following the same line of analysis as in the two-state channel, the network state $(h_1^2,h_2^2)=({s}_1,{s}_1)$ can be readily verified to be degraded with respect to states $({s}_1,{s}_2)$, $({s}_2,{s}_1)$, and $({s}_2,{s}_2)$ when ${s}_2>{s}_1$. Additionally, channel combinations $({s}_1,{s}_2)$ and $({s}_2,{s}_1)$ are also degraded with respect to state $({s}_2,{s}_2)$. When a particular transmitter's channel becomes stronger while the interfering channel remains constant, the transmitter affords to decode additional codebooks. Similarly, when a transmitter's own channel remains constant while the interfering channel becomes stronger, the transmitter can decode additional layers. This can be facilitated by decoding and removing the interfering transmitter's message, based on which the transmitter experiences reduced interference. Based on these observations, by ordering the different realizations of $h_1$ and $h_2$ in the ascending order and determining their relative degradedness, a successive decoding strategy is illustrated in Table~\ref{table_3}. In this table $A_{p,q}$ denotes the cell in the $p^{\rm th}$ row and the $q^{\rm th}$ column, and it specifies the set of codebooks $\mathcal{U}_{pq}$ to be decoded when the combined channel state is $(h_1^2,h_2^2)=({s}_q,{s}_p)$. In this table, the codebooks set to be decoded in each possible combined state is recursively related to the codebooks decoded in the weaker channels. Specifically, the state corresponding to $A_{p-1,q-1}$ is degraded with respect to states $A_{p,q-1}$ and $A_{p-1,q}$. Therefore, in the state $A_{p,q}$, the receiver decodes all streams from states $A_{p-1,q-1}$ (included in ${\cal U}_{p-1,q-1}$), $A_{p,q-1}$ (included in ${\cal U}_{p,q-1}$), and $A_{p-1,q}$ (included in ${\cal U}_{p-1,q}$). Subsequently, these are followed by decoding one additional stream from each user denoted by $W^1_{pq}$ and $W^2_{qp}$. 
When both channel coefficients have the strongest possible realizations, all the streams from both users will be decoded at the receiver.

\setlength\extrarowheight{2pt}
\def\arraystretch{1.2}
\begin{table*}[t]
\caption{Successive decoding order for the $\ell$-state MAC.}
\label{table_3}
{\footnotesize
\hfill{}
\begin{tabular}{|c||c|c|c|c|c|c|}
\hline
\diagbox{\textcolor{red}{$h^2_2$}}{\textcolor{blue}{$h^2_1$}}& ${s}_1$ & ${s}_2$ & $\ldotp\ldotp$ &${{s}_q}$ & $\ldotp\ldotp$ & ${s}_\ell$\\
\hline\hline
${s}_1$ 
&  \begin{tabular}[t]{@{}c@{}}\\ \textcolor{blue}{$W^1_{11}$} , \textcolor{red}{$W^2_{11}$}\end{tabular} 
& \begin{tabular}[t]{@{}c@{}}$\mathcal{U}_{11}$\\ \textcolor{blue}{$W_{12}^1$} , \textcolor{red}{$W_{21}^2$}\end{tabular}
& $\ldotp\ldotp$
& $\ldotp$
& $\ldotp\ldotp$
& \begin{tabular}[t]{@{}c@{}}$\mathcal{U}_{1(\ell-1)}$\\ \textcolor{blue}{$W_{1\ell}^1$} , \textcolor{red}{$W_{\ell 1}^2$}\end{tabular}\\
\hline
${s}_2$ 
& \begin{tabular}[t]{@{}c@{}}$\mathcal{U}_{11}$\\ \textcolor{blue}{$W_{21}^1$} , \textcolor{red}{$W_{12}^2$}\end{tabular}
& \begin{tabular}[t]{@{}c@{}}$\mathcal{U}_{11}$ , $\mathcal{U}_{12}$ , $ \mathcal{U}_{21} $\\\textcolor{blue}{$W_{22}^1$} , \textcolor{red}{$W_{22}^2$}\end{tabular} 
&$\ldotp\ldotp$
&$\ldotp$
&$\ldotp\ldotp$
& \begin{tabular}[t]{@{}c@{}}$\mathcal{U}_{1(\ell-1)} \; , \; \mathcal{U}_{2(\ell-1)} \; , \; \mathcal{U}_{1l}$\\ \textcolor{blue}{$W_{2l}^1$} , \textcolor{red}{$W_{l2}^2$}\end{tabular}\\ 
\hline 
$\ldotp$&$\ldotp$&$\ldotp$&$\ldotp\ldotp$&$\ldotp$ & $\ldotp\ldotp$&$\ldotp$\\
\hline
${s}_{p}$ 
& $\ldotp$
& $\ldotp$
& $\ldotp\ldotp$
& \begin{tabular}[t]{@{}c@{}}$\mathcal{U}_{(p-1)(q-1)} , \mathcal{U}_{p(q-1)} , \mathcal{U}_{(p-1)q},$\\ \textcolor{blue}{$W_{pq}^1$} , \textcolor{red}{$W_{qp}^2$}\end{tabular}
& $\ldotp\ldotp$
& $\ldotp$\\
\hline 
$\ldotp$&$\ldotp$&$\ldotp$&$\ldotp\ldotp$&$\ldotp$ &$\ldotp\ldotp$&$\ldotp$\\
\hline
${s}_{\ell}$ 
& \begin{tabular}[t]{@{}c@{}}$\mathcal{U}_{(\ell-1)1}$ \\ \textcolor{blue}{$W_{\ell 1}^1$} , \textcolor{red}{$W_{1 \ell}^2$}\end{tabular} 
& \begin{tabular}[t]{@{}c@{}}$\mathcal{U}_{(\ell-1)1},\mathcal{U}_{\ell 1},\mathcal{U}_{(\ell-1)2},$\\\textcolor{blue}{$W_{\ell 2}^1$},\textcolor{red}{$W_{2 \ell}^2$}\end{tabular} 
& $\ldotp\ldotp$
& $\ldotp$
& $\ldotp\ldotp$ 
& \begin{tabular}[t]{@{}c@{}}$\mathcal{U}_{(\ell-1)(\ell-1)}\; , \;\mathcal{U}_{\ell(\ell-1)}\; , \;\mathcal{U}_{(\ell-1)\ell}$\\ \textcolor{blue}{$W_{\ell\ell}^1$} , \textcolor{red}{$W_{\ell\ell}^2$}\end{tabular}\\ 
\hline
\end{tabular}}
\hfill{}
\end{table*}

Next,  the rate region achieved in presented in Theorem~\ref{theorem:achievable_rate_finite} for the general multi-state channel. It can be verified that the region characterized by Theorem~\ref{theorem:achievable_rate2} is subsumed by this general rate region. Similarly to the two-state channel settings, define $R^i_{uv}$ as the rate of codebook $W^i_{uv}$ for $i\in\{1,2\}$ and $u,v\in\{1,\dots,\ell\}$. Furthermore,  define $\beta_{uv}\in[0,1]$ as the fraction of the power allocated to the codebook $W^i_{uv}$, where $\sum_{u=1}^\ell\sum_{v=1}^\ell\beta_{uv}=1$. For the simplicity in notations and for emphasizing the interplay among the rates, we focus on the symmetric case in which $R_{uv}= R^1_{uv}=R^2_{uv}$.
\begin{Theorem}[\cite{Tajer18}]
\label{theorem:achievable_rate_finite}
A region of simultaneously achievable rates $$\{R_{uv}: u< v \;\; \mbox{and} \;\; u,v \in\{1,\dots,\ell\}\}$$ for an $\ell$-state two-user multiple access channel is characterized as the set of all rates satisfying:
\begin{align}
\label{R:uv} R_{uv} & \leq  \min \left\{b_1(u,v),b_2(u,v),\frac{b_3(u,v)}{2}\right \}\\
\label{R:vu} R_{vu} & \leq   \min \left\{ b_4(u,v),\frac{b_5(u,v)}{2} \right \}\\
\label{R:{vv}} R_{uv}+R_{vu} & \leq   \min \left\{b_6(u,v),b_7(u,v),\frac{b_8(u,v)}{2} \right \}\\
\label{R:5} 2R_{uv}+R_{vu} & \leq b_9(u,v)\\
\label{R:6} R_{uv}+2R_{vu} & \leq b_{10}(u,v)\\
\label{R_7} R_{uu} & \leq \min\left\{b_{11}(u),\frac{b_{12}(u)}{2}\right\}\ ,
\end{align}
where constants $\{b_i:i\in\{1,\dots,12\}\}$ are specified in Appendix~\ref{app:constants:theorem:achievable_rate_finite}.
\end{Theorem}

\subsection{MAC with Local CSIT -- Two-state Channels: Fixed Layering}
\label{sec:LCSIT_fixed}
Next, we turn to the setting in which the transmitters have {\sl local} CSI. Specifically, each channel randomly takes one of a finite number of states, and each transmitter only knows the state of its direct channel to the receiver {\sl perfectly}, along with the probability distribution of the state of the other transmitter's channel. This model was first studied in~\cite{Zou13}, in which a single-user broadcast approach is directly applied to the MAC. In this approach,  each transmitter generates two coded layers, where each layer is adapted to one of the states of the channel linking the other transmitter to its receiver. This transmission approach is followed by successive decoding at the receiver in which there exists a pre-specified order of decoding of the information layers.

This scheme assigns codebooks based on channels' strengths such that it reserves one additional information layer as the channel state gets stronger. In this scheme, the number of transmitted layers and the decoding order are fixed and independent of the actual channel state. In the two-state channel model, when a transmitter $i$ experiences the channel state ${s}_m$, it splits its message to two information layers via two independent codebooks denoted by $T^i_{m1}$ and $T^i_{m2}$. The rate of layer $T^i_{m1}$ is adapted to the {\sl  weak} channel state of the other user while the rate of layer $T^i_{m2}$ is adapted to the {\sl  strong} channel state. Thus, each transmitter encodes its information stream by two layers and adapts the power distribution between them according to its channel state. Subsequently, the receiver implements a successive decoding scheme according to which it decodes one layer from transmitter~$1$ followed by one layer from transmitter~ $2$, and then the remaining layer of transmitter~$1$, and finally the remaining layer of transmitter~$2$. This order is pre-fixed and is used in all channel states. This scheme is summarized in Table~\ref{table:Scheme_18}.

\begin{table}[!h]
\renewcommand{\arraystretch}{1}
\caption{Successive decoding scheme in \cite{Zou13}}
\label{table:Scheme_18}
\centering
\begin{tabular}{ |c||c|c|c|c| } 
\hline
 $(h_1,h_2)$ & stage 1 & stage 2 & stage 3 & stage 4 \\ 
\hline\hline
$(s_1,s_1)$ & $T^1_{11}$ & $T^2_{11}$ & $T^1_{12}$ & $T^2_{12}$\\  
\hline
$(s_1,s_2)$ & $T^1_{11}$ & $T^2_{21}$ & $T^1_{12}$ & $T^2_{22}$\\ 
\hline
$(s_2,s_1)$ & $T^1_{21}$ & $T^2_{11}$ & $T^1_{22}$ & $T^2_{12}$\\
\hline
$(s_2,s_2)$ & $T^1_{21}$ & $T^2_{21}$ & $T^1_{22}$ & $T^2_{22}$\\
\hline
\end{tabular}
\end{table}

The following theorem characterizes an outer bound on the average rate region. For this purpose, define $R_i(h_1,h_2)$ as the rate of transmitter $i$ for the state pair $(h_1,h_2)$. Accordingly,  define $\bar R_i= \mathbb{E}_{h_1,h_2}[R_i(h_1,h_2)]$ as the {\sl average} rate of transmitter $i$, where the expected value is with respect to the distributions of $h_1$ and $h_2$.
\begin{Theorem}[\cite{Zou13}] When the transmitters have local CSIT, an outer bound on the expected capacity region contains rates $(\bar R_1,\bar R_2)$ satisfying
\begin{align}
\bar R_1 & \leq q_1C(s_1,0)+(1-q_1)C(s_2,0)\\
\bar R_1 & \leq q_2C(s_1,0)+(1-q_2)C(s_2,0)\\
\bar R_1 +\bar R_2 & \leq q_1q_2 C(2s_1,0)+(q_1+q_2-2q_1q_2)C(s_1+s_2,0)+(1-q_1)(1-q_2)C(2s_2,0)\ .
\end{align}
\end{Theorem}

\begin{figure}[!t]
\centering
\captionsetup{justification=centering}
\includegraphics[height=2.2in]{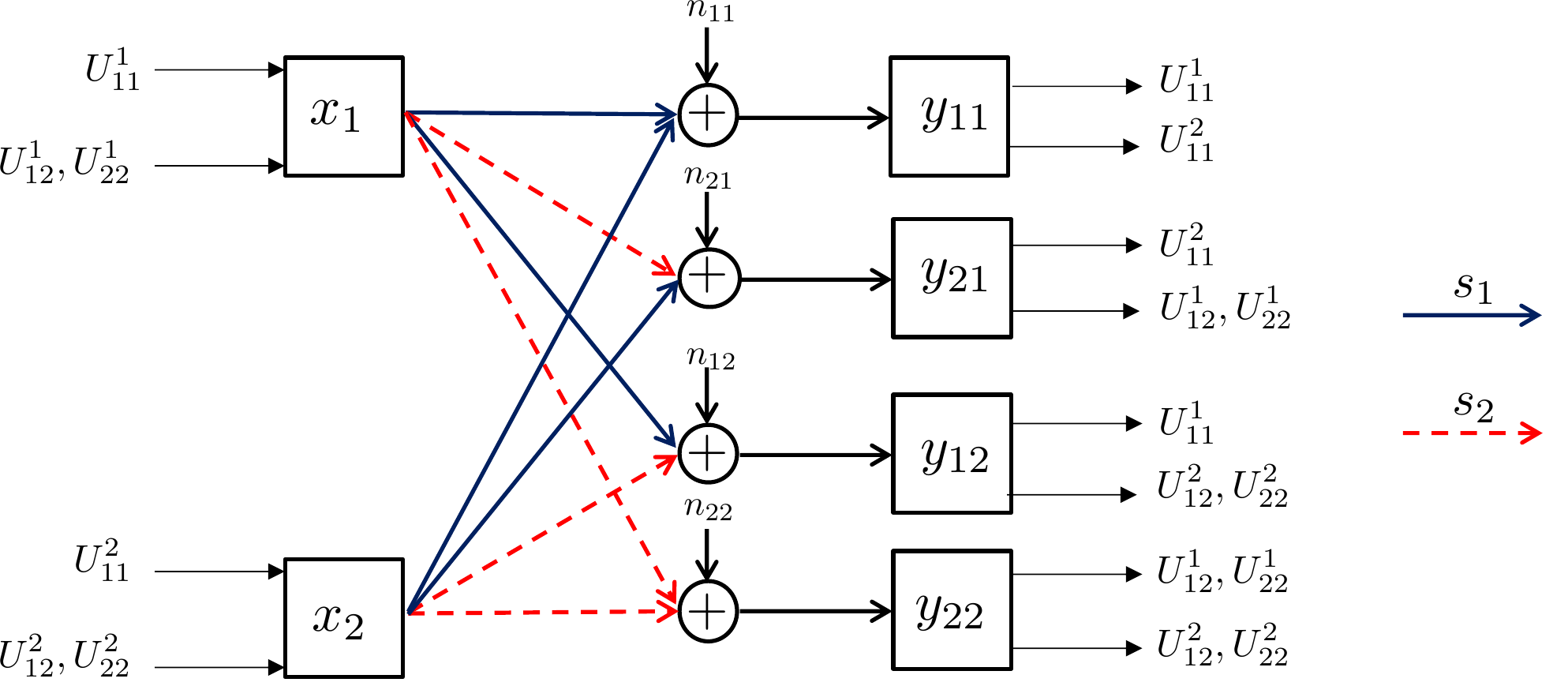}
\caption{Equivalent network for the two-user MAC (local CSIT).}
\label{fig:network2}
\end{figure}

\newpage

\subsection{MAC with Local CSIT -- Two-State Channels: State-dependent Layering}
\label{sec:LCSIT_MU}

Next, we present another scheme for the MAC with local CSIT that generalizes the scheme of Section~\ref{sec:LCSIT_fixed} via adapting information layering to the combined states of the channel. The underlying motivation guiding this generalization is that we need to account for both the direct and interfering roles that each transmitter plays. Hence, the transmission rates of different layers should be adapted to the combined state of the entire network. The major difference between this approach and that in  Section~\ref{sec:LCSIT_fixed} is that this scheme relies on the available local CSIT available to the individual transmitters such that each transmitter adapts its layers and their associated raters to the instantaneous state of the channel. This facilitates opportunistically sustaining higher rates.

\begin{figure}[h]
\centering
\includegraphics[width =4in]{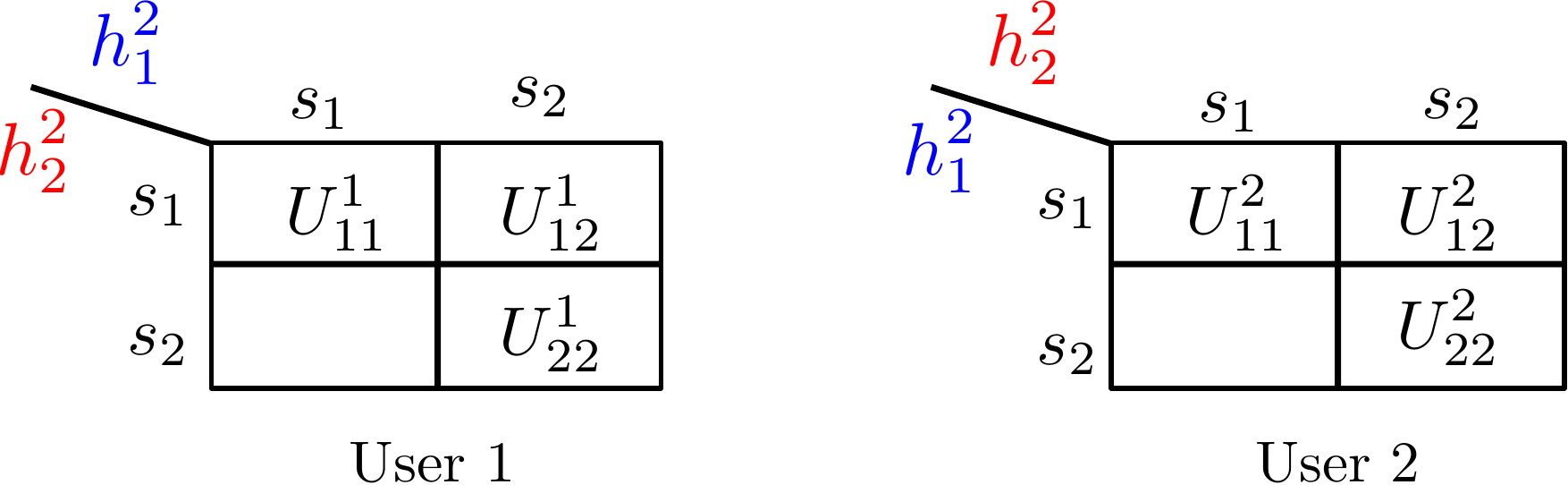}
\caption{Layering and codebook assignments.}
\label{fig:code_partial}
\end{figure}

\vspace{.1 in}
\noindent{\bf State-dependent Layering.} In this approach, each transmitter, depending on the instantaneous state of the local CSI available to it, splits its message into independent information layers. Formally, when transmitter $i \in \{1,2\}$ is in the {\sl  weak} state, it encodes its message by only one layer, which we denote by $U^i_{11}$. On the other contrary, when transmitter $i \in \{1,2\}$ is in the {\sl  strong} state, it divides its message into two information layers, which we denote by $U^i_{12}$, and $U^i_{22}$. Hence, transmitter $i$ adapts the codebook $U^i_{12}$ (or $U^i_{22}$) to the state in which the other transmitter  experiences a {\sl  weak} (or {\sl  strong}) channel. A summary of the layering scheme and the assignment of the codebooks to different network states is provided  in Fig.~\ref{fig:code_partial}. In this table, the cell associated with the state $({s}_m,{s}_n)$ for $m,n \in \{1,2\}$ specifies the codebook adapted to this state.

\vspace{.1 in}
\noindent{\bf Decoding Scheme.} A successive decoding scheme is designed based on the premise that as the combined channel state becomes stronger, more layers are decoded. Based on this, one the total number of codebooks decoded increases as one of the two channels becomes stronger. In this decoding scheme, the combination of codebooks decoded in different states is as follows (and it is summarizes in Table~\ref{table:decode_partial}):
\begin{itemize} 
\setlength\itemsep{0em}
\item {\bf State $({s}_1,{s}_1)$:} In this state, both transmitters experience {\sl  weak} states, and they generate codebooks $\{U^1_{11}, U^2_{11}\}$ according to Fig.~\ref{fig:code_partial}. In this state, the receiver jointly decodes the baseline layers $U^1_{11}$ and $U^2_{11}$.
\item {\bf State $({s}_2,{s}_1)$:} When the channel of transmitter~1 is {\sl  strong} and the channel of transmitter~2 is {\sl  weak}, three codebooks $\{U^1_{12}, U^1_{22}, U^2_{11}\}$ are generated and transmitted. As specified by Table~\ref{table:decode_partial}, the receiver jointly decodes  $\{U^{1}_{12},U^{2}_{11}\}$. This is followed by decoding the remaining codebook, i.e.,  $U^1_{22}$.
\item {\bf State $({s}_1,{s}_2)$:} In this state, codebook generation and decoding are  similar to those in the state $({s}_2, {s}_1)$, except that the roles of transmitters 1 and 2 are interchanged.
\item {\bf State $({s}_2,{s}_2)$:} Finally, when both transmitters experience {\sl  strong} channels, the receiver decodes four codebooks  in the order specified by the last row of Table~\ref{table:decode_partial}. Specifically, the receiver first jointly decodes the baseline layers $\{U^1_{12},U^2_{12}\}$, followed by jointly decoding the remaining codebooks $\{U^1_{22},U^2_{22}\}$.
\end{itemize}

\begin{table}[h]
\centering
\begin{tabular}{|c||c|c|} 
\hline
 $(h^2_1,h^2_2)$ &  stage 1 &  stage 2 \\ 
\hline\hline
$({s}_1,{s}_1)$ & $U^1_{11},U^2_{11}$ & \\ 
\hline
$({s}_2,{s}_1)$ & $U^1_{12},U^2_{11}$ & $U^1_{22}$ \\ 
\hline
$({s}_1,{s}_2)$ & $U^1_{11},U^2_{12}$ & $U^2_{22}$ \\ 
\hline
$({s}_2,{s}_2)$ & $U^1_{12},U^2_{12}$ & $U^1_{22}, U^2_{22}$ \\ 
\hline
\end{tabular}
\caption{Decoding scheme}
\label{table:decode_partial}
\end{table}

\begin{figure}[!h]
\centering
\includegraphics[width=5in]{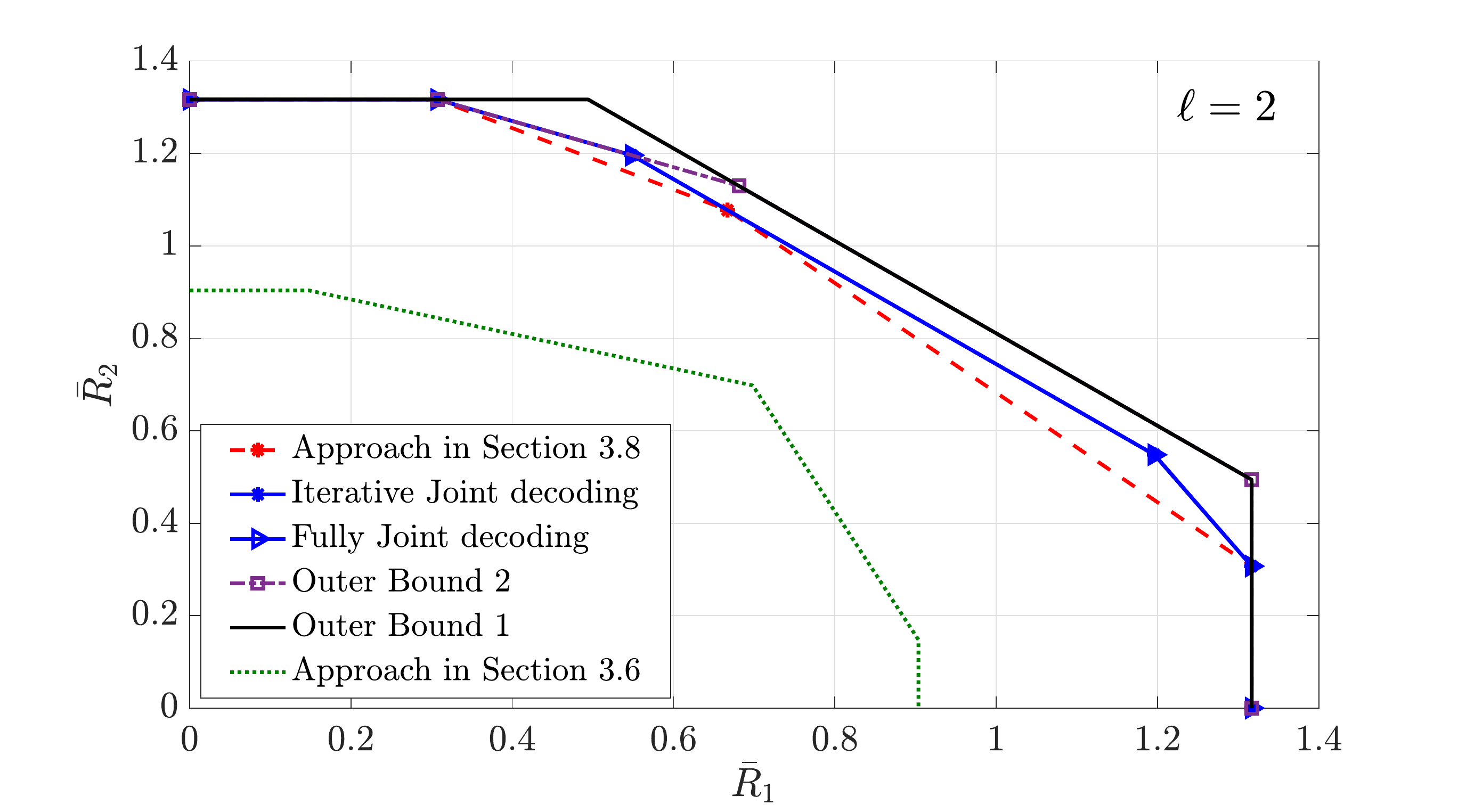} 
\caption{Average rate regions for $\ell = 2$.}
\label{fig:achievable_region_snr10}
\end{figure}

Compared to the setting without any CSIT at the transmitter (i.e., the setting discussed in Section~\ref{sec:CSIR_MU}), the key difference is that the transmitters have distinct transmission strategies when they are experiencing different channel states. Specifically, each transmitter dynamically chooses its layering scheme based on the instantaneous channel state known to it. Furthermore, the major difference with the scheme of Section~\ref{sec:LCSIT_fixed} is that this scheme adapts the number of encoded layers  proportionately to the strength of the combined channel state. Such adaptation of the number of encoded layers results in two advantages. The first one is that adapting the number of layers leads to overall fewer information layers to be generated and transmitted. This, in turn, results in decoding overall fewer codebooks and reduced decoding complexity. The second advantage pertains to providing the receiver with the flexibility to vary the decoding order according to the combined channel state. This allows for a higher degree of freedom in optimizing power allocation, and subsequently, larger achievable rate regions.
In support of these observations, the numerical evaluations in Fig.~\ref{fig:achievable_region_snr10}, the achievable rate region subsumes that of Section~\ref{sec:LCSIT_fixed}. Furthermore, as the number of channel states increases, the sum-rate gap between these two schemes becomes more significant. Finally, depending on the actual channel state, the scheme in this section decodes between $2$ and $\frac{\ell(\ell + 1)}{2}$ codebooks, whereas the scheme of Section~\ref{sec:LCSIT_fixed} always decodes $\ell^2$ codebooks.

It is noteworthy that when in the two-state channel model of Fig.~\ref{fig:network2} the channel states are ${s}_1 = 0$ and ${s}_2=1$, this model simplifies to the two-user random access channel investigated in~Section~\ref{sec:CSIR_SU}. In this special case, reserving one codebook to be decoded exclusively in each of the interference-free states, i.e., $({s}_1, {s}_2)$ and $({s}_2, {s}_1)$, enlarges the achievable rate region. Hence, it is beneficial in this special case to treat codebooks $(U^1_{22}, U^2_{22})$ as interference whenever both users are active, i.e., when the channel state is $({s}_2,{s}_2)$. In general, however, when the channel gain $s_1$ is non-zero, i.e., ${s}_1 > 0$, reserving two codebooks to be decoded exclusively in these two channel states limits the average achievable rate region. 

\vspace{.1 in}
\noindent {\bf Achievable Rate Region.} Next, we provide an inner bound  on the average capacity region. Recall that the {\sl average} rate of transmitter $i$ is denoted by $\bar R_i= \mathbb{E}_{h_1,h_2}[R_i(h_1,h_2)]$, where the expectation is with respect to the random variables $h_1$ and $h_2$. Hence, the average capacity region is the convex hull of all simultaneously achievable average rates $(\bar R_1,\bar R_2)$.  Furthermore, we define $\beta^k_{ij}\in[0,1]$ as the ratio of the total power $P$ assigned to information layer $U^k_{ij}$, where we have $$\sum_{i=1}^j \beta^k_{ij}=1$$ for all $j,k\in\{1,2\}$. The next theorem characterizes an average achievable rate region.
\begin{Theorem}[\cite{zohdy2019broadcast}]
\label{theorem_achievable_rate_partial}
For the codebook assignment in Fig.~\ref{fig:code_partial}, and the decoding scheme in Table~\ref{table:decode_partial}, for any given set of power allocation factors $\{\beta^k_{ij}\}$, the average achievable rate region $\{\bar R_1, \bar R_2\}$ is the set of all rates that satisfy 
\begin{align}
\bar R_1 & \leq q_1 C\left( {s}_1 ,{s}_2 \beta^2_{22}\right)+  q_2 \left( C\left( {s}_2 \beta^1_{12},{s}_2\beta^1_{22}+ {s}_2 \beta^2_{22}\right)\! + \!C\left({s}_2\beta^1_{22},0\right)\right),\\
\bar R_2 & \leq p_1 C({s}_1,{s}_2 \beta^1_{22}) +  p_2 \left( C\left( {s}_2 \beta^2_{12},{s}_2\beta^1_{22}+ {s}_2 \beta^2_{22}\right)\! + \! C\left({s}_2\beta^2_{22},0\right)\right), \\
\nonumber \bar R_1 + \bar R_2 & \leq q_1 p_1 C\left( 2{s}_1 ,0\right) \nonumber\\
&\quad +  q_1 p_2 C\left({s}_1 +{s}_2\beta^2_{12}+{s}_2 \beta^2_{22},0\right) \nonumber \\
&\quad +  q_2 p_1 C\left({s}_1 +{s}_2 \beta^1_{12}+{s}_2 \beta^1_{22},0\right) \nonumber \\
&\quad +  q_2 p_2 C\left({s}_2\beta^1_{12}+{s}_2\beta^2_{12}+{s}_2\beta^1_{22}+{s}_2\beta^2_{22},0\right).
\end{align} 
\end{Theorem}

Achieving the average rate region specified in this theorem requires decoding the codebooks in the order specified by Table~\ref{table:decode_partial}. Specifically, the receiver adopts a multi-state decoding scheme where in each state it decodes at most two codebooks. This decoding scheme continues until all the codebooks from both transmitters are decoded. Even though limiting the number of codebooks to be decoded at each stage is expected to result in a reduced rate region, it can be readily verified that the rate region that is achieved by employing a fully joint decoding scheme can be recovered via time-sharing among the average achievable rates corresponding to all possible decoding orders in each channel state. 

\vspace{.1 in}
\noindent {\bf Outer Bound.} Next, we provide outer bounds on the average capacity region, and we compare them with the achievable rate region specified by Theorem~\ref{theorem_achievable_rate_partial}.

\begin{figure}[!t]
\centering
\includegraphics[width=3in]{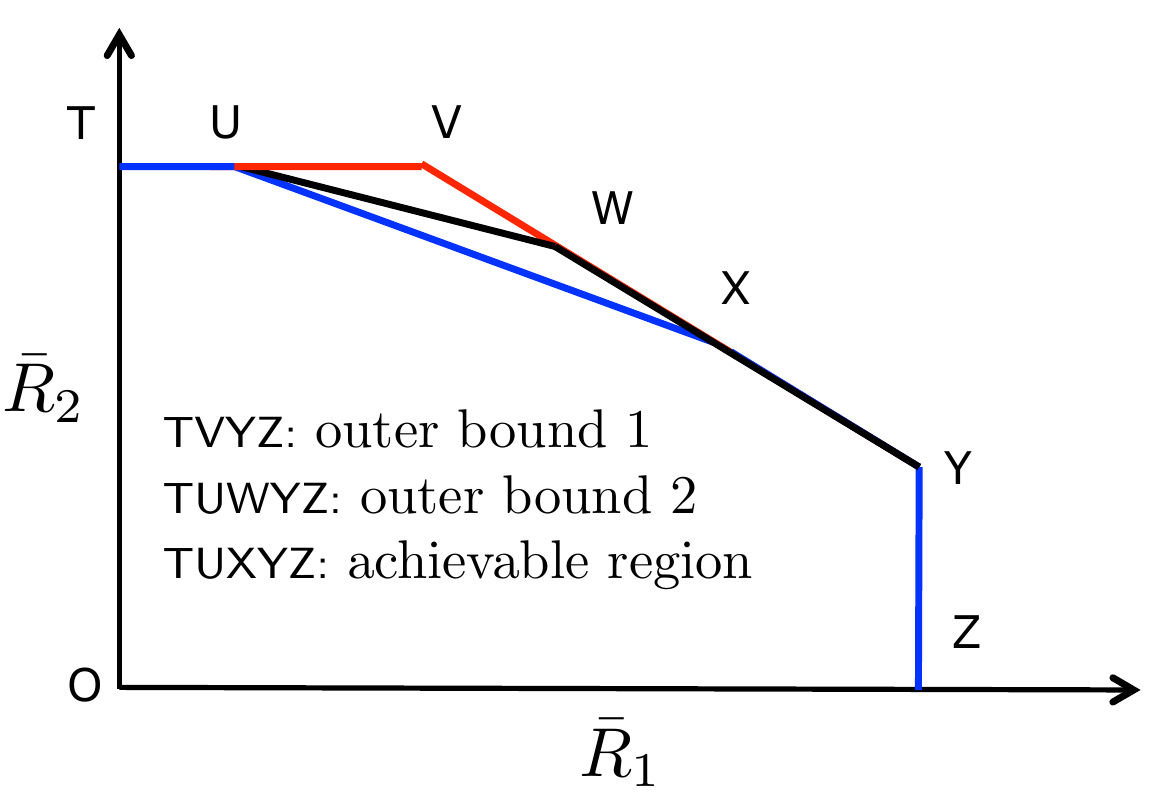}  
\caption{Outer bounds on the average achievable rate region.}
\label{fig:outer_bound_achievable_region}
\end{figure}

\vspace{.1in}
\noindent \textbf{Outer bound 1:} The first outer bound is the average capacity region corresponding to the two-user MAC in which the transmitters have complete access to the CSI~\cite{knopp1995information}. This region is specified by {\sf \footnotesize OTVYZO} in Fig.~\ref{fig:outer_bound_achievable_region}.

\vspace{.1in}
\noindent \textbf{Outer bound 2:} The second outer bound is the average capacity region of the two-user MAC with local CSI at transmitter $1$ and full CSI at transmitter $2$. Outer bound 2 is formally characterized in the following theorem.

\begin{Theorem}[\cite{zohdy2019broadcast}]
\label{theorem_achievable_rate_partialfull}
For the two-user MAC with local CSI at transmitter~1 and full CSI at transmitter~2, the average capacity region is the set of all average rates enclosed by the region {\sf \footnotesize OTUWYZO} shown in Fig.~\ref{fig:outer_bound_achievable_region}, where the corner points are specified in Appendix~\ref{sec:app_cornerpoints}.
\end{Theorem}

\noindent For the case of available local CSI at transmitter $1$ and full CSI at transmitter $2$, it can be shown that deploying the discussed layering scheme at transmitter $1$ (with local CSIT) achieves the average sum-rate capacity of Outer bound 1. This is formalized in the following theorem.
\begin{Theorem}[\cite{zohdy2019broadcast}]
\label{theorem:outer bound}
With local CSI at transmitter~1 and full CSI at transmitter~2, an average achievable rate region is the region {\sf \footnotesize OTUXYZO} shown in Fig.~\ref{fig:outer_bound_achievable_region}. The average capacity region is achieved along {\sf \footnotesize TU} and {\sf  \footnotesize YZ}, and the sum-rate capacity is achieved on {\sf \footnotesize XY}. The corner points are specified in Appendix~\ref{sec:app_cornerpoints}.
\end{Theorem}

Figure~\ref{fig:outer_bound_achievable_region} illustrates the relative representations of the inner and outer bounds on the average capacity region. Specifically,  the region specified by {\sf  \footnotesize OTVYZO} is the average capacity region of a two-user MAC with full CSI at each transmitter, which serves as Outer Bound 1 specified earlier. This region encompasses Outer Bound 2 denoted by {\sf  \footnotesize OTUWYZ}. Segments $\sf TU$ and $\sf XYZ$ of the boundary of  Outer Bound $1$ coincide with the average capacity region of the case of the two-user MAC with full CSIT. 

Figure~\ref{fig:achievable_region_snr10} demonstrates the average rate region for the two-state channel. For this region we have $P=P_1 = P_2 =10$ dB, and select the channel gains as ${s}_1=0.25$ and ${s}_2=1$. Accordingly, the channel probability parameters are set to $q_1 = p_1 = 0.5$. The main observation is that the average achievable rate region coincides with average rate region achieved when the receiver adopts joint decoding.  It can be shown that when the transmitters have local CSIT, it is possible to achieve an average sum-rate that is close to outer bound 1, and that the average the sum-rate capacity can be achieved asymptotically in the low and high power regimes. This observation is formalized in the next theorem.
\begin{Theorem}[\cite{zohdy2019broadcast}]
\label{theorem:proximity_partial_CSI}
By adopting the codebook assignment presented and setting $\beta^1_{22}=\beta^2_{22}=\frac{{s}_1}{{s}_2}$, the sum-rate capacity of a two-user MAC with full CSIT is achievable asymptotically as $P\rightarrow 0$ or $P\rightarrow\infty$. 
\end{Theorem}

\subsection{MAC with Local CSIT -- Multi-state Channels: State-dependent Layering}\label{sec:LCSIT_MU_multi}
In this section, we generalize the encoding and decoding strategy of Section~\ref{sec:LCSIT_MU}  to the general $\ell$-state channel. When the channels have $\ell$ possible states, each transmitter is allocated $\ell$ different sets of codebooks, one corresponding to each channel state. Specifically, corresponding to channel state ${s}_m$ for $m\in\{1,\dots,\ell\}$, transmitter $i$ encodes its message via $m$ information layers generated according to independent codebooks. This set of codebooks is denoted by $\W^i_m = \{U^i_{1m}, \dots, U^i_{mm}\}$.

\begin{table*}[h]
\renewcommand{\arraystretch}{1.17}
\centering
\caption{Successive decoding stages for $\ell-$state MAC with local CSIT}
\label{table:MAC_lstate_partial}
\begin{tabular}{|c||c|c|c|c|c|c|}
 \hline
 \backslashbox{$\textcolor{red}{h_2}$}{$\textcolor{blue}{h_1}$} & ${s}_1$ & ${s}_2$ & $\cdots$ & ${s}_q$ & $\cdots$ & ${s}_\ell$\\ 
 \hline\hline
 ${s}_1$ & \scalebox{.9}{\begin{tabular}[t]{@{}c@{}} $\textcolor{blue}{U^1_{11}}$ \\ $\textcolor{red}{U^2_{11}}$\end{tabular}} & \scalebox{.9}{\begin{tabular}[t]{@{}c@{}} $\textcolor{blue}{U^1_{12}, U^1_{22}}$ \\ $\textcolor{red}{\V^{2}_{11}}$ \end{tabular}} & $\cdots$ & $\cdot$ &$ \cdots$ & \scalebox{.8}{\begin{tabular}[t]{@{}c@{}} $\textcolor{blue}{U^1_{1\ell}, \dots, U^1_{\ell\ell}}$ \\ $\textcolor{red}{\V^2_{(\ell-1)1}}$ \end{tabular}}\\ 
 \hline
 ${s}_2$ & \scalebox{.9}{\begin{tabular}[t]{@{}c@{}}$\textcolor{blue}{\V^1_{11}}$ \\ $\textcolor{red}{U^2_{12}, U^2_{22}}$\end{tabular}} & \scalebox{.9}{\begin{tabular}[t]{@{}c@{}}$\textcolor{blue}{\V^1_{12}}$ \\ $\textcolor{red}{\V^2_{21}}$ \end{tabular}} &$\cdots$ &$\cdot$ & $\cdots$& \scalebox{.9}{\begin{tabular}[t]{@{}c@{}}$\textcolor{blue}{\V^1_{1\ell}}$ \\ $\textcolor{red}{\V^2_{2(\ell -1)}}$ \end{tabular}}\\ 
 \hline
 $\cdot$ & $\cdot$ & $\cdot$ & $\cdots$ & $\cdot$ & $\cdots$ & $\cdot$\\ 
 \hline
${s}_p$ & $\cdot$ & $\cdot$ & $\cdots$ &\scalebox{.9}{\begin{tabular}[t]{@{}c@{}}$\textcolor{blue}{\V^1_{(p-1)q}}$ \\ $\textcolor{red}{\V^2_{p(q-1)}}$ \end{tabular}} & $\cdots$ & $\cdot$\\ 
 \hline
 $\cdot$ & $\cdot$ & $\cdot$ & $\cdots$ & $\cdot$ & $\cdots$ & $\cdot$\\
 \hline
 ${s}_\ell$ & \scalebox{.9}{\begin{tabular}[t]{@{}c@{}}$\textcolor{blue}{\V^1_{(\ell-1)1}}$ \\ $\textcolor{red}{U^2_{1\ell}, \dots, U^2_{\ell\ell}}$\end{tabular} }& \scalebox{.9}{\begin{tabular}[t]{@{}c@{}}$\textcolor{blue}{\V^1_{(\ell -1)2}}$ \\ $\textcolor{red}{\V^2_{1\ell}}$ \end{tabular}} & $\cdots$ & $\cdot$ & $\cdots$ & \scalebox{.9}{\begin{tabular}[t]{@{}c@{}}$\textcolor{blue}{\V^1_{(\ell-1)\ell}}$ \\ $\textcolor{red}{\V^2_{(\ell-1)\ell}}$ \end{tabular}}\\ 
\hline
\end{tabular}
\end{table*}

Table~\ref{table:MAC_lstate_partial} specifies the designation of the codebooks to different combined channel states. In this table, the channels are ordered in the ascending order. In particular, varying channels for transmitter 1, the combined channel state $({s}_q,{s}_p)$ precedes all channel states $({s}_k,{s}_p)$ for all $k>q$. Similarly, for transmitter $2$ channel state $({s}_q,{s}_p)$ precedes the channel state $({s}_q,{s}_k)$, for every $k > p$. Furthermore, according to this approach, when user $i$'s  channel becomes stronger, it decodes additional codebooks. The sequence of decoding the codebooks, as shown in Table~\ref{table:MAC_lstate_partial}, is specified in three steps:
\begin{enumerate}
\item \underline{State $({s}_1,{s}_1)$:} Start with the weakest channel combination $({s}_1,{s}_1)$, and reserve the baseline codebooks $U^1_{11},U^2_{11}$ to be the only codebooks to be decoded in this state. Define $\V^i_{11}= \{U^i_{11}\}$ as the set of codebooks that the receiver decodes from transmitter $i$ when the channel state is $({s}_1,{s}_1)$.
\item \underline{States $({s}_1,{s}_q)$ and $({s}_q,{s}_1)$:} Next,  construct the first row of the table. For this purpose,  define ${\cal V}^2_{1q}$ as the set of the codebooks that the receiver decodes from transmitter $2$, when the channel state is $({s}_1, {s}_q)$. Based on this, the set of codebooks in each state can be specified recursively. Specifically, in the state $({s}_1,{s}_q)$,  decode what has been decoded in the preceding state $({s}_1,{s}_{q-1})$, i.e., the set of codebooks ${\cal V}^2_{1(q-1)}$, plus new codebooks $\{U^1_{1q},\dots, U^1_{qq}\}$. Then, construct the first column of the table in a similar fashion, except that the roles of transmitter 1 and 2 are swapped.
\item \underline{States $({s}_q,{s}_p)$ for $p,q>1$:} By defining the set of codebooks that the receiver decodes from transmitter $i$ in the state $({s}_q,{s}_p)$ by ${\cal V}^i_{qp}$, the codebooks decoded in this state are related to the ones decoded in two preceding states. Specifically, in state $({s}_q,{s}_p)$  decode codebooks ${\cal V}^1_{(p-1)q}$ and ${\cal V}^1_{p(q-1)}$. For example, for $\ell=3 $, the codebooks decoded in $({s}_2,{s}_3)$ includes those decoded for transmitter $1$ in state $({s}_2,{s}_2)$ along with those decoded for transmitter $2$ in channel state $({s}_1, {s}_3)$.
\end{enumerate}

 The decoding order in the general case is similar the one used for $\ell=2$ in Table~\ref{table:decode_partial}. In particular, in channel state $({s}_q,{s}_p)$ the receiver successively decodes $q$ codebooks from transmitter 1 along with $p$ codebooks from transmitter~2. The set of decodable codebooks in channel state $({s}_q,{s}_p)$ is related to set of codebooks decoded for transmitter $2$  in state $({s}_{q-1},{s}_p)$ and those decoded for transmitter $1$ $({s}_{q},{s}_{p-1})$.  The average achievable rate region for the codebook assignment and decoding strategy presented in this section is summarized in Theorem~\ref{partial:multi state}. Similar to the two-state channel case,  define $\beta^i_{mn}\;\in\;[0,1]$ as the fraction of power allocated to the codebook $U^i_{mn}$ such that $\sum_{m=1}^n \beta^i_{mn}=1, \; \forall n \; \in \; \{1,\dots. \ell\}$. 

\begin{Theorem}[\cite{zohdy2019broadcast}]
\label{partial:multi state}
For the codebook assignment in this section and the decoding scheme in Table~\ref{table:MAC_lstate_partial}, for any given set of power allocation factors $\{\beta^i_{mn}\}$, the average achievable rate region $\{\bar R_1, \bar R_2\}$ for the $\ell$-state channel is the set of all rates that satisfy
\begin{align}
\bar R_2& \leq \mathbb{E} [r_1(n, m)]\ ,\\
\bar R_2& \leq \mathbb{E} [r_2(n, m)]\ ,\\
\bar R_{1} + \bar R_2& \leq \mathbb{E} [\min \{r_3(n, m), r_4(n, m)\}]\ ,
\end{align}
where the functions $\{r_1(n, m), \dots, r_4(n, m)\}$, for all $m, n \in \{1, \dots , \ell\}$ are defined as follows.
\begin{align}
r_1(n, m) = & \min_{m}\sum^{\ell}_{j = 1} c_1(j,m) + c_3(j,n,m)\, ,\\
r_2(n, m) = & \min_{m}\sum^{\ell}_{j = 1} c_2(j,m) + c_4(j,n,m)\, ,\\
r_3(n, m) = & \sum_{\forall k < m} c_5(m) + c_7(m,n) + c_9(k,m,n)\, ,\\
r_4(n, m) = & \sum_{\forall k < m} c_5(m) + c_6(m,n) + c_8(k,m,n)\ , 
\end{align}
where
\begin{align}
c_1 (j,m) = & \; C(s_j \beta^1_{jj}, s_m C_2(j,m)) ,  \quad \forall j \in  \{1,\dots, \ell\}, m \in  \{j,\dots, \ell\}  \, ,\\
c_2 (j, i) = & \; C(s_j \beta^2_{jj}, s_m C_1(j,m)), \quad \forall j  \in   \{1,\dots, \ell\}\, ,\\
c_3 (j,n,m) = & \; C(s_n \beta^1_{jn}, s_n C_1(j,n) + s_j C_2(j,j)) , \quad  \forall n  \in  \{j+1,\dots, \ell\}, m  \in \{j,\dots, \ell\}\, ,\\
c_4 (j,n,m) = & \; C(s_n \beta^2_{jn}, s_j C_1(j,m) + s_n C_2(j,n))\ ,\quad \forall n \in \{j+1,\dots, \ell\}\, ,\\
c_5 (m) = & \;  C(s_m \beta^1_{mm} +s_n \beta^2_{mm}), \quad  \forall m \in \{1,\dots, \ell\}\ ,\\
c_6(m,n) = & \; C(s_m \beta^1_{mm} + s_n \beta^2_{mn}, s_n C_2(m,n)), \quad  \forall m  <  n, \forall n \in \{m+1,\dots, \ell\}\, ,\\
c_7 (m,n)= & \; C( s_n \beta^1_{mn} + s_m \beta^2_{mm}, s_n C_1(m,n)) , \quad  \forall m  < n, \forall n \in \{m+1,\dots, \ell\}\, , \\
c_8 (k,m,n) = & \;  C\left(s_m \beta^1_{km} + s_n \beta^2_{kn}, s_m C_1 (k,m) + s_n C_2 (k,n)\right), \quad  \;\;\;\;\; \forall k  < m, \forall n \in \{m,\dots, \ell\}\, ,\\
c_9 (k,m,n) = & \; C\left(s_n \beta^1_{kn} + s_m \beta^2_{km},  a_n C_1(k,n)+ s_m C_2(k,m)\right), \quad  \forall \; k < m, \forall n \in \{m,\dots, \ell\}\, ,
\end{align}
and we have defined $C_1(m,n)= 1-\sum_{i=1}^{m} \beta^1_{in}$ and $C_2(m,n)= 1-\sum_{i=1}^{m} \beta^2_{in}$, for all $ m<n$ and $n \in  \{m+1, \dots, \ell\}$.

\end{Theorem}
\begin{figure}[!t]
\centering 
\includegraphics[width=5in]{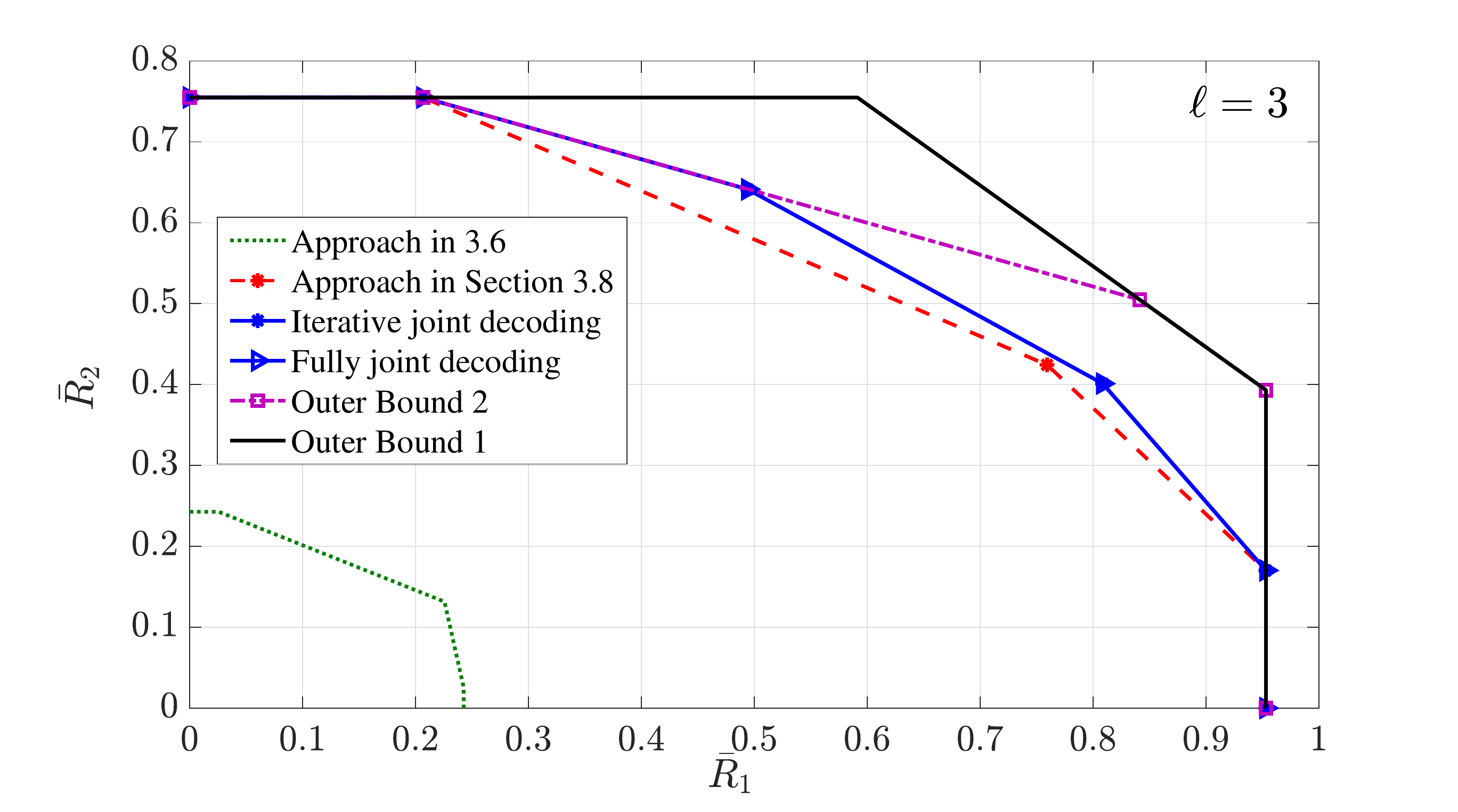} 
\caption{Average rate regions for $\ell = 3$.}
\label{fig:multi_state}
\end{figure}

\begin{figure}[!t]
\centering
\includegraphics[width=5in]{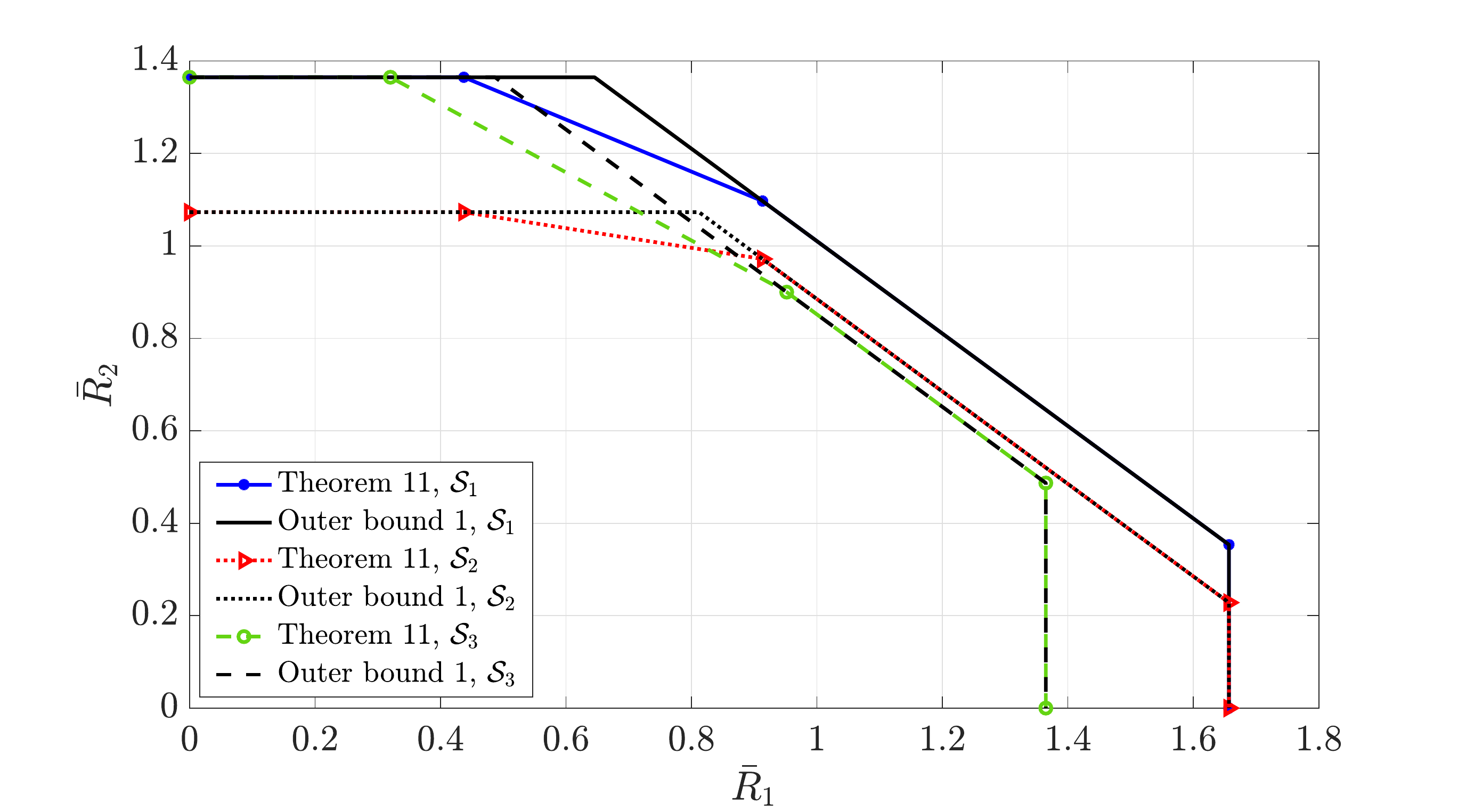} 
\caption{Average rate regions in Theorem~\ref{theorem:outer bound}.}
\label{fig:7}
\end{figure}

Figure~\ref{fig:multi_state} demonstrates the average rate region for the three-state channel, in which the channel gains are ${s}_1=0.04$, ${s}_2 = 0.25$, ${s}_3=1$, and channel probability parameters $q_1= 0.3, q_2= 0.4$ for transmitter $1$, and $p_1 = 0.6, p_2=0.1$ for transmitter $2$. Furthermore, the region in Theorem~\ref{theorem:outer bound} is evaluated in Fig.~\ref{fig:7}. Specifically, the average achievable rate region {\sf OTUXYZ} specified in Fig.~\ref{fig:outer_bound_achievable_region} is evaluated for three scenarios $\hat{\cal{S}}_1, \hat{\cal{S}}_2, \hat{\cal{S}}_3$. In all three scenarios, the average power constraint is set to $10$ dB, i.e., $P_1 = P_2 = P = 10$ dB, and the channel states are $({s}_1, {s}_2) = (0.3, 1)$. Evaluations are carried out for the symmetric setting $\hat{\cal{S}}_1$ with the probability distribution $q_1 = p_1 = 0.5$, and the asymmetric cases $\hat{\cal{S}}_1,\hat{\cal{S}}_2$ with probability distributions $q_1 =0.2, p_1 = 0.8$ and $q_1 =0.4, p_1 = 0.5$. This figures illustrates that the average capacity region of the two-user MAC with full CSIT can be partially achieved  when only one user has full CSIT.

\section{The Interference Channel}
\label{sec:IC}

\subsection{Overview}

In this section, we turn the focus to the interference channel as a key building block in interference-limited wireless networks. In this channel, multiple transmitters communicate with their designated receivers, imposing interference on one another. Designing and analyzing interference management schemes has a rich literature. Irrespective of their discrepancies, the existing approaches often rely on the accurate availability of the CSIT and CSIR. We discuss how the broadcast approach can be viewed as a distributed interference management scheme, rendering a practical approach to have effective communication in the interference channel in the face of unknown CSIT.

While the literature on assessing the communication reliability limits of the interference channel and the attendant interference management schemes is rich, a significant focus is on the channels with perfect availability of the CSIT at all transmitters. Representative known results in the asymptote of high SNR regime include the degrees-of-freedom (DoF) region achievable by interference alignment~\cite{Jafar:IT2008,maddah2008communication}. In the non-asymptotic SNR regime of particular note is the achievable rate region due to Han-Kobayashi (HK)~\cite{han1981new, chong2008han}, which is shown to achieve rates within one bit of the capacity region for the Gaussian interference channel~\cite{etkin2008gaussian}. While unknown in its general form, the capacity region is known in special cases, including the strong interference channel~\cite{carleial1975case, sato1981capacity}, the discrete additive degraded interference channel~\cite{benzel1979capacity}, certain classes of the deterministic interference channel~\cite{gamal1982capacity,cadambe2009capacity,chong2007capacity, bresler2008two}, and opportunistic communication under bursty interference, which is a form of the broadcast approach and is studied under different assumptions on the non-causal availability of the CSI at the transmitters and receivers~\cite{Villacres2018}. some other examples of interference channel and broadcast approach are found in \cite{YiSun20,YiSum20TCOM}.
There are extensive studies on circumventing the challenges associated with analyzing and optimal resource allocation over the HK region~\cite{wang2014sliding,bandemer2015optimal,tuan2017superposition, yagi2011multi, zhao2012maximum,geng2015optimalityJ,Tajer:IT2016, YiSun20}. A more detailed and thorough overview of these can be found in~\cite{KimElGamal2011}.

Interference management without CSIT has also been the subject of intense studies more recently, with more focus on the high SNR regime. Representative studies in the high SNR regime include characterizing  the DoF region for the two-user multi-antenna interference channel in~\cite{huang2012degrees, zhu2011degrees,vaze2012degree, gou2011degrees, shin2016mimo, jeon2017degrees, morales2019degrees}; blind interference alignment in~\cite{jafar2012blind, lu2013blind, lu2014blind, jafar2010exploiting, gou2011aiming,  wang2011improved, akoum2012data, wang2014degrees, castanheira2017retrospective, chen2017blind}; interference management via leveraging network topologies in~\cite{jafar2013topological, naderializadeh2014interference};  and ergodic interference channels in~\cite{morales2014blind, yang2015degrees,Akhlaghi:CL2011}. In the non-asymptotic SNR regime, the studies are more limited, and they include analysis on the capacity region of the erasure interference channel in~\cite{vahid2017binary,zhu2016layered}; the compound interference channel in~\cite{raja2009two}; ergodic capacity for the Z-interference channel in~\cite{zhu2011ergodic};  ergodic capacity of the strong and very strong interference channels in~\cite{lin2016ergodic, lin2019stochastic}; and approximate capacity region for the fast-fading channels~\cite{sebastian2015rate,sebastian2018approximate}.

In this section, conductive to relieving dependency on full CSIT, we discuss how the broadcast approach can be viewed as a distributed interference management solution for circumventing the lack of CSIT in the multiuser interference channel. One significant intuition provided by the HK scheme is that even with full CSIT, layering and superposition coding is necessary. Built upon this intuition, the broadcast approach is a natural evolution of the HK scheme. We focus on the two-user and finite-state Gaussian interference channel to convey the key ideas in rate-splitting, codebook assignments, and decoding schemes. The remainder of this section is organized as follows. This section focuses primarily on the two-user Gaussian interference channel, for which we provide a model in Section~\ref{sec:ICmodel}. We start by discussing the setting in which the receiver has full CSI, and the transmitters have only the statistical model of the CSI and review the application of the broadcast approach in this setting in Section~\ref{sec:CSIR_IC_Two} for the two-user channel and in Section~\ref{sec:CSIR_IC_Multi} for the multiuser channel. Finally, we will review the interference channel with local CSIT in Sections~\ref{sec:LCSIT_IC}.
Under the setting with local CSIT, we consider two scenarios in which each transmitter either knows the level of the interference that their respective receiver experiences, or the level of interference they impose on the unintended receiver. We discuss how the broadcast approach can be designed for each of these two scenarios.

\subsection{Broadcast Approach in the Interference Channel -- Preliminaries}\label{sec:ICmodel}
\begin{figure}[t]
\centering
\includegraphics[height=2.2in]{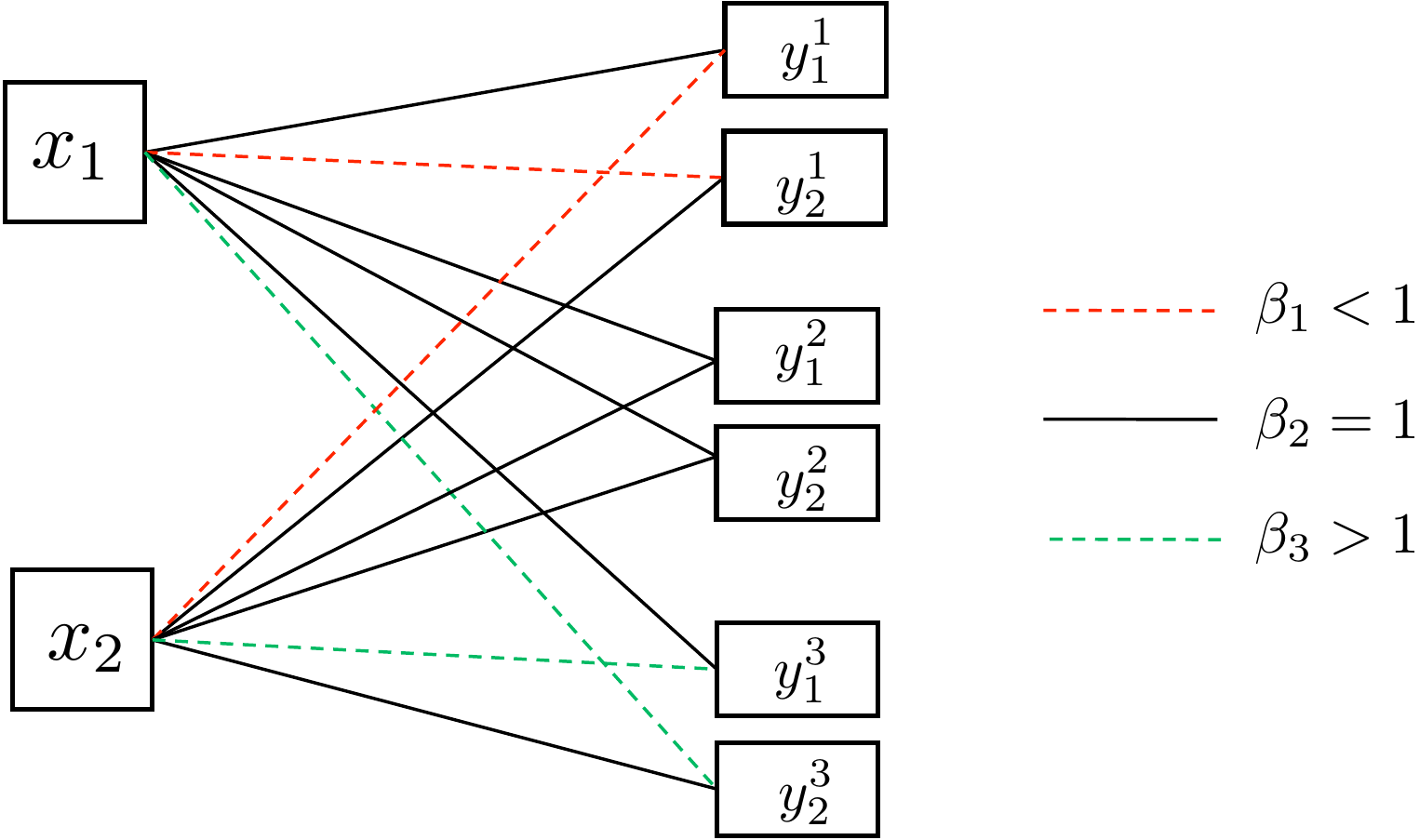}
\renewcommand{\figurename}{Fig.}
\captionsetup{justification=centering}
\caption{Equivalent network for the symmetric Gaussian interference channel ($\ell=2$).}
\label{equi_channel}
\end{figure}

Consider the two-user slowly-fading Gaussian interference channel, in which the coefficient of the channel connecting transmitter $i$ to receiver $j$ is denoted by $h^*_{ij}$ for $i,j \in \{1,2\}$. We refer to $h^*_{ii}$ and $h^*_{ij}$ as the {\sl direct} and {\sl cross} channel coefficients, respectively, $\forall\; i \neq j$. 
The signal received by receiver $i$ is denoted by
\begin{align}
\label{eq:input_output}
y^*_{i} &= h^*_{ii} \; x^*_{i} \; + \; h^*_{ij} \; x^*_{j} \;+\; n^*_{i}\ ,
\end{align}
where $x^*_{i}$ denotes the signal transmitted  by transmitter $i$, and $n^*_{i}$ accounts for the AWGN distributed according to $\mathcal{N}\left(0, N_i\right)$. The transmitted symbol $x^*_i$ is subject to the average power constraint $P^*_i$, i.e., \textcolor{black}{${\mathbb E}[|x_i^*|^2] \leq P^*_i$}. Each channel is assumed to follow a block fading model in which the channel coefficients remain constant for the duration of a transmission block of length $n$, and randomly change to another state afterward. We consider an $\ell$-state channel model in which each channel coefficient $h_{ij}^*$ randomly and independently of the rest of the channels takes one of the $\ell$ possible states $\{\sqrt{s_i}: i\in\{1,\dots,\ell\}\}$. Without loss of generality, we assume that $0 < s_1 < \dots < s_{\ell} < + \infty$. The $\ell$-state interference channel in~\eqref{eq:input_output} gives rise to an interference channel with $\ell^2$ different states. The entire channel states are assumed to be fully known to the receivers while being unknown to the transmitters. A statistically equivalent form of the $\ell$-state interference channel in~\eqref{eq:input_output} is the standard interference channel model given by~\cite{carleial1978interference, sason2004achievableJ} 
\begin{align}
\label{eq:standard_input_output}
y_{1} &= x_{1} \; + \; \sqrt{a_{1}}x_{2} \;+\; n_{1}\ , \qquad \mbox{and} \qquad y_{2} = \sqrt{a_{2}}x_{1} \; + \;  x_{2} \;+\; n_{2}\ ,
\end{align}
and the inputs satisfy \textcolor{black}{${\mathbb E}[|x_i^*|^2] \leq P^*_i$}, where we have defined

\begin{align}\label{eq:normalization}
a_{1} = \left(\frac{h^*_{12} }{h^*_{22} }\right)^2\cdot\frac{N_2}{N_1}\ ,  \qquad a_{2} = \left(\frac{h^*_{21} }{h^*_{11} }\right)^2\cdot\frac{N_1}{N_2}\ , \qquad \mbox{and} \qquad P_i = \frac{(h^*_{ii})^2}{N_i}\cdot P^*_i\ .
\end{align}
and the terms $n_1$ and $n_2$ are the additive noise terms distributed according to ${\cal N}(0,1)$. The equivalence between \eqref{eq:input_output} and \eqref{eq:standard_input_output} can be established by setting
\begin{align}\label{eq:equivalent}
y_{i} = \frac{y^*_{i}}{\sqrt{N_i}}\ , \quad\ x_{i} = \frac{h^*_{ii}} {\sqrt{N_i}}\;x^*_{i} \ , \quad n_{i} = \frac{n^*_{i}}{\sqrt{N_i}} \ .
\end{align}
Channel gains $a_{1}$ and $a_{2}$ are statistically independent, inheriting their independence from that of the channel coefficients. 
By invoking the normalization in~\eqref{eq:normalization}, it can be readily verified that the {\sl cross} channel gains $a_{i}$ take one of $K= \ell(\ell - 1) +  1$ possible states, which we denote by $\{\beta_1, \dots, \beta_K\}$. Without loss of generality we assume they are in the ascending order. For the two-state channel, the {\sl cross} channel gain takes one of the three states $\beta_1 = \frac{s_1}{s_2}, \beta_2  = 1$, and $\beta_3  =\frac{1}{\beta_1}$. Hence, the state of the network is specified by  two cross links, rendering $K^2$ states for the network. We say that the network is in the state $(\beta_{s}, \beta_{t})$ when $(a_{1}, a_{2}) = (\beta_{s}, \beta_{t})$. To distinguish different states, in the network state $(\beta_{s}, \beta_{t})$, we denote the outputs by
\begin{align}
\label{eq:standard_input_output_k}
y^s_{1} &= x_{1} \; + \; \sqrt{\beta_s}\; x_{2} \;+\; n_{1}\ , \qquad \mbox{and} \qquad y^t_{2} = \sqrt{\beta_t} \; x_{1} \; + \;  x_{2} \;+\; n_{2}\ .
\end{align}
Hence, this interference channel can be equivalently presented as a network with two transmitters and $K^2$ receiver pairs, where each receiver pair corresponds to one possible channel state. In the case of the {\sl symmetric} interference channel, we have $a_{1}= a_{2}$, and the number of possible channel combinations reduces to $K$, rending an equivalent network with two transmitters and $2K$ receivers. Figure~\ref{equi_channel} depicts such a symmetric network for the two-state channel. Finally, we define 
\begin{align}\label{eq:q}
q^s_1= \Pro (a_1=\beta_s) \qquad \mbox{and} \qquad q^s_2= \Pro (a_2=\beta_s)\ .
\end{align}

\subsection{Two-user Interference Channel without CSIT}
\label{sec:CSIR_IC_Two}

Effective interference management in the interference channel hinges on how a transmitter can balance the two opposing roles that it has as both an information source and an interferer. Striking such a balance requires designating a proper notion of degradedness according to which different realizations of the network can be distinguished and ordered. Hence, specifying an order of degradedness plays a central role in assigning codebooks and designing the decoding schemes. We adopt the same notion of degradedness that was used for the MAC with proper modifications.

When each channel has $\ell$ possible states, the cross channels take one of the $K=\ell(\ell-1)+1$. Hence, by adopting the broadcast approach, this two-user interference channel becomes equivalent to a multiuser network consisting of two transmitters and $K^2$ receivers. The transmitters and each of these receivers form a MAC, in which the receiver is interested in decoding as many information layers as possible. To this end and by following the same line of arguments we had for the MAC, we use the capacity region of the individual MACs to designate degradedness among  distinct network states. 

The network model in~\eqref{eq:standard_input_output_k} is equivalent to a collection of MACs. The MAC associated with  receiver $y^{s}_i$, for $s < k$, is degraded with respect to the MAC associated with the receiver $y_i^{k}$. Hence, receiver $y_i^{k}$ can successfully decode all the information layers that are decoded by the receivers  $\{y^1_i,\dots,y^s_i\}$. Driven by this approach to designating degradedness, each transmitter splits its message into multiple independent codebooks, where each is adapted to one combined state of the network and intended to be decoded by specific receivers. 

At receiver $y^k_i$, decoding every additional layer form transmitter $i$ directly increases the achievable rate. In parallel,  decoding each additional layer from the the interfering other transmitter indirectly increases the achievable rate by canceling a part of the interfering signal. Driven by these two observations, transmitter $i$ breaks its message into $2K$ layers denoted by $\{V_i^k, U_i^k\}_{k = 1}^K$, each serving a specific purpose. Recall that in the canonical model in~\eqref{eq:standard_input_output_k}, the direct channels remain unchanged and only the cross channels have varying states. Hence, each of the $2K$ layers of each transmitter is designated to a specific {\sl cross} channel state and receiver.

\begin{figure}[t]
\centering
\includegraphics[height=1.8in]{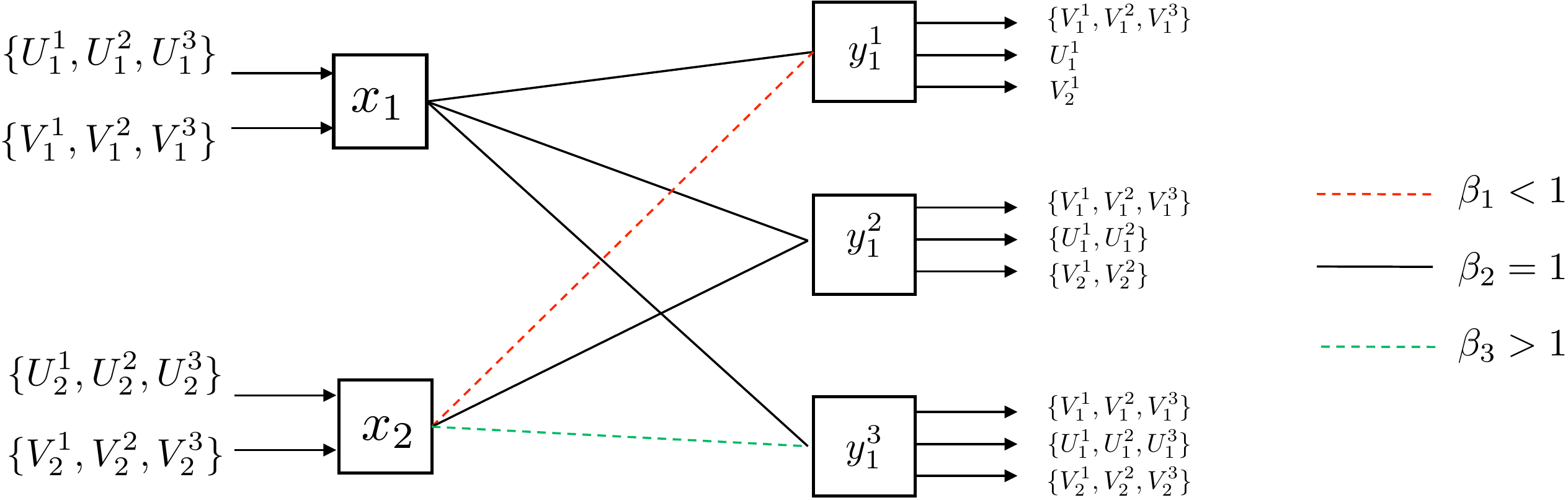}
\renewcommand{\figurename}{Fig.}
\captionsetup{justification=centering}
\caption{Codebook assignments at transmitter 1 in the two-state channel.}
\label{fig:assignment}
\vspace{-0.15in}
\end{figure}

\begin{itemize}
\item Transmitter 1 (or 2) reserves the information layer $V_1^k$ (or $V_2^k$)  for adapting it to the channel from transmitter $1$ (or 2) to the {\sl unintended} receiver $y_2^k$ (or $y_1^k$). Based on this designation,  the intended receivers $\{y_1^k\}_{k = 1}^K$ (or $\{y_2^k\}_{k = 1}^K$) will decode all codebooks $\{V_1^k\}_{k = 1}^K$ (or $\{V_2^k\}_{k = 1}^K$), and the non-intended receivers $\{y^k_2\}_{k = 1}^K$ (or $\{y^k_1\}_{k = 1}^K$) will be decoding a subset of these codebooks. The selection of the subsets depends on on channel strengths of the receivers, such that the non-intended receiver $y_2^k$ (or $y_1^k$) decodes only codebooks $\{V_1^s\}_{s = 1}^k$ (or $\{V_2^s\}_{s = 1}^k$).

\item Transmitter 1 (or 2) reserves the layer $U_1^k$ (or $U_2^k$) for adapting it to the channel from transmitter $2$ (or 1) to the {\sl intended} receiver $y_1^k$ (or $y_2^k$). Based on this designation, the unintended receivers $\{y_2^k\}_{k = 1}^K$ (or $\{y_1^k\}_{k = 1}^K$) will {\sl not} decode any of the codebooks $\{U_1^k\}_{k = 1}^K$ (or $\{U_2^k\}_{k = 1}^K$), and the intended receivers $\{y_1^k\}_{k = 1}^K$ (or $\{y_2^k\}_{k = 1}^K$) will be decoding a subset of these codebooks. The selection of these subsets depends on channel strengths of the receives such that the intended receiver $y_1^k$ (or $y_2^k$) decodes only the codebooks $\{U_1^s\}_{s = 1}^k$ (or $\{U_2^s\}_{s = 1}^k$).
\end{itemize}  
Figure~\ref{fig:assignment}  specifies how the codebooks are assigned to transmitter 1 as we as the set of codebooks decoded by each of the three receivers $\{y_1^k\}_{k = 1}^3$ associated with transmitter 1.

\subsubsection{Successive Decoding: Two-state Channel}
\label{sec:IC:NCSIT:decoding:2}

We review a successive decoding scheme for the two-state channel. This scheme will be then generalized in Section~\ref{sec:IC:NCSIT:decoding:multi}. In this decoding scheme, each codebook will be decoded by a number of receivers. Therefore, the rate of each codebook will be limited by the its most {\sl degraded} channel state. The codebooks that are not decoded by a receiver will be treated as Gaussian noise. These codebooks impose interference on the receiver, which compromises the achievable rate at the receiver. This observation guides designing a successive decoding scheme that dynamically identifies (i) the set of the receivers that decode a given codebook, and (ii) the order by which the codebooks are successively decoded by each receiver.

For formalizing this decoding strategy, denote the set of receivers that decode codebook $V_i^k$ by $\mathcal{V}_i^k$, and denote the set of the receivers that decode $U_i^k$ by  $\mathcal{U}_i^k$.  Therefore, we have
\begin{align}\label{eq:common_set}
\mathcal{V}_1^k = \{y_1^{s}\}_{s = 1}^3 \cup   \{y_2^{s}\}_{s = k}^3\ , \qquad  
\mathcal{V}_2^k = \{y_2^{s}\}_{s = 1}^3 \cup   \{y_1^{s}\}_{s = k}^3\ , \qquad  \mbox{and} \qquad {\cal U}_i^{k} = \{y_i^{s}\}_{s = k}^3\  .
\end{align}
The order of successively decoding the codebooks at receiver $y_i^k$ is specified  in Table~\ref{table:Decoding_order}.

\begin{table*}[h]
\centering
\renewcommand{\arraystretch}{0.8}
{
\caption{Successive decoding order at the receivers.}
\label{table:Decoding_order}
\centering
\begin{tabular}{|c||c|c|c|c|c|c|c|c|c|} 
\hline
Receiver & Stage 1  &  Stage 2  &  Stage 3 & Stage 4 & Stage 5 & Stage 6 & Stage 7 & Stage 8 & Stage 9\\ 
\hline\hline
$y_i^1$ & $V_i^1$ & $V_j^1$ & $V_i^2$ & $V_i^3$ & $U_i^1$ & & & & \\ 
\hline
$y_i^2$ & $V_i^1$ & $V_j^1$ & $V_i^2 $ & $V_j^2$ & $V_i^{3}$ & $U_i^1$ &  $U_i^2$ & &\\  
\hline
$y_i^3$ & $V_j^1$ & $V_i^{1}$ & $V_j^{2}$ & $V_i^{2}$ & $V_j^3$ & $V_i^{3}$ & $U_i^1$ & $U_i^2$ & $U_i^3$  \\  
\hline
\end{tabular}
}
\end{table*}

\subsubsection{Successive Decoding: $\ell$-state Channel}
\label{sec:IC:NCSIT:decoding:multi}
In this section, we generalize the successive decoding scheme to the general multi-state channels. Similarly to \eqref{eq:common_set} we define
\begin{align}\label{eq:common_set2}
\mathcal{V}_1^k = \{y_1^{s}\}_{s = 1}^K \cup   \{y_2^{s}\}_{s = k}^K\ , \quad\ \mathcal{V}_2^k = \{y_2^{s}\}_{s = 1}^K \cup   \{y_1^{s}\}_{s = k}^K\ , \quad\mbox{and} \quad {\cal U}_i^{k} = \{y_i^{s}\}_{s = k}^K\  .
\end{align}
Each of the two receivers decodes a set of the codebooks. The choice of the set depends on the channel states. Specifically, when the network state is $(\beta_q, \beta_p)$, receiver 1 decodes $K + q$ codebooks from transmitter 1 and $q$ codebooks from transmitters 2. These codebooks are decoded successively in two stages in the following order:
\begin{itemize}
\setlength\itemsep{0em}
\item \textbf{Receiver 1 -- stage 1} (Codebooks $\{V^s_i\}_{s = 1}^q$): 
Receiver 1 decodes one information layer from each transmitter in an alternating manner until all codebooks $\{V^s_1\}_{s = 1}^q$ and $\{V^s_2\}_{s = 1}^q$ are decoded. The first layer to be decoded in this stage depends on the state $\beta_q$. If $\beta_q < 1$, the receiver starts by decoding codebook $V^1_1$ from transmitter 1, then it decodes the respective layer $V^1_2$ from transmitter 2, and continues alternating between the two transmitters. Otherwise, if $\beta_q > 1$, receiver 1 first decodes $V^1_2$ from the interfering transmitter 2, followed by $V^1_1$ from transmitter 1, and continues alternating. By the end of stage 1, receiver 1 has decoded $q$ codebooks from each transmitter. 

\item \textbf{Receiver 1 -- stage 2} (Codebooks $\{V^s_1\}_{s = q+1}^K$ \& $\{U^s_1\}_{s = 1}^q$): 
In stage 2, receiver 1 carries on decoding layers $\{V^s_1\}_{s = q+1}^K$ from transmitter 1, in an ascending order of the index $s$. Finally, receiver 1 decodes layers $\{U^s_1\}_{s = 1}^q$ specially adapted to receivers $\{y^s_1\}_{s = 1}^q$, in an ascending order of index $s$. Throughout stage 2, receiver 1 has additionally decoded $K$ codebooks from its {\sl intended} transmitter 1. 

\end{itemize}
The decoding scheme at receiver 2 follows the same structure by swapping the roles of the two transmitters. The set of codebooks decoded by receiver $i$ in channel state $(\beta_q,\beta_p)$ is partly defined by the set of codebooks decoded by receiver $i$ and the set decoded by receiver $j$ in state $(\beta_{q-1},\beta_{p-1})$. The decoding scheme is summarized in Table~\ref{table:decoding_multi_state}. In this table,  the channels are ordered in the ascending order such that at receiver 1, state $(\beta_q,\beta_p)$ precedes all channel states $(\beta_k,\beta_p)$ for all $k > q$. Similarly, at receiver $2$, state $(\beta_q,\beta_p)$ precedes network state $(\beta_q,\beta_k)$, for every $k > p$. Furthermore, according to this approach, when the {\em cross} channel of receiver $i$ becomes stronger, receiver $i$ decodes additional codebooks from both transmitters. In particular, in Table~\ref{table:decoding_multi_state}, every cell contains the codebooks decoded in the combined channel state $(\beta_q, \beta_p)$ where we mark the codebooks decoded by receiver 1 in blue color, while those decoded by receiver 2 in red color. To further highlight the relationship between the decodable codebooks in different states, we denote by ${\cal C}^k_i$ the set of codebooks decoded by the receiver $i$ when $a_i = \beta_k$.

\setlength\extrarowheight{2pt}
\def\arraystretch{1.2}
\begin{table*}[!h]
\centering
{
\renewcommand{\arraystretch}{1.5}
\caption{Successive decoding for $\ell-$state channel}
\label{table:decoding_multi_state}
\vspace{-0.1in}
\begin{tabular}{|c||c|c|c|c|c|c|}
 \hline
 \backslashbox{$\textcolor{red}{a_2}$}{$\textcolor{blue}{a_1}$} & $\beta_1$ & $\beta_2$ & $\cdots$ & $\beta_q$ & $\cdots$ & $\beta_K$\\ 
 \hline\hline
 $\beta_1$ & \scalebox{.8}{\begin{tabular}[t]{@{}c@{}} $\textcolor{blue}{\{V^s_{1}\}_{s = 1}^K, U^1_{1}, V^1_2}$ \\ $\textcolor{red}{\{V^s_2\}_{s = 1}^K, U^1_{2}, V^1_1}$\end{tabular}} & \scalebox{.8}{\begin{tabular}[t]{@{}c@{}} $\textcolor{blue}{\mathcal{C}^1_1, U^2_{1}, V^2_2}$ \\ $\textcolor{red}{\mathcal{C}^1_2}$ \end{tabular}} & $\cdots$ & $\cdot$ &$ \cdots$ & \scalebox{.8}{\begin{tabular}[t]{@{}c@{}} $\textcolor{blue}{\mathcal{C}^{K-1}_1, U^K_{1}, V^K_2}$ \\ $\textcolor{red}{\mathcal{C}^{1}_2}$ \end{tabular}}\\ 
 \hline
 $\beta_2$ & \scalebox{.8}{\begin{tabular}[t]{@{}c@{}} $\textcolor{blue}{\mathcal{C}^1_1}$ \\ $\textcolor{red}{\mathcal{C}^1_2, U^2_{2}, V^2_1}$\end{tabular}} & \scalebox{.8}{\begin{tabular}[t]{@{}c@{}} $\textcolor{blue}{\mathcal{C}^1_1, U^2_{1}, V^2_2}$ \\ $\textcolor{red}{\mathcal{C}^1_2, U^2_{2}, V^2_1}$ \end{tabular}} & $\cdots$ & $\cdot$ &$ \cdots$ & \scalebox{.8}{\begin{tabular}[t]{@{}c@{}} $\textcolor{blue}{\mathcal{C}^{K-1}_1, U^K_{1}, V^K_2}$ \\ $\textcolor{red}{\mathcal{C}^{1}_2, U^2_{2}, V^2_1}$ \end{tabular}}\\ 
 \hline
 $\cdot$ & $\cdot$ & $\cdot$ & $\cdots$ & $\cdot$ & $\cdots$ & $\cdot$\\ 
 \hline
$\beta_p$ & $\cdot$ & $\cdot$ & $\cdots$ &\scalebox{.8}{\begin{tabular}[t]{@{}c@{}}$\textcolor{blue}{\mathcal{C}^{q-1}_1, U^q_{1}, V^q_2}$ \\ $\textcolor{red}{\mathcal{C}^{p-1}_2, U^p_{2}, V^p_1}$ \end{tabular}} & $\cdots$ & $\cdot$\\ 
 \hline
 $\cdot$ & $\cdot$ & $\cdot$ & $\cdots$ & $\cdot$ & $\cdots$ & $\cdot$\\
 \hline
 $\beta_K$ & \scalebox{.8}{\begin{tabular}[t]{@{}c@{}} $\textcolor{blue}{\mathcal{C}^1_1}$ \\ $\textcolor{red}{\mathcal{C}^{K-1}_2, U^{K}_{2}, V^K_1}$\end{tabular}} & \scalebox{.8}{\begin{tabular}[t]{@{}c@{}} $\textcolor{blue}{\mathcal{C}^1_1, U^2_1, V^2_2}$ \\ $\textcolor{red}{\mathcal{C}^{K-1}_2, U^{K}_{2}, V^K_1}$  \end{tabular}} & $\cdots$ & $\cdot$ &$ \cdots$ & \scalebox{.8}{\begin{tabular}[t]{@{}c@{}} $\textcolor{blue}{\mathcal{C}^{K-1}_1, U^K_{1}, V^K_2}$ \\ $\textcolor{red}{\mathcal{C}^{K-1}_2, U^K_{2}, V^K_1}$ \end{tabular}}\\  
\hline
\end{tabular}
}
\end{table*}

\subsubsection{Average Achievable Rate Region}
\label{sec:IC:achievable_rates}

In this section, we provide an overview on the average achievable rate region. The average rates of the users are specified by the rates of codebooks $\{V_i^k, U_i^k\}_{k=1}^K,$ for $i \in\{1,2\}$. These rates should satisfy all the constraints imposes by different receivers in order for them to successfully decode all their designated codebooks. Hence, the rates are bounded by the smallest achievable rates by the receivers $\mathcal{V}_i^k$ and $\mathcal{U}_i^k$. To formalize the rate regions, define $R (A)$ as the rate of codebook  $A\in \{V_i^k,U_i^k: \forall i,k\}$, and define $\gamma (A)$ as the fraction of the power $P_i$ allocated to the codebook $A\in \{V_i^k,U_i^k: \forall i,k\}$. Accordingly, define $R_{i}(s, t)$ as the total achievable rate of user $i$, when the network is in the state $(\beta_s, \beta_t)$. Finally, denote the average achievable rate at receiver $i$ by  $\bar R_i = \mathbb{E}[R_{i}(\beta_s, \beta_t)]$, where the expectation is taken with respect to the probabilistic model of the channel. Note that the the transmitters, collectively have $4K$ codebooks. Corresponding to the set ${\cal S}\subseteq \mathbb{R}_+^{4K}$, define the rate region ${\cal R}_{\rm in}({\cal S})$ as the set of all average rate combinations $(\bar R_1,\bar R_2)$ such that $R(A)\in{\cal S}$ for all $A\in \{V_i^k,U_i^k: \forall i,k\}$, i.e., 
\begin{align}\label{eq:rateregion}
{\cal R}_{\rm in}({\cal S}) = \left\{(\bar R_1,\bar R_2)\;:\;  
R(A)  \in {\cal S}\ , \quad  \forall A\in \{V_i^k,U_i^k: \forall i,k\right\}\ .
\end{align}
Furthermore, corresponding to each receiver $y^k_i$ and codebook $A \in \{U^k_i, V^k_i: \forall i,k\}$ that should be decoded by $y^k_i$,  define $R^k_i(A)$ as the maximum rate that we can sustain for codebook $A$, while being decodable by $y^k_i$. Accordingly, for user $i$, and corresponding to $s,t\in\{1,\dots,K\}$ define the rates
\begin{align}
r_i(s, t) & = \sum_{k = t+ 1}^K R^s_i(V^t_i) + \sum_{k = 1}^s R^k_i(U^k_i)  + \underset{\ell\in \{1, 2\}}{\min} \sum_{k = 1}^t R^k_{\ell}(V^k_i)\ ,
\end{align}
where the the rates $R^s_i(A)$ are defined as follows. First, define $\gamma(A)$ as the fraction of $P_i$ allocated to $A \in \{U^k_i, V^k_i: \forall k\}$, and set 
\begin{align}\label{eq:Gamma_define}
\Gamma_{v}(i, k) = \sum_{j = 1}^k \gamma(V^j_i)\, , \qquad \mbox{and} \qquad \Gamma_{u}(i,k) = \sum_{j = 1}^k \gamma(U^j_i)\, .
\end{align}
Based on these definitions, if the codebook $V_i^k$ is decoded by the receiver $y^s_i$, then we have
\begin{align}
\mbox{If $\beta_s \leq 1$}\, , \qquad \qquad  R^s_i(V^k_i) & =  C\left(\gamma(V^k_i) P_i,  (1 - \Gamma_v(i, k)) P_i + \beta_k (1 - \Gamma_v (j, s-1)) P_j\right)\, ,\\
\mbox{If $\beta_s > 1$}\, , \qquad  \qquad R^s_i(V^k_i) & =  C\left(\gamma(V^k_i) P_i,  (1 - \Gamma_v(i, k)) P_i +  \beta_k(1 - \Gamma_v (j, s)) P_j\right)\, .
\end{align}
Similarly,  if the codebook $V_i^k$ is decoded by the receiver $y^s_j$, then we have
\begin{align}
\mbox{If $\beta_s \leq 1$}\, , \qquad \qquad  
R^s_j(V^k_i) & =  C\left(\beta_s \gamma(V^k_i) P_i, \beta_k (1 - \Gamma_v(i, k)) P_i +  (1 - \Gamma_v (j, s)) P_j\right)\, ,\\
\mbox{If $\beta_s > 1$}\, , \qquad  \qquad 
R^s_j(V^k_i) & =  C\left(\beta_s\gamma(V^k_i) P_i,   \beta_k(1 - \Gamma_v(i, k)) P_i + (1 - \Gamma_v (j, s - 1)) P_j\right)\, .
\end{align}
Finally, when  codebook $U_i^k$ is decoded by the receiver $y^s_i$, then we have
\begin{align}
R^s_i(U^k_i) & = C\left(\gamma(U^k_i) P_i, \beta_k (1 - \Gamma_v(1, s)) P_j + (1 - \Gamma_v(1, K)- \Gamma_u(1, s)) P_i \right)\, .
\end{align}

\begin{Theorem}[\cite{zohdytajersh:2020}]\label{theorem:sequential}
The average achievable rate region via sequential decoding is specified by
\begin{align}
\label{eq:Ravg1}
{\cal R}_{\rm in}^{\rm seq} = \left\{(\bar R_1,\bar R_2)\;:\;  
R_i(s,t)\leq r_i(s,t) , \quad \forall i\in{1,2}\; , s, t\in\{1,\dots,K\}\right\}\ .
\end{align}
\end{Theorem}

 \begin{figure}[!t]
\centering
\includegraphics[width=5in]{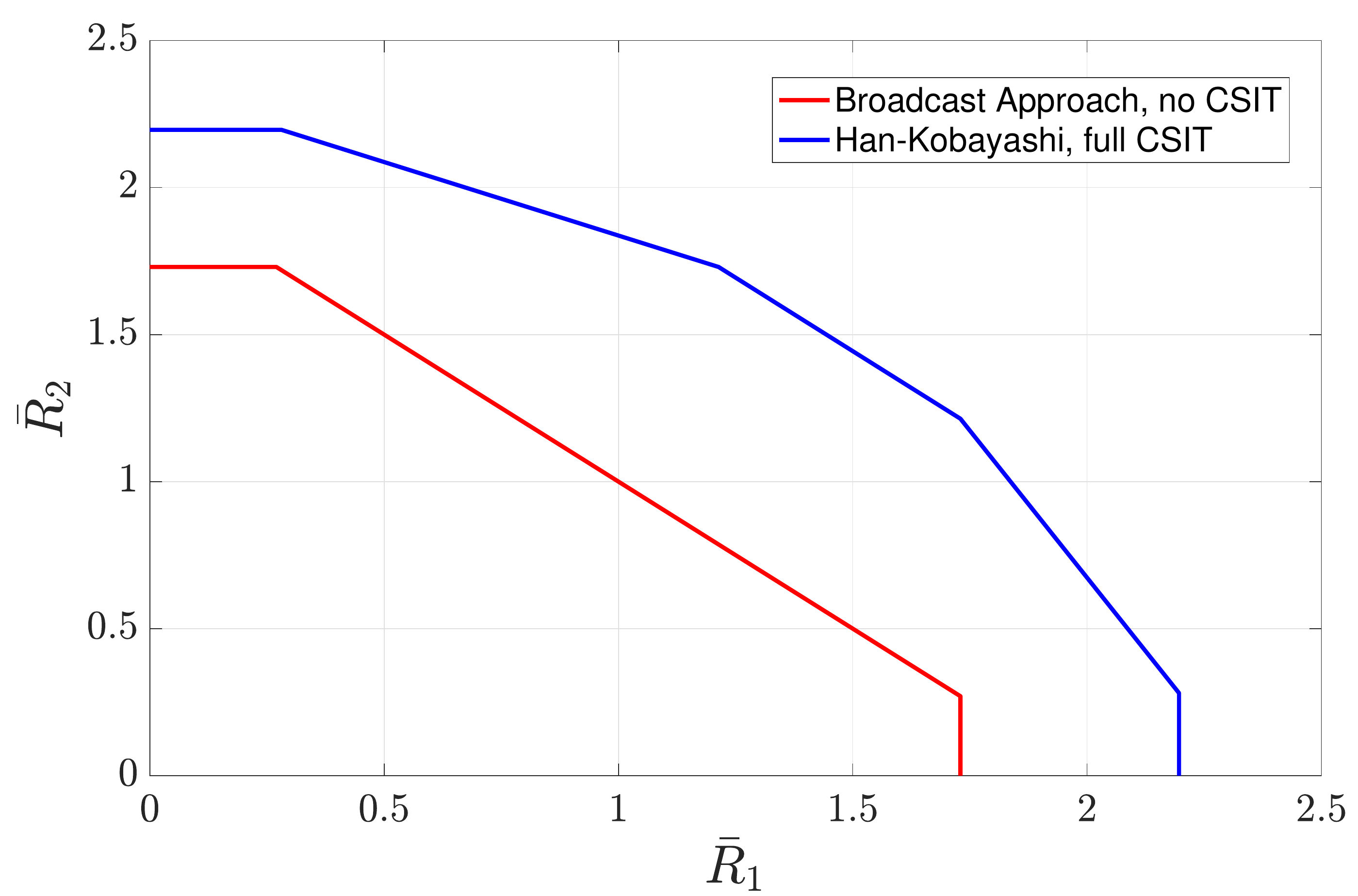}
\caption{Average rate region ($\ell = 2$).}
\label{fig:region2}
\end{figure}

 \begin{figure}[!t]
 \centering
\includegraphics[width=5in]{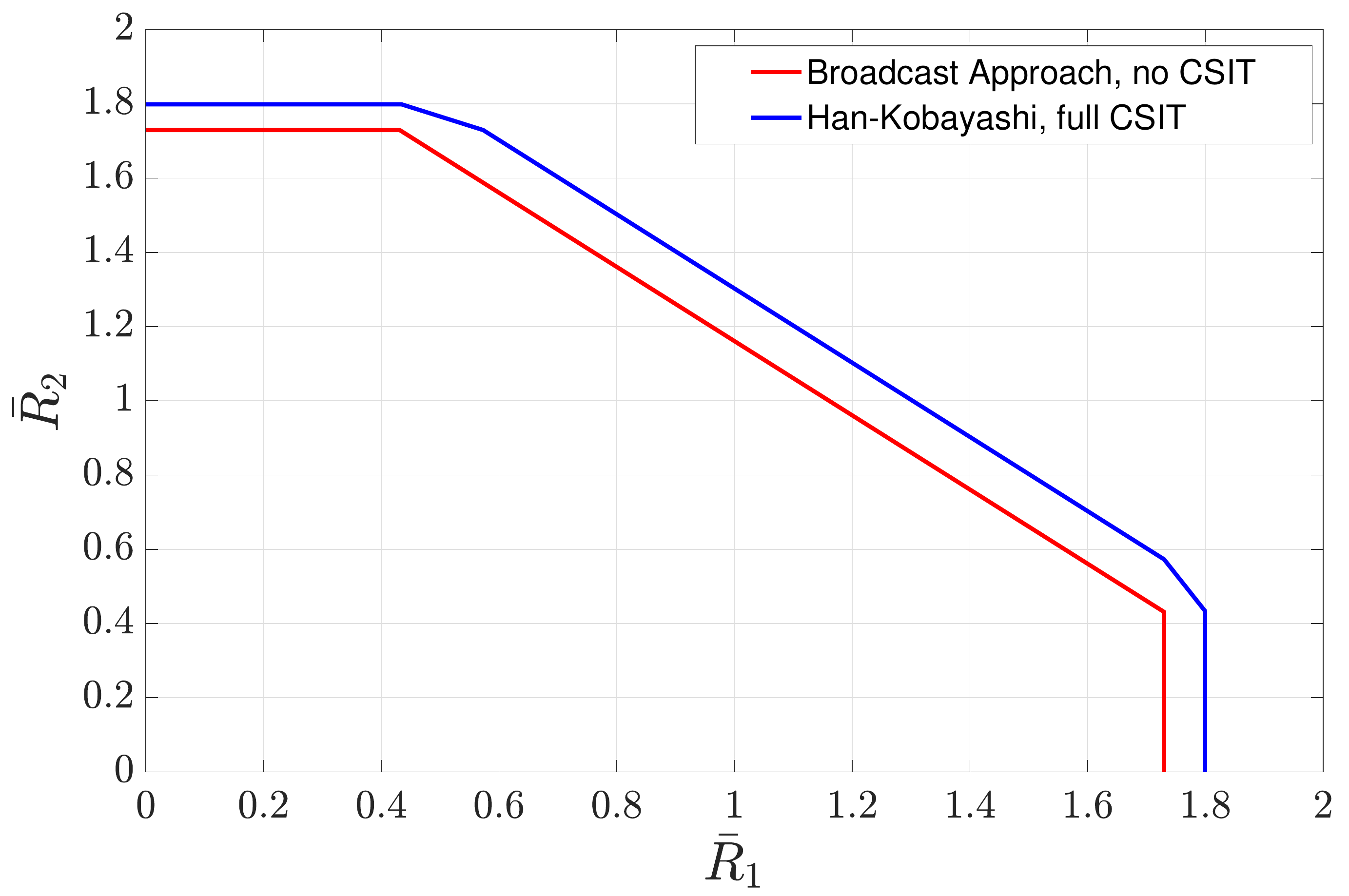} 
\caption{Average rate region ($\ell = 3$).}
\label{fig:region3Asym}
\end{figure}
The average achievable rate region characterized in Theorem~\ref{theorem:sequential} is illustrated in figures~\ref{fig:region2} and \ref{fig:region3Asym}. Figure~\ref{fig:region2} shows a general asymmetric setting for the following set of parameters $P_1 = P_2 = 10$  dB, $(q_1^1 ,q_1^2, q_1^3) = (0.5, 0.3, 0.2)$, $(q_2^1 ,q_2^2, q_2^3) = (0.3, 0.3, 0.4)$, and the weak channel state at each receiver is given by $\beta_1 = 0.5$. Similarly, for $\ell = 3$, Fig.~\ref{fig:region3Asym} evaluates the average rate region for an  asymmetric channel with $(q_i^1 ,q_i^2, q_i^3, q_i^4 ,q_i^5, q_i^6, q_i^7) = (0.1, 0.1, 0.1, 0.2, 0.1, 0.1, 0.3)$,  where the weak channel states at each receivers 1 and 2 are set as $(\beta_1 ,\beta_2, \beta_3) = (0.1, 0.2, 0.9)$ and $(\beta_1 ,\beta_2, \beta_3) = (0.1, 0.4, 0.7)$, respectively. Comparing these figures indicates that increasing the number of states reduces the gap between the two regions (broadcast approach and HK).

\subsubsection{Sum-rate Gap Analysis}\label{subsec:sumrate_gap}
Next, we provide an upper bound on the gap of the the average achievable sum-rate to the average sum-rate capacity of the interference channel in which the the transmitters have full CSI. To this end, we use the existing results on the gap between the sum-rate achievable by the HK scheme and the sum-rate capacity. to present the main ideas and for simplicity in notations, we present the results for the symmetric setting, i.e., when $a_1 =a_2 = a$, symmetric average power constraints, i.e., when $P_1 = P_2 =P$, and symmetric probabilistic models for the channels, i.e., when $q_1^s = q_2^s = q^s  = \frac{1}{3}$. Generalization of the results to the non-symmetric settings is straightforward.
 
In the symmetric settings, the channel in~\eqref{eq:standard_input_output_k} simplifies to either a weak interference channel (when $a = \beta_1)$, or to a strong interference channel  (when $a\in\{\beta_2,\beta_3\}$). To assess the average sum-rate gap, we start by analyzing the gap in the weak and strong interference regimes separately. The average of these gaps provides the average sum-rate gap. Throughout this discussions we set $\beta = \beta_3 = \frac{1}{\beta_1}$.
\begin{itemize}[leftmargin=5.5mm]
\item {\bf Weak interference:} In the weak interference regime, the capacity with full CSIT is in unknown. In this regime, in order to quantify the gap of interest, we first evaluate the gap of the sum-rate achieved by the scheme of Section~\ref{sec:IC:NCSIT:decoding:multi} to the sum-rate achieved by the HK scheme.  By  using this gap in conjunction with the known results on the gap between the sum-rate of HK and the sum-rate capacity, we provide an upper bound on the average sum-rate gap of interest.

\item {\bf Strong interference:} In the strong interference regime, the sum-rate capacity with full CSIT is known. It can be characterized by evaluating the sum-rate of the intersection of two capacity regions corresponding to two multiple access channels formed by the transmitters and each of the receivers~\cite{sato1981capacity}. 
\end{itemize}

By quantifying the two gaps above, and then aggregating them based on the probabilistic model of the channel, the following theorem establishes upper bounds on the gap between the average sum-rate achieved by the approach in Section~\ref{sec:IC:NCSIT:decoding:2} and the sum-rate capacity. The gap is characterized in two distinct regimes of transmission power, denoted by ${\rm G}_1$ and ${\rm G}_2$. Define $\bar{R}_{\rm sum}({\rm G}_i)$ as the minimum average sum-rate achievable under region ${\rm G}_i$, and denote the average sum-rate capacity with full CSIT by $C_{\rm sum}({\rm G}_i)$. Finally,  define the gap $\Delta(G_i) = C_{\rm sum}({\rm G}_i) - \bar{R}_{\rm sum}({\rm G}_i)$.


\begin{Theorem}[\cite{zohdytajersh:2020}]\label{theorem:inner_bound}
The average sum-rate achievable by the broadcast approach in Fig.~\ref{fig:assignment} and Table~\ref{table:Decoding_order} has the following gap with the sum-rate capacity of the symmetric Gaussian interference channel with full CSIT:
\begin{enumerate}
\item[(i)] For $P \in {\rm G}_1 = \left(0, \beta\right) \cup  \left(\beta (\beta^2 + \beta - 1), + \infty\right)$ we have
\begin{align}\label{eq:Delta1}
\Delta({\rm G}_1) & \leq \frac{1}{3}\left[1+   \log\left(\frac{1 + P (1+ \beta)}{1 + P (1+ \frac{1}{\beta})}\right)+\frac{1}{2}\log(2 + \beta)\right] \ .
\end{align} 

\item[(ii)] For $P \in {\rm G}_2 = \left[ \beta, \beta (\beta^2 + \beta - 1)\right]$ we have
\begin{align}\label{eq:Delta2}
\Delta({\rm G}_2) \leq \frac{1}{3}\left[\log \frac{4}{3}+\log\left(\frac{1+P(1+\beta)}{1+P/\beta+\beta}\right)+3\log\frac{(2+\beta)^2}{1+2\beta}\right]\  .
\end{align} 
\end{enumerate}
\end{Theorem}
Further analyzing the result in this theorem shows that the gap in the high SNR regime is upper bounded by a constant (for fixed channel gains).
\begin{Theorem}[\cite{zohdytajersh:2020}]\label{theorem:gap}
For any fixed network model (i.e., fixed $\beta$), when $P$ is sufficiently large, the gap between the sum-rate  capacity of the symmetric Gaussian interference channel with full CSIT and the average sum-rate achievable by the broadcast approach in Fig.~\ref{fig:assignment} and Table~\ref{table:Decoding_order} is upper bounded by
\begin{align}\label{eq:Delta1}
\Delta &\leq \frac{1}{6} \log \left(8\beta^2\cdot \frac{ \beta+2 }{ \beta+1}\right)\ .
\end{align} 
\end{Theorem}

 \begin{figure}[!t]
\centering
\includegraphics[width=5in]{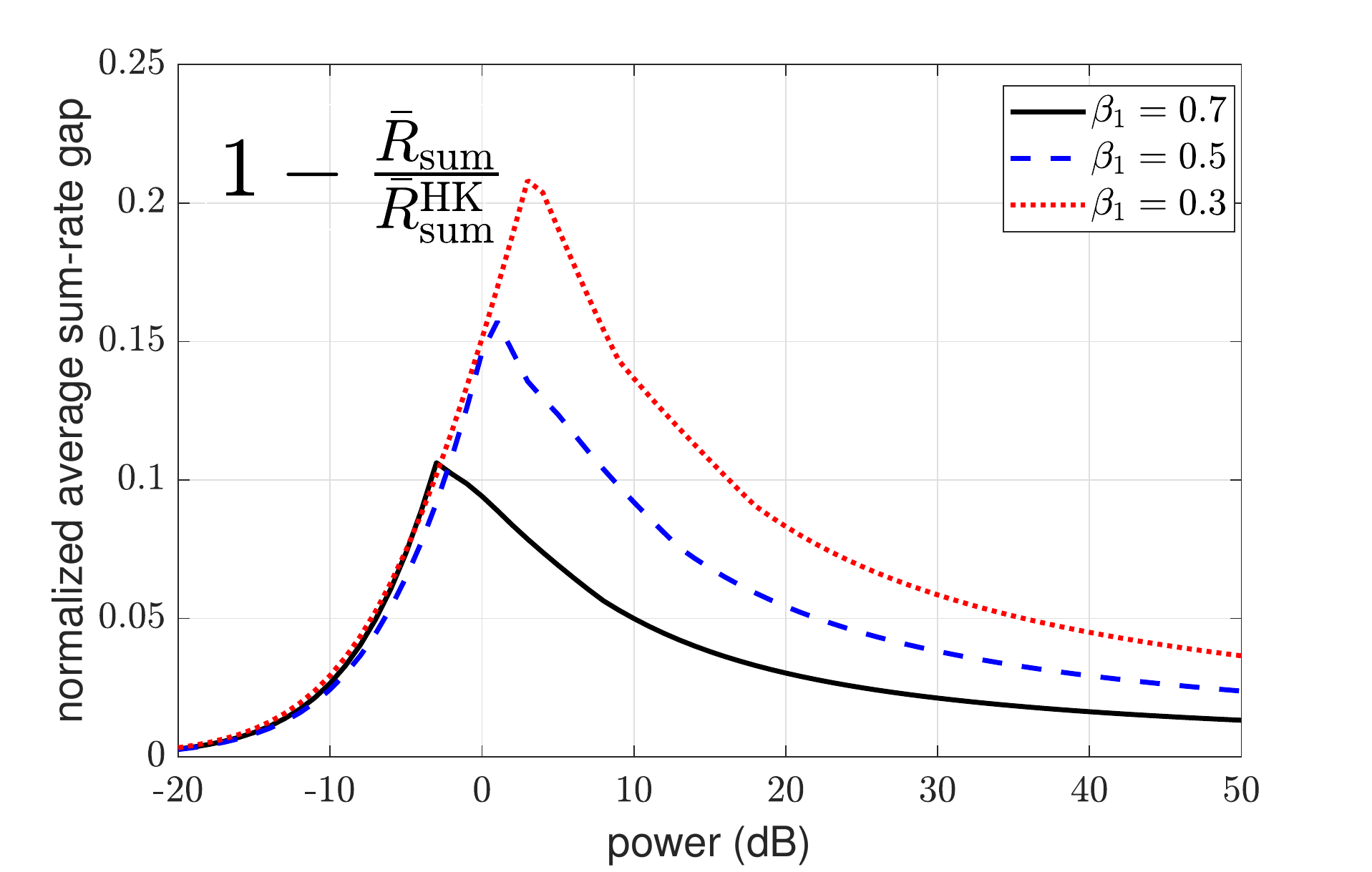} 
\caption{Sum-rate gap versus power.}
\label{fig:gap_power}
\end{figure}

\begin{figure}[!t]
 \centering
\includegraphics[width=5in]{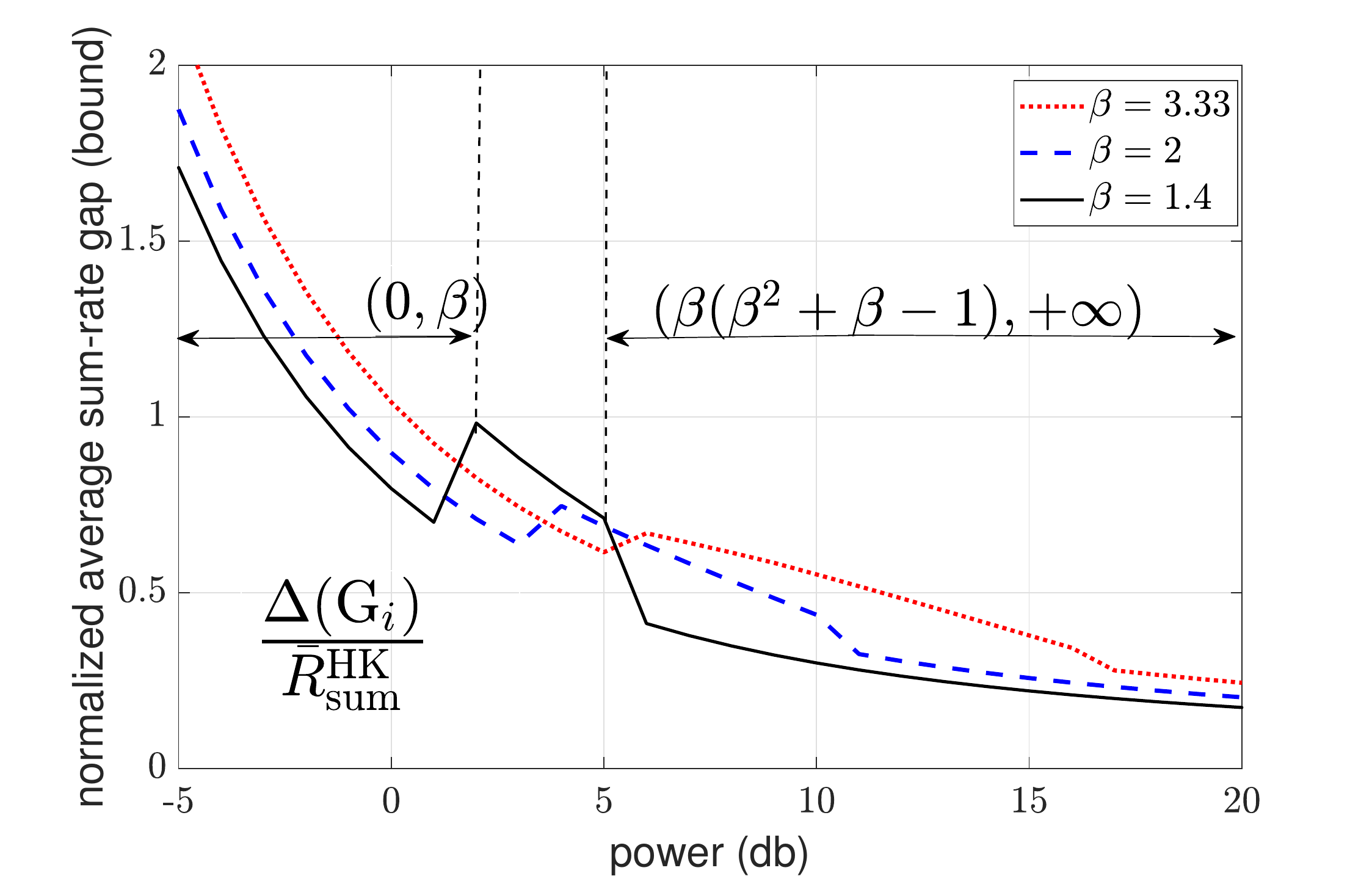} 
\caption{Bound on sum-rate gap versus power.}
\label{fig:gapbound_power}
\end{figure}

Figures~\ref{fig:gap_power} and~\ref{fig:gapbound_power} compare the maximum average sum-rate with that of the sum-rate of the HK scheme. In both schemes, the average sum-rate is maximized over all possible power allocation schemes over different codebooks. Figure~\ref{fig:gap_power} depicts the gap between the two methods normalized by the sum-rate of HK. This signifies the relative sum-rate loss that can be attributed, for the most part, to the lack of the CSI at the transmitters. It is observed that the relative loss with respect to the HK peaks for moderate power regimes, while in the small and large power regimes, it is diminishing. For the evaluations in Fig.~\ref{fig:gap_power},  a two-state channel with a symmetric channel probability model ($q^s_1 = q^s_1 = q^s$) is considered, in which $(q^1 ,q^2, q^3) = (0.3, 0.6, 0.1)$. The results follow the same trend for different values of $\beta_1$. Figure~\ref{fig:gapbound_power} evaluates the bounds on the sum-rate gaps presented in Theorem~\ref{theorem:inner_bound}. The three plots in this figure correspond to those in Fig.~\ref{fig:gap_power}. Specifically, the plots in Fig.~\ref{fig:gapbound_power} depict the  $\Delta({\rm G}_i)$ normalized by the HK sum-rate. This figure shows that the bound on the gap becomes tighter as the power increases.

\subsection{$N$-user Interference Channel without CSIT}
\label{sec:CSIR_IC_Multi}

Finally, we provide a brief overview of a representative approach to generalizing the approaches discussed thus far to the general $N$ user channel. To this end, consider a generalization of~\eqref{eq:standard_input_output} to the $N$-user channel, in which for user $m\in\{1,\dots, N\}$ we have
\begin{align}
\label{eq:standard_input_output_N}
y_{m} &= x_{m} \; + \; \sum_{i\neq m}\sqrt{a_{mi}}x_{i} \;+\; n_{m}\ .
\end{align}
Each of the channel coefficients $a_{im}$ takes one of the $L$ possible states $\{\beta_1,\dots,\beta_L\}$, order in the ascending order. The state of the network will be specified by the cross-links states, rendering a network with $ N $ transmitters and $NL^{N-1}$ receivers. In the case of the symmetric interference channel, i.e., $a_{mi}=a$, the number of network states reduces to $N$ transmitters and $NL$ receivers.

Each transmitter, to balance its impacts on the intended and unintended receivers, performs rate-splitting and adapts each information layer to one combined state of the network designated to be decoded by a specific group of receivers. In the equivalent network, each transmitter and its associated $S= L^{N-1}$ receivers form a MAC. The critical stage in specifying the broadcast approach is adopting a notion of degradedness among different such MAC. Facing a lack of a natural choice, we define an aggregate strength for receiver $y^m$ as
\begin{align}
\theta_m= \sum_{i\neq m} a_{mi}\ .
\end{align}
This metric is used to sort the $S$ receivers associated with $y^m$ in the equivalent network. Denote these receivers by $\{y_m^1,\dots,y_m^{S}\}$, which are sorted in the ascending order such that for $s<k$ receivers the $\theta_m$ value associated with the states in the channel receiver $y_i^s$ is smaller than that of receiver $y_i^k$. Hence, for $s < k$, the multiple access channel at receiver $y^{s}_i$ is degraded with respect to the channel at $y_i^{k}$. Therefore, receiver $y_i^{k}$ can successfully decode all the layers adapted to the channels with the designated receivers $\{y^1_i,\dots,y^s_i\}$. At receiver $y^k_i$, every layer decoded form transmitter $i$ directly increases the achievable rate, whereas every layer decoded from the other transmitter indirectly increases the achievable rate by canceling a part of the interfering signal. Hence, similarly to the two-user channel, transmitter $i$ splits its message into $2S$ layers denoted by $\{V_i^k, U_i^k\}_{k = 1}^{S}$ with the following designations and objectives:
\begin{itemize}
\item Transmitter $m$ adapts layer $V_m^k$ to the state of the channels linking all other transmitters to the {\sl unintended} receivers $\{y_1^k,\dots,y_{N^{L-1}}^k\}\backslash\{y^k_m\}$: while the intended receivers $\{y_m^k\}_{k = 1}^{S}$ will be decoding all codebooks $\{V_m^k\}_{k = 1}^{S}$, the non-intended receivers $\{y_1^k,\dots,y_{S}^k\}\backslash\{y^k_m\}$ decode a subset of these codebooks depending on their channel strengths. More specifically, a non-intended receiver $y_i^k$ decodes only the codebooks $\{V_m^s\}_{s = 1}^k$.
\item Transmitter $m$ adapts the layer $U_m^k$ to the state of the channels linking all other transmitters to the {\sl intended} receiver $y_m^k$: while the unintended receivers $\{y_1^k,\dots,y_{S}^k\}\backslash\{y^k_m\}$ will {\sl not} be decoding any of the codebooks $\{U_m^k\}_{k = 1}^{S}$, the intended receivers $\{y_m^k\}_{k = 1}^{S}$ decode a subset of these codebooks depending on their channel strengths. More specifically, the intended receiver $y_m^k$ decodes only the codebooks $\{U_m^s\}_{s = 1}^k$.
\end{itemize}  
As the number of users $N$ increases, the total number of codebooks per transmitter $2S=2L^{N-1}$ grows exponentially with the number of users. This renders joint decoding to have a prohibitive decoding complexity. A viable remedy is adopting the opportunistic successive decoding schemes (e.g.,~\cite{Tajer:IT2016, Gong:TCOM2012}), which are low-complexity schemes in which each receiver can dynamically identify an optimal set of codebooks to decode. 

\subsection{Two-user Interference Channel with Partial CSIT}
\label{sec:LCSIT_IC}

We now turn the attention to how the transmitters have partial information about the overall network state. Specifically, based on the model in~\eqref{eq:standard_input_output}, we consider two separate scenarios in which each transmitter either knows the interference that it causes to the unintended receiver, or the interference that its intended receiver experiences. More specifically, in Scenario 1, transmitter $i$ knows the channel state $a_j$ for $j\neq i$, while being unaware of $a_i$. In contrast, in Scenario~2, transmitter $i$ knows the channel state $a_i$ while being unaware of $a_j$ for $j\neq i$. These two scenarios and their associated broadcast approaches are discussed next.

\subsubsection{Two-user Interference Channel with Partial CSIT -- Scenario 1}

\label{sec:encoding_decodingD1}

\noindent{\bf State-dependent Adaptive Layering.} In this setting,  each transmitter controls the interference that it imposes by leveraging the partially known CSI. Concurrently, each transmitter adapts one layer to every possible channel state at its intended receiver, overcoming the partial uncertainty about the other transmitter's interfering link. Based on these observations, transmitter $i$ splits its information stream into a certain set of codebooks depending on the state of its outgoing {\em cross} channel. We denote the set of codebooks transmitted by user $i$ when $a_j=\sqrt{\beta_k}$ by $\C_i^k$. Each set $\C_i^k$ consists of $K + 1$ codebooks given by 
\begin{align}\label{eq:setC}
\C_i^k = \{ V_i^k, U^k_i(1), \dots, U^k_i(K)\} \, .
\end{align}
The layers denoted by $V^k_i$ are adapted to be successfully decoded by both receivers in all network states. Additionally, layers $U^k_i(s), \forall s \in \{1, \dots, K\}$ are specifically adapted to be opportunistically decoded by the intended receiver $y_{i}^s$ only.  
In particular, when transmitter $i$ knows the state of its outgoing channel to be $\beta_k$, it splits its information stream into $K+ 1$  layers specified in~\eqref{eq:setC} where 
\begin{itemize}
\item layer $V^k_1$ ($V^k_2$) is adapted to the {\sl cross} channel state at the {\sl unintended} receiver $y^k_{1}$ ($y^k_2$); and
\item Layer $U^k_1(s)$ ( $U^k_2(s)$) is adapted to the {\sl cross} channel state at the {\sl intended} receiver $y_2^{s}$ ($y_1^{s}$), for $s \in \{1, \dots ,K\}$.
\end{itemize}  
Note that the sets $\{\C_i^k\}_{i,k}$ contain the same number of codebooks corresponding to each user $i$ and outgoing channel state $\beta_k$. However, the motivation behind using $K$ different sets is that the power allocation among the layers in each set is distinct. Given that transmitter $i$ adapts the layers $V^k_i$ to each {\em cross} channel state at the {\em unintended} receiver, it directly imposes a constraint on the fraction of power allocated to the remaining layers $\{U^k_i(s)\}$.

\vspace{.1 in}

\noindent{\bf Successive Decoding.} Each codebook will be decoded by multiple receivers in the equivalent network formed by different receivers associated with different network states. Hence, each codebook rate will be constrained by its associated most {\sl degraded} channel state. Furthermore, any undecoded layer at a particular receiver imposes interference, which degrades the achievable rate at that receiver. Motivated by these premises, a simple successive decoding scheme can be designed that specifies (i) the set of receivers at which each layer is decoded, and (ii) the order of successive decoding order at each receiver.

In network state $(\beta_s, \beta_t)$, user $1$ transmits the superposition of all the layers in the set $\C_1^s = \{V_1^s, U_1^s(1), \dots,  U_1^s(K)\}$, while user $2$ transmits the layers in the set $\C_2^t = \{V_2^t, U_2^t(1), \dots,  U_2^t(K)\}$. Accordingly, layers $V_1^s$ and $V_2^t$ are decoded by both receivers $y_1^s$ and $y_2^t$. Further, a subset of the layers $\{U_i^s(t)\}$ from user $i$ are opportunistically decoded by the {\em intended} receiver only depending of the interfering channel state $\beta_t$ from transmitter $j \neq i$. For state $(\beta_s, \beta_t)$, we summarize the decoding order at each receiver $y_i^s$  as follows:
\begin{itemize}
\item \textbf{ Receiver $y_1^{s}$:} First, it decodes one layer from the {\em unintended} transmitter $V_2^t$ and remove it from the received signal. Secondly, it decodes the baseline layer from its {\em intended} transmitter $V_1^s$. Finally, depending on the network state $(\beta_s, \beta_t)$, it successively decodes all the layers $\{U_1^s(1), \dots, U_1^s(t)\}$.

\item \textbf{ Receiver $y_2^{t}$:} First, it decodes one layer from the {\em unintended} transmitter $V_1^s$ and remove it from the received signal. Secondly, it decodes the baseline layer from its {\em intended} transmitter $V_2^t$. Finally, depending on the network state $(\beta_s, \beta_t)$, it successively decodes all the layers $\{U_1^t(1), \dots, U_1^t(s)\}$.

\end{itemize}

\subsubsection{Two-user Interference Channel with Partial CSIT -- Scenario 2}

\noindent{\bf State-dependent Adaptive Layering.}  In contrast to Scenario~1, in this scenario,  transmitter $i$ knows $a_i$, and it is oblivious to the other channel. Lacking the extent of interference that each transmitter causes, transmitter $i$ adapts multiple layers with different rates such that the unintended receiver opportunistically decodes and removes a part of the interfering according to the actual state of the channel. Simultaneously, transmitter $i$ adapts the encoding rate of a single layer to be decoded only by its intended receiver based on the actual state of channel $a_2$. Based on this vision, transmitter $i$ splits its information stream into a distinct set of codebooks corresponding to each state of the {\em cross} channel at its intended receiver. We denote the set of codebooks transmitted by user $i$ when $a_i = \sqrt{\beta_k}$ by $\D_i^k$. Each set $\D_i^k$ consists of $K + 1$ codebooks given by 
\begin{align}\label{eq:setD}
\D_i^k = \{ V_i^k (1), \dots, V^k_i (K), U^k_i\} \, .
\end{align}
The layers denoted by $\{V^k_i (s)\}_{s = 1}^K$ are adapted to be fully decoded by transmitter $i$ and partially decoded by receiver $j$, depending on the actual network state. Contrarily, layers $U^k_i$ are specifically adapted to be decoded by the intended receiver $y_{i}^s$ only.   In particular, when transmitter $i$ knows the state of the interfering link to its {\em intended} receiver to be $\beta_k$, it splits its information stream into $K + 1$ encoded layers specified in~\eqref{eq:setD} where 
\begin{itemize}
\item Layer $V^k_1 (s)$ ($V^k_2 (s)$) is adapted to the {\sl cross} channel state at the {\sl unintended} receiver $y^s_{2}$ ($y^s_{1}$), for $s \in \{1, \dots,K\}$; and
\item Layer $U^k_1$ ($U^k_2$) is adapted to the {\sl cross} channel state at the {\sl intended} receiver $y_1^{k}$ ($y_2^{k}$).
\end{itemize}  
Similarly to the layering approach in Scenario~`11, sets $\D_i^k$ contain an equal number of codebooks for each $i\in\{1,2\}$ and $k\in\{1,\dots,K\}$. Nevertheless, power allocation schemes among the layers in each set are distinct.

\vspace{.1 in}
\noindent{\bf Successive Decoding.} Given that each codebook will opportunistically be decoded by multiple receivers, its maximum achievable rate is constrained by the most {\sl degraded} network state in which it is decoded. Similarly to that of Scenario~1, a successive decoding scheme is devised that specifies (i) the set of receivers at which each layer is decoded, and (ii) the order of successive decoding order at each receiver.

In network state $(\beta_s, \beta_t)$, user $1$ transmits the superposition of all the layers in the set $\D_1^s = \{V_1^s (1),  \dots, V_1^s(K),  U_1^s\}$, while user $2$ transmits the layers in the set $\D_2^t = \{V_2^t(1),  \dots, V_2^t(K),  U_2^t\}$. Layers $U_1^s$ and $U_2^t$ are decoded by both receivers $y_1^t$ and $y_2^s$. On the other hand, a subset of the layers $\{V_i^s (k)\}$ from user $i$ are opportunistically decoded by the {\em unintended} receiver $j$ depending on the interfering channel state. In network state $(\beta_s, \beta_t)$, we summarize the decoding order at each receiver $y_i^s$  as follows:
\begin{itemize}
\item \textbf{Receiver $y_1^{s}$:} First, it decodes one layer from the interfering signal $V^t_2(1)$. Afterwards, it decodes one layer from the intended signal $V^s_1(1)$. This receiver continues the decoding process in an alternating manner until codebooks $\{V^t_2(j)\}_{j = 1}^s$ from transmitter $2$ and codebooks $\{V^s_1(j)\}_{j = 1}^K\}$ are decoded from the intended receiver $1$ are successfully decoded. Finally, the last remaining layer fro the intended message $U^s_1$ is decoded.

\item \textbf{ Receiver $y_2^{t}$:} First, it decodes one layer from the interfering signal $V^s_1(1)$. Afterwards, it decodes one layer from the intended signal $V^t_2(1)$. This receiver continues the decoding process in an alternating manner until codebooks $\{V^s_1(j)\}_{j = 1}^s$ from transmitter $1$ and codebooks $\{V^t_2(j)\}_{j = 1}^K\}$ are decoded from the intended receiver $2$ are successfully decoded. Lastly, the last remaining layer fro the intended message $U^t_2$ is decoded.
\end{itemize}

\section{Relay Channels}
\label{sec:relay}

\subsection{Overview}
In this section, we extend the discussions to two-hop networks, in which a source and a destination communicate while being assisted by relay nodes.  In line with the key assumptionthroughout this paper (i.e., no CSIT), we assume that the transmitter and the relay(s) are oblivious to instantaneous realizations of their outgoing channels. Such settings are especially relevant when a source communicates with a remote destination, and a relay terminal is occasionally present near the source but without the source's knowledge~\cite{Katz05,Katz06,Katz07,Katz09}.

We start the discussion with a two-hop network model in Section~\ref{Sec5_1}, where the source and destination communicated through a relay node (no direct source-destination communication). In this section, we review various decode-and-forward (DF), amplify-and-forward (AF), quantize-and-forward (QF), and amplify-and-quantize-and-forward (AQF) relaying schemes and characterize their attendant average communication rates, based on the results in \cite{AsSh06TwoHop}. The work in \cite{AsAsSh07} considers the problem of communication between a single remote transmitter and a destined user while being helped by co-located users. This problem is discussed in more detail in Section \ref{Sec5_2}.   In a dynamic wireless network where a source terminal communicates with a destination, it is worth considering oblivious relaying strategies of a relay near the source transmitter in the presence of other users in the network \cite{BraginskiyAsSh12}, as discussed in Section \ref{Sec5_3}. In Section~\ref{sec:diamond}, we review the broadcast transmission schemes of the diamond channel investigated in~\cite{Zamani14}.
Motivated by addressing the distributed nature and delay sensitivity of modern communication systems, the study in~\cite{SimoneSh09} investigates a network consisting of a source-destination pair, the communication between which is assisted by multiple relays. This setting is reviewed in Section~\ref{sec:multirelay}.  Finally, motivated by the fact that in practical wireless networks, it is often difficult for each user to keep track of the relay nodes, in Section~\ref{sec:occasionally}, we review the settings in which the relays are available only occasionally.

\subsection{A Two-Hop Network}\label{Sec5_1}

Let us consider a two-hop relay fading channel \cite{AsSh06TwoHop}, where the transmitter and receiver communicate through an intermediate network node that serves as a relay. Various relaying protocols and broadcasting strategies are considered. For example, DF relaying, a simple relay with a single packet buffer, which cannot reschedule retransmissions, is first studied. DF relaying limitations in various cases give rise to consideration of other relaying techniques, such as AF, where a maximal broadcasting achievable rate is analytically derived. A QF relay, coupled with a single-level code at the source, uses codebooks matched to the received signal power and performs optimal quantization. This is simplified by an AQF relay, which performs scaling, and single codebook quantization on the input. As discussed later in this section, it is observed that the latter may be throughput-optimal on the relay-destination link while maintaining a lower coding complexity compared with the QF setting.

The work in \cite{Baghani16} concerns two-hop transmissions over relay-assisted block fading channels, assuming there is no direct link between the transmission ends and the communication is carried out by a relay. Various relaying strategies are considered in combination with the multi-layer coded transmission. The study in \cite{Akhlaghi19} optimizes the power allocation for relaying strategies in a similar two-way relay setting. The work in \cite{Attia14} considers a two-layer transmission and optimizes the power allocation for a DF relay.

In a DF \cite{JNL04} scheme, the relay decodes
the received source message; re-encodes it; and forwards the
resulting signal to the destination. Note that, since the relay must
perfectly decode the source message, the achievable rates are
bounded by the capacity of the channel between the source and the relay.
A non-regenerative relay has a different coding scheme than the
source, and it can improve, for example, the reliability of the
relay-destination transmission. The work in \cite{DENIZ05} compares
two DF protocols assuming knowledge of channel gains at the
transmitter and adhering to delay-limited capacity. Further work on user cooperation to increase diversity gains, using DF cooperation techniques over a
Rayleigh fading channel is found in \cite{YUKSEL04}.

In \cite{BOYER04}, different types of AF relay settings are studied
and general expressions for the aggregate SNR at the destination
are derived for a varying number of relaying nodes. This study 
is motivated by previous observations that AF relays
can sometimes approach or exceed the performance of their
DF counterparts \cite{JNL04}.

A QF relay implementation is considered in
\cite{Katz06} and it is shown to be superior to the DF and AF, in terms of average
throughput in the presence of a direct link and a known channel
gain on the relay-destination link, which models a two co-located
user cooperation. Practical compress-and-forward (CF) code design was
presented in \cite{LIU05} for the half-duplex relay channel. The
quantization in \cite{Katz06,LIU05} is of Wyner-Ziv (WZ) quantization
type \cite{WYNER76}, which refers to the relay quantizing its
received observation of the source symbol while relying on the side
information that is available at the destination receiver, from
the direct link. In a two-hop relay setting, the receiver has no
additional side information, and thus the quantization applied at
the relay is a standard quantization of a noisy Gaussian source
\cite{BERGER71}. Consider the following SISO channel:
\begin{align}\label{S_1}
\textbf{y}_r\; = \;  h_s \textbf{x}_s \; + \;  \textbf{n}_s ~ ,
\end{align}
where $\textbf{y}_r$ is a received vector of length $N$ at the relay, which is also the transmission block length, $ \textbf{x}_s $ is the transmitted vector, and $ \textbf{n}_s $ is the additive noise
vector, with elements that are complex Gaussian i.i.d. with zero
mean and unit variance denoted ${\mathcal{CN}}(0,1)$. $h_s$ is
the (scalar) fading coefficient, assumed to be
perfectly known at the relay and the destination receivers only.
The source transmitter has no CSI. The
power constraint at the source is given by $P_s=\bbe[|x_s|^2]$.  The channel between the relay and the destination is described by
\begin{align}\label{S_2}
\textbf{y}_d \; = \; h_r \textbf{x}_r \;+ \;  \textbf{n}_r ~ ,
\end{align}
where $ \textbf{y}_d $ is a received vector of length $N$ at the destination receiver, and $ \textbf{x}_r $ is the relay transmitted vector. $
\textbf{n}_r $ is the additive noise vector, with elements that
are complex Gaussian i.i.d. with zero mean and unit variance
denoted by ${\mathcal{CN}}(0,1)$, and $h_r$ is the (scalar) fading
coefficient. The fading coefficients $h_s$ and $h_r$ are assumed to
be perfectly known at the destination receivers only. The relay
transmitter does not possess $h_r$. The power constraint at the
relay is given by $P_r=\bbe[|x_r|^2]$.

It is assumed that the relay operates in a full-duplex mode by
receiving and transmitting on different frequency bands, realizing
a two-hop network. Furthermore, the relay is not capable of
buffering data. In the DF protocols, the relay has to forward all
the data successfully decoded immediately. Layers that were not
decoded on the path from source to destination must be rescheduled
for retransmission at the source. If the relay had packet
scheduling capabilities, the DF protocols could be improved by
letting the relay perform retransmission of layers that are not
decoded at the destination. However, this calls for distributed
scheduling control, which highly complicates the system and is
beyond the scope of this subsection.

\subsubsection{Upper Bounds}

A full CSI (FCSI) upper bound is derived for a hypothetical case that
both source and relay have perfect CSI of all links, and the
source always transmits in the maximal achievable rate over this
relay channel. This achievable rate is the minimal rate determined
by the fading gain realizations on both links. It is generally
expressed by
\begin{align}\label{S_6}
C_{\rm FCSI}=\bbe_{s_s,s_r}[ \log(1+\min(P_ss_s,P_rs_r))]\ ,
\end{align}
where $s_s=|h_s|^2$, and $s_r=|h_r|^2$. By explicitly extracting the expectation in
(\ref{S_6}) we get
\begin{align}\label{S_7}
C_{\rm FCSI}&=\int\limits_0^\infty \d\nu \int\limits_0^\infty \d\mu
f(\nu)f(\mu) \log(1+\min(P_s\nu,P_r\mu))\\
&=\int\limits_0^\infty \d\nu
\int\limits_{\frac{P_s}{P_r}\nu}^\infty \d\mu f(\nu)f(\mu)
\log(1+P_s\nu)\\
&\qquad +  \int\limits_0^\infty \d\nu f(\nu)\int\limits_{0}^{\frac{P_s}{P_r}\nu} \d\mu f(\mu)
\log(1+P_r\mu)\\
&=\int\limits_0^\infty \d\nu f(\nu)(1-F(\frac{P_s}{P_r}\nu))
\log(1+P_s\nu)\\
&\qquad + \frac{P_s}{P_r}\int\limits_0^\infty \d\nu
(1-F(\nu)) f(\frac{P_s}{P_r}\nu) \log(1+P_s\nu)\ ,
\end{align}
where $f(x)$ and $F(x)$ are the PDF and CDF of the fading gain, respectively. For a Rayleigh fading channel, the FCSI upper bound is given by
\begin{align}
\label{S_8}
C_{\rm FCSI} & =  (1+ \frac{P_s}{P_r})\int\limits_0^\infty \d\nu
e^{-(1+\frac{P_s}{P_r})\nu} \log(1+P_s\nu)\\
&= e^{\frac{P_s+P_r}{P_rP_s}}E_1\myround{\frac{P_s+P_r}{P_rP_s}}\ ,
\end{align}
where $E_1(x)$ is the exponential integral function
$E_1(x)\triangleq\int_x^\infty \d t \frac{e^{-t}}{t}$ for $x\geq 0$
\cite{Abramowitz65}. The ergodic cut-set upper bound is the minimum of the average
achievable rates on the two links (source-relay and
relay-destination). This is specified by
\begin{align}\label{S_8_1}
C_{\rm erg}=\min \{ \bbe_{s_s} [\log(1+P_ss_s)]\; ,\;  \bbe_{s_r}
	[\log(1+P_rs_r)] \}\ .
\end{align}
For Rayleigh fading channels, and similar fading gain distribution
functions $f(x)$ for the two links, the ergodic upper bound
simplifies to
\begin{align}\label{S_8_2}
C_{\rm erg}=\int\limits_0^\infty \d\nu e^{-\nu}
\log(1+P\nu)\ ,~~~~\textrm{s.t.}\qquad  P=\min(P_s,P_r)\ ,
\end{align}
which is justified by the monotonicity of the ergodic capacity as
function of $P$. A tighter upper bound on the broadcast strategy
is the broadcasting cut-set bound. This is the minimum average
broadcasting rate achievable on each of the links separately. It
is specified by
\begin{align}\label{S_8_3}
{R_{\rm bs-cutset}} =  \min\mytwist{\int\limits_0^\infty
	\d u~f_\mu(u)R_\mu(u)\; ,~\int\limits_0^\infty \d u~f_\nu(u)R_\nu(u)}\ ,
\end{align}
where $f_\nu(u)$ and $f_\mu(u)$ are the PDFs of the source-relay
and relay destination fading gains, respectively, and $R(u)$ is
the broadcasting achievable rate for a fading gain $u$. For a
Rayleigh fading channel with similar distribution on both links,
the cut-set bound is given by \cite[equation (18)]{ShitzSteiner03}
\begin{align}\label{S_8_4}
R_{\rm bs-cutset} = 2E_1(s_0)-2E_1(1)-(e^{-s_0}-e^{-1})\ ,
\end{align}
where 
\begin{align}
s_0=\frac{2}{1+\sqrt{1+4\min(P_s,P_r)}}\ .
\end{align}
 The broadcasting
cut-set bound (\ref{S_8_4}) may be achieved if the relay is
allowed to delay its data and reschedule retransmissions
independently. Furthermore, the relay has to inform the source how
many layers were decoded for every block. We do not assume such
feedback is available. The only feedback, in our channel model, is
from destination to source indicating the number of successfully decoded layers.

\subsubsection{DF Strategies}
Consider first the simple DF relaying for an outage approach with single-level coding at source and relay.
In single-level coding, the code rate from the source transmitter
to the relay is determined by the fading gain threshold selected.
For a power threshold $s_s$, the code rate is $R =
\log\myround{1+P_ss_s}$, and this same rate is transmitted from
the relay to the destination with power $P_r$, thus $R =
\log\myround{1+P_rs_r}$, and $s_r = \frac{P_s}{P_r}s_s$. Hence, the
average achievable rate from the source to the destination is
\begin{align}
\label{S_3}
R_{1,\rm ave} &= \PP(\nu>s_s)\PP(\mu>s_r)\log(1+P_ss_s)\\
&= (1-F_\nu(s_s))(1-F_\mu(s_r))\log(1+P_ss_s)\ ,
\end{align}
where $\nu$, $\mu$ are the fading gain random variables,
$F_\nu(x)$, $F_\mu(x)$ are the corresponding CDFs, and $P_s$ is the source transmission power. For a
Rayleigh fading channel, with $F_\mu(x) =F_\nu(x)=1-e^{-x}$, the
average rate is given by
\begin{align}\label{S_4}
R_{1,\rm ave}=e^{-s_s}e^{-\frac{P_s}{P_r}s_s}\log(1+P_ss_s)\ ,
\end{align}
and the maximal achievable rate is thus
\begin{align}\label{S_5}
R_{1L}=\max\limits_{s_s}
e^{-s_s}e^{-\frac{P_s}{P_r}s_s}\log(1+P_ss_s)\ .
\end{align}
Let the source perform two-level coding, and the relay has to decode as many layers as possible, depending on the fading realization. If successful in decoding both layers,
it transmits a single-level code at a rate that is the sum of
source rates. If only one layer was decoded successfully at the
relay, it encodes it into a different single-level code, which is
equal in rate to the first level of the source channel code. This
gives higher flexibility in decoding of a single layer at the
destination receiver when the channel conditions on the
source-relay link allow only one layer detection at the relay. The
channel code rate at the source is given by
\begin{align}
\label{S_13}
R_1^s & =\log(1+P_ss_{s,1}) - \log(1+(1-\alpha_s)P_ss_{s,1})\ ,\\
R_2^s & =\log(1+(1-\alpha_s)P_ss_{s,2})\ ,
\end{align}
where $0\leq \alpha_s\leq 1$, $s_{s,1}$ and $s_{s,2}$ are the
fading gain thresholds implicitly specifying the layering rates.
The rates of the single-level code at the relay are then given by
\begin{align}\label{S_14}
R_1^r & =\log(1+P_rs_{r,1})\ , \quad \textrm{s.t.} \quad  R_1^r=R_1^s\ ,\\
R_2^r & =\log(1+P_rs_{r,2})\ , \quad \textrm{s.t.} \quad R_2^r=R_1^s+R_2^s\ ,
\end{align}
where $s_{r,1}$ and $s_{r,2}$ are determined from the
rate equalities on the right hand side of (\ref{S_14}). The
overall average rate is then
\begin{align}
\label{S_15}
\nonumber {R_{2-1,\rm ave}} & =\max\limits_{s_{s,1},s_{s,2},\alpha_s}
\PP(s_{s,1}\leq\nu<s_{s,2})P(\mu>s_{r,1})R_1^s\\
&\qquad\qquad + \PP(\nu>s_{s,2})P(\mu>s_{r,2})(R_1^s+R_2^s)\\
\nonumber &=\max\limits_{s_{s,1},s_{s,2},\alpha_s}
\myround{F_\nu(s_{s,2})-F_\nu(s_{s,1})}(1-F_\mu(s_{r,1}))R_1^s\\
& \qquad\qquad + 
\myround{1-F_\nu(s_{s,2})}(1-F_\mu(s_{r,2}))(R_1^s+R_2^s)\ ,
\end{align}
where $\nu$ is the fading gain RV on the source-relay link, and
$\mu$ is the RV on the relay destination link.  This approach outperforms single-level coding at the source
and two-level coding at the relay, described in the previous
subsection. The main reason for this difference is that the outage
approach described here adapts to the source-relay channel
conditions. That is, the outage rate from the relay to the destination is
equal to the decoded rate and depends on the number of
successfully decoded layers (\ref{S_14}). However, when considering
the opposite approach (source: outage, relay: two-level), the
outage rate is fixed for all channel conditions, and if the relay
fails in its decoding, nothing is transmitted to the destination.

\subsubsection{Continuous Broadcasting DF Strategies}

\noindent {\bf{Coding Scheme I -- Source: Outage \& Relay: Continuum Broadcasting.}} In this coding scheme, the source transmitter performs single-level
coding. Whenever channel conditions allow decoding at the relay,
it performs continuum broadcasting, as described in the previous
subsection. Thus, the received rate at the destination depends on
the instantaneous channel fading gain realization on the
relay-destination link. Clearly, a necessary condition for
receiving something at the destination is that channel conditions
on the source-relay link will allow decoding. The source
transmission rate is given by
\begin{align}\label{O_1}
R_1^s=\log(1+P_ss_s)\ ,
\end{align}
and the corresponding achievable rate at the destination is given
by
\begin{align}\label{O_2}
R^r(\nu) = \int_0^\nu\frac{u\rho_r(u)\d u}{1+uI_r(u)}\ ,
\end{align}
where $I_r(\nu)$ is the residual interference distribution
function and its boundary conditions are stated in
(\ref{SISO4})-(\ref{SISO5}). The total rate transmitted in the
broadcasting link is equal to the single-level code rate of the
source-relay link, that is
\begin{align}\label{O_3}
R_1^s = \int\limits_0^\infty\frac{u\rho_r(u)\d u}{1+uI_r(u)}\ .
\end{align}
The above condition in (\ref{O_3}) states a constraint on the
optimization of the average rate. The average rate expression,
considering the transmission scheme on the two links is
\begin{align}\label{O_4}
R_{\rm ave} &= \PP(\nu > s_{s})\int\limits_0^\infty \d x f_\mu(x)
\int\limits_0^x\frac{u\rho_r(u)\d u}{1+uI_r(u)} \\
&=(1-F_\nu(s_s))\int\limits_0^\infty \d x (1-F_\mu(x))
\frac{x\rho_r(x)}{1+xI_r(x)}\ ,
\end{align}
where we have used partial integration rule. The average rate
maximization problem can now be posed as
\begin{align}
\label{O_5}
\displaystyle {R_{1-\rm bs,\rm ave}} = \left\{
\begin{array}{ll}
 \max\limits_{s_s,I_r(\nu)} & \displaystyle 
(1-F_\nu(s_s))\int\limits_0^\infty \d x (1-F_\mu(x))
\frac{x\rho_r(x)}{1+xI_r(x)}\\
\textrm{s.t.} & \displaystyle \int\limits_0^\infty\frac{u\rho_r(u)du}{1+uI_r(u)} =
\log(1+P_ss_s)
\end{array}\right. \ . 
\end{align}
As a first step in solving the maximal average rate, the residual
interference distribution $I_r(\nu)$ is found for every $s_s$.
That is
\begin{align}
\label{O_6}
{R_{1-\rm bs}(s_r)} & = \left\{
\begin{array}{ll}
\max\limits_{I_r(\nu)} & \displaystyle \int\limits_0^\infty \d x (1-F_\mu(x)) \frac{x\rho_r(x)}{1+xI_r(x)}\\
 \textrm{s.t.}
& R_1^s = \displaystyle \int\limits_0^\infty\frac{u\rho_r(u)\d u}{1+uI_r(u)}\\
\end{array}\right. \\
& = \left\{
\begin{array}{ll}
\max\limits_{I_r(\nu)} & \displaystyle \int\limits_0^\infty \d xG_1(x,I_r(x),I_r'(x))\\
 \textrm{s.t.}
& R_1^s = \displaystyle \int\limits_0^\infty \d xG_2(x,I_r(x),I_r'(x))
\end{array}\right. \ ,
\end{align}
where $I_r'(x)=\frac{\d I_r(x)}{\d x}$. The necessary condition for
extremum in (\ref{O_6}) subject to the subsidiary condition, is in
generally stated \cite{GF91}
\begin{align}\label{O_7}
G_{1,I_r} + \lambda G_{2,I_r} - \frac{d}{\d x}\myround{G_{1,I_r'} +
	\lambda G_{2,I_r'}} ~= ~0\ ,
\end{align}
where $G_{1,I_r}$ is the derivative of $G_{1}$ with respect to $I_r$, and
$G_{1,I_r'}$ is the derivative of $G_{1}$ with respect to  $I_r'$. The scalar
$\lambda$ is also known as a Lagrange multiplier, and it is
determined  from the subsidiary condition in
(\ref{O_6}). The substitution of $S_{I_r}\triangleq G_{1,I_r} + \lambda
G_{2,I_r}$, and $S_{I_r'}\triangleq G_{1,I_r'} + \lambda G_{2,I_r'}$ by
using (\ref{O_6}) results in
\begin{align}\label{O_8}
S_{I_r} &= \frac{x^2I_r'\myround{1-F_\mu + \lambda}}{(1+xI_r)^2}\ ,\\
S_{I_r'}& = \frac{-x\myround{1-F_\mu + \lambda}}{1+xI_r}\ ,\\
\frac{\d S_{I_r'}}{\d x}&=\frac{(x^2I_r'-1)\myround{1-F_\mu +
		\lambda}}{(1+xI_r)^2}\ .
\end{align}
Substituting  the expressions in (\ref{O_8}) into the extremum
condition in (\ref{O_7}) yields a general solution for the
residual interference, as function of the Lagrange multiplier $\lambda$. This is summarized in the following proposition.
\begin{Proposition}
The relay broadcasting residual power distribution function $I_r(x)$ that maximizes the expected rate over the two-hop wireless fading channel \eqref{S_1}-\eqref{S_2} is given by
\begin{align}\label{O_9}
I_r(x)=\mycase{
	\begin{array}{ll}
	P & 0\leq x \leq x_0\\
	&\\
	\frac{1-F_\mu(x)+\lambda-xf_\mu(x)}{f_\mu(x)x^2} & x_0\leq x \leq x_1\\
	&\\
	0 & x \geq x_1
	\end{array}}\ ,
\end{align}
where $x_0$ and $x_1$ are determined from the boundary conditions
$I_r(x_0)=P$ and $I_r(x_1)=0$, respectively. The scalar $\lambda$
is determined from the subsidiary condition in (\ref{O_6}). 
\end{Proposition}
When
considering a Rayleigh flat fading channel for the relay
destination link, i.e., $F_\mu(x)=1-\exp(-x)$, the residual
interference distribution gets the following form
\begin{align}\label{O_10}
I_r(x)= \frac{\lambda}{e^{-x}x^2}+\frac{1}{x^2}-\frac{1}{x}\ ,
\qquad \textrm{for}\quad  x_0\leq x \leq x_1\ ,
\end{align}
and the condition $I_r(x_1)=0$ provides
\begin{align}\label{O_11}
x_1 = 1-W_L(-\lambda e)\ ,
\end{align}
where $W_L(x)$ is the Lambert W-function, also called the omega
function, and it is the inverse of the function $f(W)=We^W$.
Interestingly, the subsidiary condition with (\ref{O_10}) as the
solution for $I_r(x)$ yields a simplified expression
\begin{align}
\label{O_12}
R_T&=\int\limits_{x_0}^{x_1}\frac{u\rho_r(u)\d u}{1+uI_r(u)} \\
&=2\log(x_1)-x_1 - (2\log(x_0)-x_0)\\
&= 2\log(1-W_L(-\lambda e)) - 1+W_L(-\lambda e)\\
&- 2\log(x_0)+x_0\ ,
\end{align}
where (\ref{O_11}) is used for substitution of $x_1$. Using the
subsidiary condition (\ref{O_6}), i.e., $R_T=R_1^s$, the solution
of $x_0$ as function of $\lambda$ is
\begin{align}
\label{O_13}
x_0 = & -2W_L \myround{-0.5e^{\log(1-W_L(-\lambda e)) - 0.5+0.5W_L(-\lambda e) - 0.5R_1^s}}\ .
\end{align}
Finally, $\lambda$ can be found by solving $I_r(x_0)=P$. Thus, all initial conditions are satisfied, the solution
for $\lambda$ is obtained by numerically solving the nonlinear
equation specified by $I_r(x_0)=P$. The maximal rate
$R_{1-\rm bs,\rm ave}$ is then obtained by searching numerically over
$s_s$ and evaluating $R_{1-\rm bs,\rm ave}$ for all $s_s$ in the search. \vspace{.1 in}

\noindent {\bf{Coding Scheme II --  Source: Continuum Broadcasting, Relay: Outage.}} In this coding scheme, the source transmitter performs continuum
broadcasting, as described in the previous subsection. The relay
encodes the successfully decoded layers into a single-level block
code. Thus, the rate of each transmission from the relay depends on
the number of layers decoded. For a fading gain realization $\nu$
on the source-relay link the decodable rate at the relay is
\begin{align}\label{S_16}
R^s(\nu) = \int_0^\nu\frac{u\rho_s(u)\d u}{1+uI_s(u)} \ .
\end{align}
This is also the rate to be transmitted in a single-level coding
approach, yielding
\begin{align}\label{S_17}
R_1^r(\nu)=\log(1+P_rs_r(\nu))\ ,
\end{align}
where $s_r(\nu)$ is the fading gain threshold for decoding at the
destination. In order to ensure equal source and relay
transmission rates, it is required that $R_1^r(\nu)=R^s(\nu)$. The
average rate is then given by
\begin{align}
\label{S_18}
{R_{\rm bs-1,ave}}
&= \max\limits_{I_s(x)} \int\limits_0^\infty \d x
\PP(\mu\geq s_r(x))f_\nu(x) R^s(x)\\
&=\max\limits_{I_s(x)} \int\limits_0^\infty \d x (1-F_\mu(s_r(x)))
f_\nu(x) \int_0^x\frac{u\rho_s(u)\d u}{1+uI_s(u)}\\
&=\max\limits_{I_s(x)} \int\limits_0^\infty \d x e^{-x}e^{-s_r(x)}
\int_0^x\frac{u\rho_s(u)\d u}{1+uI_s(u)}\ ,
\end{align}
where a Rayleigh fading distribution is assumed on the last
equality, and
\begin{align}\label{S_19}
s_r(\nu) = \frac{1}{P_r} \myround{\exp\myround{{\int\limits_0^\nu \d x
			\frac{x\rho_s(x)}{1+xI_s(x)}}} -1}\ .
\end{align}
As may be noticed from (\ref{S_19}), the functional subject to
optimization in (\ref{S_18}) does not have a localization property
\cite{GF91}, and thus, it does not have a standard Euler-Lagrange
equation for an extremum condition.\vspace{.1 in}

\noindent {\bf{Coding Scheme III --  Source and Relay: Continuous Broadcasting.}} In this scheme, both source and relay perform the optimal continuum
broadcasting. The source transmitter encodes a continuum of layered
codes. The relay decodes up to the maximal decodable layer. Then it
retransmits the data in a continuum multi-layer code matched to
the rate that has been decoded last. In this scheme, the source
encoder has a single power distribution function, which depends
only on a single fading gain parameter. The relay uses a power distribution that depends on the two fading gains on the source-relay and the
relay-destination links.

In general, the source channel code rate as a function of the fading
gain is the same one specified in (\ref{S_16}). The rate
achievable at the destination is then given by
\begin{align}
\label{B_1}
R^s(\nu,\mu) = \int_0^\mu\frac{u\rho_r(\nu,u)\d u}{1+uI_r(\nu,u)}\ .
\end{align}
The maximal average rate is then specified by
\begin{align}
{R_{\rm bs-bs,ave}}=\left\{
\begin{array}{ll}
 \max\limits_{I_s(\nu),I_r(\nu,\mu)} & 
\displaystyle \int\limits_0^\infty \d x\int\limits_0^\infty \d y
f_\nu(x)f_\mu(y) \int\limits_0^y \frac{u\rho_r(x,u)du}{1+uI_r(x,u)}\\
 \textrm{s.t. } &\displaystyle \int\limits_0^\infty \frac{u\rho_r(x,u)\d u}{1+uI_r(x,u)}  \leq R^s(v)
\end{array}\right. \ ,
\end{align}
which may be simplified into
\begin{align}
\label{B_3}
{R_{\rm bs-bs,ave}}=\left\{
\begin{array}{ll}
 \max\limits_{I_s(\nu),I_r(\nu,\mu)} & 
\displaystyle\int\limits_0^\infty \d x\int\limits_0^\infty \d y
f_\nu(x)(1-F_\mu(y))  \frac{y\rho_r(x,y)}{1+yI_r(x,y)}\\
 \textrm{s.t. } & \displaystyle\int\limits_0^\infty \frac{u\rho_r(x,u)\d u}{1+uI_r(x,u)} \leq R^s(v)
\end{array}\right. \ .
\end{align}
In order to present an Euler-Lagrange equation here, the
subsidiary condition in (\ref{B_3}) still has to be brought to a
functional form, and then it could be solved with the aid of the
Lagrange multipliers.

\subsubsection{AF Relaying}

Consider a relay that cannot decode and encode the data, but rather it can only
amplify the input signal. The channel model to consider here is
the same one specified in (\ref{S_1})-(\ref{S_2}). However,
it may be assumed that the relay can estimate the input signal power and
amplify the signal (without distortion) by a factor that ensures
maximal transmission $P_r$ from the relay. In such a case, the
amplification coefficient is given by
\begin{align}\label{AF_1}
\gamma=\sqrt{\frac{P_r}{P_s|h_s|^2+1}} \ .
\end{align}
The equivalent received signal at the destination can be specified
by
\begin{align}\label{AF_2}
\textbf{y}'_d = \frac{\gamma h_rh_s}{\sqrt{\gamma^2|h_r|^2+1}}
\textbf{x}_s + \textbf{n}'_r ~\  ,
\end{align}
where $\textbf{n}'_r \sim {\mathcal{CN}}(0,1)$ and the original
source signal is multiplied by a factor, which represents an
equivalent fading coefficient with power
\begin{align}\label{AF_3}
s_{\rm b} = \frac{\gamma^2 s_rs_s}{\gamma^2s_r+1} =
\frac{P_rs_rs_s}{P_rs_r + P_ss_s + 1} ~\ ,
\end{align}
where $s_r = |h_r|^2$ and $s_s = |h_s|^2$, and we have used the
amplification factor definition from (\ref{AF_1}) for explicitly
stating the equivalent fading gain. The CDF of the equivalent
fading gain $s_b$ is then given by
\begin{align}\label{AF_4}
F_{s_{\rm b}}(x) = \PP(s_b<x) = \int\int\limits_{\mathcal{R}}\d x_s\d x_r
f_{s_s}(x_s)f_{s_r}(x_r)\ ,
\end{align}
where $${\mathcal{R}} = \mytwist{x_r,x_s\in[0,\infty) \; \Big| \; 
	\frac{P_rx_rx_s}{P_rx_r + P_sx_s + 1}\leq x }\ .$$ 
	When assuming
a Rayleigh fading channel, i.e.,  $f_{s_r}(x_r)=e^{-x_r}$ and
$f_{s_s}(x_s)=e^{-x_s}$, we have 
\begin{align}\label{AF_5}
F_{s_{\rm b}}(x) &= 1- \int\limits_{\frac{P_s}{P_r}x}^\infty \d x_r
\int\limits_{\frac{x(1+P_rx_r)}{x_rP_r-xP_s}}^\infty \d x_s
e^{-x_s}e^{-x_r} \\
&= 1-\int\limits_{\frac{P_s}{P_r}x}^\infty \d x_r
e^{-x_r-{\frac{x(1+P_rx_r)}{x_rP_r-xP_s}}}\ ,
\end{align}
which does not lend itself to a closed-form expression. In a broadcast approach, the transmitter performs continuous code
layering, matched to the equivalent single fading gain RV. Using
the equivalent channel model (\ref{AF_2}) and using the results of
\cite{ShitzSteiner03}, the average received rate is given by
\begin{align}\label{AF_7}
R_{\rm bs,AF,ave} = \max\limits_{I(x)} \int\limits_0^\infty
\myround{1-F_{s_{\rm b}}(x)}\frac{-xI'(x)}{1+xI(x)}\ ,
\end{align}
where the optimal residual interference distribution $I_{\rm opt}(x)$
is given by \cite{ShitzSteiner03}
\begin{align}\label{AF_8}
I_{\rm opt}(x)=\mycase{
	\begin{array}{ll}
	P & 0\leq x \leq x_0\\
	\frac{1-F_{s_{\rm b}}(x)-xf_{s_{\rm b}}(x)}{f_{s_{\rm b}}(x)x^2} & x_0\leq x \leq x_1\\
	0 & x \geq x_1
	\end{array}}\ ,
\end{align}
where $x_0$ and $x_1$ are determined from the boundary conditions
$I_r(x_0)=P_s$ and $I_r(x_1)=0$, respectively. The maximal achievable rate is provided by the following proposition.
\begin{Proposition}
The maximal achievable expected rate of a two-hop AF-relay network is explicitly given by
\begin{align}
\label{AF_10}
{R_{\rm bs,AF,ave}} =  \int\limits_{x_0}^{x_1} \d x \mat{
	\frac{2(1-F_{s_{\rm b}}(x))}{x} +
	\frac{(1-F_{s_{\rm b}}(x))f_{s_{\rm b}}'(x)}{f_{s_{\rm b}}(x)}}\ ,
\end{align}
where the CDF $F_{s_{\rm b}}(x)$ is specified in (\ref{AF_5}), and thus the
corresponding PDF is given by
\begin{align}
\label{AF_9}
f_{s_{\rm b}}(x)  =  \frac{d}{\d x}F_{s_{\rm b}}(x)
= \int\limits_{\frac{P_s}{P_r}x}^\infty \d x_r
\frac{P_rx_r(1+P_rx_r)}{(P_sx-P_rx_r)^2}e^{-x_r-{\frac{x(1+P_rx_r)}{x_rP_r-xP_s}}}\ .
\end{align}
\end{Proposition}
Finally, $R_{\rm bs,AF,ave}$ (\ref{AF_10}) can be obtained via a
numerical integration.

\subsubsection{AQF Relay and Continuum Broadcasting}
Now, let the source encoder perform  continuum layering, and the
relay, as before, amplifies its input signal, quantizes it with
average distortion $D$, optimally in an MSE sense.  The
destination tries to  first decode the quantized signal $u_q$. Upon
successful decoding, it decodes the multi-level code up to the
highest layer possible, depending on the fading gain on the source
relay link. In this setting, we  consider  single-level quantization. In broadcasting, it may
be assumed that part of the original signal cannot be decoded. Therefore, it is modeled as additive Gaussian noise. The quantized
signal, after suitable amplification and forward channel
conversion, as a function of the source data, is given
by
\begin{align}\label{BAQ_1}
u_q = \beta\gamma h_sx_{s,s} +\beta\gamma h_sx_{s,I} + \beta\gamma
n_s+ n_q'\ ,
\end{align}
where $n_q'$ is the equivalent quantization noise distributed
according to ${\mathcal{CN}}(0,\beta D)$, $\beta=1-\frac{D}{P_r}$,
$\gamma=\sqrt{\frac{P_r}{P_s\nu_s+1}}$ with $\nu_s=|h_s|^2$, and
$x_{s,I}$ represents the residual interference in the decoded
signal. Consider a power distribution $\rho(\nu_s)$ which is the
source power distribution as function of the fading gain. Then the
incremental rate associated with a fading $\nu_s$ is
\begin{align}\label{BAQ_2}
\d R(\nu_s)=\frac{\gamma^2\nu_s\rho(\nu_s)\beta^2\d\nu_s}{\gamma^2 +
	\beta D + \gamma^2\nu_sI(\nu_s)\beta^2}\ ,
\end{align}
which simplifies after substituting $\gamma$ and some algebra
\begin{align}\label{BAQ_25}
\d R(\nu_s)=\frac{\nu_s\rho(\nu_s)\d\nu_s}{1+D_\beta + \nu_s(I(\nu_s)
	+ P_sD_\beta)}\ ,
\end{align}
where the $D_\beta\triangleq \frac{D/P_r}{1-D/P_r}$. Thus, the
average rate attainable, when $u_q$ is successfully decoded, is
\begin{align}\label{BAQ_3}
R_{\rm ave}&=\int\limits_0^\infty \d\nu_s f(\nu_s)\int\limits_0^{\nu_s}
\d R(u)\\
&= \int\limits_0^\infty
(1-F(\nu_s))\frac{\nu_s\rho(\nu_s)\d\nu_s}{1+D_\beta +
	\nu_s(I(\nu_s) + P_sD_\beta)}\\
&=\int\limits_0^\infty (1-F(\nu_s))
\frac{\nu_s\rho_N(\nu_s)\d\nu_s}{1+\nu_sI_N(\nu_s)}\ ,
\end{align}
where the first equality is obtained by solving the integral in
parts. The following relationships follow from the definitions of
the normalized power distribution and residual interference:
\begin{align}\label{BAQ_4}
\rho_N(\nu_s) &  \triangleq \frac{\rho(\nu_s)}{1+D_\beta}\ ,\\
I_N(\nu_s)  & \triangleq \frac{I(\nu_s) + D_\beta P_s}{1+D_\beta}\ ,
\end{align}
that satisfy $\rho_N(\nu_s)=-I_N'(\nu_s)$. For a given
average distortion $D$, $D_\beta$ is also explicitly determined,
and the maximal average rate $R_{\rm ave}$ is achieved for
\begin{align}\label{BAQ_5}
\rho_N(\nu_s) & =\frac{2}{\nu_s^3}-\frac{1}{\nu_s^2}\ ,\\
I_N(\nu_s) & =\frac{1}{\nu_s^2}-\frac{1}{\nu_s}\ ,
\end{align}
on the range of $\nu_s \in [\nu_0,\nu_1]$, where the boundary
conditions are $I_N(\nu_0)=P_s$ and $I_N(\nu_1)=0$. Thus,
the range of the optimal solution is
\begin{align}\label{BAQ_6}
\nu_0 & = \frac{2}{1+\sqrt{1+4P_s}}\ ,\\
\nu_1 & = \frac{2}{1+\sqrt{1+4\frac{P_sD_\beta}{1+D_\beta}}}\ .
\end{align}
This rate is attainable only when the compressed signal may be
decoded at the destination. Otherwise, an outage event occurs, and
nothing can be restored from the original signal. Evidently, the
event of outage depends only on the relay-destination link. Hence,
the average achievable rate for the broadcast-amplify-quantize
(BAQ) approach is formalized in the next proposition.

\begin{Proposition}
	In the system model described by
		(\ref{S_1})-(\ref{S_2}), with $\nu_s$ known to relay and
		destination, and $\nu_r$ known to destination only, the maximal
		average attainable rate in a BAQ
		scheme is specified by
		\begin{align}\label{BAQ_7}
		R_{\rm BAQ,ave} = \max\limits_{D}~ \overline{P}_{\rm out}\cdot
		\int\limits_0^\infty (1-F(\nu_s))
		\frac{\nu_s\rho_N(\nu_s)\d\nu_s}{1+\nu_sI_N(\nu_s)}\ ,
		\end{align}
		where the complementary outage probability is
		\begin{align}\label{BAQ_8}
		\overline{P}_{\rm out} = \PP\myround{\log\frac{P_r}{D} \leq
			\log(1+\nu_rP_r)}\ .
		\end{align}
\end{Proposition}
The complementary outage probability for a Rayleigh fading channel
reduces (\ref{BAQ_8}) into $\overline{P}_{\rm out} =
e^{-\frac{1}{D}+\frac{1}{P_r}}$. Computing  $R_{\rm BAQ,ave}$
can be directly pursued, while optimizing the selection of the
average distortion $D$, and directly computing the average rate
for every $D$. 
\begin{figure}[htb]
	\centering
	\includegraphics[width=5in]{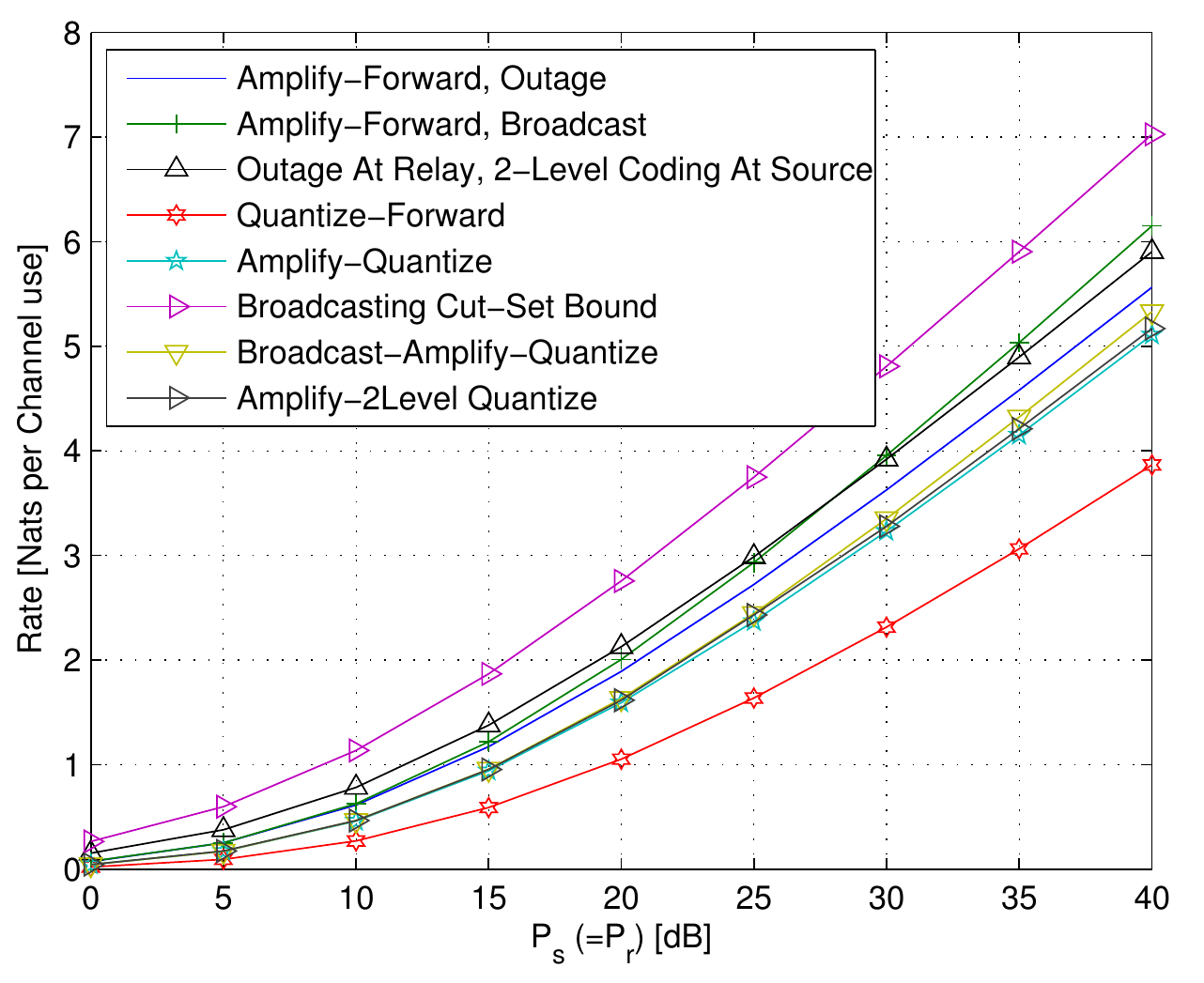}
	\caption{Achievable average rates, for $P_r=P_s$, and for various relaying protocols and
		broadcasting strategies.}\label{fig:rates_all}
\end{figure}

Figure \ref{fig:rates_all} demonstrates the maximal attainable expected
rates for the various relaying protocols. The numerical results correspond to Rayleigh fading channels on
both source-relay and relay-destination links. A comparison of all relaying
protocols (DF, AF, QF, AQF) is provided for equal SNR on both links, i.e., $P_r=P_s$. As
may be noticed, the broadcasting for AF relay has the highest
throughput gains for high SNRs.
The AQF scheme with two levels of refinement at
the relay, shows only a small gain in the overall expected throughput. This questions the possible benefits of higher levels
of successive refinement at the relay, when the source performs only
single-level coding.

\subsection{Cooperation Techniques of Two Co-located Users}\label{Sec5_2}

\begin{figure}[htb]
	\begin{center}
		\includegraphics[width=2.5in]{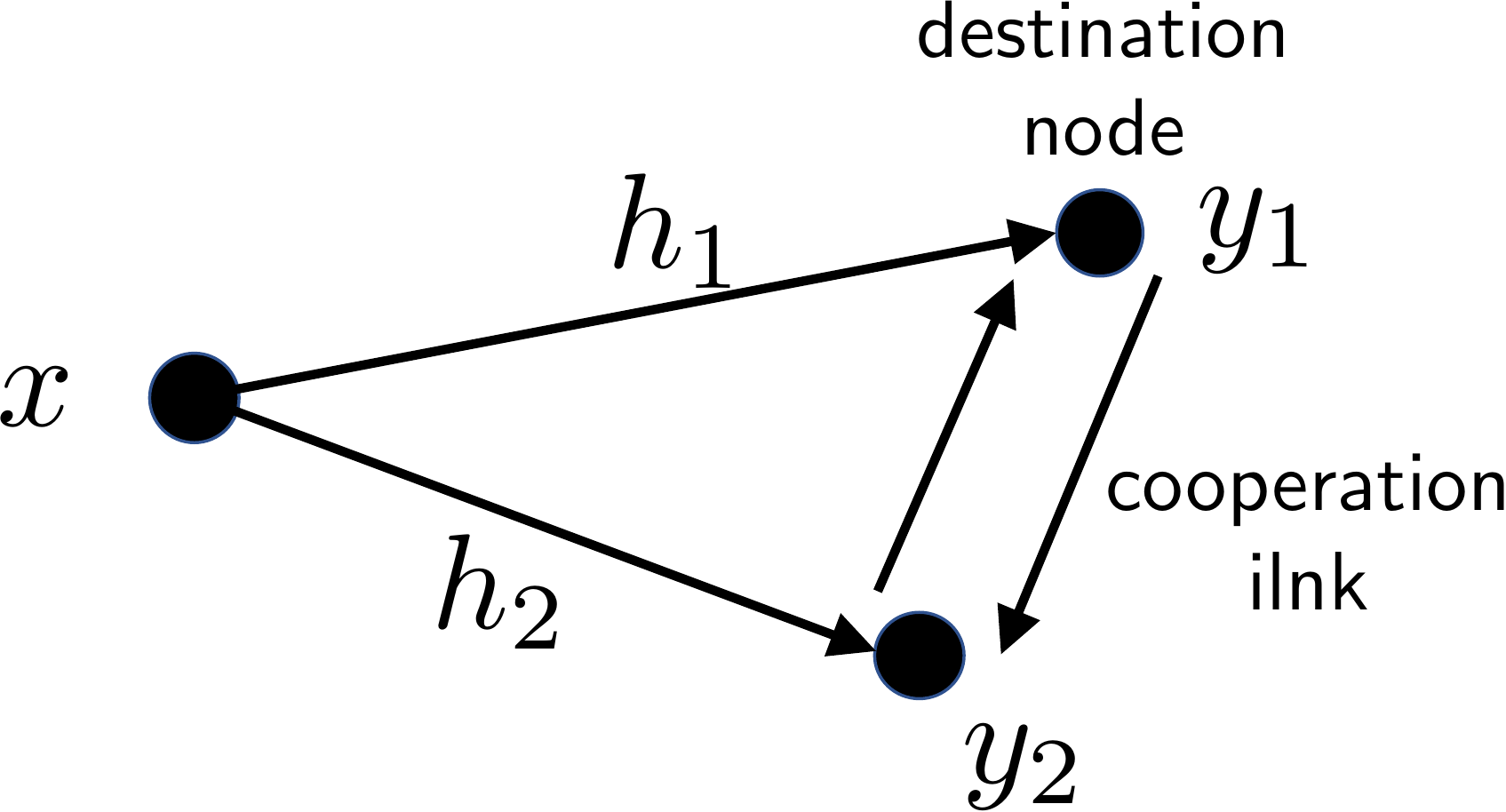}
	\end{center}
	\caption{A schematic diagram of a source transmitter and two co-located users - destination and a helping node, performing multi-session cooperation.}\label{fig:system}
\end{figure}

The work in \cite{AsAsSh07} considers the problem of communication between a single remote transmitter and a destination while being helped by co-located users, over an independent block Rayleigh-fading channel, as depicted in Fig.~\ref{fig:system}. The users' colocation nature allows cooperation, enabling a higher communication rate from the transmitter to the destination. The transmitter has no CSI, while receivers have access to perfect CSI. Under this channel model, cooperation between co-located users for a transmitter using a broadcast approach achieves higher expected rates. This is directly explained by the fundamentals of the broadcast approach, where the better the channel quality, the more layers that can be successfully decoded. The cooperation between the users is performed over AWGN channels, under a relay power constraint with unlimited bandwidth. Three cooperation techniques are considered: AF, CF, and DF. For the case of a relaxed decoding delay constraint, these techniques are extended by the broadcast approach. The high efficiency is obtained from multi-session cooperation as each session allows decoding more layers. Interestingly, closed-form expressions for infinitely many AF sessions and recursive expressions for the more complex CF can be derived.

The first cooperation strategy is based on the AF
relaying by a network user to the destination user over the
cooperation with the following variations.
\begin{enumerate}
	\item \textit{Naive AF} - A helping node scales its input and relays it to the destined user, who jointly decodes the relay signal and the direct link signal.
	\item \textit{Separate preprocessing AF} - A more efficient form of single-session AF is a separate preprocessing approach in which the co-located users exchange the values of the estimated fading gains, and each individually decodes the layers up to the smallest fading gain. The helping user removes the decoded common information from its received signal and performs AF on the residual signal to the destined user. 
	\item \textit{Multi-session AF} - Repeatedly separate preprocessing is followed by a transmitting cooperation information at both helper and destination nodes (on orthogonal channels). The preprocessing stage includes individual decoding of the received information from the direct link and previous cooperation sessions. Along the cooperation sessions, transmission of the next block already takes place. It means that multi-session cooperation introduces additional decoding delays \textbf{without any impact on the throughput}. For this purpose, multiple parallel cooperation channels are assumed. For incorporating practical constraints on the multi-session approach, the total power of multi-session cooperation is restricted to $P_r$. This is identical to the power constraint in single-session cooperation.
\end{enumerate}

In the limit of infinitely
many sessions, the multi-session cooperation channel capacity is
$C_{\rm coop}=P_r$. The other cooperation schemes (naive AF, and
separate preprocessing AF) cannot efficiently use unlimited
bandwidth. Single-session wide-band AF means duplicating the AF
signal while proportionally reducing its power. This has no
gain over narrow-band cooperation. Therefore a narrow-band
cooperation channel is used for these two schemes, with
$C_{\rm coop}=\log(1+P_r)$.  Another set of cooperative strategies based on the WZ \cite{WYNER76} CF relaying are:
\begin{enumerate}
	\item \textit{Naive CF} - A helping node performs WZ-CF over the cooperation link. The destination informs the relay of its fading gain realization prior to the WZ compression. The destination performs optimal joint decoding of the WZ compressed signal forwarded over the cooperation link, and its own copy of the signal from the direct link. 
	\item \textit{Separate preprocessing CF} - Each
	user decodes independently up to the highest common decodable layer. Then WZ-CF cooperation takes place on the residual signal by WZ coding. 
	\item \textit{Multi-session CF} - Multi session cooperation, as described for AF, is carried out in conjunction with
	successive refinement WZ \cite{SM04} CF relaying. 
\end{enumerate}
To analyze these models, consider the following SIMO channel
\begin{align}\label{2_1}
\textbf{y}_i = h_i \textbf{x}_s + \textbf{n}_i \ ,~~~~~i\in\{1,2\}\ ,
\end{align}
where $\textbf{y}_i$ is a received vector by user $i$, with length $L$, which is the transmission block length. The length $L$ is assumed to be sufficiently large that transmission rates close to the
mutual information are reliably decoded. $ \textbf{x}_s $ is the
original source transmitted vector, and $ \textbf{n}_i $ is the additive
noise vector, with complex Gaussian i.i.d. 
zero-mean and unit variance ${\mathcal{CN}}(0,1)$, and
$h_i$ is the (scalar) fading coefficient, which is perfectly known at the $i-th$ receiver. The fading
$h_i$ is distributed according to the Rayleigh distribution $h_i
\sim {\mathcal{CN}}(0,1)$, and it remains constant for the duration of
every transmission block (adhering to a block fading channel). It means that the two users have equal average SNR, which is
realistic due to their colocation. Nevertheless, the results may be
extended to the case of unequal average SNRs in a straightforward way.
Receivers being co-located may also suggest channel realization
correlation ($h_1$ and $h_2$). In the case of such correlation, the
cooperation gains are expected to be smaller since even the joint
decoding channel capacity decreases. We assume, for simplicity of
analysis, fully independent fading channel realizations. The cooperation link between the users are modeled by AWGN
channels as follows:
\begin{align}\label{2_2}
\textbf{y}_{2,1}^{(k)} & = \textbf{x}_1^{(k)} + \textbf{w}_1^{(k)} \ ,\\
\textbf{y}_{1,2}^{(k)} & = \textbf{x}_2^{(k)} + \textbf{w}_2^{(k)}\ ,
\end{align}
where $\textbf{y}_{2,1}^{(k)}$ is the length $L$ helper's received
cooperation vector from the destination ($i=1$), on
the $k^{\rm th}$ cooperation link, and vise-versa for
$\textbf{y}_{1,2}^{(k)}$. $\textbf{x}_i^{(k)}$ is the cooperation
signal from user $i$, on the $k^{\rm th}$ cooperation link, and
$\textbf{w}_i$ is the noise vector with i.i.d. elements distributed
according to ${\mathcal{CN}}(0,1)$. On a single-session cooperation
$k=1$, and the power of $x_i^{(1)}$ is limited by
$\bbe[\myround{|x_i^{(1)}|^2}]\leq P_r$ (for $i=1,2$). On a $K$-session cooperation there are $K$ orthogonal cooperation channels available for each user with a total power constraint $P_r$. The power constraint here is specified by 
\begin{align}
\bbe\left[\myround{\sum\limits_{k=1}^K|x_i^{(k)}|^2}\right]\leq P_r\ .
\end{align}
Hence, $K$ is also the bandwidth expansion that results from
the multi-session cooperation. It is assumed that the next block's receive can be performed
while transmitting a cooperation message of previous blocks, that is, a full-duplex receiver. Cooperation is
without interference, as orthogonal channels are assumed
for this purpose. Naturally, the link capacity of a single-session narrow-band
cooperation over the AWGN channel defined in (\ref{2_2}) is given by $C_{\rm coop,NB} = \log(1+P_r)$.

In the limit of $K\rightarrow\infty$ with a power constraint for
multi-session cooperation, the cooperation link capacity is given by
\begin{equation}\label{eq:PR1_1coop}
C_{\rm coop,WB} = \int\limits_0^\infty \d R(s)=\int\limits_0^\infty \log(1+\rho(s)ds) =\int\limits_0^\infty
\rho(s)\d s = P_r\ ,
\end{equation}
where the fractional rate of a session $s$ is $\d R(s)$ and $dR(s)=\log(1+\rho(s)ds)$. The fractional power
at the $s^{\rm th}$ session is $\rho(s)$. The multi-session power
constraint implies $$\int\limits_0^\infty\rho(s)\d s=P_r\ ,$$ which
justifies the last equality in (\ref{eq:PR1_1coop}).

\subsubsection{Lower and Upper Bounds}
To evaluate the benefit of cooperation among receivers in
a fading channel following the model described in
(\ref{2_1})-(\ref{2_2}), we provide upper and lower
bounds on relevant figures of merit. There are three types of bounds relevant to our channel
model. The first is the outage capacity, which is the ultimate
average rate achievable using a single-level code (without
multi-layer coding). The second one is the achievable broadcasting rate, which refers to a
scheme using a continuous broadcast approach. The last one is the ergodic
capacity, which gives the ultimate upper bound on average rates by
averaging maximal rates obtained with full transmitter CSI.

The lower bounds are obtained by considering no-cooperation. That is a single transmitter-receiver
 pair with no cooperating user. Therefore, all lower bounds are simple SISO fading channel capacities. Upper bounds refer to the case where a co-located helping node is available, and the two users can perform optimal joint decoding of their received signals. In all cases, the bounds relate to a Gaussian block fading channel, adhering to
 (\ref{2_1})-(\ref{2_2}). 

\vspace{.1 in}
\noindent {\bf{Outage lower bound.}} The single-layer coding expected rate is 
\begin{align}\label{03_.5}
R_{\rm outage,LB} =
\max\limits_{u_{\rm th}>0}\left\{(1-F(u_{\rm th}))\log(1+u_{\rm th}P_s)\right\}\ ,
\end{align}
where the optimal threshold $u_{\rm th}$ that maximizes (\ref{03_.5})
is given by $u_{\rm th,opt} = \frac{P_s-W(P_s)}{W(P_s)P_s}$. The
function $W(x)$ is the Lambert-W function.

\vspace{.1 in}\noindent{\bf{Broadcasting lower bound.}}
This bound is based on a SISO block fading
channel, with receive CSI. The maximal expected
broadcasting rate \cite{ShitzSteiner03}, for a Rayleigh
fading channel is
\begin{align}\label{03_1}
R_{\rm bs,LB} = e^{-1}-e^{-s_0} + 2E_1(s_0) - 2E_1(1)\ ,
\end{align}
where $s_0 = 2/(1+\sqrt{1+4P_s})$, and $E_1(x)$ is the exponential
integral function.

\vspace{.1 in}\noindent{\bf{Ergodic lower bound.}}
Ergodic capacity of a general SIMO channel with $m$ receiver
antennas is \cite{T99}
\begin{align}\label{03_5}
C_{\rm erg}(m) = \int\limits_0^{\infty} u^{m-1}e^{-u}
\log(1+P_su)\d u,~~~m\in\mathbb{N}\ ,
\end{align}
which simplifies for a SISO channel into
\begin{align}\label{03_5_a}
C_{\rm erg,LB} = C_{\rm erg}(1)=e^{1/P_s}E_1(1/P_s)\ .
\end{align}

\vspace{.1 in}\noindent{\bf{Outage upper bound.}}
Fully cooperating users bound is derived similarly to
(\ref{03_.5}), with $F_{\rm UB}(u)$ as the fading gain distribution
function.

\vspace{.1 in}\noindent{\bf{Broadcasting upper bound.}}
The broadcasting upper bound is a two receive antenna block fading channel. The expected broadcasting rate for a Rayleigh fading channel \cite{ShitzSteiner03} is 
\begin{align}\label{03_3}
R_{\rm bs,UB} =& s_1e^{-s_1}-e^{-s_1}-3E_1(s_1)  - (s_0e^{-s_0}-e^{-s_0}-3E_1(s_0))\ ,
\end{align}
where $s_0$ and $s_1$ are determined by the boundary conditions
$I_{\rm UB}(s_0)=P_s$ and $I_{\rm UB}(s_1)=0$, respectively. The residual
interference $I_{\rm UB}(x)$ is given by $I_{\rm UB}(x) = (1+x-x^2)/x^3$.

\begin{figure}[tb]
	\begin{center}
		\includegraphics[width = 5in]{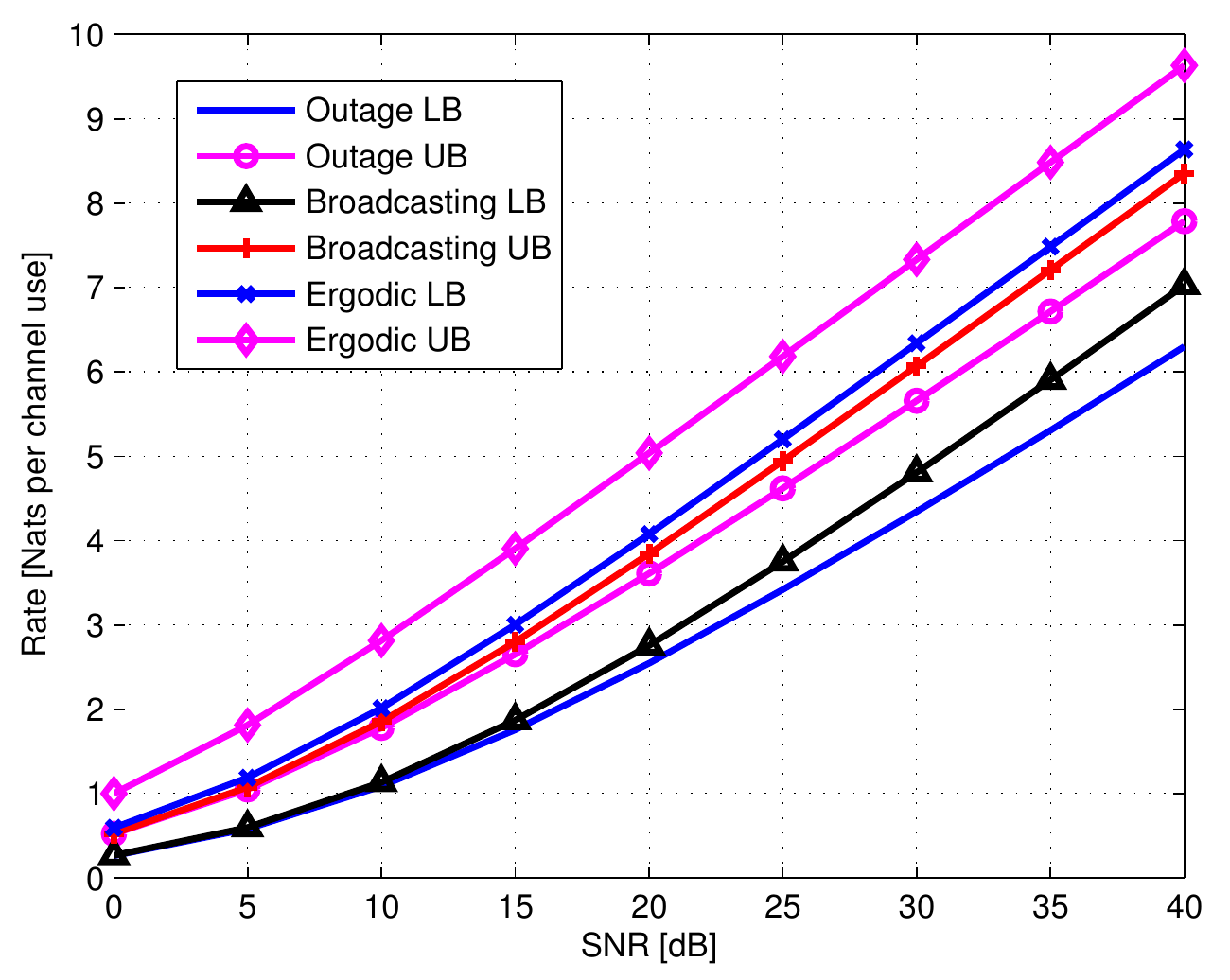}
	\end{center}
	\caption{Ranges of the average rates for both outage and broadcast
		approaches, over the cooperation channel, which were calculated
		using these approaches for either single antenna user (LB) or two
		antennas user (UB). The corresponding rate-range for an ergodic
		channel from (\ref{03_5_a}) and (\ref{03_6}) is also given for
		comparison.}\label{fig:out_bs_bounds}
\end{figure}

\vspace{.1 in}\noindent{\bf{Ergodic upper bound.}}
Ergodic bound for two receive antennas SIMO fading channel is 
$C_{\rm erg}(2)$ in (\ref{03_5}),
\begin{align}\label{03_6}
C_{\rm erg,UB} &=  C_{\rm erg}(2) = 1+e^{1/P_s}E_1(1/P_s)-1/P_s e^{1/P_s}E_1(1/P_s)\ .
\end{align}
Figure \ref{fig:out_bs_bounds} exemplifies the upper and lower
bounds for two cooperating users.

\vspace{.1 in}\noindent{\bf{Single-session cut-set upper bound.}}
Another upper bound considered is the classical cut-set bound of the
relay channel \cite{Cover}. This bound may be useful for single-session cooperation, where the capacity of the cooperation link is
rather small. Using the relay channel definitions in (\ref{2_1})-(\ref{2_2}), and assuming a single cooperation session $K=1$, the cut-set bound for a
Rayleigh fading channel is given by
\begin{align}\label{03_4}
C_{\rm cut-set} &=
\sup\limits_{p(x_s),p(x_2)}\!\!\!\!\!\min\Big\{I(x_s;y_1|h_1)+I(x_2;y_{1,2}) \; , \;   I(x_s;y_1,y_2|h_1,h_2)\Big\}\\
&= \min\Big\{ C_{\rm erg}(1)+C_{\rm coop}\; ,~ C_{\rm erg}(2) \Big\}\ ,
\end{align}
where the ergodic capacity $C_{\rm erg}(m)$ is given by (\ref{03_5}),
and the terms $C_{\rm erg}(1)$ and $C_{\rm erg}(2)$ are specified in (\ref{03_5_a}) and (\ref{03_6}), respectively. The cut-set bound is tight only when $C_{\rm erg}(1)+C_{\rm coop}\leq C_{\rm erg}(2)$, since otherwise the
cut-set bound coincides with the ergodic upper bound $C_{\rm erg,UB}$ in
(\ref{03_6}).

\subsubsection{Naive AF Cooperation}
In this AF strategy, the helping node scales its input
signal to the relaying power $P_r$, and relays the signal to
the destination user. The destination received signal at the
destination, after AF relaying is
\begin{align}\label{3_1}
\textbf{y}_{b} = \mat{
	\begin{array}{c}
	\textbf{y}^{(1)}_{1,2}\\
	\textbf{y}_{1}
	\end{array}
}=\mat{
	\begin{array}{c}
	\alpha h_2\textbf{x}_s + \alpha\textbf{n}_2 + \textbf{w}_2\\
	h_1\textbf{x}_s + \textbf{n}_1
	\end{array} }
=\mat{
	\begin{array}{c}
	(\sqrt{\beta} \textbf{x}_s + \widetilde{\textbf{w}}_2)\cdot\sqrt{1+\alpha^2}\\
	h_1\textbf{x}_s + \textbf{n}_1
	\end{array}
}\ ,
\end{align}
where $\textbf{y}_{b}$ is the signal to be decoded at the
destination, and $\alpha$ scales the relay output
to $P_r$. Hence, $\alpha=\sqrt{\frac{P_r}{1+P_s s_2}}$, and
$s_i = |h_i|^2$. The normalized noise vector
$\widetilde{\textbf{w}}_2$ has i.i.d. elements distributed
${\mathcal{CN}}(0,1)$. Hence, the normalized signal gain after the
scaling of user $i=2$ is
\begin{align}\label{3_2}
\beta = \frac{P_r s_2}{1+P_s s_2 + P_r}.
\end{align}
The achievable rate as a function of the channel fading gains is
given by the following mutual information
\begin{align}\label{3_3}
I(x_s;\textbf{y}_{b}|h_1,h_2)=\log(1+P_s s_b) =
\log\left(1+P_s\left(s_1+\frac{P_r s_2}{1+P_s s_2 +
	P_r}\right)\right),
\end{align}
where $s_b=s_1+\beta$, and therefore, the equivalent fading $s_b$ is the broadcasting variable. The CDF of $s_b$ \cite{ShitzSteiner03} is
\begin{align}\label{3_4}
F_{s_{\rm b}}(x)=\PP(s_b\leq x) =\int\limits_0^{\infty}\d u
f_{s_1}(u)\int\limits_0^{\max\myround{0,x-\frac{P_ru}{1+P_su+P_r}}}\d v
f_{s_2}(v)\ ,
\end{align}
where $f_{s_i}(u)$ is the PDF of $s_i$. The CDF of $s_b$, for a Rayleigh fading
channel, is 
\begin{align}\label{3_5}
F_{s_{\rm b}}(x)=\begin{cases}
\begin{array}{ll}
0 & x\leq 0\\
1-e^{-\frac{(1+P_r)x}{P_r-P_sx}}
- \!\!\!\!\!  \displaystyle \int\limits_0^{\frac{(1+P_r)x}{P_r-P_sx}}\d u\cdot
e^{-u-x+\frac{P_ru}{1+P_su+P_r}}
&  0\leq x< \frac{P_r}{P_s} \\
1-\displaystyle \int\limits_0^{\infty}\d u\cdot e^{-u-x+\frac{P_ru}{1+P_su+P_r}}
& x\geq \frac{P_r}{P_s}
\end{array}\ .
\end{cases} 
\end{align}
The corresponding PDF $f_{s_{\rm b}}(x)$ is given by
\begin{align}\label{3_5_1}
f_{s_{\rm b}}(x)=\begin{cases}
0 x\leq 0\\
\displaystyle \int\limits_0^{\frac{(1+P_r)x}{P_r-P_sx}}\d u\cdot
e^{-u-x+\frac{P_ru}{1+P_su+P_r}} &
0\leq x< \frac{P_r}{P_s} \\
\displaystyle \int\limits_0^{\infty}\d u\cdot e^{-u-x+\frac{P_ru}{1+P_su+P_r}} &
x\geq \frac{P_r}{P_s}
\end{cases}
\end{align}
This provides the single-layer and broadcasting expected rates for
the naive AF. The transmitter performs broadcasting optimized on fading gain random variable $s_b$ from~(\ref{3_4}). The maximal expected rate is expressed as follows
\begin{align}\label{S_5}
R_{\rm bs,ave} = \max\limits_{I(u)}\int\limits_0^{\infty} \d u
(1-F_{s_{\rm b}}(u)) \frac{\rho(u)u}{1+I(u)u}\ ,
\end{align}
where $F_\nu(u)$ is the fading gain CDF. 
\begin{Proposition}
The optimal power distribution, which maximizes the broadcasting achievable expected rate for naive AF cooperation is given by
\begin{align}\label{S_7}
I_{\rm NAF}(u)=\mycase{
	\begin{array}{ll}
	P_s & u< u_0\\
	\frac{1-F_{s_{\rm b}}(u)-u\cdot f_{s_{\rm b}}(u)}{u^2f_{s_{\rm b}}(u)} & u_0\leq u\leq u_1\\
	0 & u> u_1
	\end{array}}\ ,
\end{align}
where $u_0$ and $u_1$ are obtained from the boundary conditions
$I_{\rm opt}(u_0)=P_s$ and $I_{\rm opt}(u_1)=0$, respectively. The equivalent fading gain distribution $F_{s_{\rm b}}(x)$ and $f_{s_{\rm b}}(x)$ are specified in \eqref{3_5} and \eqref{3_5_1}, respectively.
\end{Proposition}
The broadcasting gain is compared to the single-layer coding
under the same fading gain distribution. Using the equivalent SISO channel model, which is governed by $s_{\rm b}$ with CDF $F_{s_{\rm b}}(u)$ in (\ref{3_5}),
the optimal power allocation for naive-AF can be specified
following (\ref{S_7}). Note that $I_{\rm NAF}(u)$ is non-increasing,
starting from $P_s$ at $u=0$. The average rate is explicitly given
by
\begin{align}\label{3_13}
R_{\rm NAF} =\!\! \int\limits_0^\infty\! \d x \mat{ \frac{2(1-F_{s_{\rm b}}(x))}{x}
	+ \frac{(1-F_{s_{\rm b}}(x))f_{s_{\rm b}}'(x)}{f_{s_{\rm b}}(x)}}.
\end{align}
The first derivative of the PDF of $s_b$ is denoted by
$f_{s_{\rm b}}'(x)$.

\subsubsection{AF with Separate Preprocessing}
In this section, every node tries to decode independently as
many layers as possible. Then both
users exchange the index of the highest layer successfully decoded.
Every node re-encodes the decoded data of each layer up to the lowest common index and removes it from the
original received signal. The helping node scales the result to
power $P_r$ and relays it over the cooperation link to the
destination. This improves on the naive AF, as the
cooperation is more efficient, though it requires the helping node to be aware of the source codebook and be able to decode its transmission.
Like the naive AF, this is a single-session $K=1$ cooperation. The received signal at the helping node can be expressed as
\begin{align}\label{3_14}
\boldsymbol{y}_2 = h_2(\boldsymbol{x}_{s,D} +
\boldsymbol{x}_{s,I}) + \boldsymbol{n}_2,
\end{align}
where $\boldsymbol{x}_{s,D}$ is the part of the source data
successfully independently decoded by helping node $i=2$. The coded layers not decoded independently $\boldsymbol{x}_{s,I}$ are actually the residual
interference signal.

When $s_1\geq s_2$,  the decoded data in
$\boldsymbol{x}_{s,D}$ include layers up to 
$s_2$. This reflects in the residual interference power $I(s)$,
where $s$ is the fading gain equivalent. The residual signals at both sides (before a cooperation session) are 
\begin{align}\label{3_15}
\boldsymbol{y}_{1,I} = h_1\boldsymbol{x}_{s,I(s_2)} +
\boldsymbol{n}_1\ ,\\
\boldsymbol{y}_{2,I} = h_2\boldsymbol{x}_{s,I(s_2)} +
\boldsymbol{n}_2\ .
\end{align}
It may be shown, similarly to AF derivation, that the
equivalent fading gain, after AF relaying $y_{2,I}$,
is (\ref{3_17}). Generally speaking, the helping node removes only
common information from its input signal and relays the scaled residual
signal to the destination. The destination user receives
a relayed residual signal, containing only its undecoded layers
when $s_2\geq s_1$. Otherwise, the helping node transmits its scaled residual
interference, including some layers that could be independently decoded
by the destination. The equivalent fading gain observed by the
destination and its distribution are stated in the following
proposition.

\begin{Proposition}\label{propAFone}
In an \textit{AF with separate preprocessing
				}cooperation strategy, with a single cooperation session $K=1$
				with a limited power $P_r$, the highest decodable layer is associated with an
				equivalent fading gain determined by
				\begin{align}\label{3_17}
				s_a = s_1 + \frac{P_rs_2}{1+s_2\cdot\max(I(s_1),I(s_2))+P_r}\ ,
				\end{align}
				with the following CDF for a Rayleigh fading channel
				\begin{align}\label{3_21}
				F_{s_a}(x) = \int\limits_0^{\phi_1^{-1}(x)}  [\exp{(-2u)}-
				\exp\myround{-u-\phi_2(u)}
				- \exp\myround{-u-\phi_3(u)} ]\d u\ ,
				\end{align}
				where
				\begin{align}\label{3_21_a}
				\phi_1(u) & =u+\frac{uP_r}{1+uI(u)+P_r} \ ,\\
				\phi_2(u) & = \max\myround{u, x-\frac{uP_r}{1+uI(u)+P_r} } \ ,\\
				\phi_3(u) & =\max\myround{u, \phi_4(x-u)}\ , \\
	\label{3_20}
				\phi_4(x-u) & = \begin{cases}
				\frac{(1+P_r)(x-u)}{P_r-I(u)(x-u)} &
				P_r-I(u)(x-u)>0 \\
				\infty & P_r-I(u)(x-u) \leq 0
				\end{cases}\ .
				\end{align}
\end{Proposition}			
Additional  details are available in \cite{AsAsSh07}.

\subsubsection{Multi-Session AF with Separate Preprocessing}
Next, we discuss $K$ multi-session AF with separate preprocessing
per session. The total power allocation per transmitted codeword for all sessions
corresponding to its decoding is $P_r$, in the limit of $K=\infty$. In this
approach, common layers are subtracted before every AF session by
both users. After every AF relaying, each node attempts to
decode more layers using all received AF signals so far and its
own received signal. It should be emphasized that the
multi-session is performed over parallel orthogonal channels in
such a way that the source transmission is block-wise continuous.
For example, during the $k^{\rm th}$ cooperation session of the
$1^{st}$ transmitted block (from the source), the first
cooperation session for the $k-1$ transmitted block takes place.
As the overall multi-session power is limited to $P_r$, at every
block epoch, the total power of $P_r$ is used.

As parallel channels are used for cooperation, with
infinitesimal power $\rho(s)$ allocated per channel, 
this wide-band cooperation link's capacity is the capacity of the corresponding
parallel channel. The power allocation is
$\int_0^\infty \rho(s)\d s = P_r$ under the constraint of $P_r$. The fractional rate per sub-band is then $\d R(s) = \log(1+\rho(s)\d s)=\rho(s)\d s$, \cite{V90}. Therefore,
the average capacity of this wide-band cooperation link, regardless
of the actual power allocation density, is $C_{\rm coop}=P_r$~(\ref{eq:PR1_1coop}). Notice that we use AF, which cannot
effectively use such capacity increase in single-session
cooperation ($P_r>\log(1+P_r)$). This capacity is
available in two directions: relay-destination and
destination-relay. It is required that information is exchanged in
both directions. Otherwise, multi-session cooperation becomes
inefficient, and unidirectional transmission, of only the relay to
the destination, will not gain from multi-session relaying. In the case of unlimited sessions, the scalar equivalent fading gain
can be derived for a given broadcasting power allocation $I(s)$.

\begin{Proposition}\label{prop_ms}
In a \textit{multi-session AF
					($K\rightarrow\infty$, cooperation power constraint $P_r$) with
					separate preprocessing} cooperation strategy, the highest decodable
				layer is associated with an equivalent fading gain determined by
				\begin{equation}\label{eq:PR1_1}
				s_{\rm ms} = \mycase{
					\begin{array}{ll}
					s_a^* & s_1\geq s_2\\
					s_b^* & s_1<s_2\\
					\end{array}\ ,
				}
				\end{equation}
				where $s_{\rm b}^*$ is the solution of 
				\begin{equation}\label{eq:PR_15}
				\int_{s_2}^{s_b^*}\frac{s_1}{(s_1+s_2-\sigma)^2}[1+s_1I(\sigma)]\d\sigma=P_r\ ,
				\end{equation}
				and by using $s_{\rm b}^*$,
				\begin{equation}\label{eq:PR_18}
				s_a^* = s_1 + s_2\cdot \frac{Z(s_b^*)}{1+Z(s_b^*)}\ ,
				\end{equation}
				where
				\begin{equation}\label{eq:PR_17}
				Z(s)=\int_{s_2}^{s}\frac{1+s_1I(\sigma)}{(1+s_2I(\sigma))}\frac{s_1}{(s_1+s_2-\sigma)}\d\sigma\ .
				\end{equation}
\end{Proposition}			
Similarly, achievable rates are obtained for naive CF and CF with separate preprocessing \cite{AsAsSh07}.

\subsubsection{Multi-Session Wyner-Ziv CF}
In this cooperation scheme, both nodes can quantize and compress their received
signals and exchange the result via a cooperation session. The compression is performed by the  WZ \cite{WYNER76} algorithm using side information at the decoder. For this to be performed, several definitions are required. Notice that each WZ compression step can use all information collected in the previous sessions in the form of side information. Define
$$\boldsymbol{\hat{y}}_1^{(k)}=\boldsymbol{y}_1+\boldsymbol{n}_{c,1}^{(k)}\ ,$$
where $\boldsymbol{n}_{c,1}^{(k)}$ is independent of $\boldsymbol{y}_1$, as the compressed
signal that is transmitted from $i=1$ to the co-located user $i=2$.
We refer the reader to \cite{SM04}, for successive Wyner-Ziv
overview. Here, we deal with the case where the message that is
transmitted in each session has better side information than the previous session since more layers are decoded. Furthermore, the second session can use the information sent by all the previous sessions
in order to improve performance. Since the power that is used by
each session is a control parameter, rather than a fixed parameter,
the use of an auxiliary variable that is transmitted during a session, but decoded only at the next session is superfluous (due to
the better side information, declared as $V$ in \cite{SM04}). Next,
using \cite{SM04}, the following Markov chain is defined, where
unlike \cite{SM04}, we are interested in independent averaged
distortion, rather than plain averaged distortion. The main feature
here is that the compression noise $\boldsymbol{n}_{c,i}^{(k)}$
should decrease from iteration to iteration, ending up with a
sequence of degraded channels $\boldsymbol{\hat{y}}_i^{(k)}$, following the
Markov chain:
\begin{align}
\boldsymbol{y}_2-\boldsymbol{x}_s-\boldsymbol{y}_1-\boldsymbol{\hat{y}}_1^{(k)}-\boldsymbol{\hat{y}}_1^{(k-1)}-\dots-\boldsymbol{\hat{y}}_1^{(1)}\ ,\\
\boldsymbol{y}_1-\boldsymbol{x}_s-\boldsymbol{y}_2-\boldsymbol{\hat{y}}_2^{(k)}-\boldsymbol{\hat{y}}_2^{(k-1)}-\dots-\boldsymbol{\hat{y}}_2^{(1)}\ .
\end{align}
The equivalent fading gains after every iteration of the
multi-session cooperation are stated in the following proposition.
\begin{Proposition}
	 The achievable rate in the multi-session
				with separate preprocessing and successive refinement WZ is given in
				a recursive form for the $k^{\rm th}$ session,
				\begin{align}
				R_{WZ}^{(k)}=\bbe_{s_{\rm ms}^{(k)}}\left[\log(1+s_{\rm ms}^{(k)}P_s) \right]\ ,
				\end{align}
				where
				\begin{equation}\label{eq:WZ0101}
				s_{\rm ms}^{(k)} = \mycase{
					\begin{array}{ll}
					s_a^{(k)} & s_1\geq s_2\\
					s_b^{(k)} & s_1<s_2\\
					\end{array}
				}\ ,
				\end{equation}
				and
				\begin{align}
				s_a^{(k)}=s_1+\frac{s_2}{1+(\sigma_2^{(k)})^2}\ ,\label{eq:WZ_achv1_msNA}\\
				s_b^{(k)}=s_2+\frac{s_1}{1+(\sigma_1^{(k)})^2}\ ,\label{eq:WZ_achv2_msNA}
				\end{align}
				and
					\begin{equation}
					\label{eq:50}
					\left(\sigma_j^{(k)}\right)^2=\left(\sigma_j^{(k-1)}\right)^2
					\frac{1+s_jI(s^{(k-1)})+s_{3-j}I(s^{(k-1)})}{(1+s_{3-j}I(s^{(k-1)}))\left[1+\delta_j^{(k)}\left(1+\left(\sigma_j^{(k-1)}\right)^2\right)\right]+s_jI(s^{(k-1)})(1+\delta_j^{(k)})}\ ,
					\end{equation}
				where $\sigma_j^{(k)}$ is specified in (\ref{eq:50}) for $j=1,2$,
				and $\delta_j^{(k)}$ is the fractional power assigned to user
				$j$ for the $k^{\rm th}$ cooperation session.
\end{Proposition}

\begin{figure}[tb]
	\begin{center}
		\includegraphics[width = 5in]{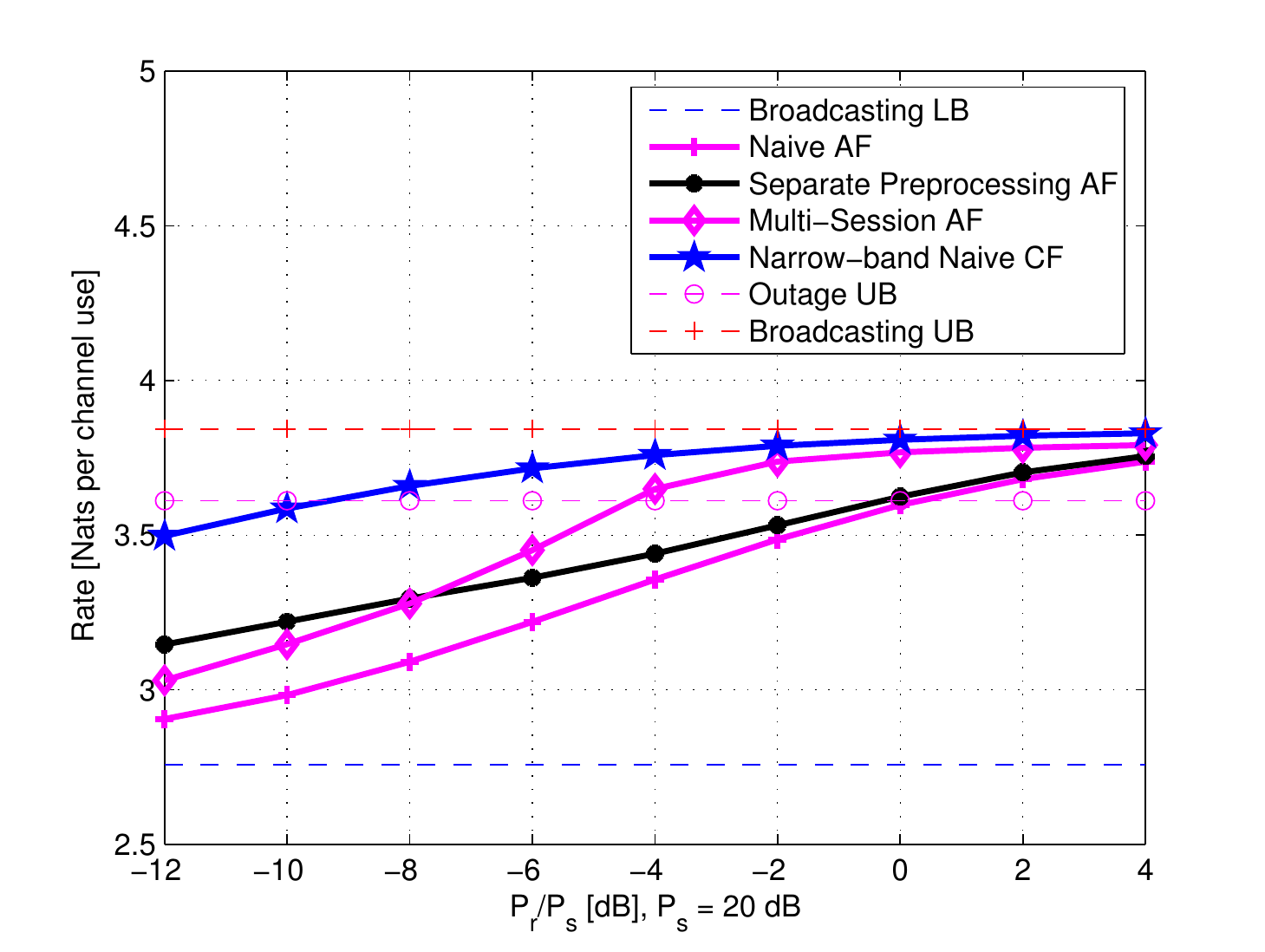}
	\end{center}
	\caption{Broadcast approach: average rates of Naive AF, AF with separate preprocessing, multi sessions AF and narrow-band (NB) naive CF compared to upper and lower bounds, as function of the channels quality ratio $\frac{P_r}{P_s}$. ($P_s=20$ dB).}\label{fig:bs_af_20}
\end{figure}

Figure \ref{fig:bs_af_20} compares the variation of the average rates versus the cooperation link
quality ($P_r/P_s$)
achieved by the naive AF, separate preprocessing AF, multi-session AF,
and narrow-band naive CF.  It is observed that the gains of separate preprocessing AF over the naive approach increase with decreasing $P_r/P_s$. For $P_s=20$ dB, both approaches achieve
gains over the outage upper bound for $P_r/P_s \geq 0$ dB. For moderate to
high $P_s$ and $P_r$, the multi-session AF approximates the
broadcasting upper bound. Again, the naive CF outperforms all other
approaches and approximates the broadcasting upper bound even on a
wider range of $P_r$ values.

\subsection{Transmit Cooperation Techniques}\label{Sec5_3}

Relaying strategies of a relay close to the source transmitter are considered in \cite{BraginskiyAsSh12}. The source-relay channel is assumed to be a fixed gain AWGN due to their colocation, while the source-destination and the relay-destination channels are subject to a block flat Rayleigh fading. A perfect CSI is assumed only at the respective receivers. With the expected throughput as a performance measure, \cite{BraginskiyAsSh12} incorporates a two-layer broadcast approach into a cooperative strategy based on the DF scheme, referred to as SDF. The SDF strategy's achievable rate expressions are derived under the broadcast approach for multiple settings, including single-user MISO and the general relay setting using successive decoding technique, both numerically and analytically.

The system consists of a source terminal $s$ communicating with a destination receiver, denoted by $d$. The multi-terminal network may consist of a helping terminal $r$. The helping terminal is occasionally present, and when available, it is near the source. However, the source is not aware of the relay's existence or availability. This model is motivated by the nature of wireless sensor networks. In such networks, numerous sensors intended to gather some information from the environment are deployed over a limited area. The sensors usually transmit information to a control point, which may have high processing capabilities. The dense deployment, along with autonomous functionality required from each sensor, leads to the concepts of collocation and obliviousness of cooperation among sensors.

The information is transmitted over a shared wireless medium where transmission received by the destination is subject to block flat Rayleigh fading. The fading coefficients between the source and the destination denoted by $h_{s}$, and between the relay and the destination denoted by $h_{r}$, are modeled by two independent zero-mean unit variance complex Gaussian RVs and are assumed to remain constant over a transmission block of $N$ symbols, with $N$ large enough to allow Shannon theoretic arguments to hold. Since the source and the relay are physically collocated, the channel gain between the two is assumed to be $\sqrt{Q} e^{j\theta } $, where $Q$ is a fixed power gain (known to all), and $\theta$ is a random phase uniformly distributed RV over $\left[-\pi ,\pi \right)$, which is assumed fixed during a transmission block of $N$ symbols and independent from one block to the next.

During the transmission period of one fading block, the relay (if one exists) can assist the source in relaying the message to the destination. Unaware of the relay's presence, the source assumes that in the worst case, it is the only active transmitter, optimizing its transmission for the SISO channel. When the relay exists, the received signals at the relay and the destination at time $n$, $n=1,\dots, N$, are modeled by
\begin{align} \label{ZEqnNum819106}
y_{r} \left(n\right) & =\sqrt{Q} x_{s} \left(n\right)+n_{r} \left(n\right)\ ,\\
y_{d} \left(n\right) & =h_{s} x_{s} \left(n\right)+h_{r} x_{r} \left(n\right)+n_{d} \left(n\right)\ ,
\end{align}
where $y_{r}(n)$ and $y_{d}(n)$ are the received signals at the relay and destination, respectively. The signals $x_s(n)$ and $x_r(n)$ designate the source and relay transmitted signals, respectively. The AWGN samples are denoted by $n_{r}(n), n_{d}(n)$ and they are distributed as ${\mathcal{CN}}(0,1)$.
Without a helping relay, the received signal at the destination is given by
\begin{equation} \label{ZEqnNum541574}
y_{d} \left(n\right)=h_{s} x_{s} \left(n\right)+n_{d} \left(n\right).
\end{equation}
For brevity, the fading gains are denoted by $\nu _{s} =\left|h_{s} \right|^{2} $ and $\nu _{r} =\left|h_{r} \right|^{2} $ each of which is exponentially distributed with unit mean. 

\subsubsection{Single-layer Sequential Decode-and-Forward (SDF)}
In the SDF strategy \cite{Katz05,Katz09}, the source transmits a single layer coded signal at the rate $R$. The relay (if present) remains silent while trying to decode the single-layer message. Once it can decode the message (after accumulating enough mutual information), it starts transmitting the message, acting as a second transmit antenna. If it is unable to decode the message before the block ends, it remains silent throughout the block, and no further cooperation occurs. The term sequential decode-and-forward emphasizes that the relay first decodes the entire message and only then starts sending its codeword. The mutual information at the relay is $\log \left(1+P_{s} Q\right)$, which means that a relay will decode a rate $R$ message for $R\leq \log \left(1+P_{s} Q\right)$. Define $\varepsilon$ as the fractional time within the transmission block which the relay uses to decode the message, i.e.,  $\varepsilon \buildrel\Delta\over= \min \left(1,\frac{R}{\log \left(1+P_{s} Q\right)} \right),\bar{\varepsilon }=1-\varepsilon $.  The expected throughput for a Rayleigh fading channel is expressed by 
\begin{equation} \label{ZEqnNum556876}
R_{\rm ave}^{\rm SDF} =R\cdot \left\{
\begin{array}{ll} 
{e^{{\frac{-e^{R} -1}{P_{s} }} }+\displaystyle\int_{0}^{\frac{e^{R} -1}{P_{s} } }\exp\left({-\frac{\left(e^{\frac{R-\varepsilon \log \left(1+\nu P_{s} \right)}{\bar{\varepsilon }} } -1-\nu P_{s} \right)}{P_{r} } }\right) \exp({-\nu }) \d\nu} & {R\leq\log \left(1+P_{s} Q\right)} \\ 
e^{{\frac{-e^{R} -1}{P_{s} }} } & { R>\log \left(1+P_{s} Q\right)} \end{array}\right. \ .
\end{equation}
The expected throughput $R_{\rm ave}^{\rm SDF}$ for $R>\log \left(1+P_{s} Q\right)$ is also equal to the achievable rate without a relay, which serves as the oblivious cooperation lower bound.

\subsubsection{Continuous Broadcasting}
Consider the problem of oblivious relaying where the transmitter performs continuous layering. It is assumed that when the relay exists, it first decodes the entire message from the source and then starts its transmission. Under a collocation assumption, the relay decoding time may be negligible compared to the transmission block duration. This setting of negligible relay decoding delay is called informed SDF. The informed SDF protocol assumes that the helping relay, when available, is informed of the transmitted packets in advance. Thus, when a relay is available, it helps throughout the transmission block.

Denote the power density at the transmitter by $\rho _{s} \left(s\right)$ and its corresponding residual interference function by $I_{s} \left(s\right)$, where $I_{s} \left(s_{0} \right)=P_{s} $ and $I_{s} \left(s_{1} \right)=0$. The layering power density at the relay is denoted by $\rho _{r} \left(s\right)$. The relay residual interference function $I_{r} \left(s\right)$ maximizing the expected throughput in presence of a helping relay is the subject for optimization. The relay power constraint is $I_{r} \left(s_{0} \right)=P_{r} $. As the optimization problem does not lend itself to a closed-form solution for a general power distribution $I_{r}(s)$, a suboptimal $I_{r} \left(s\right)$ is proposed. Consider a relay power distribution of the form
\begin{equation}\label{eq11qq}
	I_{r} \left(s\right)=\frac{P_{r} }{P_{s} } I_{s} \left(s\right)\ .
\end{equation}
The selection of such a power distribution (\ref{eq11qq}) may be analytically analyzed using a single-variable function as a subject for optimization via the calculus of variations. Any general selection of $I_s(s)$ and $I_r(s)$ requires optimizing two functionals, and does not seem to have a closed-form analytical solution. This general problem remains a subject for further research.
Under the power allocation in (\ref{eq11qq}) the equivalent fading gain of the combined source and relay signals takes the form of $s_{\rm eq}\triangleq \nu _{s} +\frac{P_{r} }{P_{s} } \nu _{r}$. The CDF of $s_{\rm eq}$ is thus
\begin{equation}
	F_{s_{\rm eq}}\left(s\right)=\mycase{
		\begin{array}{ll}
			F\left(s\right)=1-e^{-s} -se^{-s} & a=1\\
			&\\
			1+\frac{e^{-s} }{a-1} +\frac{ae^{-\frac{s}{a} } }{1-a} & \textrm{otherwise}
	\end{array}}\ ,
\end{equation}
where $a\triangleq \frac{P_{r} }{P_{s} }$. It is clear that the expected throughput may be directly computed, as $I_s(s)$ is the source optimal power allocation \cite{ShitzSteiner03}, and the relay uses the mentioned $I_{r} \left(s\right)=\frac{P_{r} }{P_{s} } I_{s} \left(s\right)$. We call this setting \textit{relay broadcasting}. In order to evaluate the oblivious cooperation gain, the achievable expected throughput can be compared to the $2\times 1$ MISO setting, where a single source with two antennas transmits using a continuous layering coded signal. This serves as a tight upper bound and is termed \textit{MISO broadcasting}.

\subsubsection{Two Layer SDF - Successive Decoding}

Previous subsections presented achievable rates for the single-layer and for the continuous broadcasting approaches. This subsection focuses on a practical layering approach, involving only two coded layers. Two coded layers are incorporated within the SDF schemes, and achievable rates are studied. Lower and upper bounds are derived first, and then achievable rates are formulated. More details are available in \cite{BraginskiyAsSh12}. The general problem can be formulated by considering a transmitter using a two-layer coding approach with a power per layer defined by $\alpha P_s$ and $\bar{\alpha } P_s$ , where $\bar{\alpha } \triangleq 1-\alpha$. Accordingly, the rate per layer is defined by
\begin{align} \label{ZEqnNum798160}
		R_1 & =\log \left(1+\frac{\eta _{1} \alpha P_{s} }{1+\eta _{1} \bar{\alpha } P_{s} } \right)\ ,\\
		R_2 & =\log \left(1+ \eta _{2} \bar{\alpha} P_{s} \right)\ ,
\end{align}
where $\eta_1<\eta_2$ can be interpreted as equivalent fading gains for reliable decoding of the $i^{\rm th}$ layer. In oblivious relaying, the source power allocation per layer, defined by $\alpha$, is set such that the expected throughput is maximized without a relay. When a helping relay is available, the source keeps using the power allocation $\alpha$, while the relay allocates $\beta P_r$ and $\bar{\beta}P_r$ for the first and second layer, respectively. Under SDF relaying, the relay has to decode the message before transmitting it. The relay fractional decoding time, $\varepsilon^i_r$ of the $i^{\rm th}$ layer, is
\begin{align} \label{ZEqnNum654105}
	\varepsilon _{r}^{1} & \triangleq  \min \left(1,\frac{R_{1} }{\log \left(1+\frac{Q\alpha P_{s} }{1+Q\bar{\alpha }P_{s} } \right)} \right)\ ,\\
 \label{ZEqnNum654105_22}
	\varepsilon _{r}^{2} & \triangleq\ \min \left(1,\max \left(\varepsilon _{r}^{1} ,\frac{R_{2} }{\log \left(1+Q\bar{\alpha }P_{s} \right)} \right)\right)\ ,
\end{align}
where $\varepsilon _{r}^{i}$ specifies the fractional time for the relay to gain sufficient mutual information to decode the $i^{\rm th}$ layer. Note that due to successive decoding, the second layer decoding cannot be shorter than its preceding layer. Using the fractional decoding times, it is required to derive the mutual information at the destination for each of the layers. When the relay requires more time to decode the second layer, it may begin allocating all its power $P_r$ for the first layer. Only once the second layer decoding is complete does the relay begin transmitting using $\beta P_r$ and $\bar{\beta}P_r$ allocated power per layer. The mutual information for decoding the first layer is given by
\begin{align} \label{ZEqnNum366232_0}
	I^{\rm SDF,1} & =\varepsilon _{r}^{1} \log \left(1+\frac{\nu _{s} \alpha P_{s} }{1+\nu _{s} \bar{\alpha }P_{s} } \right)+\left(\varepsilon _{r}^{2} -\varepsilon _{r}^{1} \right)\log \left(1+\frac{\nu _{s} \alpha P_{s} +\nu _{r} P_{r} }{1+\nu _{s} \bar{\alpha }P_{s} } \right)\\ 
	& \qquad +\left(1-\varepsilon _{r}^{2} \right)\log \left(1+\frac{\nu _{s} \alpha P_{s} +\nu _{r} \beta P_{r} }{1+\nu _{s} \bar{\alpha }P_{s} +\nu _{r} \bar{\beta }P_{r} } \right)\ ,
\end{align}
where $\nu_s$ and $\nu_r$ are the fading gain realizations of the source-destination and the relay-destination links, respectively. The coefficients $\varepsilon _{r}^{1}, \varepsilon _{r}^{2}$ are the relative time for the relay to gain sufficient mutual information to decode the first layer and second layer, respectively. The mutual information associated with the second layer is
\begin{equation} \label{ZEqnNum366232}
	I^{\rm SDF,2} =\varepsilon _{r}^{2} \log \left(1+\nu _{s} \bar{\alpha }P_{s} \right)+\left(1-\varepsilon _{r}^{2} \right)\log \left(1+\nu _{s} \bar{\alpha }P_{s} +\nu _{r} \bar{\beta }P_{r} \right)\ .
\end{equation}
The expected throughput achievable at the destination, with a helping relay, can be computed by using (\ref{ZEqnNum798160})-(\ref{ZEqnNum366232}), to obtain
\begin{multline} \label{ZEqnNum230171}
	R_{\rm ave}^{\rm BSDF} =R_{1} \cdot \PP\left[\left(I^{SDF,1} >R_{1} \right)\cap \left(I^{SDF,2} <R_{2} \right)\right]\\+\left(R_{1} +R_{2} \right)\cdot \PP\left[\left(I^{SDF,1} >R_{1} \right)\cap \left(I^{SDF,2} >R_{2} \right)\right] \ ,
\end{multline}
which can be maximized over $\alpha ,\beta ,\eta _{1} ,\eta _{2} $. We assume that $\varepsilon _{r}^{1} =\varepsilon _{r}^{2} $, implying simplex relay. This means the relay transmits only after completing the decoding of both layers.

A {\bf lower bound} for the achievable rate of oblivious relaying is considered here. In an oblivious setting, the maximal expected throughput without a helping relay is called a direct transmission rate. This rate serves as the lower bound to achievable rates for the relay channel.
\begin{Proposition}
The oblivious relaying lower bound, i.e., single user direct transmission expected throughput is
\begin{equation} \label{ZEqnNum736241}
	R_{\rm ave}^{\rm BSU} = R_{1} \PP\left(\eta _{1} <\nu _{s} <\eta _{2} \right)+\left(R_{1} +R_{2} \right)\PP\left[\left(\nu _{s} >\eta _{1} \right)\cap \left(\nu _{s} >\eta _{2} \right)\right]\ ,
\end{equation}
where $R_1,~R_2$ are the two-layers' rate allocation, and $\eta _{1}, \eta _{2}$ are the fading gain threshold for decoding the first layer and second layer, respectively.
\end{Proposition}
The expected rate $R_{\rm ave}^{\rm BSU}$ can be optimized over $\alpha ,\eta _{1} ,\eta _{2} $ to maximize (\ref{ZEqnNum736241}), and provide a tight lower bound. In an oblivious setting, it remains to optimize the relay layering power allocation, i.e., $\beta$, to maximize $R_{\rm ave}^{\rm BSDF}$ in (\ref{ZEqnNum230171}).

The MISO achievable rates serve as {\bf upper bounds}, reflecting full cooperation among transmitters. As the relay and source might have different power allocations, it is required to study the problem of MISO layering with individual power constraints per antenna. Consider first a sub-optimal approach where the same fractional power allocation per layer is used per antenna. In our setting this means $\alpha=\beta$ in (\ref{ZEqnNum366232_0})-(\ref{ZEqnNum366232}), i.e., the first layer power allocation of the source and the relay is $\alpha P_s$ and $\alpha P_r$, respectively. The expected rate then, similarly to (\ref{ZEqnNum230171}), becomes  
\begin{multline} \label{ZEqnNum454614}
	{R_{\rm ave}^{\rm BMISO} =} R_{1} \PP\left[\left(\log \left(\frac{1+Y}{1+\bar{\alpha }Y} \right)>R_{1} \right)\cap \left(\log \left(1+\bar{\alpha }Y\right)<R_{2} \right)\right]\\+\left(R_{1} +R_{2} \right)\PP\left[\left(\log \left(\frac{1+Y}{1+\bar{\alpha }Y} \right)>R_{1} \right)\cap \left(\log \left(1+\bar{\alpha }Y\right)>R_{2} \right)\right]\ ,
\end{multline}
where $Y \triangleq P_s\nu_s + P_r\nu_r$. For a Rayleigh fading channel the CDF of $Y$ is given by
\begin{equation} \label{ZEqnNum226384}
	F_{Y} \left(u\right)=\left\{\begin{array}{ll} \frac{1}{P_{r} -P_{s} } \left(P_{r} e^{-\frac{u}{P_{r} } } -P_{s} e^{-\frac{u}{P_{s} } } \right)& P_{s} \ne P_{r}  \\ \left(1+\frac{u}{P_{s} } \right)e^{-\frac{u}{P_{s} } } & P_{s} =P_{r}  \end{array}\right.\ .
\end{equation}
Now, consider the more general setting for the MISO layering problem, where source and relay layering power distribution is not necessarily equal, i.e., $\alpha\neq\beta$. The following result derived via explicit evaluation of the decoding probabilities quantifies the average throughput achievable by letting the relay use an independent power allocation.

\begin{Proposition}[\cite{BraginskiyAsSh12}]
For a $2\times 1$ MISO, a channel model described by (\ref{ZEqnNum819106}) and independent predetermined power allocation coefficients $\alpha ,\beta $, the average throughput is given by
\begin{equation} \label{ZEqnNum357278}
	R_{\rm ave}^{\rm BVMISO} =\left\{\begin{array}{ll}
		R_{1} \frac{ke^{-\eta _{1} }}{k-1} +R_{2} \left[\frac{e^{-\nu _{s_2} -k\left(\eta _{1} -\nu _{s_2} \right)} }{k-1} + \frac{ne^{-\eta _{2} } - e^{-\nu _{s_2} -n\left(\eta _{2} -\nu _{s_2} \right)} }{n-1} \right] & 1-e^{R_{1} } \bar{\beta }\leq 0 \\
		{R_{1} \frac{ke^{-\eta _{1} } - e^{-k\eta _{1} }}{k-1} +R_{2} \left[\frac{e^{-\nu _{s_1} -k\left(\eta _{1} -\nu _{s_1} \right)} -e^{-k\eta _{1} } }{k-1} + \frac{ne^{-\eta _{2} } -e^{-\nu _{s_1} -n\left(\eta _{2} -\nu _{s_1} \right)} }{n-1} \right]} & 1-e^{R_{1} } \bar{\beta }>0
	\end{array}\right.\ ,
\end{equation}
where $n\buildrel\Delta\over= \frac{\bar{\alpha }P_{s} }{\bar{\beta }P_{r} },~k\buildrel\Delta\over= \frac{\alpha P_{s} }{\left(\beta +\eta _{1} P_{s} \left(\beta -\alpha \right)\right)P_{r} } $, and where
\begin{align}\label{EQdef1}
	\nu _{s_1} & \triangleq  \left\{\begin{array}{ll} 0 & \frac{\bar{\alpha }\eta _{2} }{\bar{\beta }} >\frac{\alpha \eta _{1} }{\beta +\eta _{1} P_{s} \left(\beta -\alpha \right)}  \\ \\
	& \\
	\displaystyle \frac{-\alpha \eta _{1} \bar{\beta }+\bar{\alpha }\eta _{2} \left(\beta +\eta _{1} P_{s} \left(\beta -\alpha \right)\right)}{\bar{\alpha }\left(\beta +\eta _{1} P_{s} \left(\beta -\alpha \right)\right)-\alpha \bar{\beta }} & \mbox{\rm otherwise} \end{array}\right. \ ,\\
\label{EQdef2}
	\nu _{s_2} & \triangleq  \frac{-\alpha \eta _{1} \bar{\beta }+\bar{\alpha }\eta _{2} \left(\beta +\eta _{1} P_{s} \left(\beta -\alpha \right)\right)}{\bar{\alpha }\left(\beta +\eta _{1} P_{s} \left(\beta -\alpha \right)\right)-\alpha \bar{\beta }}\ .
\end{align}
\end{Proposition}
It is evident from the above proposition that the relay's power allocation has a crucial effect on the achievable rate, and a powerful relay does not guarantee a high achievable rate unless equipped with appropriate power allocation. For an equal layering power allocation, i.e., $\alpha=\beta$, (\ref{ZEqnNum357278}) reduces to (\ref{ZEqnNum454614}) as $\nu _{s_1} =0$. A step in determining an optimal power allocation for the MISO is taken in the following proposition, which establishes the optimal power allocation for an asymptotic source power and a constant ratio of source to relay powers.

\begin{Proposition}[\cite{BraginskiyAsSh12}] 
For a $2\times 1$ MISO setting satisfying $P_{s} \to \infty$, $P_r \to \infty$,  and $\frac{P_{s} }{P_{r} } =c$ under the channel model described by (\ref{ZEqnNum819106}), the equal power allocation is optimal.
\end{Proposition}

\subsection{Diamond Channel}
\label{sec:diamond}

Next, we review the two-hop transmission from a source to destination via two parallel full-duplex relay channel, which is investigated in~\cite{Zamani14}. Similarly to the general theme of this paper, the transmitter and the relays are oblivious to their forward links to their next hops, while being aware of their backward channel from the previous one.
\begin{figure} [h]
\centering
\includegraphics[width=2 in]{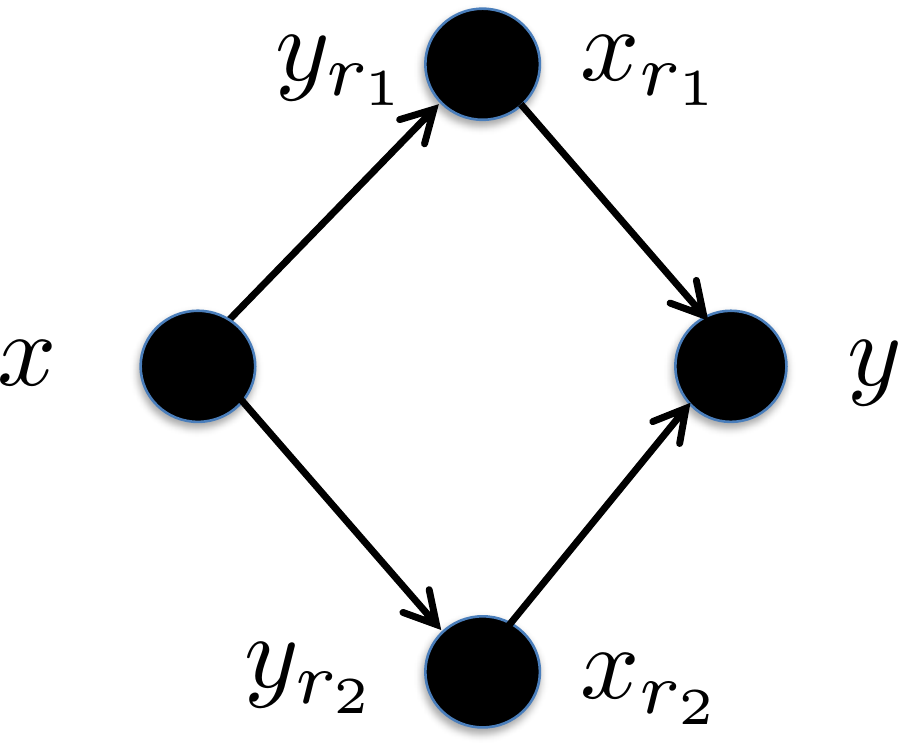}
\caption{The diamond channel.}
\label{fig:chapter5_diamond}
\end{figure}
As shown in Fig.~\ref{fig:chapter5_diamond}, the transmitter sends a signal $x$, and it is received by both relay nodes. The signal received by relay $i\in\{1,2\}$ is given by
\begin{align}
y_{r_i} \; = \; h_{r_i} x \; + \;  n_i\ ,
\end{align}
where $h_{r_i}$ follows a Rayleigh fading process and $n_i$ accounts for the AWGN. The signal received by the receiver from the concurrent transmissions by the relays is
\begin{align}
y \; = \; h_1 x_{r_1} \; + \; h_2 x_{r_2} \; + \; n \ ,
\end{align}
where $h_i$ follows a Rayleigh fading process and $n$ is the AWGN. The relays can be in the half- or full-duplex modes. Accordingly, the channel gains are defined as $s_i=|h_i|^2$ and $s_{r_i}=|h_{r_i}|^2$. 

A relevant metric to assess the broadcast approach's performance is the average rate that can be sustained reliably between the source and the destination, maximized over all possible allocations of power across different information layers at the transmitter and the relays. Each overall channel realization is the combination of the realizations of four distinct and independently varying channels. A relevant notion of degradedness in the channel can be specified based on the source-destination rate that the channel can support. Based on this, channel realizations are rank-ordered based on the aggregate rate they support. The transmitter designates one layer per realization, and the receiver at each channel realization decodes all the layers designated to that realization and those designated to the weaker ones. This strategy is next reviewed under different relaying strategies.

\subsubsection{Decode-and-Forward}
In this scheme, a transmitter generates $K$ information layers denoted by $\{x_1,\dots, x_K\}$, which are adapted to $K$ discrete channel gains. The first baseline layer is designed to be decoded by the relays when the gain of the channels linking the source to the relays is at least $s_1$, i.e., relay $i$ decodes $x_1$ when $s_{r_i} \geq a_1$. Similarly, in general,  layer $k$ is designed to be decoded by the relays when $s_{r_i} \geq a_k$. Hence, the fraction of the power allocation to layer $k\in\{1,\dots,K\}$ is denoted by $\gamma_k$. The incremental rate allocated to layer $k$ is
\begin{align}
R_k=\log\left(1+\frac{\gamma_ka_k}{1+\sum_{j=k+1}^K\gamma_ia_k}\right)\ .
\end{align}
Each relay starts decoding the information layers from the baseline layer 1 up to the layer that its actual channel realization affords. This results in the two relay nodes decoding a different number of information layers. Denote the number of layers decoded by relay 1 and relay 2 by $M_1$ and $M_2$, respectively. Relay $i$ then superimposes all the $M_i$ information layers and allocates $\alpha_{ij}$ fraction of its power to layer $j\in\{1,\dots, M_i\}$, with the rest being allocated power 0, i.e., $\alpha_{ij}=0$ for $j\in\{M_i+1,\dots, K\}$. Hence, the message transmitted by relay $i$ is
\begin{align}
x_{r_i} \; = \;  \sum_{j=1}^{K} \sqrt{\alpha_{ij}} x_j\ .
\end{align}
Since each relay is oblivious to the channel and power distribution of the other relay, due to the symmetry involved, it is assumed that power distributions are the same in both schemes.  It is shown in~\cite[Theorem 2]{Zamani14} that if power distribution across the layers is identical in both relays, then the relay signals must be uncorrelated in order to achieve the maximum expected rate. Hence, each relay's code construction is based on assuming a  similar power distribution for the other relay. The two relays adopt a transmission scheme that mimics the space-time block codes, implemented in a distributed way. Specifically, consider the time-slotted transmission in which the signal transmitted by relay $i\in\{1,2\}$ at time $t$ is denoted by $X_i(t)$. At time $t$, transmitter 1 transmits $\sum_{j=1}^{K}\sqrt{\alpha_{1j}} x_j(t)$ and transmitter 2 transmits $\sum_{j=1}^{K}\sqrt{\alpha_{2j}} x_j(t+1)$. Subsequently, at time $t+1$, transmitter 1 transmits $-\sum_{j=1}^{K}\sqrt{\alpha_{1j}} x^*_j(t+1)$ and transmitter 2 transmits $\sum_{j=1}^{K}\sqrt{\alpha_{2j}} x^*_j(t)$. Hence, the received signal at the destination is
\begin{align}\label{eq:diamond}
\left[
\begin{array}{c}
y(t)\\
-y^*(t+1)
\end{array}
\right]
= \sum_{i=1}^K 
\left[
\begin{array}{cc}
h_1\sqrt{\alpha_{1i}} & h_2\sqrt{\alpha_{2i}} \\
-h_2^*\sqrt{\alpha_{2i}} & h_1^*\sqrt{\alpha_{1i}}
\end{array}
\right]
\left[
\begin{array}{c}
x_i(t)\\
x_i(t+1)
\end{array}
\right]+
\left[
\begin{array}{c}
n(t)\\
-n^*(t+1)
\end{array}
\right]\ .
\end{align}
By capitalizing on the structure, the destination decouples the received signal into two parallel streams of signals by post-multiplying the received vector on the left side of \eqref{eq:diamond} by
\begin{align}
\left[
\begin{array}{cc}
h_1^*\sqrt{\alpha_{1i}} & - h_2\sqrt{\alpha_{2i}} \\
h_2^*\sqrt{\alpha_{2i}} & h_1\sqrt{\alpha_{1i}}
\end{array}
\right]\ .
\end{align}
Based on this approach, the interference power imposed when decoding layer $i$ by the destination is
\begin{align}
I_i = (s_1\alpha_{1i}+s_2\alpha_{2i})\sum_{j=i+1}^K (s_1\alpha_{1j}+s_2\alpha_{2j})\ .
\end{align}
Therefore, the probability of successfully decoding layer $k$ at the destination is
\begin{align}
{\sf P}_k \; = \;  \PP\left(\frac{s_1\alpha_{1ki}+s_2\alpha_{2k}}{1+\sum_{j=k+1}^K (s_1\alpha_{1j}+s_2\alpha_{2j})} \geq \frac{\gamma_ka_k}{1+\sum_{i=k+1}^K\gamma_ia_k}\right)\ .
\end{align}
Hence, the expected achievable rate is 
\begin{align}
 R_{\rm ave}=\sum_{k=1}^K {\sf P}_k R_k=\sum_{k=1}^K {\sf P}_k\log\left(1+\frac{\gamma_ka_k}{1+\sum_{j=k+1}^K\gamma_ia_k}\right)\ .
\end{align}
An optimal allocation of power across different layers can be found by maximizing the average sum-rate. A toy example showing the details and some steps involved is available in~\cite{Zamani14}.

\subsubsection{Amplify-and-Forward}
In this relaying mode, only the destination node decodes the information layers and the relay nodes only amplify what they receive. To coherently decode the signals, the destination deploys a distributed space-time code permutation along with the threshold-based ON/OFF power scheme, which is known to improve the performance of AF relaying. In this scheme, relay $i$ will remain silent if the channel gain $s_{r_i}$ is smaller than a pre-specified threshold $a_{\rm th}$. Otherwise, each relay completes its transmission in two consecutive time slots. In time $t$, relay $1$ transmits $c_1y_{r_1}(t)$ and relay 2 transmits  $c_2y_{r_2}(t+1)$. In time slot $t+1$, relay 1 transmits $-c_1y^*_{r_1}(t+1)$ and relay 2 transmits $c_2y^*_{r_2}(t)$, where coefficients $c_1$ and $c_2$ are selected properly to satisfy the  power constraints. At the destination, the received vector
\begin{align}
\left[
\begin{array}{c}
y(t)\\
-y^*(t+1)
\end{array}
\right]
\end{align}
is multiplied by
\begin{align}
\left[
\begin{array}{cc}
h_{r_1}h_1c_1 & h_{r_2}h_2c_2  \\
-h^*_{r_2}h^*_2c_2 & h^*_{r_1}h^*_1c_1  \\
\end{array}
\right]^{\sf H}\ ,
\end{align}
transforming the channel into two parallel channel, yielding the following mutual information between the transmitter and the receiver
\begin{align}
I(x;y) = \log\left(1+\frac{s_{r_1}s_1c_1^2+s_{r_2}s_2c_2^2}{1+s_1c_1^2+s_2c_2^2}\cdot P\right)\ .
\end{align}
Hence, the average rate of this channel can be found by averaging $I(x;y)$ over the distributions of all the channel gains involved.  Subsequently, the maximum average rate can be found by averaging the mutual information across different realizations of the channels.

\subsection{Multi-relay Networks}
\label{sec:multirelay}

Motivated by addressing the distributed nature and delay sensitivity of modern communication systems, the study in~\cite{SimoneSh09} investigates a network consisting of a source-destination pair, the communication between which is assisted by $M_T$ relays. The source is connected to the relays via a broadcast channel, and the relays have orthogonal channels to the destination. There is no direct link between the transmitter and the receiver. The signal received by relay $i\in\{1,\dots, M_T\}$ during time slot $t\in\{1,\dots, n\}$ is given by 
\begin{align}
y_{r_i} (t) = x(t) + n_i(t)\ ,
\end{align}
where $x(t)$ and $n_i(t)$ represent the transmitted signal by the source and the AWGN. The channel between each relay and the destination has a finite capacity denoted by $C$. Furthermore, the relays will have a non-ergodic failure profile, and it is assumed that at any given time, a random number of relays denoted by $M\in\{M_0,\dots, M_T\}$ are functioning, while communication by the rest is erased for the entire duration of a transmission. $M_0$ denotes the minimum number of relays that are functioning at any given time, and define $\bp = [p_{M_0},\dots, p_{M_T}]$ as the probability mass function of $M$. 

The success/erasure model of the relay-destination links provides a context for defining degradedness among different realizations. Specifically, a realization that has $m$ functioning relay-destination links will be considered degraded with respect to the one with $n>m$ functioning links. Based on this, the transmitter splits its messages into $M_T-M_0+1$ independently generated information layers $\{W_{M_0},\dots, W_{M_T}\}$ with rates $\{R_{M_0},\dots, R_{M_T}\}$. When there are $M=m$ active relay-destination links, the destination decodes information layers $\{W_{M_0},\dots, W_{M_T}\}$, rendering a total rate of $R^T_m=\sum_{i=1}^mR_i$. Subsequently, the average rate in the channel is
\begin{align}
R_{\rm ave} = \sum_{m=M_0}^{M_T} p_m R^T_m\ .
\end{align}
Two distinct settings will be discussed next: the {\em oblivious relays} setting, in which which the relays are oblivious to the codebooks used by the source, and the {\em non-oblivious relays} settings in which the relays are informed about the codebooks used by the source.

\subsubsection{Oblivious Relays}

In this setting, a relay performs a stochastic mapping from the message set to a codeword.  This stochastic mapping depends on a random key $F\in \cal F$ that is revealed to the destination, but it is unknown to the relays. By appropriately choosing a probabilistic model for $F$, it is possible to model a scenario in which the signal transmitted by the source is i.i.d. over the codeword elements. At each relay, being oblivious to the codebook $F$, the relay maps its received sequence to an index in the set $\{1,\dots, 2^{nC}\}$. Finally, the destination uses the relays' indexes, the knowledge of the codebook $F$, and the actual number of active relay-destination links, decode the layers associated with the number of active relay-destination links. By restricting the input to be Gaussian, it is shown in~\cite{SimoneSh09} that the average capacity of the channel is upper bounded by
\begin{align}\label{eq:Cmax_relay}
C_{\rm ave} \leq  \max \sum_{m=M_0}^{M_T} p_m \sum_{i=1}^mR_i\ ,
\end{align}
where 
\begin{align}
R_m = \frac{1}{2}\log\left(1+\frac{m\beta_mP}{1+m\sigma^2_m+mP\sum_{k=m+1}^{M_T}\beta_k}\right)\ .
\end{align}
The maximization in~\eqref{eq:Cmax_relay} is taken with respect to parameters $\beta_m\geq 0$, which satisfy
\begin{align}
\sum_{m=M_0}^{M_T} =1\ ,
\end{align}
and $\sigma^2_m$ is defined as
\begin{align}
\sigma^2_m=\left(\frac{1}{m}+P\right)\left(2^{2mC}-1\right)^{-1}\ .
\end{align}
Motivated by the structure of this upper bound, \cite{SimoneSh09} proposes a broadcast approach and single-description compression at the relays. In this approach, each relay sends over the relay-destination link a single index (description), which is a function of its received signal. The compression/decompression scheme is inspired by the technique used in ~\cite{ishwar} for robust distributed source coding in a CEO problem. The technique works by performing random binning at the agents, as is standard in distributed compression. Moreover, the test channel (i.e., equivalent compression noise) and binning rate are selected so that the receiver can recover with high probability the compressed signals on the $M$ active links irrespective of the realized value of $M$ as long as it is $M \geq M_0$ (as guaranteed by assumption). In other words, the design of the compression scheme targets the worst-case scenario of $M = M_0$. Notice that, should more than $M_0$ links be active $M > M_0$, the corresponding compressed signals would also be recoverable at the receiver, since, by the design of the binning rate, any subset of $M_0$ descriptions can be decompressed \cite{ishwar}. After decompression is performed, the receiver uses all the $M$ signals obtained from the relays to decode the codewords up to the $M^{\rm th}$ layer (that is, the layers with rates $R_m$ with $M_0 \leq  m \leq M$). Under this transmission scheme, the achievable average rate is
\begin{align}
R_m \leq \frac{1}{2}\log\left(1+\frac{m\beta_mP}{1+\sigma^2+mP\sum_{k=m+1}^{M_T}\beta_k}\right)\ ,
\end{align}
where $\sigma^2$ satisfies
\begin{align}
\frac{1}{2}\log \left[\left(1+\frac{M_0P}{1+\sigma^2}\right)^{\frac{1}{M_0}}\left(1+\frac{1}{\sigma^2}\right)\right] \leq C\ .
\end{align}
This broadcast approach can be further developed to couple the broadcast coding approach with multi-description,  rather than single-description, compression at the relays. The idea follows the work in \cite{chenberger}, which focused on the CEO problem. Accordingly, each relay shares the $nC$ bits it can convey to the destination between multiple descriptions of the received signal to the decoder. The basic idea is that different descriptions are designed to be recoverable only if certain connectivity conditions are met (that is, if the number of functioning links $M$ is sufficiently large). This adds flexibility and robustness to the compression strategy.
To simplify the presentation, the analysis is focused on the two-agent case ($M_T = 2$). In this approach, the two relays send two descriptions: a basic one to be used at the destination in case the number of active links turns out to be $M = 1$ and a refined one that will be used only if $M=2$. In this setting, the achievable average rate is
\begin{align}
R_1 & \leq \frac{1}{2}\log\left(1+\frac{\beta P}{1+(1-\beta)P+\sigma_1^2+\sigma_2^2}\right)\ ,\\
R_2 & \leq \frac{1}{2}\log\left(1+\frac{2(1-\beta)P}{1+\sigma_2^2}\right)\ ,
\end{align}
with any power allocation factor $\beta$ and $\sigma^2_1$ and $\sigma^2_2$ that satisfy
\begin{align}
\frac{1}{2}\log \left(1+\frac{P+1}{\sigma_1^2+\sigma_2^2}\right)+\frac{1}{4}\log\left(\frac{(\sigma_1^2+\sigma_2^2)^2(2P+\sigma^2_2+1)(\sigma_2^2+1)}{(2P+\sigma_1^2+\sigma_2^2+1)(\sigma_1^2+\sigma_2^2+1)\sigma_2^4)}\right)\leq C\ .
\end{align}

\subsubsection{Oblivious Agents}

Next, we briefly review the model in which the agents are informed about the codebook used at the source. As shown in~\cite{SimoneSh09}, the average capacity for this setting is upper bounded by
\begin{align}
C_{\rm ave} \leq \sum_{m=M_0}^{M_T} p_m \min\left\{\frac{1}{2}\log(1+mP)\; ,\; mC\right\}\ .
\end{align}
This result follows directly from cut-set arguments, where the first term in the $\min$ follows by considering the cut between source and relays, and the second depends on the cut from relays to the destination.

As for an achievable strategy, a generalization of the single-description approach for the setting of the oblivious relay can be constructed in a straightforward way. In this scheme, the source uses broadcast coding with Gaussian codebooks. However, on top of the $M_T - M_0 +1$ layers considered earlier, the source superimposes a further layer carrying a common message, denoted by $W_0$, with rate $R_0$, to be decoded by all relays and then forwarded to the destination. The destination is considered to recover such a message at all times, that is, as long as the number of active links $M$ satisfies $M \geq M_0$. For this purpose, each agent reserves a rate of $R/M_0$ on its outgoing links to send an index computed as a random function of the decoded $W_0$. It can be easily seen that, even though the agents are unaware of which links are currently active, the receiver will be able to recover $W_0$ with vanishing probability of error as $n\rightarrow\infty$. The extra layer carrying $W_0$ is decoded first by the agents and canceled, and the rest of coding/decoding takes place as for the broadcast approach with a single-description scheme with the caveat that now the remaining link capacity to forward compression indices is $C - R/M_0$. Under this scheme, the average rate that can be achieved is given by
\begin{align}
R_{M_0} & \leq \tilde R_{M_0} + R_0\ ,\\
R_m & \leq \frac{1}{2}\log\left(1+\frac{m\beta_mP}{1+\sigma^2+mP\sum_{k=m+1}^{M_T}\beta_k}\right)\  \qquad\mbox{for} \quad M_0+1,\dots, M_T\ .
\end{align}
where
\begin{align}
R_0=\frac{1}{2} \log\left(1+\frac{\beta_0P}{1+(1-\beta_0)P}\right)\ ,
\end{align}
and $\sigma^2$ satisfies
\begin{align}
\frac{1}{2}\log \left[\left(1+\frac{M_0P(1-\beta_0)}{1+\sigma^2}\right)^{\frac{1}{M_0}}\left(1+\frac{1}{\sigma^2}\right)\right] \leq C-\frac{R_0}{M_0}\ .
\end{align}

\subsection{Occasionally Available Relays}
\label{sec:occasionally}

Finally, we consider the impact of uncertainty in the network topology on transmission. This is motivated by the fact that in practical wireless networks, it is often difficult for each user to keep track of neighboring terminals, potentially assisting in the transmission of its information. This is especially pronounced in high-mobility networks. One immediate implication of this setting is in the IEEE 802.11 WLAN protocol using occasional relay terminals is explored. Mobile users that are far away from an access point can suffer from low uplink rates. Occasional relaying terminals between the mobile users and the access point receive the transmitted packets and relay them to the access point. When relays do not exist, then the direct links are used, albeit at a lower rate.

This setting is studied in~\cite{Katz09}, which considers communication between a source and a destination where occasionally there might be a relay node in close proximity of the source, and assisting it without its knowledge (i.e., the source is oblivious to the existence of the relay node). The destination, on the other hand, is aware of the existence of the relay node. When the relay exists, the source-relay channel is considered to be of a constant quality (due to the proximity), and the source-destination and relay-destination channels undergo block fading. All channels are known only to their associated receivers, and they are otherwise unknown to other nodes.

Hence, in this setting, the transmitter's uncertainty is due to a combination of channel uncertainty and relay existence uncertainty. Furthermore, the combination of these factors can be used for adopting a natural notion of channel degradedness. Specifically, we can use the throughput of the channel as a metric based on which different channel realizations and relay existence scenarios can be rank-ordered. By leveraging this notion of degradedness, the transmitter generates the codebooks, one corresponding to each possible realization, ensued by superposition coding for transmission. At the destination node, the receiver uses the information about the actual realization of the channel and the relay's existence and decodes all the codebooks assigned to this realization and all the weaker ones, treating the rest as noise.

\section{Communications Networks}
\label{sec:networks}

\subsection{Overview}
Previous sections discussed point-to-point communication, the MAC, the interference channel, and the relay channel. This section considers a broader span of communication networks with multiple communicating nodes and different cooperation levels. Only a limited number of examples are covered in detail, and an outlook of additional relevant problems is provided in Section~\ref{sec:outlook}.

We review the application of the broadcast approach to four different aspects of modern communication networks. First, we focus on cellular communication. Specifically, the case of uplink communications is studied in \cite{AsLupuKatzSh12} where the broadcast approach is studied in conjunction with multiuser detection for randomly spread direct sequence (DS) code-division multiple access (CDMA). This is discussed in more detail on Section \ref{Sec6_1}. In networks, it may be commonly required to minimize the distortion of the source information rather than maximize the expected rate. For fading channels combining the broadcast approach with successive refinement source coding allows minimization of expected distortion. This aspect is discussed in Section \ref{Sec6_2}.  Successive refinement as combined with the broadcast approach gives idea beyond the basic setting, and was recently used for a multi-user downlink with layered cooperation among users \cite{KimPark20}.  The broadcast approach for the information bottleneck channel is studied in~\cite{AsSh20_Bottleneck,SteinerShamai2020,steinerSh2020bottleneck}, and it is  discussed on Section \ref{Sec6_3}. Finally, the design of the broadcast approach for transmitters with harvested energy is discussed in Section \ref{Sec6_4}. There are indeed many additional network related works which are worth noting but cannot be reviewed in details in this section, such as thsoe in~\cite{Liang14,Liang12,Tulino09, Simone09,park2013multilayer,ParkSimeoneSahinShamai2014, Park_2019,SimoneSomek11,ZouLiang15,ZouLiang18,Karaksik13, ROY_6613623,Huleihel15}.

\subsection{Multi-User MAC Broadcasting with Linear Detection}\label{Sec6_1}

A cellular system where macrocells are overlaid with femtocells is studied in \cite{SimoneSh10}. Each femtocell is served by a home base station that is connected to the macrocell base station via an unreliable network access link, such as a digital subscriber line (DSL) followed by the Internet. A scenario with a single macrocell and a single femtocell is considered first, and it is then extended to include multiple macrocells and femtocells, both with standard single-cell processing and multicell processing (or network MIMO). Two main issues are addressed for the uplink channel: (i) interference management between femto and macrocells; and (ii) robustness to uncertainties on the quality of the femtocell access link. The problem is formulated in information-theoretic terms, and inner and outer bounds are derived to achievable per-cell sum-rates for outdoor and home users. Overall, the analysis lends evidence to the performance advantages of sophisticated interference management techniques, based on joint decoding and relaying, and of robust coding strategies via the broadcast approach.

The work in \cite{AsLupuKatzSh12} considers the problem of multiuser detection for randomly spread
DS-CDMA over flat
fading channels. The analysis focuses on the case of many users and large spreading sequences such that their ratio, which is the system load, is kept fixed. Spectral efficiency of practical linear detectors such as match-filter and decorrelator employing successive interference cancellation (SIC) at the receiver is derived. This is used to extend the notion of the strongest users detectors for SIC receivers. The strongest users detectors system design relies on an outage approach where each user transmits in a single layer (fixed rate), and only users experiencing good channel conditions may be reliably decoded, while the other users are not decoded. In  \cite{AsLupuKatzSh12}, iterative SIC decoding is studied, and it is shown that for equal power users, the optimal rate allocation, for maximizing the expected spectral efficiency, is equal rates for all users. This outage approach analysis is extended for a multilayer coding broadcast approach per user. The expected sum-rate, under iterative decoding with linear multiuser detectors, is optimized, and the optimal layering power distribution is obtained. For small system loads, the achievable spectral efficiency with the continuous broadcast approach and a linear matched filter detector exhibits significant gains over the single-layer coding approach.

Multiuser wireless communication systems using CDMA have been studied and
implemented in recent years. The results on the asymptotic distribution
of singular values of certain random matrices allowed the analysis
of randomly spread direct sequence CDMA
\cite{TSEHANLY99,VerduShamai99,ShamaiVerdu01}. In those multiple
access channels, random and independent signature waveforms are
assigned to the network subscribers.

In \cite{VerduShamai99}, the sum-rate capacity per chip was analyzed
for a non-fading channel, the number of users $K$ is taken to the
limit ($K\rightarrow\infty$), and the spreading sequence length $N$
is also large ($N\rightarrow\infty$). The system load, which is also
the number of users per chip is kept fixed, i.e.,
\begin{align}\label{intro_eq1}
	\beta =\frac{K}{N}\ .
\end{align}
The main conclusions from the results in \cite{VerduShamai99} are
that for low $\beta$, the linear multiuser detectors (e.g., 
decorrelator and linear MMSE detectors)
have near-optimal spectral efficiency. For any $\beta$ and
$E_b/N_0$, the match-filter multiuser detector is far from optimal.
The spectral efficiency of the linear detectors, except for the
matched filter, grows unbounded with $E_b/N_0$, for a given $\beta$.
The work in \cite{ShamaiVerdu01} extended these results to the case
where every user experiences a flat fading channel. The sum-rate
capacity is an ergodic capacity, which is achievable for fast fading
channels, where every transmitted block experiences sufficiently
many fading realizations to approximate ergodicity. Otherwise, a
framework of outage capacity may better characterize the expected
performance. The channel model with slow fading, where a fading
remains fixed throughout a transmission block, is considered in
\cite{shamaiZaidel02}, where an outage probability is equivalent to
the fraction of undecoded users, providing a framework for strongest
users detection. In this work, it is assumed that all users transmit
at equal rates and equal power, regardless of their individual
fading realizations. In such a case, the receiver can no longer
guarantee reliable decoding for all active users. In this case, the
receiver ranks all active users by their received powers and decodes
the transmissions of the largest number of users, for which decoding
is successful. The system design can be done such that a fraction of
undecodable users (FUU) is defined, and this dictates the fixed rate
to be used by all active users. The total achievable sum-rate is
referred to as the outage capacity. The FUU can be optimized such
that the average sum rate is maximized.

In \cite{AsLupuKatzSh12}, the sum-rate capacity of linear detectors with
SIC receivers is studied for different
types of detectors. Different approaches to rate
allocation and multi-stage decoders with SIC are considered. Interestingly, it turns out that with
iterative SIC decoding, equal rate allocation achieves the highest
average spectral efficiency. In iterative decoding, the receiver
decodes as many users as possible and performs SIC every iteration.
The effective system load is reduced after every SIC iteration,
increasing the multiuser detector efficiency. Thus, more users with
worse channel conditions can be decoded. Moreover, by letting every
user employ multi-layer coding, the expected spectral efficiency may increase further.

The multi-access channel  combined with the broadcast approach \cite{ShitzSteiner03}
in its continuous layering form was first analyzed in \cite{SH00}.
Some MAC outage approaches and MIMO multi-layering schemes were
studied in \cite{SteinerShamai2007}. In \cite{Minero:ISIT07}, a simple two
state multi-access channel with two users is studied, where it is
shown that superposition coding is optimal, and the sum-rate
capacity per layer is derived. A random-access (non-fading) channel
is also a special case of the MAC. Achievable rates over this
channel are studied with superposition coding in
\cite{Minero:ISIT07,GOLDSMITH04}. An alternative practical approach
is to use variable-rate coding over the MAC
\cite{CAIRE04_VARIABLERATE}.\\
The main results of \cite{AsLupuKatzSh12} may be summarized  as follows:
\begin{enumerate}
	\item formulation of ergodic bounds for systems with random spreading DS-CDMA over fading channels, employing SIC receivers;
	\item derivation of the expected spectral efficiency achievable with equal rate allocation per user, and iterative SIC decoding. It is also shown that equal rate allocation maximizes the expected spectral efficiency;
	\item derivation of the expected spectral efficiency for the case of multi-layer coding taken to the limit of many layers (continuous broadcast approach);
	\item analysis of a multi-layer coding where parallel decoders are used, without employing SIC;
	\item analysis of a more complicated setting, including a multi-layer coded transmission with iterative SIC decoding. It is shown that, like in the single-layer case, the expected spectral efficiency is maximized for equal rate allocation per user. Furthermore, the optimal layering power allocation function, which maximizes the expected spectral efficiency, is obtained for the matched-filter and decorrelator detectors. The case of broadcasting with MMSE and optimal detectors under iterative SIC decoding remains an open problem.
\end{enumerate}

\subsubsection{Channel Model}\label{sec:model}

We describe the channel model and the basic assumptions. 
Consider the following system model
\begin{align}\label{eqCDMA_Strongest_1}
	\mathbf{y} =  \mathbf{V}\mathbf{H} \mathbf{x} + \mathbf{n}\ ,
\end{align}
where $\mathbf{x}=[x_1,...,x_K]$ is a vector of length $K$. An
individual term $x_k$ is a sample of a layered coded signal of the
$k^{\rm th}$ user, and $\{x_k\}$ are i.i.d. and distributed according to ${\mathcal{CN}}(0,P)$, where $P$ sets the power
constraint per user. $\mathbf{V}$ is an $[N\times K]$ signature
matrix with elements i.i.d. distributed according to
$v_{i,j}\sim{\mathcal{CN}}(0,\frac{1}{N})$, and $\mathbf{n}$ is,
without loss of generality, a normalized AWGN vector
$\mathbf{n}\sim{\mathcal{CN}}(0,I_N)$. The channel matrix
$\mathbf{H}$ is a diagonal matrix
$\mathbf{H}=\textrm{diag}(h_1,h_2,...,h_K)$ of fading gains. The empirical distribution of $\{s_k\}\triangleq\{|h_k|^2\}$ converges
almost surely to a distribution $Q(s)$ such that $\bbe_{Q}[s] = 1$. The channel
matrix $\mathbf{H}$ remains fixed throughout a transmission block,
which corresponds to a slowly fading channel model. Note that, since
the additive noise is normalized we have $\p= P$.

The energy per bit to noise spectral density ratio is used for
evaluating the spectral efficiency and for comparing different
strategies, and it is defined as
\begin{align}\label{intro_eq2}
	\frac{E_b}{N_0} = \frac{\beta }
	{R_{\rm sum}}\; \p\ ,
\end{align}
where $R_{\rm sum}$ is the total spectral efficiency, i.e., the sum-rate
in bits per second per Hertz. The system load $\beta$ is defined
in (\ref{intro_eq1}).

\subsubsection{Strongest Users Detection - Overview and Bounds}\label{sec:overview}

Motivated by practical
considerations, the decoding of strongest users on block fading channels
is studied in \cite{shamaiZaidel02}. This study assumes that
all users transmit at equal rates and equal powers, regardless of
their individual fading realizations. In such a case, the receiver
can no longer guarantee the reliable decoding of all active users. Thus,
the receiver ranks all active users by their received powers and
decodes the transmissions of the largest number of users, for which
decoding is successful. The system design can be optimized to a fixed
FUU, which dictates the rate to be
used by all active users. The maximal achievable sum-rate is referred
to as the outage capacity. It is obtained by optimizing the FUU such that the
average sum rate is maximized.

The ergodic spectral efficiency for the fading CDMA channel model in
(\ref{eqCDMA_Strongest_1}) is given by \cite{ShamaiVerdu01}
\begin{align}\label{eqCDMA_Strongest_2}
	C_{\rm erg}(\beta,\p) = \beta \bbe_s\mat{\log(1+s\eta(\beta) \p) }\ ,
\end{align}
where $\eta(\beta)$ is the multiuser detector efficiency, which
depends on the detector type (e.g, matched filer, decorrelator, MMSE), and is
a function of the system load $\beta$, and the SNR. The expectation
is taken with respect to the fading gain distribution $Q(s)$. For
the completeness of this presentation, the multiuser detector
efficiency is specified for each relevant detector. The detector
efficiency of a matched filter is \cite{ShamaiVerdu01}
\begin{align}\label{eqCDMA_Strongest_3}
	\eta_{\rm mf}(\beta) = \frac{1}{1+\beta \p}\ .
\end{align}
The detector efficiency of a decorrelator receiver is
\begin{align}\label{eqCDMA_Strongest_4}
	\eta_{\rm dec}(\beta) = \max\{0, 1-\beta\}\ ,
\end{align}
and for an MMSE detector, $\eta_{\rm mmse}(\beta)$ satisfies the
following equation
\begin{align}\label{eqCDMA_Strongest_5}
	\eta_{\rm mmse}(\beta)  + \beta E_s\mat{
		\frac{s\eta_{\rm mmse}(\beta)\p}{1+s\eta_{\rm mmse}(\beta)\p} } = 1\ .
\end{align}
The expectation here is taken with respect to the fading gain distribution $Q(s)$.
For a Rayleigh fading channel, the expectation is explicitly
expressed as
\begin{align}\label{eqCDMA_Strongest_6}
	\bbe_s\mat{ \frac{s\eta_{\rm mmse}\p}{1+s\eta_{\rm mmse}\p} } =
	1-\frac{E_1\myround{\frac{1}{\eta_{\rm mmse}\p}}
	}{\eta_{\rm mmse}\p}\exp\myround{\frac{1}{\eta_{\rm mmse}\p}}\ ,
\end{align}
where $E_1(x)$ is the exponent integral function. 

\vspace{.1 in}
\noindent {\bf Upper Bound.}  It is well-known that the optimum multiuser detector capacity is
also equal to the ergodic successive decoding sum-rate capacity with an MMSE
detector, according to the mutual information chain rule \cite{Cover}.
Thus the ergodic capacity, obtained with an optimum detector, can be expressed by the ergodic SIC MMSE detection capacity \cite{ShamaiVerdu01}
\begin{align}\label{eqCDMA_Strongest_2_1_1_1}
	C_{\rm opt}(\beta,\p) =  \bbe_s \mat{\int\limits_0^\beta \log\myround{
		1+s\cdot\eta_{\rm mmse}(z)\cdot \p} ~\d z}\ .
\end{align}

\begin{figure}[h]
	\centering
	\includegraphics[width=5in]{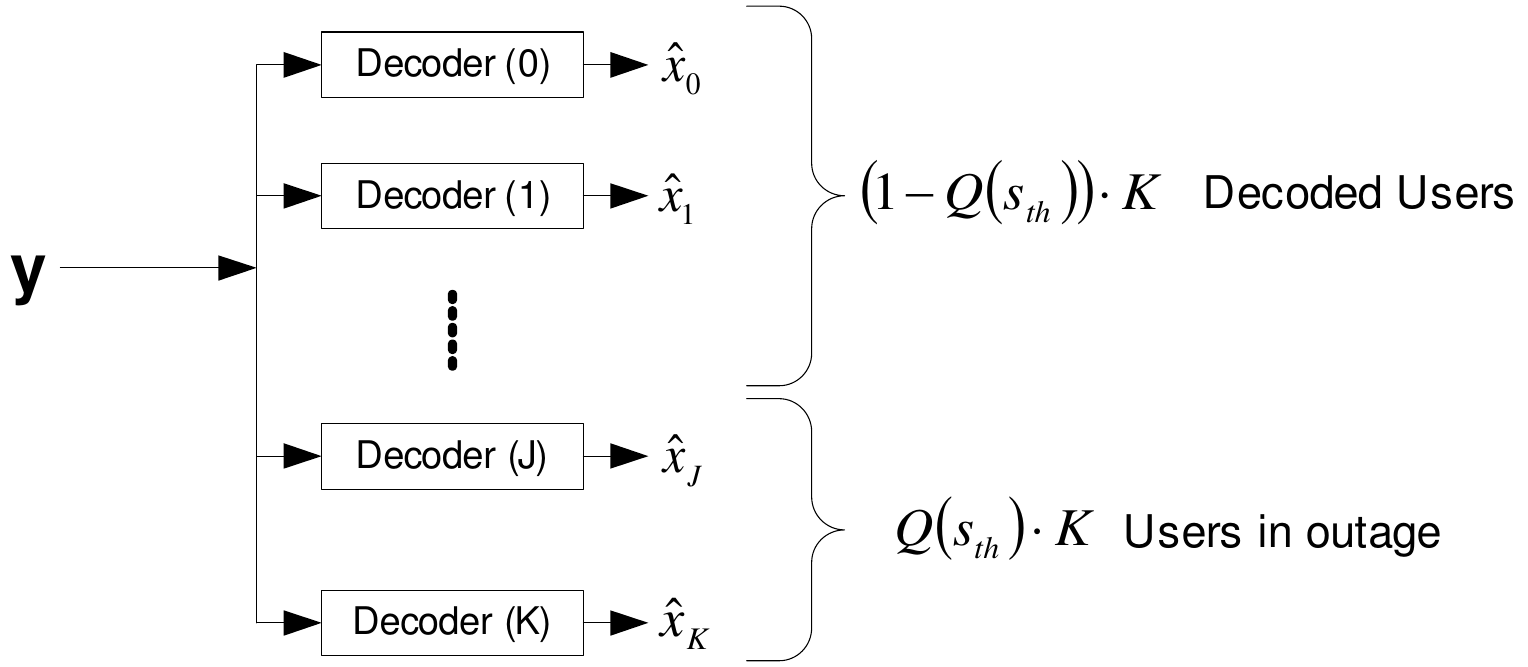}
	\caption{Schematic description of a parallel multiuser decoder, without SIC.}\label{fig:decoders_parallel}
\end{figure}

\noindent {\bf Strongest users detection.} It refers to the practical case where all users transmit at a fixed rate, via single-layer coding. The
adequate channel model here is the block fading channel, where a
fixed fading realization throughout the block for each user is
observed. Thus all users, experiencing fading gains smaller than a
threshold $s_{\rm th}$, will not be reliably decoded. This is demonstrated in Fig.~\ref{fig:decoders_parallel}, where a fraction of users, corresponding to $Q(s_{\rm th})$, is in an outage, and all other users are reliably decoded. The average achievable sum-rate for outage decoding is given by
\cite{shamaiZaidel02}
\begin{align}\label{eqCDMA_Strongest_7}
	C_{\rm out}(\beta,\p) = \beta (1-Q\myround{s_{\rm th}})
	{\log(1+s_{\rm th}\eta(\beta) \p) }\ ,
\end{align}
where $Q(s_{\rm th})$ is the probability of outage corresponding to the
fraction of users that cannot be reliably decoded. The multiuser
detector efficiency $\eta(\beta)$ is specified in equations
(\ref{eqCDMA_Strongest_3})-(\ref{eqCDMA_Strongest_5}) for the
underlying linear detectors.

In parallel decoding schemes, the decoding latency may be small. However, there is an inherent spectral efficiency loss due to decoding every user in the presence of interference from all other users. An SIC decoder attempts decoding the users one by one, where after every successful decoding, a reconstructed signal associated with the decoded user is subtracted from the input signal. The procedure continues until the last user is decoded. Consider the case that each user transmits over a fading channel. Such a
channel model was studied in \cite{ShamaiVerdu01}, where the
detectors considered were an optimal detector, MMSE, decorrelator,
and MF. The derivations in \cite{AsLupuKatzSh12} extend the results for the SIC receiver
strategy.

For a given system load $\beta$, the ergodic sum-rate is specified
in (\ref{eqCDMA_Strongest_2}). This sum-rate is an upper bound
since its achievability requires fast feedback from the receiver
to all users. With SIC decoding, the ergodic sum-rate is given by
\begin{align}\label{eqStrong_SIC_10}
	C_{\rm SIC,erg}(\beta,\p) = \bbe_s\mat{\int\limits_{0}^{\beta}\d z~
		{\log\myround{1+s\cdot\eta\myround{z} \p} } }\ .
\end{align}
The ergodic sum-rate for an MF-SIC detector is derived in the same lines
as for the non-fading case, yielding
\begin{multline}\label{eqStrong_SIC_11}
	C_{\rm SIC,MF}(\beta,\p) = \bbe_s\left[\myround{s+\beta
		+\frac{1}{\p}}\log(1+\p(s+\beta))\right. \\- 
		\left.(\frac{1}{\p}+s)\log(1+s\p)-
	\myround{\beta +\frac{1}{\p}}\log(1+\beta \p)\right] \ .
\end{multline}
The sum-rate capacity, for an SIC decorrelator detector, is also
available as a function of the fading gain distribution
\begin{multline}\label{eqStrong_SIC_11_1}
	C_{\rm SIC,Dec}(\beta,\p) = \bbe_s\mat{\myround{1
		+\frac{1}{s\p}}\log(1+s\p)\right. \\
		\left. -\beta~ - \myround{1-\beta
		+\frac{1}{s\p}}\log(1+s\p(1-\beta)) ~}\ .
\end{multline}
For an MMSE detector, the sum-capacity cannot be given in a closed-form, and it is computed using $\eta(z)$ given in
(\ref{eqCDMA_Strongest_6}), plugged into the ergodic capacity
expression in (\ref{eqStrong_SIC_10}).

\subsubsection{Broadcast Approach with Strongest Users Detection - (NO SIC)} 
If we let every user transmit a
continuously layered coded signal, then the number of decodable
layers per user directly depends on the experienced fading level.
Consider here parallel decoding, where the receiver decodes all
users in parallel up to the highest reliably decoded layer. Thus the
achievable rate, averaged over all possible fading realizations, is
given by
\begin{align}\label{eqCDMA_Strongest_7__}
	R_{\rm bs}(\beta,\p) &= \lim\limits_{K,N,J\rightarrow\infty,~
		\frac{K}{N}\rightarrow\beta,~ J(s)/K\rightarrow q(s) }
	\int\limits_0^\infty \d s \frac{J(s)}{N} \int\limits_0^s \d u
	\frac{u\eta\ 
		\rho(u)}{1+u\eta I(u)}\ ,
\end{align}
where $J(s)$ is the number of decoded users at fading level $s$, and
where the broadcasting rate, derived in (\ref{SISO6}), is modified
here by the detector efficiency $\eta$. The expected sum rate is
simplified into
\begin{align}\label{eqCDMA_Strongest_8}
	R_{\rm bs}(\beta,\p) =\int\limits_0^\infty \d s ~q(s) \int\limits_0^s \d u
	\frac{u\eta \rho(u)}{1+u\eta I(u)} = \int\limits_0^\infty \d s(1-Q(s))
	\frac{s\eta \rho(s)}{1+s\eta I(s)}\ .
\end{align}
It can be shown that the optimal power distribution, which maximizes
$R_{\rm bs}(\beta,\p)$, is like in (\ref{SISO14}), where the detector
efficiency $\eta$ scales the power distribution
\begin{align}\label{SISO14_1_}
	I(x) = \mycase{
		\begin{array}{cl}
			\p &~ x<x_0\\
			\frac{1-Q(x)-x\cdot q(x)}{x^2q(x)\eta} &~ x_0\leq x\leq x_1 \\
			0 &~\mbox{else}
	\end{array}}~,
\end{align}
where $x_0$ is determined by $I(x_0)= \p$, and $x_1$ by $I(x_1)=0$.

\subsubsection{SIC Broadcast Approach Upper Bound}
In order to characterize an achievable rate via layering, the power
distribution for layering should be optimized for every subset of
users, and their corresponding residual interference must be
accounted for in the stages of the SIC, as described above for the
outage case. Such an analytical analysis for the broadcast approach
seems to be intractable. Therefore, an upper bound significantly
tighter than the ergodic upper bound is provided.

The upper bound of the broadcast approach is simply the broadcast
approach combined with SIC, where optimal layering is performed for
every subset of users. It is assumed that at any decoding stage,
there is no residual interference from previous SIC stages. Although
interference from undecoded layers of early stages does exist, this
assumption allows full derivation and optimization of a continuous
broadcast approach. Under this simplified setting, the layering
sum-rate with SIC is given by
\begin{align}\label{eqStrong_SIC_17}
	C_{\rm SIC,BS}(\beta,\p) = \int\limits_0^\beta \d z \int\limits_0^\infty
	\d s~(1-Q(s))\frac{s\eta(z)\rho(s)}{1+s\eta(z)I(s)}\ ,
\end{align}
where the inner integral is the average achievable rate for a given
system load $z$. The maximization of this average rate is given in
(\ref{eqCDMA_Strongest_8}), with an optimal power distribution
specified in (\ref{SISO14_1_}). For a Rayleigh fading channel, this
maximal average rate can be expressed more explicitly as
\begin{align}\label{eqStrong_SIC_18}
	C_{\rm SIC,BS}(\beta,\p) = \beta( e^{-1}-2E_1(1))+ \int\limits_0^\beta
	\d z \myround{ 2E_1(S_0(z))-e^{-S_0(z)} }\ ,
\end{align}
where $S_0(z)=2/\myround{1+\sqrt{1+4\p\eta(z)} }$.
Since this broadcasting upper bound does not provide an achievable
expected rate, the analysis of
continuous broadcasting, which follows assumes equal rates with iterative decoding.

\subsubsection{Broadcast Approach with Iterative SIC}

\begin{figure}[h]
	\centering
	\includegraphics[width=5in]{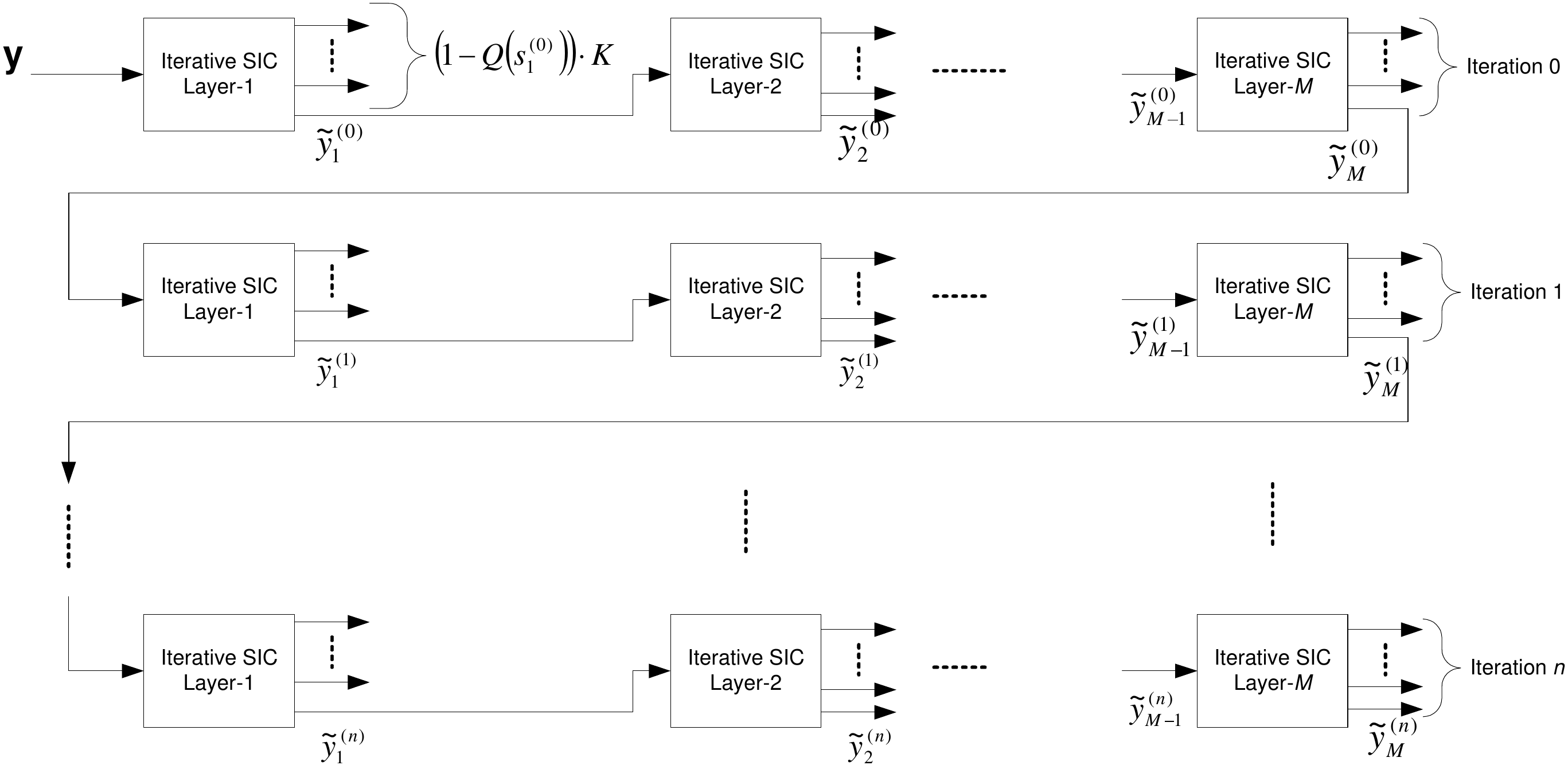}
	\caption{Schematic description of an iterative successive interference cancellation multiuser decoder for multi-layer coded transmission with $M$ layers. Every iteration includes multiple sub-iterations of iterative-SIC decoding per code layer.}\label{fig:decoders_SIC_iter_Broadcast}
\end{figure}

A broadcast approach is employed by all transmitting users, where all users transmit with the same rate and layering power distribution. The decoder applies iterative SIC decoding. The main idea in the decoding scheme is to apply an iterative SIC per layer. The decoding process is illustrated in Fig.~ \ref{fig:decoders_SIC_iter_Broadcast}. Every iteration includes $M$ stages of iterative SIC decoding, where $M$ is the number of coded layers. The first stage attempts iterative SIC decoding of the first layer of all users. Then the next stage performs iterative SIC decoding for the group of users for which decoding of the first layer was successful. This continues until the last layer decoding is done. Then the second iteration continues similarly.

The continuous layering characterizes the highest achievable sum-rate for the broadcast approach, i.e., the number of
layers is unlimited. Every layer is associated with a fractional
rate and power allocation, as described in the broadcast approach overview. The achievable rates and overall performance strongly depend on the
transmission scheme and the decoding strategy. The decoding strategy
which is adopted here is the multi-round iterative decoding.  The maximal achievable average rate can be expressed by the following optimization problem
\begin{align}\label{eqStrong_SIC_BS2_40_xx}
	R_{\rm sum,bs} = \max\limits_I ~ R_{\rm sum,bs}(I)\ ,
\end{align}
where the achievable continuous layering rate $R_{\rm sum,bs}(I)$ is given by
\begin{align}\label{eqStrong_SIC_BS2_40}
	R_{\rm sum,bs}(I) = \beta \int\limits_0^\infty
	~\d s(1-Q(s))\frac{s\eta(G)\rho(s)}{1 + s\eta(G)I(s)} \triangleq
	\int\limits_0^\infty \d s J(s,I,I')\ ,
\end{align}
where $G$ corresponds to the remaining layers per user, which induce
the mutual interference
\begin{align}\label{eqStrong_SIC_BS2_41}
	G\triangleq \frac{\beta}{\p}\int\limits_0^\infty Q(s)\rho(s)\d s
	\triangleq \int\limits_0^\infty \d s Z(s,I,I')\ ,
\end{align}
where $I(s)=\int_s^{\infty}\d u~\rho(u)$.

The optimization of
(\ref{eqStrong_SIC_BS2_40}) with respect to the residual interference
constraint in (\ref{eqStrong_SIC_BS2_41}) can be solved by fixing
the interference parameter $G$ to an arbitrary value such that $0<
G\leq \beta$. For such $G$, the optimization in
(\ref{eqStrong_SIC_BS2_40}) is a standard variational problem with a
residual interference constraint on top of the power constraint
$I(0)=\p$. The optimization problem is, therefore
\begin{align}\label{eqStrong_SIC_BS2_42}
		\max\limits_I ~\int\limits_0^\infty \d s J(s,I,I') \qquad \textrm{s.t.} \qquad G\geq \int\limits_0^\infty \d s Z(s,I,I')\ .
\end{align}
We can write the Lagrangian form
\begin{align}\label{eqStrong_SIC_BS2_43}
	L = \int\limits_0^\infty \d s J(s,I,I') ~+~ \lambda\myround{ G-
		\int\limits_0^\infty \d s Z(s,I,I') }\ .
\end{align}
The Euler-Lagrange condition for extremum can be derived, and the
optimal layering power distribution can be expressed in a closed-form, as summarized in the next proposition.

\begin{Proposition}[\cite{AsLupuKatzSh12}]
The optimal power distribution, which
			maximizes the expected sum-rate of a continuous broadcast approach
			(\ref{eqStrong_SIC_BS2_42}), with matched-filter multiuser detection
			and iterative SIC decoding, is achieved from
			\begin{align}\label{eqStrong_SIC_BS2_44}
				I(s) = \mycase{
					\begin{array}{ll}
						\p & s<s_0\\
						\dfrac{-\p+\sqrt{\p^2~+~\dfrac{4\lambda\beta(1-Q(s))\p}{\eta(G)s^2Q'(s)}}}{2\lambda\beta}-\dfrac{1}{s\eta(G)} & s_0 \leq s\leq s_1\\
						0 & s>s_1
				\end{array}}\ ,
			\end{align}
			where $s_1$ is the smallest fading gain for which $I(s_1)=0$, and the
			left boundary condition on $s_0$ satisfies $I(s_0)=\p$. The
			Lagrangian multiplier $\lambda$ is obtained by an equality for the
			residual interference constraint (\ref{eqStrong_SIC_BS2_41}), as specified by
			\begin{align}\label{eqStrong_SIC_BS2_44_xx1}
				\int\limits_{s_0}^{s_1}Q(s)I'(s)~\d s = -G\frac{\p}{\beta}\ .
			\end{align}
\end{Proposition}
The decoding algorithm for a decorrelator multiuser detector is
similar. In the continuous setting, the detector efficiency is
updated according to the number of users for which all layers are
decoded. This is the reason the upper boundary of the power
distribution is a subject for optimization. The solution is
obtained by solving the corresponding variable endpoint variational
optimization problem.

It is assumed here that the optimal solution for the power
distribution lies on a single continuous interval $[s_0,s_1]$. The
extension to multiple continuous intervals may be done as in
\cite{Tian08}. The average achievable rate with a decorrelator
detector, in its general form, is
\begin{equation}\label{eqStrong_SIC_BS2_100}
	R_{\rm bs,decorr} = \beta \int\limits_{s_a^+}^{s_b^-}\d s
	(1-Q(s))\frac{s\rho(s)\eta\myround{ \beta Q(s_b) }}{ 1 + s
		I(s)\eta\myround{ \beta Q(s_b) } } + (1-Q(s_a))R_0(s_a) +
	(1-Q(s_b))R_1(s_b)\ ,
\end{equation}
where $\eta(x)=1-x$, $I(s_0)=\p$, $I(s_1)=0$, and the rate of
the first layer is
\begin{align}\label{eqStrong_SIC_BS2_101}
	R_0(s_a) = \beta \log\myround{ 1 + \frac{s_a\eta\myround{\beta
				Q(s_b)}(\p-I(s_a^+)) }{ 1 + s_a\eta\myround{\beta Q(s_b)}I(s_a^+) }
	}\ ,
\end{align}
where $I(s_a^-)=\p$, and $I(s_a^+)$ is the remaining power
allocation for the continuous and last layers. The last layer is
allocated
\begin{align}\label{eqStrong_SIC_BS2_102}
	R_1(s_b) = \beta \log\myround{ 1 + s_b\eta\myround{\beta
			Q(s_b)}\myround{ I(s_b^-) - I(s_b^+) }  }\ ,
\end{align}
where $I(s_b^+)=0$. Thus the discontinuity in $I(s)$, can be in
$s_a$, and $s_b$. Define the functional subject for optimization,
from (\ref{eqStrong_SIC_BS2_100}), by
\begin{align}\label{eqStrong_SIC_BS2_103}
	G(s_b,s,I,I') = \beta (1-Q(s))\frac{-sI'(s)\eta\myround{ \beta Q(s_b)
	}}{ 1 + s I(s)\eta\myround{ \beta Q(s_b) } }\ .
\end{align}
The following variable end point variational optimization problem is
solved following
\begin{align}\label{eqStrong_SIC_BS2_104}
	R_{\rm bs,decorr} =\left\{
	\begin{array}{ll}
	\max\limits_{s_a,s_b,I} & 	\int\limits_{s_a^+}^{s_b^-}\d s G(s_b,s,I,I')
	+ (1-Q(s_a))R_0(s_a) + (1-Q(s_b))R_1(s_b) \\
	{\rm s.t.} & I(s_a^-)=\p \\
	& I(s_b^+)=0\ 
	\end{array}\right .\ .
\end{align}
The optimal power distribution is formulated in the next
proposition. 
\begin{Proposition}
The expected sum-rate for continuous layering
			per user, with a {decorrelator} multiuser detector and
			iterative SIC decoding, is given by
			\begin{equation}\label{eqStrong_SIC_BS2_100-prop}
				R_{\rm bs,decorr} = \beta \int\limits_{s_a^+}^{s_b^-}\d s
				(1-Q(s))\frac{-sI'_{\rm opt}(s)\eta\myround{ \beta Q(s_b) }}{ 1 + s
					I_{\rm opt}(s)\eta\myround{ \beta Q(s_b) } } + (1-Q(s_a))R_0(s_a) +
				(1-Q(s_b))R_1(s_b)\ ,
			\end{equation}
			where $\eta(x) = 1-x$, and the optimal layering power distribution
			is given by
			\begin{align}\label{eqStrong_SIC_BS2_105prop}
				I_{\rm opt}(s) = \mycase{
					\begin{array}{cl}
						\p &~ s\leq s_a^{-}\\
						\frac{1-Q(s)-s\cdot Q'(s)}{s^2Q'(s)\eta(\beta Q(s_b))} &~ s_a^+\leq s\leq s_b^- \\
						0 &~ else
				\end{array}}\ ,
			\end{align}
			and the interval for continuous layering satisfies
			\begin{multline}\label{eqStrong_SIC_BS2_106prop}
				\int\limits_{s_a^+}^{s_b^-}\d s \frac{\partial
					G(s_b,s,I_{\rm opt},I'_{\rm opt})}{\partial s_b} +
				G(s_b,s_b,I_{\rm opt},I'_{\rm opt})  \\+ \frac{ \partial} {\partial
					s_b}\mat{  (1-Q(s_b))R_1(s_b) + (1-Q(s_a))R_0(s_a) }
				= 0\ ,
		\end{multline} 
		and
		\begin{align}\label{eqStrong_SIC_BS2_107prop}
			-G(s_b,s=s_a,I_{\rm opt},I'_{\rm opt})  + \frac{ \partial} {\partial
				s_a}\myround{  (1-Q(s_a))R_0(s_a) }
			= 0\ .
		\end{align}
		\label{prop:decorrBS}
\end{Proposition}

\begin{figure}[tb]
	\centering
	\includegraphics[width = 5in]{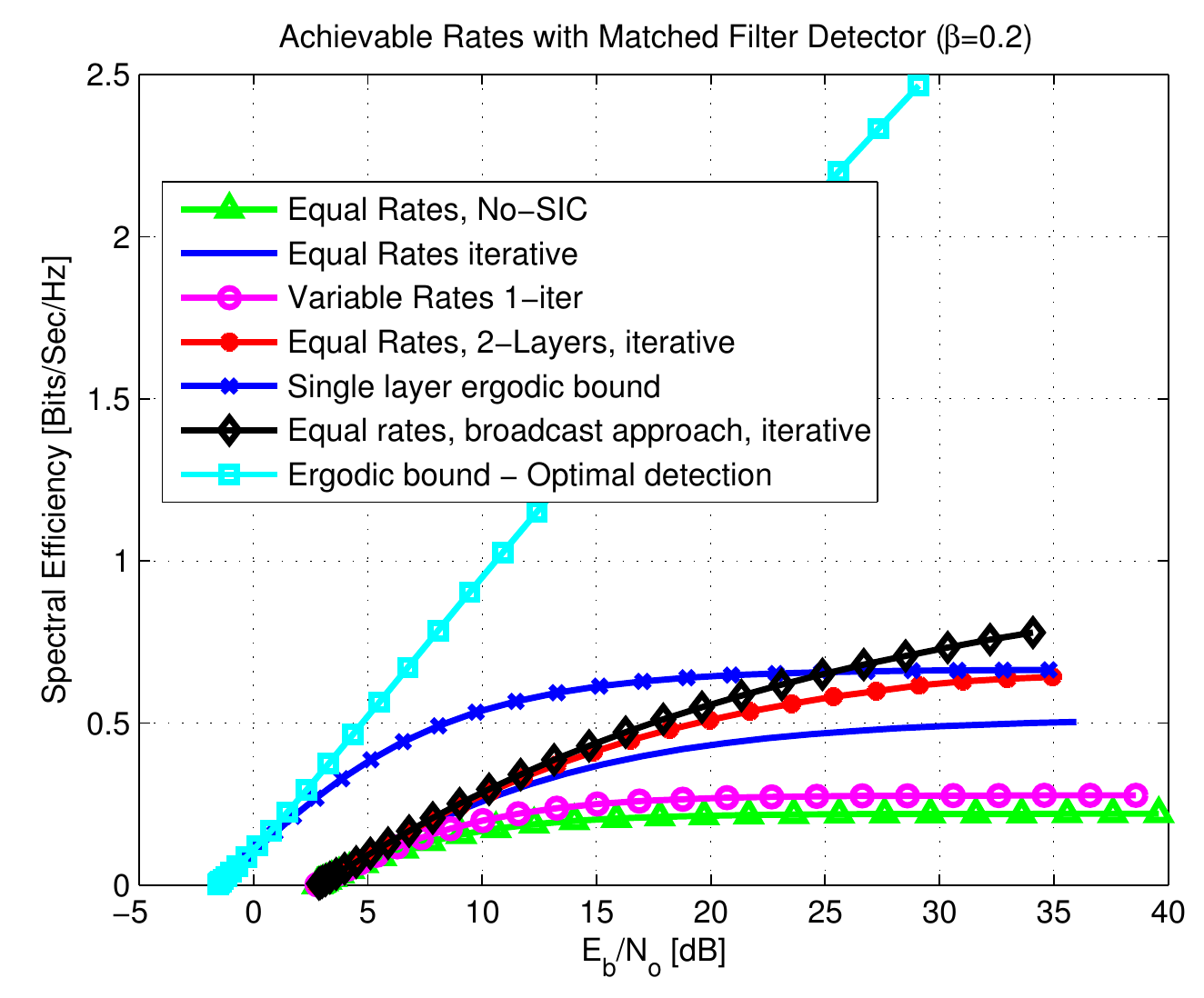}
	\caption{Expected sum-rate for a Rayleigh fading channel,
		with different transmission and decoding strategies, based on a
		matched filter multiuser detector ($\beta=0.2$). 
	}\label{fig:mf0.2}
\end{figure}

\begin{figure}[tb]
	\centering
	\includegraphics[width = 5in]{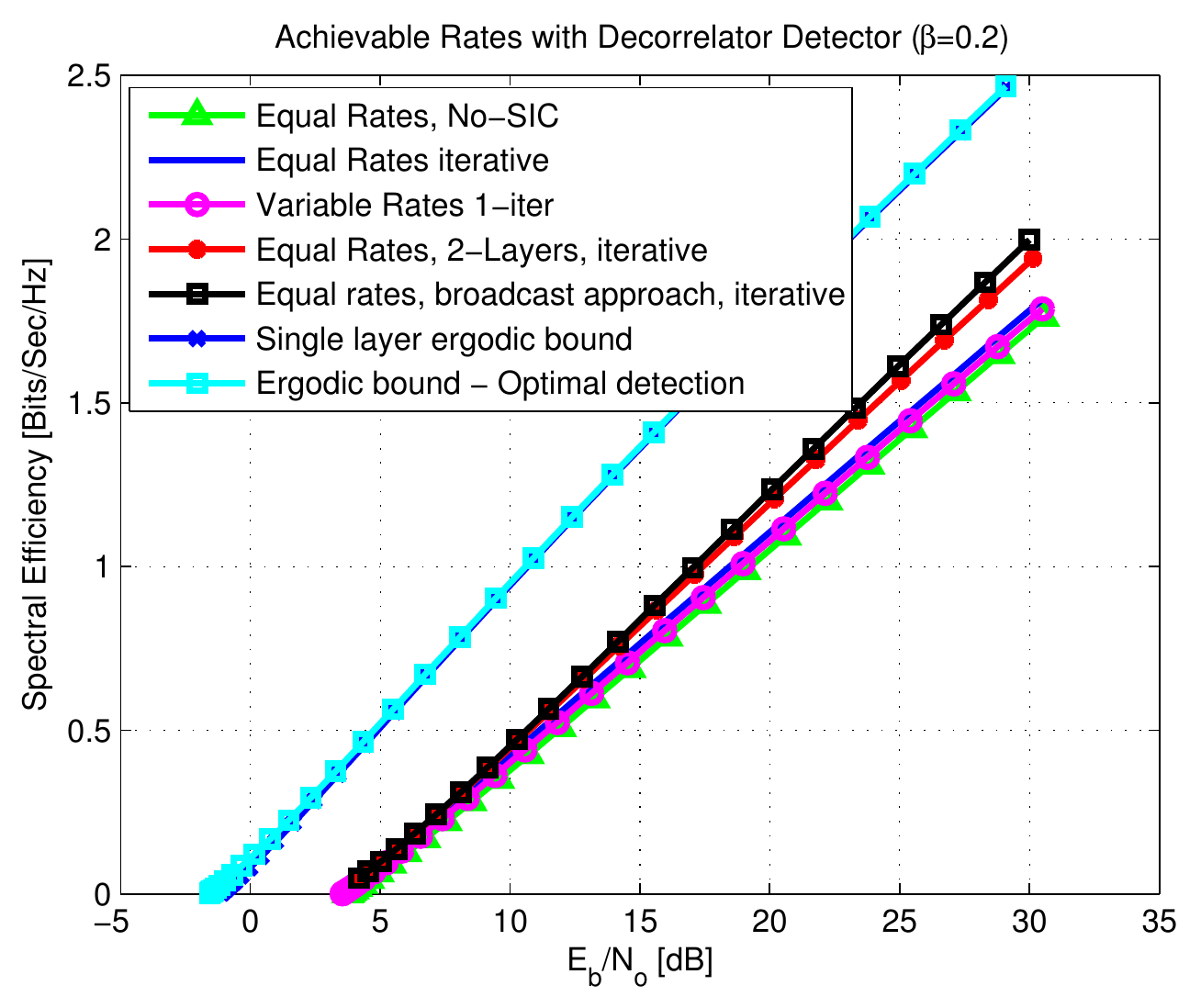}
	\caption{Expected sum-rate throughput for a Rayleigh fading channel,
		with different transmission and decoding strategies, based on a
		decorrelator multiuser detector ($\beta=0.2$). 
	}\label{fig:decorr0.2}
\end{figure}

From the numerical results in \cite{AsLupuKatzSh12}, and Fig.~\ref{fig:mf0.2}, the broadcast approach with multi-round iterative SIC decoding offers a significant spectral efficiency gain over the single-layer coding
strategies. The gain is especially noticeable for the lower system
loads. Interestingly, for $\beta\leq 0.2$, it can be noticed that the
spectral efficiency of the broadcast approach exceeds the MF
single-layer ergodic bound at high $E_b/N_0$ (Fig.~\ref{fig:mf0.2}). For a single user setting, the ergodic bound is always an upper bound which cannot be exceeded. However, in our multiuser setting, an MF detector is used
for the ergodic bound, and the MF detection is information lossy. In
the broadcast approach, the MF detection is performed over and over
for every layer according to the iterative decoding scheme. Hence,
the broadcast approach with iterative decoding may outperform the
optimum single-layer coding with transmitter channel side
information (ergodic bound). Generally speaking, this result should
not be limited to small system loads. Any non-zero slope of the
broadcast approach is sufficient for exceeding the MF single-layer
ergodic bound. However, the crossing level will be at very high
$E_b/N_0$ values. Figure \ref{fig:decorr0.2} demonstrates the
achievable rates for a Rayleigh fading channel with a decorrelator
based multiuser detection, with different transmission and decoding
strategies.

It can be concluded that unequal transmission rate assignment is a practical strategy, as the
base-station, aware of all its users, can take care of rate
allocation. This is in contrast to ergodic bounds, where the users
must transmit at rates matching their experienced fading
realizations (which is an impractical assumption). In SIC with
the strongest users and single-iteration detection, the subsets are
ordered regardless of the instantaneous channel realizations. Therefore, such an assignment can be done once for every new
subscriber. It was shown in this work that for single-iteration
decoding, unequal rate allocation maximizes the spectral efficiency
since the decoding order is fixed. However, for iterative SIC
decoding, the decoding order is no longer fixed, and it is shown
that equal rate allocation maximizes the expected sum-rate.

It is worth noting that systems employing decorrelator detection
can significantly gain from using SIC at system loads close to
$\beta=1$. For such system loads, single-user detection is
interference-limited, and therefore, the achievable rate can be
infinitesimally small. While with SIC, only the users decoded first
transmit at low rates. Gradually, the effective system load for
decoding is reduced, and higher spectral efficiency can be achieved
for other users, resulting in higher sum-rates.

The single-layer analysis was extended to a multi-layer coding
broadcast approach per user. The expected sum-rate, under iterative
decoding with linear multiuser detectors, is optimized, and the
optimal power distribution is obtained (for a decorrelator and an MF
detector). The achievable spectral efficiency for a linear matched
filter detector shows significant gains over the single-layer coding
approach. The interesting observation here is that the expected
spectral efficiency exceeds the single-layer ergodic sum-capacity.
The ergodic bound assumes that every user transmits at a rate
matched to its decoding stage and channel realization. For a single-user setting, the ergodic bound is always an upper bound for the
broadcast approach. However, in our multiuser setting, an MF detector
is used for the ergodic bound, and the MF detection is information
lossy. In the broadcast approach, the MF detection is performed over
and over for every layer according to the iterative decoding scheme.
Therefore, the broadcast approach can provide spectral efficiencies
exceeding those of a single layer coding with channel side
information, when an MF detector is used.

\subsection{The Broadcast Approach for Source-Channel Coding}\label{Sec6_2}

In networks, it may be commonly required to minimize the distortion of the source information rather than maximize the expected rate.
This broadcast approach is useful in a variety of
applications, and it matches the successive refinement (SR) source
coding approach \cite{Koshelev:80,successiveCover1991,RI99} and later works \cite{RI99,NgTian07,Tian08,Ng09,NgTian12}. That is, the more information rate is provided, the less average distortion is evident in the reconstructed source. On a wireless fading channel, in order to minimize the
expected distortion at the receiver, it is essential to find the
optimal power allocation in the broadcast strategy and this is
indeed our focus in this section. This cross-layer design approach was, in fact,
already suggested in \cite{ShitzSteiner03}.

The broadcast-SR approach facilitates to achieve via coding the basic features
of analog communications, that is the better the channel, the
better the performance (say measured by received SNR (MMSE)).
Furthermore, that is without the transmitter knowing the state channel realization, see applications as referenced in the book \cite{DuhamelKieffer09}.

The initial effort on this problem was  made in
\cite{Sesia:05}, where the broadcast strategy coupled with SR source
coding was compared with several other schemes. The optimization
problem was formulated by discretizing the continuous fading states,
and an algorithm was devised when the source coding layers are
assumed to have the same rate. This algorithm, however, does not
directly yield the optimal power allocation when the fading states
are discrete and pre-specified, nor does it give a closed-form
solution for the continuous case. This problem is also considered in \cite{EtemadiJafarkhani:06}, which provides
an iterative algorithm by separating the optimization problem into
two sub-problems. The study in \cite{Ng09} provides a
recursive algorithm to compute the optimal power allocation for $M$
fading states, with worst-case complexity of $O(2^M)$. Furthermore, by
directly taking the limit of the optimal solution for the discrete
case, a solution was given for the continuous case optimal power
allocation, under the assumption that the optimal power allocation
is concentrated in a single interval. Similar problems were
considered in  \cite{CaireNarayanan:05,Gunduz:06} in the high
SNR regime from the perspective of distortion exponent. Successive refinement, as combined with the broadcast approach, gives idea beyond the basic setting and was 
used in \cite{KimPark20}.

The work in \cite{Tian08} proposes a new algorithm that can compute in linear time, i.e., of
$O(M)$ complexity, the optimal power allocation for the case with
$M$ discrete fading states. Furthermore, it provides a derivation of the
continuous case optimal power allocation solution by the classical
variational method \cite{GF91}. Both the algorithm and
the derivation rely on an alternative representation of the Gaussian
broadcast channel capacity, which appeared in \cite{TS02}. The
dual problem of minimizing power consumption subject to a given
expected distortion constraint is also discussed.

\subsubsection{SR with Finite Layer Coding}
Finite-layer coding can be matched to a finite number of fading states, the $M$ possible
power gains in an increasing order $s_1< s_2<...<s_M$ are
distributed according to a probability mass function $p_i$ such that
$\sum_{i=1}^M p_i=1$. The transmitter has an average power
constraint $P$, and if power $P_i$ is allocated to the $i^{\rm th}$ layer
in the broadcast strategy, the $i^{\rm th}$ layer channel rate $R_i$ is
given by
\begin{align}
	R_i=\frac{1}{2}\log\left(1+\frac{s_i P_i}{1+s_i\sum_{j=i+1}^MP_j}\right)=\frac{1}{2}\log\left(1+\frac{ P_i}{1/s_i+\sum_{j=i+1}^MP_j}\right)\ .
	\label{eqn:rate}
\end{align}
From the second expression in \eqref{eqn:rate}, the
equivalence to broadcast on a set of channels with different noise
variances is clear. Let $n_i\triangleq1/s_i$, which implies $n_1>
n_2> ...> n_M$ are the equivalent noise power on the channels. The
layers corresponding to smaller values of $s_i$ (and larger values of
$n_i$) will be referred to as the lower layers, which is consistent
with the intuition that they are used to transmit the more protected
base layers of the SR source coding. Since the Gaussian source is successively refinable
\cite{successiveCover1991}, the receiver with power gain $s_i$ can thus
reconstruct the source within distortion
\begin{align}
	D_i=\exp\left(-2b\sum_{j=1}^i R_j\right)\ ,
	\label{eqn:distortion}
\end{align}
where $b$ is the bandwidth expansion coefficient. Combining (\ref{eqn:rate}) and (\ref{eqn:distortion}), the problem we wish to solve is essentially the following minimization over the power allocation $(P_1,P_2,...,P_M)$:
\begin{align}
	\label{eqn:originalOpt} 
\left\{
\begin{array}{ll}
	\min &  \sum_{i=1}^M p_i
	\left(\prod_{j=1}^i\left(1+\frac{
		P_j}{1/s_j+\sum_{k=j+1}^MP_k}\right)\right)^{-b}\\
		{\rm s.t.} & P_i\geq 0,\quad i=1,\dots ,M\\
	&\displaystyle \sum_{i=1}^M P_i\leq P
\end{array}	\right. \ .
\end{align}
When the fading state is continuous, the density of the power gain
distribution is then given by $f(s)$, which is assumed to be
continuous and differentiable almost everywhere. In this case, the
goal is then to find a power allocation density function $P(s)$, or
its cumulative function, which minimizes the expected distortion, see more details in \cite{Tian08}.

\subsubsection{The Continuous SR-Broadcasting}
We next turn our attention to the case of a continuum of layers, which
is, in fact, the case considered in \cite{ShitzSteiner03}. To
facilitate understanding, we first provide a less technical derivation
under the assumption that the optimal power allocation concentrates
on a single interval of the power gain range, and show that this is
indeed true for some probability density function $f(s)$. This
simple derivation provides important intuitions for the general
case, based on which a more general derivation is then given and
some properties of the solution are subsequently discussed. For
simplicity, we first assume $f(s)$ has support on the entire
non-negative real line $[0,\infty)$. Later it is shown that this
assumption can be relaxed. The optimization problem can be reformulated as follows. Define
\begin{align}
	I(i)=\exp\left(\sum_{j=1}^i 2R_j\right)\ .
\end{align}
We take the number of layers to infinity and the constraint becomes
an integral equation, where we  convert back to the power gain $s$
instead of noise power $n$, and it is clear we can replace the
inequality by equality without loss of optimality
\begin{align}
	\int\limits_0^\infty I(s)\frac{1}{s^2}\d s = \int\limits_0^\infty
	\frac{\exp\myround{2R(s)}}{s^2}\d s=P\ ,
\end{align}
where $R(s)$ is the cumulative rate associated with a fading gain
$s$. The term to be optimized is given by
\begin{align}
	\overline{D}(I) = \int\limits_0^\infty f(s)\exp(-2bR(s))\d s =
	\int\limits_0^\infty \frac{f(s)}{I(s)^b}\d s\ .
\end{align}
Note the additional condition that $I(s)$ has to be monotonically non-decreasing, and it should satisfy the boundary conditions $I(0)=1$. Ignoring the positivity constraint $I'(s)\geq 0$ for now, take
\begin{align}
	J(s,I,I')=\frac{f(s)}{I^b(s)}, \quad G(s,I,I')=\frac{I(s)}{s^2}\ .
\end{align}
Hence the optimization problem can be written as
\begin{align}
\min & \quad \int_{0}^\infty J(s,I,I') \d s \quad 
{\rm s.t. }  \quad \int_{0}^\infty G(s,I,I') ds=P\ .
\end{align}
Next, we assume there is a unique interval $[s_1,s_2]$ for which
power allocation is non-zero. Under this assumption, the objective
function reduces to
\begin{align}\label{toAppA1}
	\overline{D}(I)=\int_0^{\infty}J(s,I,I')\d s=\int_{s_1}^{s_2}
	\frac{f(s)}{I(s)^b}\d s+F(s_1)+\frac{1-F(s_2)}{I(s_2)^b}\ ,
\end{align}
where $F(s)$ is the CDF of the fading gain random variable, i.e.,
$F(s)=\int_{0}^s f(r)\d r$, and the constraint becomes
\begin{align}\label{toAppA2}
	P(I)=\int_0^{\infty}G(s,I,I')ds=\int_{s_1}^{s_2}
	I(s)\frac{1}{s^2}\d s+\frac{I(s_2)}{s_2}-\frac{1}{s_1}=P\ .
\end{align}
Then we can write the Lagrangian form $L(I)=\overline{D}(I)+\lambda
(P(I)-P)$. To find the extremal solution, we consider an increment
$q(s)$, and the increment of the Lagrangian functional is given by
$\Delta(q)=L(I+q)-L(I)$. By taking an increment $q(s)$ with $q(s_1)=q(s_2)=0$ as well as
$q(s)=0$ for $s\notin [s_1,s_2]$, then the Euler-Lagrange equation (pages
42-50 in \cite{GF91})  requires
\begin{align}
	J_I+\lambda G_I-\frac{\d}{\d s}[J_{I'}+\lambda G_{I'}]=0\ ,
\end{align}
with
\begin{align}
	J_I=\frac{-bf(s)}{I^{b+1}(s)}\ ,\quad G_I=\frac{1}{s^2},\quad J_{I'}=G_{I'}=0\ ,
\end{align}
which further simplifies to
\begin{align}
	I(s)=\left(\frac{bf(s)s^2}{\lambda}\right)^{1/(b+1)}\ .
	\label{eqn:Is}
\end{align}
At this point, it is clear that for $I'(s)\geq 0$ to be true, which
is necessary for $I(s)$ to be a valid solution, $f(s)s^2$ should
have non-negative derivative in any interval such that
(\ref{eqn:Is}) holds. In fact, for any interval that a positive rate is
allocated to, $f(s)s^2$ should have strictly positive derivative
such that $I(s)$ is strictly increasing. If there is only one
interval over the support of $f(s)$ where $f(s)s^2$ has strictly
positive derivative, then the single interval solution assumption is
indeed true. Now, since $q(s_2)$ can be arbitrary, at this variable
end (pages 25-29 in \cite{GF91}) a necessary condition
for an extremum is
\begin{align}
	\label{eqn:simples2}
	\frac{-b(1-F(s_2))}{I(s_2)^{b+1}}+\lambda\frac{1}{s_2}=0\ ,
\end{align}
which gives
\begin{align}
	\lambda=\frac{bs_2(1-F(s_2))}{I(s_2)^{b+1}}\ .
\end{align}
Since $I(s_1)=1$,
$\lambda=bf(s_1)s^2_1$, with the expression of $I(s)$ gives one
boundary condition
\begin{align}
	\label{eqn:boundarys2}
	1-F(s_2)=f(s_2)s_2\ .
\end{align}
The lower bound $s_{1}$ is determined by the power constraint, from
which we have
\begin{align}
	\int_{s_{1}}^{\infty}\frac{I(s)}{s^2}\d s=\int_{s_{1}}^{s_2}\left(\frac{f(s)}{f(s_1)s^2_1}\right)^{1/(b+1)}s^{-2b/(b+1)}\d s+\frac{1}{s_2}\left(\frac{f(s_2)s_2^2}{f(s_1)s^2_1}\right)^{1/(b+1)}=P+\frac{1}{s_1}\ ,
	\label{eqn:boundarys1}
\end{align}
where in the second equation we split the integral into two parts
partitioned by $s=s_{2}$. Hence, the unique extremal solution is
\begin{align}
	I(s)=\left(\frac{f(s)s^2}{f(s_1)s^2_1}\right)^{1/(b+1)}\ ,
	\label{eqn:constraint}
\end{align}
with the boundary conditions specified by (\ref{eqn:boundarys2}) and (\ref{eqn:boundarys1}). To find the corresponding power allocation, define
$T(s)=\int_{s}^{\infty} P(r)\d r$. We derive from
(\ref{eqn:constraint}) that 
\begin{align}
		T(s)&=\int\limits_s^\infty \frac{I(r)}{I(s)}\d r-\frac{1}{s}
		=\left(\frac{f(s_{2}){s^2_{2}}}{f(s)s^2}\right)^{1/(b+1)}\frac{1}{s_2}+\int_{s}^{s_{2}}\left(\frac{f(r)r^2}{f(s)s^2}\right)^{1/(b+1)}\frac{1}{r^2}\d r-\frac{1}{s}\ .
\end{align}
Through some basic calculation, it can be shown that this is
the same solution as that in \cite{Ng09}. Thus, the limit of the
optimal solution of the discrete case in \cite{Ng09} indeed
converges to the extremal solution derived through the classical
variational method. Furthermore, the variational method derivation
directly asserts that $f(s)s^2$ has a non-negative derivative for
any positive power allocation interval. This condition was, however,
lacking in the derivation in \cite{Ng09}.

We consider the average achievable distortion for a SISO Rayleigh
fading channel, with CDF $F(s) = 1-\exp\myround{-\frac{s}{\overline{s}}}$,
where $\overline{s}$ is the expected fading gain power. For this
distribution, the optimal power allocation is single interval
continuous, and zero outside the interval $[s_1, s_2]$. That can be
immediately observed from $f(s)s^2$ by taking its first derivative
\begin{align}\label{eq_r_2}
	\frac{\d}{\d s}f(s)s^2 =
	\left(\frac{2s}{\overline{s}}-\frac{s^2}{\overline{s}^2}\right)
	\exp\myround{-\frac{s}{\overline{s}}}\ ,
\end{align}
where $\frac{\d}{\d s}f(s)s^2\geq 0$ on a single interval
$s\in[0,2\overline{s}]$. Then the upper bound $s_2\in
[0,2\overline{s}]$ is determined by (\ref{eqn:boundarys2}), which reduces to
\begin{align}\label{eq_r_3}
	\exp\myround{-\frac{s_2}{\overline{s}}} = \frac{s_2}{\overline{s}}
	\exp\myround{-\frac{s_2}{\overline{s}}}\ ,
\end{align}
yielding $s_2 = \overline{s}$. Solving (\ref{eqn:boundarys1}) gives the other boundary value $s_1$, denoted by $s_{1,\rm opt}$; the condition (\ref{eqn:boundarys1}) does not lead to an analytical expression, but can be solved numerically. Then, the general expression
for $I(s)$ for the Rayleigh fading channel is given by
\begin{align}\label{eq_r_5}
	I(s) = \mycase{\begin{array}{ll} 1 & s\leq s_{1,\rm opt}\\
			\myround{\frac{s^2}{s^2_{1,\rm opt}}\exp{\myround{-\frac{s-s_{1,\rm opt}}{\overline{s}}}}
			}^{1/(b+1)} & s_{1,\rm opt}<s\leq \overline{s}\\
			\myround{\frac{\overline{s}^2}{s^2_{1,\rm opt}}\exp{\myround{-\frac{\overline{s}-s_{1,\rm opt}}{\overline{s}}}}
			}^{1/(b+1)} & s> \overline{s}
	\end{array}}.
\end{align}

\begin{figure}[tb]
	\unitlength=1in
	\begin{center}
		\includegraphics[width = 5in]{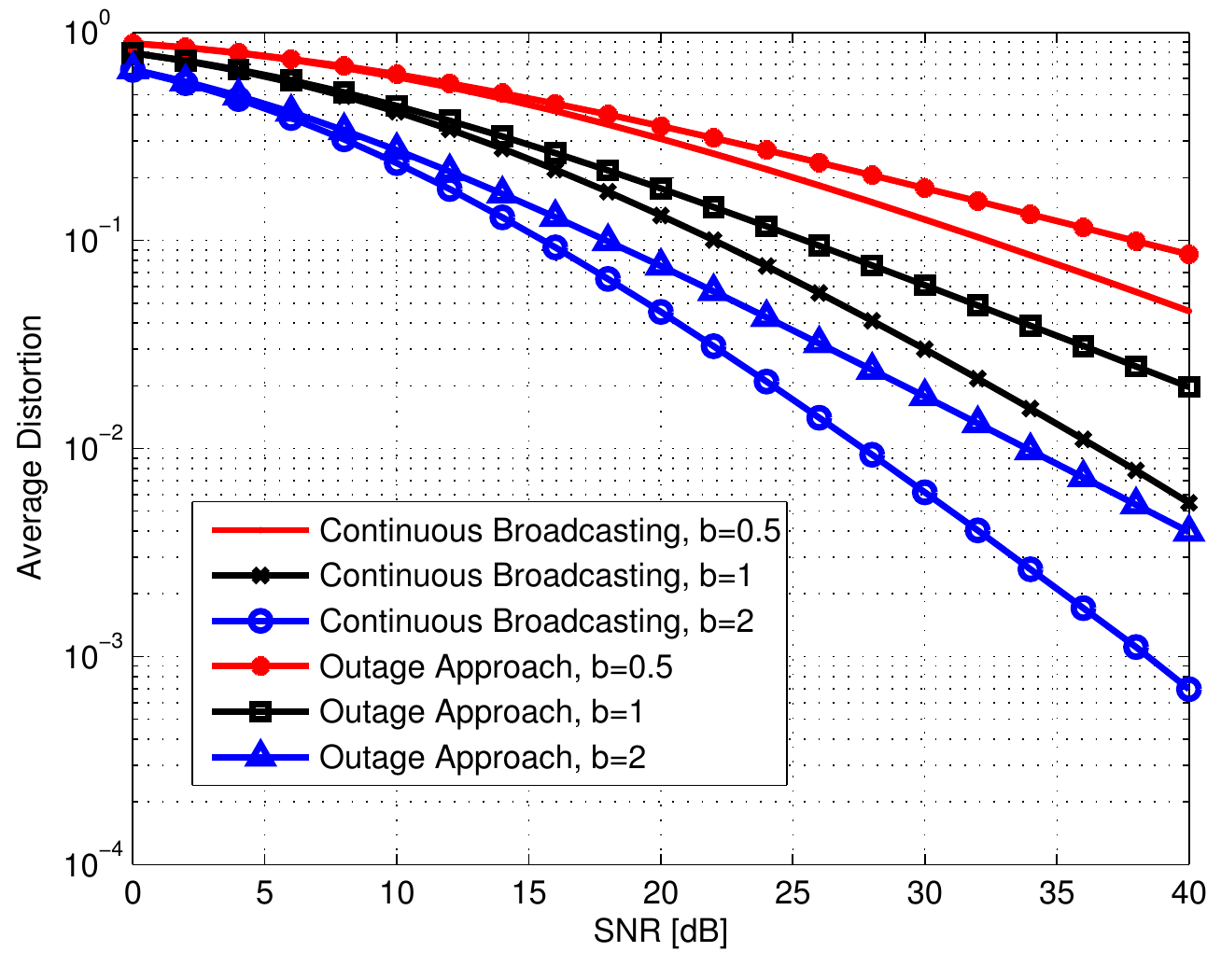}
	\end{center}
	\caption{Minimal average distortion, a comparison of outage approach and broadcast approach, for bandwidth expansions  $b=0.5, 1, 2$.}\label{figDist}
\end{figure}

In Fig.~\ref{figDist}, the average distortion bounds for Rayleigh
fading channels are demonstrated for three different values of
bandwidth expansion values ($b=0.5,1,2$). For every bandwidth
expansion, the minimal average distortion of the outage approach and
broadcast approach are compared. It can be noticed that the smaller
$b$ is, the larger is the broadcast gain, which can be defined as
the SNR gain of the broadcast approach over the outage approach for
the same average distortion value. Thus, the benefit of the broadcast
approach compared to the outage directly depends on the system
design parameter $b$.

\subsection{The Information Bottleneck Channel}\label{Sec6_3}

An interesting setting is the information bottleneck channel, the objective of which is the efficient transmission of data over a wireless block fading channel that is connected to a limited capacity reliable link. This setting is known as the bottleneck channel \cite{AsSh20_Bottleneck}. Two main broadcast approaches are considered for the bottleneck channel in \cite{AsSh20_Bottleneck}. The first is an oblivious approach, where the sampled noisy
observations are compressed and transmitted over the bottleneck
channel without having any knowledge of the original information
codebook. This is compared to a decode-forward (non-oblivious) approach, where the sampled noisy data is decoded,
and whatever is successfully decoded is reliably transmitted over
the bottleneck channel. This work is extended for an uncertain
bottleneck channel capacity setting in \cite{SteinerShamai2020}, where the transmitter is not
aware of the available backhaul capacity per transmission and knows only its distribution. In both settings, it is possible
to analytically describe the optimal
continuous layering power distribution that maximizes the
average achievable rate in closed-form expressions. The topic is covered in more details \cite{steinerSh2020bottleneck}.

\begin{figure}[]
	\centering
	\includegraphics[width=4.5 in]{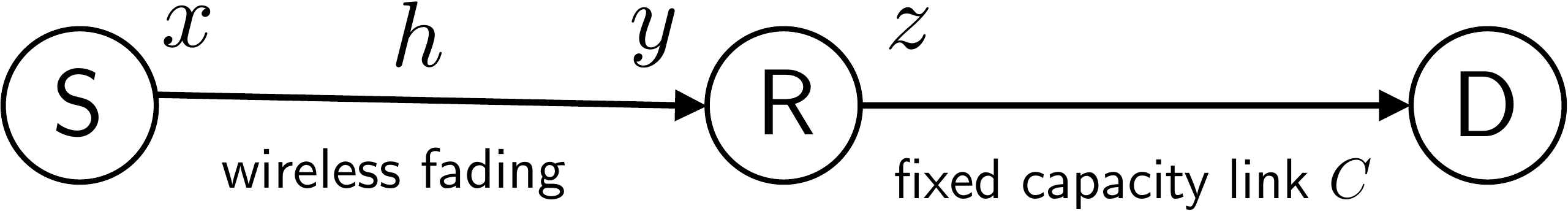}
	\caption{Information bottleneck fading channel system model block diagram.}
	\label{fig:model}
\end{figure}

The Gaussian bottleneck problem is depicted in Fig.~\ref{fig:model}. Consider a Markov chain of a random variable triplet $x-y-z$, related according to 
\begin{equation}\label{eqChannelModel}
	y = h\cdot x + n\ ,
\end{equation}
where $x$ and $n$ are i.i.d, with $n\sim {\cal N}(0,1)$. The fading gain is $s=|h|^2$ fixed per transmission block. The SNR is $P\cdot s$, where $P$ is the transmission power $\bbe[X^2]=P$. The fading gain $s$ distribution is known to transmitter and receiver as the broadcast approach \cite{ShitzSteiner03} discussed earlier. The output $z$ of the bottleneck channel is a compressed version of the received signal $y$ under the bottleneck channel capacity $C$ constraint. The optimization problem can be formalized as
\begin{equation}\label{maxEq}
	\max\limits_{\PP(z|y), \PP(x)~{\rm s.t.} I(y;z)\leq C} ~ I(x;z)\ .
\end{equation}
If $x$ is Gaussian, then it is clear \cite{Tishby2004,tishby99information} that $y-z$ is also a Gaussian channel. Therefore, the maximization result of \eqref{maxEq} is
\begin{equation}\label{maxCompressionRate}
	C_{\rm Obliv} = I(x;z) = \frac{1}{2}\log\myround{ \frac{  1+P|h|^2 }  {1+P|h|^2\cdot \exp(-2C)}} \ ,
\end{equation}
which is a direct result of the rate-distortion approach. The output of the relay $y$ may be represented by quantizing its input 
\begin{equation}\label{eqRD}
	z=y+m\ ,
\end{equation}
where $P|h|^2+1$ is the variance of $y$ of the channel model \eqref{eqChannelModel}. The quantization noise variance, denoted by $m$, is obtained by $I(z;y) = C$, i.e.,
\begin{equation}\label{eqVar}
	\bbe[m^2] = \frac{P|h|^2+1}{\exp(2C)-1}\ .
\end{equation}
The problem underhand is the reliable transmission rate from $x$ to destination with an oblivious relay that uses the bottleneck channel to send compressed versions of its input, without knowledge of the transmitter codebook.

For a DF non-oblivious relay, the relay can decode its input and then send the decoded data under bandwidth limitation $C$ over the bottleneck channel $y-z$. Hence, the minimum of two capacities provides the achievable transmission rate 
\begin{equation}
	C_{\rm DF}=\min \left\{ \frac{1}{2}\log(1+P|h|^2),C \right\}\ .
\end{equation}
An alternative setting that generalizes the current model is a variable availability of the bottleneck capacity, which is common in cellular uplink transmission. This can be due to changing traffic loads over time on the network \cite{ROY_6613623}. This means that the relay-destination bottleneck channel capacity $C$ is a random variable. The source transmitter knows its distribution; however, like in the wireless fading channel, feedback to the transmitter is not available due to the capacity variability dynamics. If the relay perfectly knows, per received codeword, the bottleneck currently available capacity, it can adapt its data rate. However, if the relay has no access to the capacity per codeword, it may use successive refinement source coding \cite{Tian08} matched to the capacity distribution.

Consider a fading wireless link to $y$, where $s=|h|^2$ is a unit variance block fading gain. It is assumed to change independently between codewords and remains fixed over a single codeword. The channel model of $z$ is expressed by its block fading gain as
\begin{equation}
	z = \sqrt{{\rm FPR}_{\rm eq}} x + n\ ,
\end{equation}
where $n$ is a unit variance Gaussian noise. The equivalent fading gain ${\rm FPR}_{\rm eq}$ is given by
\begin{equation}\label{snreq}
	{\rm FPR}_{\rm eq} = \frac{s(1-\exp(-2C))}{1+s\cdot P\cdot \exp(-2C)}\ ,
\end{equation}
which is directly obtained from \eqref{eqVar}. It may be observed that ${\rm FPR}_{\rm eq}$ is finite for $s\geq 0$, and at the limit of $s\rightarrow\infty$ becomes
\begin{equation}\label{snreq_limit}
	\lim\limits_{s\rightarrow\infty} {\rm FPR}_{\rm eq} = (\exp(2C)-1)/P\ ,
\end{equation}
and the ergodic capacity of the bottleneck fading channel is  
\begin{align}\label{erg_Obliv}
	C_{\rm Obliv,Erg} &= \bbe_s\left[\frac{1}{2}\log(1+P\cdot {\rm FPR}_{\rm eq})\right]  \\
	&=   \bbe_s\left[ \frac{1}{2} \log\left( 1+ \frac{s\cdot P\cdot (1-\exp(-2C))}{1+s\cdot P\cdot \exp(-2C)} \right) \right]\ . \label{eqn111}
\end{align}
The continuous broadcasting approach solution is rather straightforward here. The channel model here can be expressed by equivalent fading gain $\nu={\rm FPR}_{\rm eq}$ from (\ref{snreq}), which depends on the bottleneck channel capacity $C$ and the distribution of the channel fading gain $s$. In this bottleneck channel with oblivious relaying, the broadcast approach is optimized for a fading distribution $F_\nu(u)$ of (\ref{snreq}). Obtaining optimal power distribution can be derived directly. Clearly for high bottleneck channel capacity $C\rightarrow\infty$, then ${\rm FPR}_{\rm eq}\rightarrow s$.

A DF relay (non-oblivious approach) can decode the received signal $y$, and reliably convey to the destination the decoded data under capacity limit $C$. An ergodic upper bound of the bottleneck fading channel $C_{\rm DF,Erg}$, is not achievable for a block fading channel, as the transmitter has no CSI. It is beneficial to transmit a multi-layer coded signal for this channel model. The DF non-oblivious ergodic capacity is expressed as
\begin{equation}\label{eq_erg}
	C_{\rm DF,Erg} = \bbe_s\left[\min\left\{C,~\frac{1}{2}\log(1+sP) \right\}\right]\ ,
\end{equation}
where a single fading realization is assumed per transmission and decoding of a single codeword, for the slowly fading channel.

The continuous broadcasting approach for the non-oblivious DF approach can be optimized in the following way. A transmitted signal $x$ is multi-layer coded in a continuum of layers. The received signal $y$ is decoded layer-by-layer in a successive decoding manner. All the successfully decoded layers with a total rate up to $C$, the bottleneck channel capacity, can be reliably conveyed over the bottleneck channel. The optimization goal is to maximize the average transmitted rate over the bottleneck channel in this block fading channel model. We formulate here the optimization of power density distribution function $\rho_{\rm opt}(u)$ so that the average transmission rate is maximized under the bottleneck channel capacity constraint.

\begin{Proposition}\label{thm:RbsNonObliv}
	For the non-oblivious block fading bottleneck channel, the total \emph{expected average achievable rate of the broadcast approach} is obtained by the following residual power distribution function
	\begin{align}\label{prop_1}
		I_{\rm opt}(u) =\left\{
			\begin{array}{ll}
 \arg \max\limits_{I(u)} & 	 \frac{1}{2} \displaystyle \int\limits_0^{\infty} \d u
		(1-F_s(u)) \frac{\rho(u)u}{1+I(u)u} \\
\mbox{\rm s.t.}		&  \displaystyle\int\limits_0^{\infty} \d u \frac{\rho(u)u}{1+I(u)u} \leq C\ .
	\end{array}\right.\ , 
	\end{align}
	where $F_s(u)$ is the CDF of the fading gain random variable, and $C$ is the bottleneck channel capacity. The optimal power allocation $I_{\rm opt}(u)$ is given by
	\begin{align}\label{prop_2}
		I_{\rm opt}(u)=\mycase{
			\begin{array}{ll}
				P & u< u_0\\
				\frac{1-F_s(u)+\lambda_{\rm opt}-u\cdot f_s(u)}{u^2f_s(u)} & u_0\leq u\leq u_1\\
				0 & u> u_1
		\end{array}}\ ,
	\end{align}
	where $\lambda_{\rm opt}\geq 0$ is a Lagrange multiplier specified by
	\begin{align}\label{prop_3}
		\lambda_{\rm opt} = -u_1\cdot f_s(u_1) -1+F_s(u_1)\ ,
	\end{align}
	and for any $\lambda_{\rm opt}>0$,  
	\begin{align}\label{prop_4}
		u_1^2 \cdot f_s(u_1) = \exp(2C)\cdot u_0^2 \cdot f_s(u_0)\ .
	\end{align}	
\end{Proposition}

\begin{figure}[]
	\centering
	\includegraphics[width = 5in]{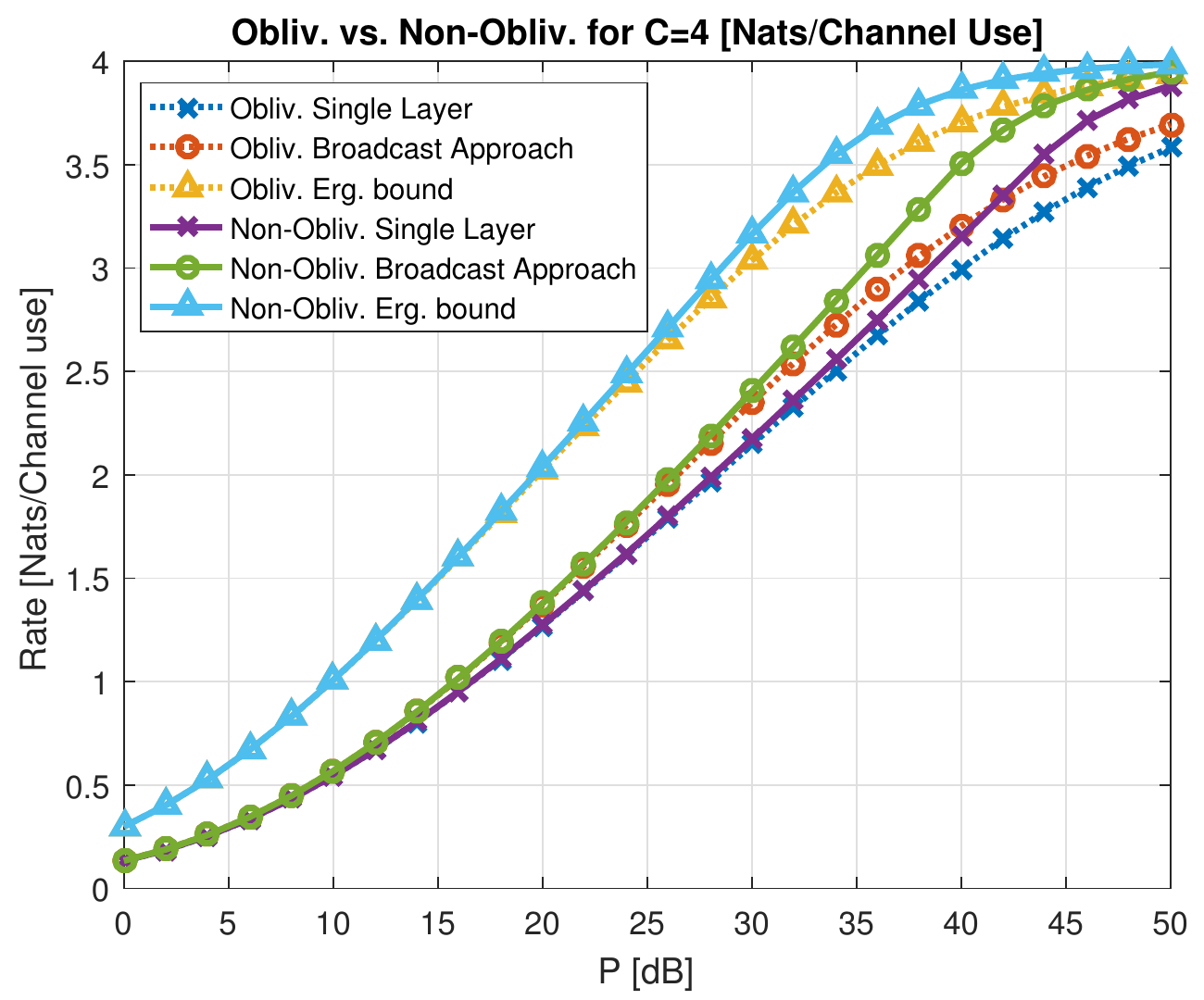}
	\caption{Oblivious vs. Non-Oblivious single layer coding and broadcast approach compared to the ergodic capacity, for bottleneck channel capacity of $C=4$ [Nats/Channel Use].}
	\label{fig:OblivVsNonObliv}
\end{figure}
Figure \ref{fig:OblivVsNonObliv} demonstrates the achievable rates with a non-oblivious approach as compared to an oblivious approach for a bottleneck channel capacity $C=4$ (Nats/Channel use). It can be observed here that in the high SNR region, the gain of the broadcast approach compared to single-layer coding is higher with a non-oblivious approach.

\subsubsection{Uncertainty of Bottleneck Capacity}
A common case in cellular uplink is a variable availability of backhaul capacity. This may be the result of variable loads on the network over time. Traffic congestion of internet data may lead to changing the availability levels of the backhaul \cite{ROY_6613623}. On the bottleneck channel, this means that the relay-destination link capacity $C$ is a random variable. It may be assumed that the transmitter is aware of the average capacity and its distribution. However, like the wireless fading channel, the capacity variability dynamics may not allow time for feedback to the transmitter. The following subsection considers the case that the relay is fully aware of the current bottleneck available capacity for each received codeword.

Consider a bottleneck channel with discrete capacity levels represented by $N$ random  capacity values $\{C_i\}_{i=1}^{N}$, such that $C_1\leq C_2 \leq \cdots \leq C_N$ with corresponding probabilities $\{p_{b,i}\}_{i=1}^{N}$, such that $p_{b,i}\geq 0$ and $\sum_{i=1}^{N}p_{b,i}=1$. The average capacity of the bottleneck channel is 
\begin{align}\label{avg_capacity_1}
	C_{\rm ave} = \sum_{i=1}^{N}p_{b,i}C_i\ .
\end{align}	
The broadcast approach can be derived here for an oblivious relay setting and under an equivalent fading gain distribution. Since the transmitter is not aware of the bottleneck capacity per codeword, and only knows its distribution, the following optimization flow is used for the continuous broadcast approach optimization. 
The combined equivalent channel viewed by the transmitter  is
\begin{equation}\label{snreqCV}
	{\rm FPR}_{\rm eq}(s,C_b) = \frac{s(1-\exp(-2C_b))}{1+s\cdot P\cdot \exp(-2C_b)}\ , \quad \mbox{and}\quad  s=|h|^2\ .
\end{equation}
The continuous broadcast approach is optimized for a fading distribution $F_\mu(u)$, where $\mu={\rm FPR}_{\rm eq}(s,C_b)$ is the equivalent channel gain depending on the fading gain realization $s$, and bottleneck channel capacity $C_b$ available per codeword. The CDF of this fading gain is
\begin{equation}\label{snreqvpdf}
	F_\mu(u)=\sum_{i=1}^{N}p_{b,i} F_s \myround{\frac{u}{1-(1+Pu)\exp(-2C_i)}}\ .
\end{equation}
The main result here is expressed in the following proposition
\begin{Proposition}	
	The power distribution, which maximizes the
	expected rate over the oblivious bottleneck channel, is
	\begin{align}\label{eq:general_Ix1}
		I(x) = \mycase{
			\begin{array}{cl}
				\frac{1-F_\mu(x)-x\cdot f_\mu(x)}{x^2f_\mu(x)} & ,~x_0\leq x\leq x_1 \\
				0 & ,~\mbox{\rm else}
		\end{array}}\ ,
	\end{align}
	where $x_0$ is determined by $I(x_0)=P$, and $x_1$ by
	$I(x_1)=0$. Furthermore, the broadcasting rate is expressed as function of the ${\rm FPR}_{\rm eq}$ distribution $F_\mu(u)$
	\begin{align}\label{eq5Mult_BS}
		R_{\rm opt}(s) =\mycase{
			\begin{array}{ll}
				0 & s< x_0\\
				\log(s/x_0)+\frac{1}{2} \log\left( \frac{f_\mu(s)}{f_\mu(x_0)} \right) & x_0\leq s\leq x_1\\
				\log(x_1/x_0)+\frac{1}{2} \log\left( \frac{f_\mu(x_1)}{f_\mu(x_0)} \right) & s > x_1
		\end{array}}\ .
	\end{align}	
\end{Proposition}	
The derivation of this optimization is based on the analysis in \cite{ShitzSteiner03} for characterizing the power distribution under an equivalent channel model that includes the relayed signal after compression to a rate which matches the bottleneck channel capacity.
The channel model for the relayed signal $z$ can be expressed by its block fading gain, under an oblivious approach. Specifically,
\begin{equation}
	z= \sqrt{{\rm FPR}_{\rm eq}} \cdot x + n\ ,
\end{equation}
where $n$ is a unit variance Gaussian noise, and ${\rm FPR}_{\rm eq}(s,C_b)$ is specified in (\ref{snreqCV}). More details can be found in \cite{steinerSh2020bottleneck}. It is interesting to note here that although the relay can perform successive refinement source coding matched to backhaul capacity distribution, it does not help and cannot increase the expected achievable rate if the relay is informed with
the available capacity per codeword.

An interesting problem arises when the wireless channel is fast fading, and bottleneck channel capacity is random. That is, the fading $h$ \eqref{eqChannelModel} changes independently (i.i.d.) for every channel use. For long codewords, the ergodic nature of the channel can be utilized per transmitted codeword. 
Evidently, under a non-oblivious DF relay, the relay decodes the transmission, and then whatever possible is conveyed through the backhaul.
The interesting part is the oblivious processing. Here the relay should
also convey the fading realizations $h$ and received signal $y$ to the destination, the best possible way. Hence, $h$ plays the role of the source to be conveyed with successive refinement. Furthermore, note that even if all $y$ is provided to the destination, unavailable fading realization vector $h$ makes the capacity behave as $\log \log({\sf SNR})$, as in the i.i.d. channel with unavailable fading at the transmitter and receiver. This problem is analyzed for a known bottleneck capacity in \cite{CaireShamai18}.

\subsection{Transmitters with Energy Harvesting}\label{Sec6_4}

As the last model in this section, we review the channel model of~\cite{zohdytajerharvesting} in which the transmitter relies on an exogenous energy harvesting unit as its only source of energy. Energy harvesting has been evolving rapidly as a promising alternative to systems with lifetime-limited batteries. Communication systems empowered by energy harvesting units rely on ambient sources, which facilitate potentially perpetual sources of power~\cite{lu2015wireless2,lu2015wireless,panatik2016energy,jabbar2010rf}. Specifically, the recent advances in both the theory and implementation of energy harvesting circuitry has facilitated the growth in various wireless domains, e.g., ad-hoc networks~\cite{huang2013spatial}, wireless body networks~\cite{zhang2010energy}, wireless sensor networks \cite{nishimoto2010prototype}, and radio frequency identification systems~\cite{sudevalayam2011energy}, which constitute the main technologies that the IoT relies on.

By relying on harvested energy, the transmitter faces two sources of randomness due to the fading and energy arrival processes. The transmitter knows only the statistical descriptions of these processes while remaining oblivious to the actual realizations of both. We review the optimal distribution of power across information layers and over time in order to maximize the average rate that can be reliably sustained. An interesting observation is that allocation of power across layers and over time can be decoupled into two independent power allocation tasks, one specifying the allocation over time, and the second one optimizing the available power at any given time across different layers. Furthermore, both sub-problems can be solved optimally (under proper assumptions on the fading process).

To lay the context, consider transmission over a slowly-fading Gaussian channel. The channel undergoes block fading, where the fading gain is constant over a block of $n$ channel uses and changes independently across blocks. The block length $n$ is assumed to be sufficiently long such that under the given delay constraints (finite transmission duration), one codeword can be reliably transmitted to the receiver. The input-output relationship across $B$ fading blocks is given by
\begin{equation}\label{eq:channelmodel}
y_b^i = h_b \cdot x_b^i + n_b^i\ , \; \forall\; i\in \n\ ,\;b\in\B\ ,
\end{equation} 
where $x_b^i$ and $y_b^i$ are the transmitted and received symbols at time $i$ in block $b$, $h_b$ is the fading coefficient in block $b$, and $n_b^i$ accounts for the AWGN distributed according to ${\cal N}_{\mathbb{C}}(0,1)$. Denote the channel gains by $s_b= |h_b|^2$, for $b\in\B$, and denote the CDF of $s_b$, known to the transmitter, by $F_b:\mathbb{R}_+\rightarrow [0,1]$. Accordingly,  denote the associated PDF by $f_b:\mathbb{R}_+\rightarrow\mathbb{R}_+$. Finally, set $p_b^i=\mathbb{E}[|x_b^i|^2]$ as the transmission power at time $i$ in block $b$, and define $p_b$ as the aggregate power used in block $b$, i.e., 
\begin{align}\label{eq:power}
p_b= \sum_{i=1}^n\mathbb{E}[|x_b^i|^2]=\sum_{i=1}^np_b^i\ .
\end{align}
Let $g^i_b$ denote the amount of energy harvested during time slot $i$ of block $b$. Accordingly, corresponding to each block $b$ define the vectors $\bp_b= [p_b^1,\dots,p_b^n]^T$ and $\bg_b= [g_b^1,\dots,g_b^n]^T$.
The transmitter is equipped with a battery whose capacity order-dominates that of the amount of harvested energy. This induces a set of power consumption constraints according to which the amount of energy consumption up to each time instant cannot exceed the harvested energy up to that point. Specifically, by defining
\begin{align}\label{eq:1}
\bone_i=[\underset{i}{\underbrace{1,\dots,1}},\underset{n-i}{\underbrace{0,\dots,0}}]^T\ ,\quad\forall i\in\n\ ,
\end{align}
corresponding to each pair $b\in\B$ and $i\in\n$ 
\begin{equation}\label{eq:energy_constraint}
\sum_{j=1}^{b-1}\bone^{T}_n\cdot\bp_j \;+\; \bone^{T}_i\cdot\bp_b  \;\leq \;  \sum_{j=1}^{b-1}\bone^{T}_n\cdot\bg_j \;+\; \bone^{T}_i\cdot\bg_b \ .
\end{equation}
Based on this approach, when the aggregate transmission power over any transmission block is $p$ and the actual channel gain during that block is $s= |h|^2$,  define $\rho(p,s)$ as the density of the power allocated to the information layer indexed by~$s$. Hence, the amount of power allocated to realization $s$ is $\rho(p,s)\d s$, and the amount of interference power imposed on the receiver designated to the channel realization with gain $s$ is
\begin{equation} \label{eq:interference}
 I \left( p, s \right) = \int_{s}^{\infty} \rho \left( p, u \right) \d u \ .
 \end{equation}
To satisfy the power constraint for the aggregate power split across different layers, the following condition must be satisfied.
\begin{align}\label{eq:initial}
I(p,0)=p\ .
\end{align}
Based on such power allocation and interference terms, the average rate over all possible fading realizations within one transmission block is 
\begin{align}\label{eq:rate_average_b}
\Rb (p_b) & = \int_0^\infty [1-F_b(s) ]\; \dfrac{s \cdot \rho_b \left(p_b, s\right)}{1 + s\cdot  I_b \left(p_b, s\right)}\; \d s\ , 
\end{align}
Sum-rate optimization is constrained with the energy availability constraints in~\eqref{eq:energy_constraint} and the aggregate power allocation constraint $I_b(p_b,0)=p_b$. Hence, the optimal allocation of power across information layers and over time is the solution to the following problem, which involves a stochastic guarantee on meeting the power constraints.
\begin{align}\label{eq:Q}
\R
= \left\{
\begin{array}{lll}
\max\limits_{\{\bp_b\},\{\rho_b(p_b,s)\}} & \sum\limits_{b=1}^B \Rb(p_b) & \\
\quad {\rm s.t.} & \PP\left( \sum\limits_{j=1}^{b-1}\bone^{T}_n\cdot\bp_j \;+\; \bone^{T}_i\cdot\bp_b  \;\leq \;  \sum\limits_{j=1}^{b-1}\bone^{T}_n\cdot\bg_j \;+\; \bone^{T}_i\cdot\bg_b \right)\geq \eta \, , & \forall \;b,i \\
&I_b(p_b,0)=p_b\, , &  \forall \;b\\
& \bp_b\succeq  0\, , & \forall\; b
\end{array}\right.\ .
\end{align}

\subsubsection{Optimal Power Allocation Densities}

Based on~\eqref{eq:Q}, for any given set of power allocation terms $\{\bp_b:b\in\B\}$, the set of optimal densities can be found as the solution to
\begin{align}\label{eq:Q_tilde}
\P(\bp_1,\dots,\bp_B)= \left\{
\begin{array}{lll}
\!\!\!\underset{\{\rho_b(p_b,s)\}}{\max} & \!\!\!\sum\limits_{b=1}^B \Rb(p_b) & \\
\quad{\rm s.t.} &  \!\!\!I_b(p_b,0)=p_b\, , &  \forall\; b
\end{array}\right. .
\end{align}
By noting the expansions of $I_b(p,s)$ and $\Rb(p)$ in \eqref{eq:interference} and \eqref{eq:rate_average_b}, respectively, we have
\begin{align}\label{eq:rate_b2}
\Rb(p)= - \int_0^\infty [1-F_b(s) ]\; \cdot\dfrac{s \cdot \frac{\partial I_b(p,s)}{\partial s}}{1 + s\cdot  I_b \left(p, s\right)}\; \d s\ .
\end{align}
Based on this characterization,
for a given power allocation over time $\{\bp_b:b\in\B\}$, we have
\begin{equation}\label{InterferenceResiduals}
I_b(p,s)= 
\begin{cases}
\dfrac{1 - F_b(s)}{s^2 f_b (s)} - \dfrac{1}{s}\ , &  \quad \ell_b \leq s \leq u_b\\
0\ , &  \quad \mbox{\rm otherwise}
\end{cases}\ ,
\end{equation}
where $\ell_b$ and $u_b$ can be determined uniquely by solving 
\begin{align}\label{eq:lu}
I_b(p_b,\ell_b)\; = \; p_b \qquad\mbox{and}\qquad I_b(p_b,u_b) \; = \;  0\ .
\end{align}
The analysis directly follows the same line of arguments as in the setting without an energy harvesting transmitter. Based on the characterization of interference residual functions $\{I_b:b\in\B\}$, the optimal rate over the fading block $b\in\B$ at the fading state $s$ is
\begin{align}
\label{eq:rate2}
R_b(p_b,s) \; = \;
\begin{cases}
0 & \mbox{\rm for}\quad s<\ell_b\\
\ln\frac{s^2f_b(s)}{\ell_b^2f_b(\ell_b)} & \mbox{\rm for}\quad \ell_b\leq s \leq u_b\\
\ln\frac{u_b^2f_b(u_b)}{\ell_b^2f_b(\ell_b)} & \mbox{\rm for}\quad u_b \leq s
\end{cases}\ ,
\end{align}
Subsequently, the average transmission rate over the fading block $b$ with aggregate power $p_b$ is
\begin{align}
\label{eq:rate_average_b2}
R_b(p_b) \; = \;   \ln \frac{u_b^2 f_b(u_b)}{\ell_b^2 f_b(\ell_b)} - \int_{\ell_b}^{u_b} \left[\frac{2}{s}+\frac{f'_b(s)}{f_b(s)}\right]\cdot F_b(s)\; \d s \ .
\end{align}
which can be used to show the interesting property that for any  continuous CDF, $R_b(p_b) $, it is non-decreasing and strictly concave in $p_b$~\cite{zohdytajerharvesting}.

\subsubsection{Optimal Power Allocation over Time}
\label{sec:OptimalPolicy} 

Next, based on the given allocation of power across information layers and leveraging the key properties of $R_b(p_b)$, i.e., concavity and being non-decreasing, optimal power distribution over time can be delineated. For this purpose, we present the solution to the following problem studied in~\cite{tajersequentialopt}, which is a more general problem that subsumes both problems $\R$ its special case. 
\begin{align}\label{eq:Q_bar1}
\Q(\bgamma) = \left\{
\begin{array}{lll}
\underset{\{p_b\}}{\max} & \sum\limits_{b=1}^B w_b(p_b) & \\
{\rm s.t.} & \sum\limits_{i=1}^{b}p_i  \leq  \gamma_b  & \forall\; b \\
& p_b \geq 0 & \forall\; b
\end{array}\right. \ ,
\end{align}
where $\bgamma = [\gamma_1,\dots,\gamma_B]$ and $w_b:\mathbb{R}_+\rightarrow\mathbb{R}_+$ is \textsl{strictly concave} and \textsl{non-decreasing}  in $p_b$.  Based on the expressions for $\Rb (p)$, the sum-rate over block $b$ depends on the power vector $\bp_b$ only through its sum $p_b$, defined in~\eqref{eq:power}. This implies that instead of enforcing the energy availability constraints in~\eqref{eq:energy_constraint}, we can equivalently enforce a constraint only on the aggregate power in each block. Hence, by defining 
\begin{equation}\label{eq:gamma}
\gamma_b=\sum_{i=1}^b\bone^{T}_n\cdot\bg_i\ ,
\end{equation}
the linear constraints in $\R$ can be equivalently stated as the linear constraint in $\Q$. The detailed steps of solving the problem~\eqref{eq:Q_bar1} analytically and the attendant performance guarantees are discussed in details in~\cite{tajersequentialopt}, a summary of which is provided next.

 In order to facilitate different steps in the analysis,  define the following \textsl{auxiliary} problem, solving which is instrumental to characterizing the properties of interest. Corresponding to each pair $i$ and $j$ such that $1\leq i< j\leq B$  define
\begin{align}\label{eq:Q_bar}
\Q_{i\rightarrow j}(\bgamma) = \left\{
\begin{array}{lll}
\underset{\{p_b\}}{\max} & \sum\limits_{b=i}^j w_b(p_b) & \\
{\rm s.t.} & \sum\limits_{b=i}^{j}p_b  = \gamma_j-\gamma_i  \\
 & p_b \geq 0, \qquad  \forall\; b \in\{i,\dots,j\}
\end{array}\right. \!\!\!\! ,
\end{align}
which has a unique globally optimal solution since the utility function is strictly concave.

{\renewcommand{\arraystretch}{1}{
\begin{center}
{\small
\begin{minipage}[t]{3.4 in}
\rule{\linewidth}{0.3mm}\vspace{0.0in}
{
{\bf  Algorithm~1 - Computing $\bp$}}\vspace{-.1in}\\
\rule{\linewidth}{0.3mm}\vspace{.1in}
{ {
\begin{tabular}{ll}
1:& {\bf set} $\gamma_b$ according to \eqref{eq:gamma} $\forall \;b\in\B$.\\
2:& {\bf initialize} $\: d=0$ and $u_{0}=0$,\\
3:& \textbf{\bf while} $u_d\;\leq \; B-1$\\
4: & \quad $d\leftarrow d+1$\\
5:&  \quad \textbf{\bf set} $\A_d= \{u_{d-1}+1,\dots, B\}$\\
6: & \quad  {\bf for} $b\in \A_d$ \\
7: & \quad \quad set $\by^{d,b}$ as the solution to $\Q_{u_{d-1}\rightarrow b}(\bgamma)$\\
8:& \quad \quad set $q^{d,b}= \min\left\{\frac{\der w_i}{\der y}(y^{d,b}_i):i\in\{u_{d-1}+1,\dots,b\}\right \}$\\
 9: & \quad  {\bf end for}\\
 10:& \quad $u_{d}= \arg\underset {b\in \A_d}{\max}\;\;q^{d, b}$ (if not unique select the smallest)\\
 11:& \quad $v_{d}= \underset{b\in \A_d}{\max}\;\;q^{d, b} $\\
 12:& \quad $\bz^{d}= \by^{d,u_{d}} $\\
 13:& {\bf end while}\\
 14:& {\bf for} $i\in\{1,\dots,d\}$\\
 15:& \quad {\bf for} $b\in\D_i=\{u_{i-1}+1,\dots, u_i\}$\\
 16:& \quad \quad $p_b\; = \; z_b^{i}$\\
 17 :& \quad{\bf end for} \\
 18 :& {\bf end for}
\end{tabular}}}\\
\rule{\linewidth}{0.3mm}
\end{minipage}}
\end{center}
}\vspace{-0.2in}
\subsubsection{Grouping the Constraints}
\label{sec:grouping}
The auxiliary term $\tilde\bp$ has a pivotal role in establishing the properties of $\bp$. Corresponding to $\bp$ we define the auxiliary vector $\tilde\bp$ by slightly modifying Algorithm~1. Specifically, by modifying line 1 to initiate the values of $\gamma_b$ according to $\gamma_b = \sum_{\ell=1}^b p_\ell$. This modified version of Algorithm~1 \textsl{successively} partitions the set of constraints $\{\sum_{\ell=1}^bp_\ell\; \leq \; \gamma_b\}$ into $d$ disjoint subsets of constraints. Specifically, it returns time indices $u_0 < u_1< \dots< u_d< B$, and partitions the set $\B$ into $d$ disjoint sets 
\begin{align}\label{eq:D}
\D_i = \{u_{i-1}+1,\dots, u_i\} \ ,\qquad \mbox{for}\quad i\in\{1,\dots,d\}\ .
\end{align}
Furthermore, this algorithm computes the metrics $\{v_i:i\in\{1,\dots,d\}\}$ and assigns $v_i$ to the set $\D_i$. Once these sets are known, solving $\Q$ reduces to solving a collection of smaller problems in the form of $\Q_{u_{i-1}\rightarrow u_i}(\bgamma)$ defined in~\eqref{eq:Q_bar}. The properties of $\tilde\bp$ are formalized next. 
\begin{Theorem}\label{lemma:constraint}
Given $\bp$ as the optimal solution to $\Q$, vector $\tilde \bp$ satisfies all the constraints of $\Q$. Furthermore, the vector $\tilde\bp$ satisfies $\sum_{b=1}^B w_b(\tilde p_b) \geq \sum_{b=1}^B w_b(p_b)$ and the equality holds if and only if $\bp=\tilde\bp$. 
\end{Theorem}
This establishes the optimality of $\tilde\bp$ generated by modifying Algorithm~1.
\begin{Theorem}\label{theorem:alg1_optimal}
If $\bp$ is the optimal solution to the problem $\Q$, then $\tilde\bp$ generated by modifying Algorithm~1 is also optimal. Uniqueness of $\bp$ indicates $\tilde\bp=\bp$. 
\end{Theorem}

\subsubsection{Dominant Constraints}
\label{sec:dominant}
By leveraging the results in the previous subsection, which partition the set of constraints into a collection of $d$ disjoint constraint sets, additional properties for these sets of constraints can be concluded. Specifically, in each of the given $d$ sets, it can be shown that one constraint holds with equality, which we refer to as the \textsl{dominant} constraint. These $d$ dominant constraints are the only constraints needed to characterize the optimal solution $\bp$. This property is formalized in the following theorem. 

\begin{Theorem}\label{lemma:v}
Under the optimal solution $\bp$, all the inequality constraints with indices included in $\{u_m:m\in\{1,\dots,d\}\}$ hold with equality. Furthermore, the sequence $\{v_1,v_2,\dots,v_d\}$ is strictly decreasing.
\end{Theorem}

We remark that the set of indices $\{u_i:i\in\{1,\dots, d\}\}$ and measures $\{v_i: i\in\{1,\dots, d\}\}$ have significant physical meanings in power allocation. The elements of $\{u_i:i\in\{1,\dots, d\}\}$ specify the time instances at which all the resources arrived by that time instance are entirely consumed. Furthermore, the second part of Lemma~\ref{lemma:v} establishes a connection among the derivative measures $q^{d,b}$ and $v_d$ defined in Algorithm~1. In particular, the measures $\{v_i:i\in\{1,\dots, d\}\}$ are the derivatives of the utility functions at the optimal solution $\bp$ over time.

\subsubsection{Optimality of Algorithm 1}
\label{sec:optimal}

So far we have shown that if we modify Algorithm 1 such that instead of initializing the terms $\gamma_b$ as defined in~\eqref{eq:gamma} we initialize them based on $\bp$, then the output will be in fact the optimal solution $\bp$. Next we show that initiating Algorithm 1 with either $\gamma_b = \sum_{\ell=1}^b p_\ell$ or according to~\eqref{eq:gamma} yields the same output. The underlying insight is that closer scrutiny of Algorithm~1 shows that this algorithm depends on $\bp$ primarily for determining the metrics $\{v_i: i\in\{1,\dots, d\}\}$ and their associated constraint indices $\{u_i:i\in\{1,\dots, d\}\}$. By invoking the result of Lemma~\ref{lemma:v}, we next show that for determining the sets $\{v_i:i\in\{1,\dots, d\}\}$ and $\{u_i:i\in\{1,\dots, d\}\}$, alternatively, we can also initialize $\gamma_b$ according to~\eqref{eq:gamma}, based on which we can show that the outcome of
Algorithm 1 will be, in fact, the optimal solution $\bp$. This observation is formalized in the following theorem.
\begin{Theorem}\label{theorem:alg2_optimal}
By setting $\gamma_b$ according to \eqref{eq:gamma}, Algorithm~1 generates the optimal solution of $\Q$.
\end{Theorem}

\section{Outlook}
\label{sec:outlook}

We conclude this survey by providing an outlook for some of the key open or uninvestigated research directions.

\vspace{.1 in}
\noindent {\bf Single-user MIMO channel.} Designing an optimal broadcast approach for the general MIMO channel is still an open problem since the MIMO channel is a non-degraded broadcast channel \cite{Chong14,Chong18}. Its capacity region is known for multiple users with private messages  \cite{Weingarten06}, and for two users with a common message \cite{GengNair14}. However, a complete characterization of the broadcast approach requires the full solution of the most general MIMO broadcast channel with a degraded message set \cite{KornaerMarton77}, which is not yet available (infinite number of realizations, for $H$ with Gaussian components), and hence suboptimal ranking procedures were considered. Various degraded message sets and transmission schemes with sub-optimal ranking at the receiver are studied in \cite{ShitzSteiner03,AsSh04,BustinPaySh13}. Formulation of the general MIMO broadcasting with degraded message sets and the optimization of the layering power distribution, which maximizes the expected rate, is stated in \eqref{eq17_1}-\eqref{eq17_2}. This optimization problem does not lend itself to a closed-form solution and remains an open problem for future research. The framework analyzed in \cite{romero2020rate}, which uses rate-splitting and binning, may be useful for the general broadcast problem with degraded message sets.
It is shown in \cite{GohariNair2020} that a tight upper bound might be obtained for the two users broadcast channel by adding an auxiliary receiver. Generalizing this work for multiple users may provide an efficient tool for obtaining outer bounds in general and on the MIMO broadcast approach.

The capacity region of a compound multiple-antenna broadcast channel is characterized under a particular degradedness order of users in \cite{Weingarten09}. The channel considered there has two users, where each user has a finite set of possible realizations. This again suggests that there is much room for further research to fully characterize the broadcast approach for the general MIMO channel. The majority of contributions discussed so far have considered Gaussian distribution for transmitted signals. It may be of interest to apply the broadcast approach to finite-input signals \cite{WuXiao2018}, or even binary-input channels \cite{GengNair2013}. This facilitates analyzing more practical settings and, in turn, obtaining tighter achievable bounds with the broadcast approach.

\vspace{.1 in}
\noindent {\bf Binary-dirty paper coding.} DPC has a pivotal role in Gaussian broadcast transmissions. Owing to its optimality for some settings (e.g., MIMO broadcast channel~\cite{Weingarten06}), an interesting research direction is investigating the performance or operation gains (e.g., rate and latency) of using DPC instead of superposition coding in the settings discussed in sections~\ref{sec:BCApproach}-\ref{sec:networks}. From a broader perspective, binning techniques facilitate DPC to be effective beyond Gaussian channels. In particular, Marton's general capacity region relies on the basic elements of binning~\cite{Marton1979}, in the context of which the classical Gelfand-Pinsker~\cite{GelfandPinsker80} strategy can be interpreted as a vertex point~\cite{GelfandPinsker80}. The Gelfand-Pinsker strategy in the Gaussian domain becomes DPC~\cite{Costa1983,GelfandPinsker80}. The study in~\cite{Somekh} addresses both binning and superposition coding aspects in a unified framework.  Furthermore, this study also investigates mismatched decoding, which can account for the imperfect availability of the CSI at the receivers. It is also noteworthy that throughout the paper, we primarily focused on the notion of physically degraded channels and rank-ordering them based on their degradedness. Nevertheless, it is important to investigate less restrictive settings, such as less-noisy channels~\cite{KornerMarton1975,ElGamal1979,KimElGamal2011}.

\vspace{.1 in}
\noindent {\bf Secrecy.} When considering the broadcast approach, it is
natural to look also at secrecy in communications. Such an approach not only involves determining which decoded messages depend on the channel state, but it also involves determining those that are required to be kept secret \cite{Liang12,Liang14,ZouLiang15,Hyadi}. This can be designed as part of the multi-layer broadcast approach.

\vspace{.1 in}
\noindent {\bf Latency.} There are various aspects in which delay constraints in communications may impact the system design, some of which were discussed in Section~\ref{sec:BCApproach}. There exists significant room for incorporating fixed-to-variable channel coding and variable-to-variable channel coding in the broadcast approach. In a way, this is a combination of variable-to-fixed coding (broadcast approach) and fixed-to-variable coding (that is, Fountain-like schemes). For example, some applications allow decoding following multiple independent transmission blocks, as considered in \cite{YEH01}, and studied by its equivalent channel setting, which is the MIMO parallel channel \cite{Kfir:IZS2020}. Queuing theory can be used to analyze the expected achievable latency, as in \cite{AsSh10}. An interesting observation is that layering often offers higher latency gains than throughput gains.
The problem of resource allocation for delay minimization, even under a simple queue model as in \cite{AsSh10}, remains an open problem for further research. Similarly, a generalization of the queue model with parallel queues associated with multiple streams, each with a different arrival random process and a different delay constraint, is an important direction to investigate.

\vspace{.1 in}
\noindent {\bf Connection to I-MMSE.}  It is well-known that the scalar additive Gaussian noise channel has the single crossing point property between the MMSE in the estimate of the input given the channel output. This property also provides an alternative proof to the capacity region of the scalar two-user Gaussian broadcast channel~\cite{Guo13}. This observation is extended to the vector Gaussian channel \cite{BustinPaySh13} via information-theoretic properties on the mutual information, using the I-MMSE relationship, a fundamental connection between estimation theory and information theory shown in~\cite{Guo13}. An interesting future direction is investigating the impact of I-MMSE relation on the broadcast approach.

\vspace{.1 in}
\noindent {\bf Information Bottleneck.} Another interesting setting is the information bottleneck channel. In this channel model, a wireless block fading channel is connected to a  reliable channel with limited capacity, referred to as
the bottleneck channel \cite{AsSh20_Bottleneck,SteinerShamai2020}. In these studies, it is assumed that the transmitted signal is Gaussian, which made it possible to describe the optimal
continuous layering power distribution in closed-form expressions. Extensions beyond Gaussian have both practical and theoretical significance.

One may consider the bottleneck channel setting, as depicted in Fig.~ \ref{fig:model}, where the transmitted signal is not necessarily Gaussian. Define the random variable triplet $x-t-z$ that form a Markov chain, and are related according to \eqref{eqChannelModel}, i.e., $y = x + n$, where $x$ and $n$ are independent random variables,
with $n\sim N(0,1)$ being real Gaussian with a unit variance. The transmitted signal $x$ distribution is subject to optimization, and SNR$=P\cdot s$, and $P$ are the transmission power $\bbe[x^2]=P$. The bottleneck channel output $z$ is a compressed version of $y$ adhering to a limited capacity of the bottleneck channel $C$, i.e., $I(y;z)\leq C$. It is of interest to maximize $I(x;z)$, with a maximizing probability that is not necessarily Gaussian,
see for example \cite{SAND08}. It is conjectured in \cite{Zaidi20} that the optimal distribution maximizing $I(x;z)$ is discrete.
The bottleneck channel may also consist of multiple independent relays connected through digital links to the destination, creating a distributed bottleneck. This setting is the CEO problem with logarithmic loss \cite{Aguerri18IZS,Ugur2018}, and under this setting, the broadcast approach for multi-access channels \cite{Tajer18} becomes very beneficial. With other loss functions, e.g., MMSE, the problem falls within source quality via broadcasting. Hence, the distributed bottleneck can also be viewed as source-channel coding problems with a distortion performance measure, as discussed in Section \ref{Sec6_2}. A model with two relays, known as the diamond channel, is also interesting and relevant. In the oblivious non-fading case, the optimal transmission and relay compression, together with joint decompression at the receiver, are known and characterized in \cite{sholomoDiamond19}. For the non-oblivious diamond channel, only upper bounds \cite{Michaelsholomo19}, and achievable rates of the type discussed in~\cite{Urbanke19} are available. It may also be interesting to consider the setting of recent work \cite{ZouhairBataineh20} and extend it to the case that no CSIT is available and consider a broadcast approach strategy for each user. Another possible direction is extending \cite{Karaksik13} to scenarios in which the variable backhaul links capacities $\{C_i\}$ are available only at the destination. Adapting the broadcast MIMO approach for the vector bottleneck channel \cite{ShitzSteiner03,6875354} is another important direction.

\vspace{.1 in}
\noindent {\bf Implementation.} The actual implementation of the broadcast approach, in general, is a rich topic for further research. Evidently, as it is mainly associated with layered decoding, this can be done by a variety of advanced coding and modulation techniques such as the low-density parity-check (LDPC) codes and turbo codes. The work in \cite{Gong:TCOM2011} considers LDPC implementation in conjunction with rate-splitting (no CSIT) in the interference channel, and \cite{Barak2008} provides bounds on LDPC codes over an erasure channel with variable erasures. Polar codes can be directly adopted for implementing the broadcast approach as their decoding is based on successive cancellations, and hence they naturally fit in the broadcast approach. Its efficiency has been demonstrated in the general broadcast channel \cite{Golea15,Mondelli14}, and further its
ability to work on general channels without adapting
the transmitter to the actual channel \cite{Mondelli17} demonstrates
the special features that are central to the broadcast approach.
Furthermore, its applicability to multiple description \cite{Bhatt17}
make it a natural candidate that can be used for
implementing joint source-channel coding via a broadcast approach.
Polar codes may also be used to practically address the variable-to-variable rate channel coding, as it is suitable for variable-to-fixed channel coding as well as fixed-to-variable
channel coding, as demonstrated in \cite{Li2016} for rateless codes. Power allocation across different information layers in special cases is investigated in~\cite{Boyle}, and there is room for further generalizing the results.

\vspace{.1 in}
\noindent {\bf Finite blocklength.} This paper focuses primarily on the asymptotically long transmission blocks. It is also essential to analyze the broadcast approach in the non-asymptotic block length regime. In such regimes, one could compromise the distribution of rates (asymptotic regime) with second-order descriptions, or even random coding error exponents, as there is a tradeoff between the error exponent rate of a finite block and the maximum rate. The practical aspects of communication under stringent finite blocklength constraints are discussed in~\cite{Mary}.

\vspace{.1 in}
\noindent {\bf Identification via channels.} The identification problem introduced in~\cite{ahlswede1989} is another case of a state-dependent channel. Its objective is communicating messages over a channel to select a {\em group} of messages at the receiver. This is in contrast to Shannon's formulation in which the objective is selecting {\em one} message. Many of the challenges pertinent to state-dependent channels and the lack of CSIT that appear in Shannon's formulation are relevant for the identification problem as well. Recent studies on the identification via channels without the CSIT include~\cite{pereg}.

\vspace{.1 in}
\noindent {\bf Mixed-delay constraints.} One major challenge in modern communication systems is heterogeneity in data type and their different attendant constraints. One such constraint pertains to latency, where different data types and streams can face various delay constraints. The broadcast approach investigated for addressing mixed-delay constraints in the single-user channel~\cite{CohenSteinerShamai12}, can be further extended to address this problem in more complex settings (e.g., soft handoff in cellular systems~\cite{Nikbakht_2020} and C-RAN uplink~\cite{Nikbakht2019}) while facing the lack of CSIT and in the context of fixed-to-variable channel coding~\cite{Verdu10variable-ratechannel} and fountain codes~\cite{Qureshi}.

\vspace{.1 in}
\noindent {\bf Source coding.} Another application is source coding with successive refinement where side information at the receiver (Wyner-Ziv) can be different, e.g., another communications link that might provide information and its quality is not known at the transmitter \cite{Kaspi94}. Another possible extension is
the combination of successive refinements and broadcast approach~\cite{Tian08}.

\vspace{.1 in}
\noindent {\bf Caching.}  In cooperative communication, it is common that relay stations perform data caching \cite{ParkSimeone16,KarasikSimeone18}, and the transmitter has no information about what is being cached. This random aspect of the amount and location (for multi-users) of cashing might play an interesting role in a broadcast approach for such a system.

\vspace{.1 in}
\noindent {\bf Algebraic structured codes.} The information-theoretic analyses of the networks reviewed in this paper generally are based on unstructured code design. In parallel to unstructured codes, there is rich literature on the structured design of codes with a wide range of applications to multi-terminal communication (e.g., multiple access and interference channels) and distributed source coding. A thorough recent overview of algebraic codes is available in~\cite{Pradhan:FnT2021}.

\vspace{.1 in}
\noindent {\bf Networking.} All different settings and scenarios discussed in this article play important roles in communication networks. As a network's size and complexity grow, the users cannot be all provided with the complete and instantaneous state of the networks. Specifically, in the future wireless systems (e.g., 6G), cell-based hierarchical network architectures will be dispensed with~\cite{Giordani}. In such networks, acquiring the CSI at the transmitters will be impossible, in which case the broadcast approach will be effective in circumventing the lack of CSIT. Furthermore, network coding can be incorporated in the broadcast approach, as it can account for latency, general wireless impediments (e.g., fading), and various network models, e.g., the relay, broadcast, interference, and multiple-access channels~\cite{Lee2010}.

\vspace{.1 in}
Finally, we highlight that the broadcast approach's hallmark is that it enables communication systems to adapt their key communication performance metrics (e.g., data rate, service latency, and message distortion) to the actual realizations of the communication channels. Such a feature is especially important as the size, scale, and complexity of the communication systems grow, rendering the instantaneous acquisition of channel realizations at the transmitters costly, if not prohibitive altogether. Adapting communication to unknown channels is an inherent property of communication systems in the pre-digital (analog) era, facilitating the mainstream adoption of broadcasting technologies for distributing audio and video contents. The broadcast technology instates this property in digital communication systems as well.

\vspace{6pt} 




\funding{The work of A. Tajer has been supported in part by the U.S. National Science Foundation under the grant ECCS-1933107. The work of S. Shamai has been supported by the
European Union's Horizon 2020 Research And Innovation Programme,
grant agreement no. 694630, and the WIN consortium via the Israel
minister of economy and science.}



\newpage
\abbreviations{The following abbreviations are used in this manuscript:

\noindent 
\begin{tabular}{@{}ll}
AF & amplify-and-forward \\
AQF & hybrid amplify-quantize-and-forward\\
AWGN & additive white Gaussian noise\\
BCC & broadcasting coherent combining\\
BIR & broadcasting incremental redundancy\\
CC & coherent combining\\
CDF & cumulative distribution function\\
CDMA & code-division multiple access\\
CF & compress-and-forward \\
CSI & channel state information\\
CSIT & channels state information at the transmitter sites\\
DC & delay constrained \\\
DF & decode-and-forward \\
DoF & Degrees-of-freedom\\
DS & direct-sequence \\
DSL & digital subscriber line\\
FCSI & full CSI\\
FUU & fraction of undecodable users  \\
HARQ & hybrid automatic retransmission request\\
HK & Han-Kobayashi\\
IR & incremental redundancy\\
LTSC & Long-term static channel\\
MAC & multi-access channel\\
MF & matched filter\\
MIMO & multiple input multiple output\\
MISO & multiple input single output\\
MLC & multi-level coding\\
MMSE & minimum mean squared-error\\
NDC & non-delay constrained \\
OAR & outage approach retransmission\\
PDF & probability distribution function\\
PET & priority encoding transmission\\
QF & quantize-and-forward\\
RV & random variable\\
SDF & sequential decode and forward\\
SIC & successive interference cancellation\\
SINR & signal to interference and noise ratio\\
SISO & single input single output\\
SIMO & single input multiple output\\
SNR & signal-to-noise ratio\\
SR & successive refinement\\

\end{tabular}

}
\appendixtitles{yes} 
\appendix

\section{Constants of Theorem~\ref{theorem:achievable_rate_finite}}
\label{app:constants:theorem:achievable_rate_finite}

\begin{align}
\nonumber & b_1(u,v)  =   \min_{j\in J_1}\left\{ C\big ( {s_v \beta_{uv}\; , \; s_jB_{1}(j,u,v)+s_vB_{2}(j,u,v)}\big) \right \}\ ,\\
\label{eq:b2} & b_2(u,v)  = C\big ( {s_v \beta_{uv} \; , \; (s_v+s_\ell)B_{3}(u,v)}\big)\ ,\\
\label{eq:b3} & b_3(u,v)   =  C\big ( {2s_v \beta_{uv}\; , \; 2s_vB_{3}(u,v)} \big)\ ,\\
\label{eq:b4} & b_4(u,v)  = C\big ( {s_u \beta_{vu}\; , \;  s_\ell B_{4}(u,v)+s_uB_{5}(u,v)}\big)\ ,\\
\label{eq:b5} & b_5(u,v)  =     C\big ( 2s_v \beta_{vu} \; , \;  2s_vB_3(u,v) \big)\ ,\\
\nonumber  & b_6(u,v)  = \min_{(j,k)\in J_2} \{ C\big (s_j \beta_{vu}+ s_k\beta_{uv}\; , \;   s_jB_{6}(k,u,v)+s_kB_{7}(k,u,v)\big) \}\ ,\\
\label{eq:b7} & b_7(u,v)  = C\big ( {s_v (\beta_{uv}+\beta_{vu}) , (s_v+s_\ell)B_{3}(u,v)}\big)\ ,\\
\label{eq:b8} & b_8(u,v)  =    C\big ( {2s_v (\beta_{uv}+\beta_{vu})\; , \; 2s_vB_{3}(u,v)} \big)\ ,\\
\nonumber & b_9(u,v)  = \min_{(j,k)\in J_3} \{ C\big (s_j(\beta_{uv}+\beta_{vu})+s_k\beta_{uv}\; , \;   (s_j+s_k)B_{3}(u,v)\big) \}\ ,\\
\nonumber & b_{10}(u,v)  = \min_{j,k\in J_3}\{ C\big ( s_j(\beta_{uv}+\beta_{vu})+s_k \beta_{vu}\; , \; (s_j+s_k)B_{3}(u,v))\big ) \}\ ,\\
\label{eq:b11} & b_{11}(u)  = C\big ( {s_u\beta_{uu}\; , \; (s_u+s_\ell)B_8(u,u)}\big)\ ,\\
\label{eq:b12} & b_{12}(u)  = C\big ({2s_u\beta_{uu} \; , \; 2s_uB_8(u,u)}\big)\ ,
\end{align}
and
\begin{align}
\label{eq:B1} B_{1}(j,u,v) & = 1 - \sum_{n=1}^j\sum_{m=1}^{v-1}\beta_{mn}-\sum_{n=1}^u\beta_{vn}\ ,\\
\label{eq:B2} B_{2}(j,u,v) & = 1 - \sum_{n=1}^{v-1}\sum_{m=1}^j\beta_{mn}-\sum_{n=1}^u\beta_{nv}\ ,\\
\label{eq:B3} B_{3}(u,v) & = 1 - \sum_{n=1}^{v-1}\sum_{m=1}^{v-1}\beta_{mn}-\sum_{n=1}^u\beta_{vn}-\sum_{n=1}^u\beta_{nv}\ ,\\
\label{eq:B4} B_{4}(u,v) & = 1 -  \sum_{n=1}^{v-1}\sum_{m=1}^u\beta_{mn}-\sum_{n=1}^u\beta_{nv} \ ,\\
\label{eq:B5} B_{5}(u,v) & = 1 - \sum_{n=1}^{u}\sum_{m=1}^{v-1}\beta_{mn}-\sum_{n=1}^u\beta_{vn}\ ,\\
\label{eq:B6} B_{6}(j,u,v) & = 1 -  \sum_{n=1}^j\sum_{m=1}^{v-1}\beta_{mn}-\sum_{n=1}^u\beta_{vn} \ ,\\
\label{eq:B7} B_{7}(j,u,v) &  = 1 -  \sum_{n=1}^{j}\sum_{m=1}^{v-1}\beta_{nm}-\sum_{n=1}^u\beta_{nv}\\
\label{eq:B8}  B_{8}(u,v) &  = 1 -  \sum_{n=1}^u\sum_{m=1}^v\beta_{mn}\ .
\end{align}

\section{Corner Points in Figure \ref{fig:outer_bound_achievable_region}}
\label{sec:app_cornerpoints}
The coordinates of the corner points of Fig.~\ref{fig:outer_bound_achievable_region} are specified as follows
\begin{align}
&{\sf T}: (0,b_1), \quad  {\sf U}:(b_2,b_1),  \quad {\sf V}: (b_7,b_1),\quad {\sf W}:(b_3,b_4) ,\nonumber\\
& {\sf X}:  (f_1,f_2) ,\quad  {\sf Y}: (b_5,b_6),  \quad   {\sf Z}:  (b_5,0),  
\end{align}
where we have defined
\begin{align}
b_1 & = p_1\;C(s_1,0)+p_2\;C(s_2,0)\, ,\\
b_2 & = 	q_1\;C\left(s_1,s_2\right)+q_2\;C\left(s_2,s_2\right)\, ,\\
b_3 & = q_1\; \rho_{i^*} +q_2 \hat{\rho}_{j^*}\, ,\\
b_4 & = p_1 \mu_{i^*}+ p_2 \hat{\mu}_{j^*}\, ,\\
b_5 & = q_1\;C(s_1,0)+q_2\;C(s_2,0)\, ,\\
b_6 & = p_{11}\;C\left(s_1,s_1\right)+p_{12}\;C\left(s_2,s_1\right) + p_{21}\;C\left(s_1,s_2\right)  +p_{22}\;C\left(s_2,s_2\right)\, ,\\
b_7 & = p_{11}\;C\left(s_1,s_1\right)+p_{21}\;C\left(s_2,s_1\right)+ p_{12}\;C\left(s_1,s_2\right)  +p_{22}\;C\left(s_2,s_2\right)\, ,\\
f_1 & = q_1 C\left(s_1,0\right)  + q_2\left[ C\left(s_2 \beta^1_{12},s_1 + s_2 \beta^1_{22}\right)
+\;C(s_2\beta^1_{22},0)\right]\, ,\\
f_2 & = p_{11} C(2s_1,0) +(p_{12}+p_{21}) C(s_1+s_2,0) + p_{22} C(2s_2,0)-f_1\, ,
\end{align}
and we have defined $i^* = \arg\max_{i} \mu_i$ and $j^* = \arg\max_{j} \hat{\mu}_j$ for
\begin{align}
\mu_1 &= \;p_1\; C(s_1, 0) + p_2 [C(s_1+s_2, 2s_1)  + C(s_1, 0)]\, ,\\
\mu_2 &= \;p_1\; [C(2s_1, s_1 + s_2) +C(s_2, 0)] +  p_2 C(s_2, 0)\, ,\\
\hat{\mu}_1 &= \;p_1\; C(s_1, 0) + p_2 [C(2s_2, s_1 + s_2)  + C(s_1, 0)]\, ,\\
\hat{\mu}_2 &= \;p_1\; [C(s_1 + s_2, 2 s_2) +C(s_2, 0)] +  p_2 C(s_2, 0)\, ,\\
\rho_1  &=  C(s_1,s_1)\ ,\\
 \rho_2 & = C(s_1, s_2)\, ,\nonumber \\
\hat{\rho}_1  &= C(s_2,s_1)\ , \\
\hat{\rho}_2 &= C(s_2, s_2)\, .
\end{align}

\reftitle{References}

%

\begin{thebibliography}{100}
\providecommand{\url}[1]{#1}
\csname url@samestyle\endcsname
\providecommand{\newblock}{\relax}
\providecommand{\bibinfo}[2]{#2}
\providecommand{\BIBentrySTDinterwordspacing}{\spaceskip=0pt\relax}
\providecommand{\BIBentryALTinterwordstretchfactor}{4}
\providecommand{\BIBentryALTinterwordspacing}{\spaceskip=\fontdimen2\font plus
\BIBentryALTinterwordstretchfactor\fontdimen3\font minus
  \fontdimen4\font\relax}
\providecommand{\BIBforeignlanguage}[2]{{%
\expandafter\ifx\csname l@#1\endcsname\relax
\typeout{** WARNING: IEEEtran.bst: No hyphenation pattern has been}%
\typeout{** loaded for the language `#1'. Using the pattern for}%
\typeout{** the default language instead.}%
\else
\language=\csname l@#1\endcsname
\fi
#2}}
\providecommand{\BIBdecl}{\relax}
\BIBdecl

\bibitem{Burnashev}
M.~V. Burnashev, ``{Data transmission over discrete channel with feedback:
  Random transmission time},'' \emph{Problemy Peredachi Informatsii}, vol.~12,
  no.~4, pp. 10--30, 1976.

\bibitem{Tchamkerten}
A.~Tchamkerten and E.~Telatar, ``Variable length coding over an unknown
  channel,'' \emph{IEEE Transactions on Information Theory}, vol.~52, no.~5,
  pp. 2126--2145, May 2006.

\bibitem{Shayevitz}
O.~Shayevitz and M.~Feder, ``{Achieving the empirical capacity using feedback:
  Memoryless additive models},'' \emph{IEEE Transactions on Information
  Theory}, vol.~55, no.~3, pp. 1269--1295, March 2009.

\bibitem{PPV:ISIT2010}
Y.~Polyanskiy, H.~V. Poor, and S.~Verd\'u, ``Variable-length coding with
  feedback in the non-asymptotic regime,'' in \emph{Proc. IEEE International
  Symposium on Information Theory}, Austin, TX, June 2010.

\bibitem{Tyagi}
H.~Tyagi and P.~Narayan, \emph{Excursions in Harmonic Analysis}, 2013, vol.~1,
  ch. State-dependent channels: Strong converse and bounds on reliability
  function, pp. 461--477.

\bibitem{Verdu10variable-ratechannel}
S.~Verd\'u and S.~Shamai~(Shitz), ``Variable-rate channel capacity,''
  \emph{IEEE Transactions on Information Theory}, vol.~56, no.~6, pp.
  2651--2667, June 2010.

\bibitem{SH98}
E.~Biglieri, J.~Proakis, and S.~Shamai~(Shitz), ``Fading channels:
  Information-theoretic and communication aspects,'' \emph{IEEE Transactions on
  Information Theory}, vol.~44, no.~6, pp. 2619--2692, October 1998.

\bibitem{TelatarOpp}
S.~Shamai~(Shitz) and E.~Telatar, ``Some information-theoretic aspects of
  decentralized power control in multiple access fading channels,'' in
  \emph{Proc. Information Theory and Networking Workshop}, Metsovo, Greece,
  June 1999.

\bibitem{Sharif}
M.~Sharif and B.~Hassibi, ``Delay considerations for opportunistic scheduling
  in broadcast fading channels,'' \emph{IEEE Transactions on Wireless
  Communications}, vol.~6, no.~9, pp. 3353--3363, September 2007.

\bibitem{asadi}
A.~Asadi and V.~Mancuso, ``A survey on opportunistic scheduling in wireless
  communications,'' \emph{IEEE Communications Surveys and Tutorials}, vol.~15,
  no.~4, pp. 1671--1688, Fourth Quarter 2013.

\bibitem{Zhao}
Q.~Zhao and B.~M. Sadler, ``A survey of dynamic spectrum access,'' \emph{IEEE
  Signal Processing Magazine}, vol.~24, no.~3, pp. 79 -- 89, May 2007.

\bibitem{Tanab}
M.~E. Tanab and W.~Hamouda, ``Resource allocation for underlay cognitive radio
  networks: {A} survey,'' \emph{IEEE Communications Surveys and Tutorials},
  vol.~19, no.~2, pp. 1249 -- 1276, Second Quarter 2016.

\bibitem{OZ98}
L.~Ozarow, S.~Shamai~(Shitz), and A.~Wyner, ``Information-theoretic
  considerations for cellular mobile radio,'' \emph{IEEE Transactions on
  Vehicular Technology}, vol.~43, no.~2, pp. 359--378, May 1994.

\bibitem{HanlyTse}
S.~V. Hanly and D.~N.~C. Tse, ``Multiaccess fading channels - {Part II:
  D}elay-limited capacities,'' \emph{IEEE Transactions on Information Theory},
  vol.~44, no.~7, pp. 2816--2831, November 1998.

\bibitem{LiJindalGoldsmith}
L.~Li, N.~Jindal, and A.~Goldsmith, ``Outage capacities and optimal power
  allocation for fading multiple-access channels,'' \emph{IEEE Transactions on
  Information Theory}, vol.~51, no.~4, pp. 1326--1347, April 2005.

\bibitem{narasimhan}
R.~Narasimhan, ``Individual outage rate regions for fading multiple access
  channels,'' in \emph{Proc. IEEE International Symposium on Information
  Theory}, Nice, France, June 2007, pp. 1571--1575.

\bibitem{Haghi}
A.~Haghi, R.~Khosravi-Farsani, M.~Aref, and F.~Marvasti, ``The capacity region
  of fading multiple access channels with cooperative encoders and partial
  {CSIT},'' in \emph{Proc. IEEE International Symposium on Information Theory},
  Austin, TX, June 2010, pp. 485--489.

\bibitem{DasNarayan}
A.~Das and P.~Narayan, ``Capacities of time-varying multiple-access channels
  with side information,'' \emph{IEEE Transactions on Information Theory},
  vol.~48, no.~1, pp. 4--25, January 2001.

\bibitem{jafar}
S.~Jafar, ``Capacity with causal and noncausal side information: {A} unified
  view,'' \emph{IEEE Transactions on Information Theory}, vol.~52, no.~12, pp.
  5468--5474, December 2006.

\bibitem{Kfir:IZS2020}
K.~M. Cohen, A.~Steiner, and S.~Shamai~(Shitz), ``On the broadcast approach
  over parallel {MIMO} two-state fading channel,'' in \emph{Proc. IEEE
  International Zurich Seminar on Information and Communication}, Zurich,
  Switzerland, February 2020.

\bibitem{KornaerMarton77}
J.~{K\"orner} and K.~{Marton}, ``General broadcast channels with degraded
  message sets,'' \emph{IEEE Transactions on Information Theory}, vol.~23,
  no.~1, pp. 60--64, January 1977.

\bibitem{NairGamal09}
C.~{Nair} and A.~El~Gamal, ``The capacity region of a class of three-receiver
  broadcast channels with degraded message sets,'' \emph{IEEE Transactions on
  Information Theory}, vol.~55, no.~10, pp. 4479--4493, October 2009.

\bibitem{ShitzSteiner03}
S.~Shamai~(Shitz) and A.~Steiner, ``{A broadcast approach for a single-user
  slowly fading MIMO channel},'' \emph{IEEE Transactions on Information
  Theory}, vol.~49, no.~10, pp. 2617--2635, October 2003.

\bibitem{CO72}
T.~M. Cover, ``Broadcast channels,'' \emph{IEEE Transactions on Information
  Theory}, vol.~18, no.~1, pp. 2--14, January 1972.

\bibitem{Shitz97broadcast}
S.~Shamai~(Shitz), ``A broadcast strategy for the {G}aussian slowly fading
  channel,'' in \emph{Proc. IEEE International Symposium on Information
  Theory}, Ulm, Germany, June 1997, p. 150.

\bibitem{bergergibson}
T.~Berger and J.~D. Gibson, ``Lossy source coding,'' \emph{IEEE Transactions on
  Information Theory}, vol.~44, no.~6, pp. 2693--2723, October 1998.

\bibitem{WWZ}
J.~K. Wolf, A.~D. Wyner, and J.~Ziv, ``Source coding for multiple
  descriptions,'' \emph{Bell System Technical Journal}, vol.~59, pp.
  1417--1426, 1980.

\bibitem{AsSh08_1}
A.~{Steiner} and S.~Shamai~(Shitz), ``The broadcast approach in communications
  systems,'' in \emph{Proc. IEEE Convention of Electrical and Electronics
  Engineers in Israel}, Eilat, Israel, December 2008.

\bibitem{successiveCover1991}
W.~H.~R. {Equitz} and T.~M. {Cover}, ``Successive refinement of information,''
  \emph{IEEE Transactions on Information Theory}, vol.~37, no.~2, pp. 269--275,
  March 1991.

\bibitem{RI99}
B.~Rimoldi, ``Successive refinement of information: {C}haracterization of the
  achievable rates,'' \emph{IEEE Transactions on Information Theory}, vol.~40,
  no.~1, pp. 253--259, January 1994.

\bibitem{NgTian07}
C.~T.~K. {Ng}, C.~{Tian}, A.~J. {Goldsmith}, and S.~Shamai~(Shitz), ``Minimum
  expected distortion in {G}aussian source coding with uncertain side
  information,'' in \emph{Proc. IEEE Information Theory Workshop}, Solstrand,
  Norway., July 2007, pp. 454--459.

\bibitem{Tian08}
C.~{Tian}, A.~{Steiner}, S.~Shamai~(Shitz), and S.~N. {Diggavi}, ``{Successive
  Refinement Via Broadcast: Optimizing Expected Distortion of a Gaussian Source
  Over a {G}aussian Fading Channel}aussian fading channel,'' \emph{IEEE
  Transactions on Information Theory}, vol.~54, no.~7, pp. 2903--2918, July
  2008.

\bibitem{Ng09}
C.~T.~K. {Ng}, D.~{Gunduz}, A.~J. {Goldsmith}, and E.~{Erkip}, ``Distortion
  minimization in {G}aussian layered broadcast coding with successive
  refinement,'' \emph{IEEE Transactions on Information Theory}, vol.~55,
  no.~11, pp. 5074--5086, November 2009.

\bibitem{NgTian12}
C.~T.~K. {Ng}, C.~{Tian}, A.~J. {Goldsmith}, and S.~Shamai~(Shitz), ``Minimum
  expected distortion in {G}aussian source coding with fading side
  information,'' \emph{IEEE Transactions on Information Theory}, vol.~58,
  no.~9, pp. 5725--5739, 2012.

\bibitem{DuhamelKieffer09}
P.~Duhamel and M.~Kieffer, \emph{{Joint Source-Channel Decoding. A Cross-Layer
  Perspective with Applications in Video Broadcasting over Mobile and Wireless
  Networks}}.\hskip 1em plus 0.5em minus 0.4em\relax Academic Press, 2009.

\bibitem{TR96}
M.~Trott, ``Unequal error protection codes: {T}heory and practice,'' in
  \emph{Proc. IEEE Information Theory Workshop}, Haifa, Israel, June 1996.

\bibitem{BO99}
S.~Boucheron and M.~R. Salamatian, ``About priority encoding transmission,''
  \emph{IEEE Transactions on Information theory}, vol.~46, no.~2, pp. 609--705,
  March 2000.

\bibitem{AL96}
------, ``Priority encoding transmission,'' \emph{IEEE Transactions on
  Information theory}, vol.~42, pp. 1737--1744, November 1996.

\bibitem{Woyach}
K.~Woyach, K.~Harrison, G.~Ranade, and A.~Sahai, ``Comments on unknown
  channels,'' in \emph{Proc. IEEE Information Theory Workshop}, Lausanne,
  Switzerland., September 2012.

\bibitem{Cover}
T.~M. Cover and J.~A. Thomas, \emph{Elements of Information Theory},
  2nd~ed.\hskip 1em plus 0.5em minus 0.4em\relax John Wiley \& Sons, 2006.

\bibitem{TS02}
D.~N.~C. Tse, ``Optimal power allocation over parallel \textsc{G}aussian
  broadcast channels,'' in \emph{Proc. IEEE International Symposium on
  Information Theory}, Ulm, Germany, June 1997, p.~27.

\bibitem{li98_1}
L.~Li and A.~Goldsmith, ``Capacity and optimal resource allocation for fading
  broadcast channels. {I}: Ergodic capacity,'' \emph{IEEE Transactions on
  Information Theory}, vol.~47, no.~3, pp. 1102--1127, March 2001.

\bibitem{li98_2}
------, ``Capacity and optimal resource allocation for fading broadcast
  channels. {II}: Outage capacity,'' \emph{IEEE Transactions on Information
  Theory}, vol.~47, no.~3, pp. 1103--1127, March 2001.

\bibitem{VI02}
P.~Viswanath and D.~N.~C. Tse, ``Sum capacity of the multiple antenna broadcast
  channel,'' in \emph{Proc. IEEE International Symposium on Information
  Theory}, Lausanne, Switzerland, July 2002.

\bibitem{SV02}
S.~Vishwanath, N.~Jindal, and A.~Goldsmith, ``Duality, achievable rates and
  sum-rate capacity of {G}aussian {MIMO} broadcast channel,'' \emph{IEEE
  Transactions on Information Theory}, vol.~49, no.~10, pp. 2658 -- 2668,
  October 2003.

\bibitem{KR02}
G.~Kramer, S.~Vishwanath, S.~Shamai~(Shitz), and A.~Goldsmith,
  ``Information-theoretic issues concerning broadcasting,'' in \emph{Proc. IEEE
  Workshop on Signal Processing for Wireless Communications}, Rutgers
  University, NJ, October 2002.

\bibitem{CA01_1}
G.~Caire and S.~Shamai~(Shitz), ``On the achievable throughput of a
  multi-antenna \textsc{G}aussian broadcast channel,'' \emph{IEEE Transactions
  on Information theory}, vol.~49, no.~7, pp. 1691 -- 1706, July 2002.

\bibitem{YU01}
W.~Yu and J.~Cioffi, ``The sum capacity of a \textsc{G}aussian vector broadcast
  channel,'' in \emph{Proc. IEEE International Symposium on Information
  Theory}, Lausanne, Switzerland, July 2002.

\bibitem{GA80}
A.~El~Gamal, ``Capacity of the product and sum of two unmatched broadcast
  channels,'' \emph{Problemy Peredachi Informatsii}, vol.~16, no.~1, pp. 3--23,
  January-March 1980.

\bibitem{Weingarten06}
H.~{Weingarten}, Y.~{Steinberg}, and S.~Shamai~(Shitz), ``The capacity region
  of the {G}aussian multiple-input multiple-output broadcast channel,''
  \emph{IEEE Transactions on Information Theory}, vol.~52, no.~9, pp.
  3936--3964, September 2006.

\bibitem{SE03}
S.~Sesia, G.~Caire, and G.~Vivier, ``Broadcasting a common source:
  Information-thoeretic results and system challenges,'' in \emph{Proc. IEEE
  International Symposium on Information Theory}, Monte Verita, Switzerland,
  February 2003.

\bibitem{FED01}
N.~Shulman and M.~Feder, ``Source broadcasting with unknown amount of receiver
  side information,'' in \emph{Proc. IEEE Information Theory Workshop},
  Banglore, India, October 2002.

\bibitem{SC97}
P.~Schramn, ``Multilevel coding with independent decoding on levels for
  efficient communications on static and interleaved fading channels,'' in
  \emph{Proc. IEEE Personal, Indoor and Mobile Radio Communications}, Helsinki,
  Finland, September 1997, pp. 1196--1200.

\bibitem{SC98}
D.~Schill and J.~Huber, ``On hierarchical signal constellations for the
  {G}aussian broadcast channel,'' in \emph{Proc. International Conference on
  Telecommunications}, Porto Carras, Greece, June 1998.

\bibitem{SC99}
D.~Schill, D.~Yuan, and J.~Huber, ``Efficient broadcasting using multilevel
  codes,'' in \emph{Proc. Information Theory and Networking Workshop}, Metsovo,
  Greece, June 1999, p.~72.

\bibitem{SA96}
M.~Sajadieh, F.~R. Kschischang, and A.~Leon-Garcia, ``Analysis of two-layered
  adaptive transmission systems,'' in \emph{Proc. IEEE Vehicular Technology
  Conference}, Atlanta, Georgia, April 1996, pp. 1771--1775.

\bibitem{Takesh01}
Y.~Liu, K.~Lau, C.~Takeshita, and M.~Fitz, ``Optimal rate allocation for
  superposition coding in quasi-static fading channels,'' in \emph{Proc. IEEE
  International Symposium on Information Theory}, Lausanne, Switzerland, July
  2002, p. 111.

\bibitem{V90}
A.~J. Viterbi, ``Very low rate conventional codes for maximum theoretical
  performance of spread-spectrum multiple-access channels,'' \emph{IEEE Journal
  on Selected Areas in Communications}, vol.~8, no.~4, pp. 641--649, May 1990.

\bibitem{GF91}
I.~Geldfand and S.~Fomin, \emph{Calculus of Variations}.\hskip 1em plus 0.5em
  minus 0.4em\relax Mineola, New-York: Courier Corporation, 2000.

\bibitem{AvestimehrTse07}
A.~S. {Avestimehr} and D.~N.~C. {Tse}, ``Outage capacity of the fading relay
  channel in the low-{SNR} regime,'' \emph{IEEE Transactions on Information
  Theory}, vol.~53, no.~4, pp. 1401--1415, April 2007.

\bibitem{BustinSh15}
R.~{Bustin}, R.~F. {Schaefer}, H.~V. {Poor}, and S.~Shamai~(Shitz), ``An
  i-{MMSE} based graphical representation of rate and equivocation for the
  {G}aussian broadcast channel\,'' in \emph{Proc. IEEE Conference on
  Communications and Network Security}, Florence, Italy, September 2015, pp.
  53--58.

\bibitem{AsSh08}
A.~{Steiner} and S.~Shamai~(Shitz), ``Achievable rates with imperfect
  transmitter side information using a broadcast transmission strategy,''
  \emph{IEEE Transactions on Wireless Communications}, vol.~7, no.~3, pp.
  1043--1051, 2008.

\bibitem{AsSh08_2}
------, ``Multi-layer broadcasting hybrid-{ARQ} strategies for block fading
  channels,'' \emph{IEEE Transactions on Wireless Communications}, vol.~7,
  no.~7, pp. 2640--2650, July 2008.

\bibitem{ShenLiuFitz08}
C.~{Shen}, T.~{Liu}, and M.~P. {Fitz}, ``Aggressive transmission with {ARQ} in
  quasi-static fading channels,'' in \emph{Proc. IEEE International Conference
  on Communications}, Shanghai, China, May 2008, pp. 1092--1097.

\bibitem{AsSh08_4}
A.~{Steiner} and S.~Shamai~(Shitz), ``Multi-layer broadcast hybrid-{ARQ}
  strategies,'' in \emph{Proc. IEEE International Zurich Seminar on Information
  and Communication}, Zurich, Switzerland, 2008, pp. 148--151.

\bibitem{T99}
E.~Telatar, ``Capacity of multi-antenna {G}aussian channels,'' \emph{European
  Transactions on Telecommunications}, vol.~10, no.~6, pp. 585--595, November
  1999.

\bibitem{GengNair14}
Y.~{Geng} and C.~{Nair}, ``The capacity region of the two-receiver {G}aussian
  vector broadcast channel with private and common messages,'' \emph{IEEE
  Transactions on Information Theory}, vol.~60, no.~4, pp. 2087--2104, April
  2014.

\bibitem{Chong14}
H.~{Chong} and Y.~{Liang}, ``The capacity region of the class of three-receiver
  {Gaussian MIMO} multilevel broadcast channels with two-degraded message
  sets,'' \emph{IEEE Transactions on Information Theory}, vol.~60, no.~1, pp.
  42--53, January 2014.

\bibitem{Chong18}
------, ``On the capacity region of the parallel degraded broadcast channel
  with three receivers and three-degraded message sets,'' \emph{IEEE
  Transactions on Information Theory}, vol.~64, no.~7, pp. 5017--5041, 2018.

\bibitem{AsSh04}
A.~{Steiner} and S.~Shamai~(Shitz), ``{Hierarchical coding for a MIMO
  channel},'' in \emph{Proc. IEEE Convention of Electrical and Electronics
  Engineers in Israel}, Tel-Aviv, Israel, September 2004, pp. 72--75.

\bibitem{BustinPaySh13}
R.~{Bustin}, M.~{Payaro}, D.~P. {Palomar}, and S.~Shamai~(Shitz), ``On {MMSE}
  crossing properties and implications in parallel vector {G}aussian
  channels,'' \emph{IEEE Transactions on Information Theory}, vol.~59, no.~2,
  pp. 818--844, February 2013.

\bibitem{MO79}
A.~Marshall and I.~Olkin, \emph{Inequalities: Theory of Majorization and Its
  Applications}.\hskip 1em plus 0.5em minus 0.4em\relax Academic Press, New
  York, 1979.

\bibitem{SH00}
S.~Shamai~(Shitz), ``A broadcast approach for the multiple-access slow fading
  channel,'' in \emph{Proc. IEEE International Symposium on Information
  Theory}, Sorrento, Italy, June 2000, p. 128.

\bibitem{SteinerShamai2007}
A.~Steiner and S.~Shamai~(Shitz), ``Multi-layer broadcasting over a block
  fading {MIMO} channel,'' \emph{IEEE Transactions on Wireless Communications},
  vol.~6, no.~11, pp. 3937 --3945, November 2007.

\bibitem{TG95}
E.~Telatar and R.~G. Gallager, ``Combining queueing theory with information
  theory for multiaccess,'' \emph{IEEE Selected Areas in Communications},
  vol.~13, pp. 963--969, August 1995.

\bibitem{HAJEK98}
A.~Ephremides and B.~Hajek, ``Information theory and communication networks: An
  unconsummated union,'' \emph{IEEE Transactions on Information Theory},
  vol.~44, no.~3, pp. 2416--2434, July 1998.

\bibitem{GALLAGER85}
R.~G. Gallager, ``A perspective on multiaccess channels,'' \emph{IEEE
  Transactions on information theory}, vol.~31, no.~2, pp. 124--142, March
  1985.

\bibitem{YooLiuShamai2012}
J.~W. Yoo, T.~Liu, S.~Shamai~(Shitz), and C.~Tian, ``Worst-case
  expected-capacity loss of slow-fading channels,'' \emph{IEEE Transactions on
  Information Theory}, vol.~59, no.~6, pp. 3764--3779, June 2013.

\bibitem{IDO01}
I.~Bettesh and S.~Shamai~(Shitz), ``Optimal power and rate control for minimal
  average delay: The single-user case,'' \emph{IEEE Transactions on Information
  Theory}, vol.~52, no.~9, pp. 4115--4141, September 2006.

\bibitem{AsSh10}
A.~{Steiner} and S.~Shamai~(Shitz), ``On queueing and multilayer coding,''
  \emph{IEEE Transactions on Information Theory}, vol.~56, no.~5, pp.
  2392--2415, 2010.

\bibitem{WOLFF89}
R.~W. Wolff, \emph{Stochastic Modeling and the Theory of Queues}.\hskip 1em
  plus 0.5em minus 0.4em\relax Englewood Cliffs, NJ: Perentice-Hall, 1989.

\bibitem{IDO02}
I.~Bettesh, ``Information and network theory aspects of communication systems
  in fading enviornment,'' Ph.D. dissertation, Technion -- Israel Institute of
  Technology, 2003.

\bibitem{kleirock_v2}
L.~Kleirock, \emph{Queueing Systems Volume 2: Theory}.\hskip 1em plus 0.5em
  minus 0.4em\relax New-York: John Wiley, 1975.

\bibitem{CohenSteinerShamai12}
K.~M. Cohen, A.~Steiner, and S.~Shamai~(Shitz), ``The broadcast approach under
  mixed delay constraints,'' in \emph{Proc. IEEE International Symposium on
  Information Theory}, Cambridge, MA, July 2012, pp. 209 --213.

\bibitem{Nikbakht2019}
H.~{Nikbakht}, M.~{Wigger}, W.~{Hachem}, and S.~Shamai~(Shitz), ``Mixed delay
  constraints on a fading {C-RAN} uplink,'' in \emph{Proc. IEEE Information
  Theory Workshop}, Visby, Sweden, August 2019.

\bibitem{Nikbakht_2020}
H.~Nikbakht, M.~A. Wigger, and S.~Shamai~(Shitz), ``Multiplexing gains under
  mixed-delay constraints on {W}yner's soft-handoff model,'' \emph{Entropy},
  vol.~22, no.~2, p. 182, February 2020.

\bibitem{YEH01}
P.~A. Whiting and E.~M. Yeh, ``Broadcasting over uncertain channels with
  decoding delay constraints,'' \emph{IEEE Transactions on Information Theory},
  vol.~52, no.~3, pp. 904--921, March 2006.

\bibitem{Cover1998CommentsBroadcast}
T.~M. {Cover}, ``Comments on broadcast channels,'' \emph{IEEE Transactions on
  Information Theory}, vol.~44, no.~6, pp. 2524--2530, October 1998.

\bibitem{zohdy2019broadcast}
M.~Zohdy, A.~Tajer, and S.~Shamai~(Shitz), ``Broadcast approach to multiple
  access with local {CSIT},'' \emph{IEEE Transactions on Communications},
  vol.~67, no.~11, pp. 7483--7498, August 2019.

\bibitem{Tajer18}
S.~{Kazemi} and A.~{Tajer}, ``Multiaccess communication via a broadcast
  approach adapted to the multiuser channel,'' \emph{IEEE Transactions on
  Communications}, vol.~66, no.~8, pp. 3341--3353, August 2018.

\bibitem{Costa1983}
M.~Costa, ``Writing on dirty paper,'' \emph{IEEE Transactions on Information
  Theory}, vol.~29, no.~3, pp. 439--441, May 1983.

\bibitem{CohenISIT2012}
A.~Cohen and A.~Lapidoth, ``Generalized writing on dirty paper,'' in
  \emph{Proc. IEEE International Symposium on Information Theory}, Lausanne,
  Switzerland, July 2002.

\bibitem{ahlsewde}
R.~Ahlswede, ``Multi-way communication channels,'' in \emph{Proc. IEEE
  International Symposium on Information Theory}, Hong kong, China, June 1971,
  pp. 103--105.

\bibitem{KimElGamal2011}
Y.-H. Kim and A.~El~Gamal, \emph{Network Information Theory}.\hskip 1em plus
  0.5em minus 0.4em\relax Cambridge, UK: Cambridge University Press, 2012.

\bibitem{CemalSteiberg}
Y.~Cemal and Y.~Steinberg, ``The multiple-access channel with partial state
  information at the encoders,'' \emph{IEEE Transactions on Information
  Theory}, vol.~51, no.~11, pp. 3992--4003, Nov. 2005.

\bibitem{SenComoYuksel}
N.~Sen, G.~Como, S.~Yuksel, and F.~Alajaji, ``On the capacity of memoryless
  finite-state multiple access channels with asymmetric noisy state information
  at the encoders,'' in \emph{Proc. Annual Allerton Conference on
  Communication, Control and Computing}, Monticello, IL, September 2011, pp.
  1210--1215.

\bibitem{BasherShiraziPermuter}
U.~Basher, A.~Shirazi, and H.~H. Permuter, ``Capacity region of finite state
  multiple-access channels with delayed state information at the
  transmitters,'' \emph{IEEE Transactions on Information Theory}, vol.~58,
  no.~6, pp. 3430--3452, June 2012.

\bibitem{SenAlajajiYuksel}
N.~{\c S}en, F.~Alajaji, S.~Yiiksel, and G.~Como, ``Multiple access channel
  with various degrees of asymmetric state information,'' in \emph{Proc. IEEE
  International Symposium on Information Theory}, Cambridge, MA, July 2012, pp.
  1697--1701.

\bibitem{SenAlajajiYiikselComo:2013}
N.~{\c S}en, F.~Alajaji, S.~Y{\"u}ksel, and G.~Como, ``Memoryless multiple
  access channel with asymmetric noisy state information at the encoders,''
  \emph{IEEE Transactions on Information Theory}, vol.~59, no.~11, pp.
  7052--7070, November 2013.

\bibitem{LapidothSteinberg:2013}
A.~Lapidoth and Y.~Steinberg, ``The multiple-access channel with causal side
  information: Double state,'' \emph{IEEE Transactions on Information Theory},
  vol.~59, no.~3, pp. 1379--1393, March 2013.

\bibitem{LapidothSteinberg:2013com}
------, ``The multiple-access channel with causal side information: Common
  state,'' \emph{IEEE Transactions on Information Theory}, vol.~59, no.~1, pp.
  32--50, January 2013.

\bibitem{LiSimeoneYener:2013}
M.~Li, O.~Simeone, and A.~Yener, ``Multiple access channels with states
  causally known at transmitters,'' \emph{IEEE Transactions on Information
  Theory}, vol.~59, no.~3, pp. 1394--1404, March 2013.

\bibitem{kotagiri2008multiaccess}
S.~P. Kotagiri and J.~N. Laneman, ``Multiaccess channels with state known to
  some encoders and independent messages,'' \emph{EURASIP Journal on Wireless
  Communications and Networking}, no.~1, 2008.

\bibitem{lapidoth2010multiple}
A.~Lapidoth and Y.~Steinberg, ``The multiple access channel with two
  independent states each known causally to one encoder.'' in \emph{Proc. IEEE
  International Symposium on Information Theory}, Austin, TX, June 2010, pp.
  480--484.

\bibitem{PermuterShamaiSomekh:2011}
H.~H. Permuter, S.~Shamai~(Shitz), and A.~Somekh-Baruch, ``Message and state
  cooperation in multiple access channels,'' \emph{IEEE Transactions on
  Information Theory}, vol.~57, no.~10, pp. 6379--6396, October 2011.

\bibitem{Wang:2012}
I.~H. Wang, ``Approximate capacity of the dirty multiple-access channel with
  partial state information at the encoders,'' \emph{IEEE Transactions on
  Information Theory}, vol.~58, no.~5, pp. 2781--2787, May 2012.

\bibitem{EmadiKhormujiSkoglundAref:2014}
M.~J. Emadi, M.~N. Khormuji, M.~Skoglund, and M.~R. Aref, ``Multi-layer
  {Gelfand-Pinsker} strategies for the generalised multiple-access channel,''
  \emph{IET Communications}, vol.~8, no.~8, pp. 1296--1308, May 2014.

\bibitem{MonemizadehBahmaniHodtaniSeyedin:2014}
M.~Monemizadeh, E.~Bahmani, G.~A. Hodtani, and S.~A. Seyedin, ``Gaussian doubly
  dirty compound multiple-access channel with partial side information at the
  transmitters,'' \emph{IET Communications}, vol.~8, no.~12, pp. 2181--2192,
  2014.

\bibitem{sreekumar2015distributed}
S.~Sreekumar, B.~K. Dey, and S.~R.~B. Pillai, ``Distributed rate adaptation and
  power control in fading multiple access channels,'' \emph{IEEE Transactions
  on Information Theory}, vol.~61, no.~10, pp. 5504--5524, October 2015.

\bibitem{EmadiZamanighomiAref:2012}
M.~J. Emadi, M.~Zamanighomi, and M.~R. Aref, ``Multiple-access channel with
  correlated states and cooperating encoders,'' \emph{IET Communications},
  vol.~6, no.~13, pp. 1857--1867, Sep. 2012.

\bibitem{PermuterWeissmanChen:2009}
H.~H. Permuter, T.~Weissman, and J.~Chen, ``Capacity region of the finite-state
  multiple-access channel with and without feedback,'' \emph{IEEE Transactions
  on Information Theory}, vol.~55, no.~6, pp. 2455--2477, June 2009.

\bibitem{Minero:ISIT07}
P.~Minero and D.~N.~C. Tse, ``A broadcast approach to multiple access with
  random states,'' in \emph{Proc. IEEE International Symposium on Information
  Theory}, Nice, France, June 2007, pp. 2566--2570.

\bibitem{KazemiTajer2017}
S.~Kazemi and A.~Tajer, ``A broadcast approach to multiple access adapted to
  the multiuser channel,'' in \emph{Proc. IEEE International Symposium on
  Information Theory}, Aachen, Germany, June 2017, pp. 883--887.

\bibitem{Zou13}
S.~Zou, Y.~Liang, and S.~Shamai~(Shitz), ``Multiple access channel with state
  uncertainty at transmitters,'' in \emph{Proc. IEEE International Symposium on
  Information Theory}, Istanbul, Turkey, July 2013, pp. 1466--1470.

\bibitem{Cao2007}
J.~Cao and E.~M. Yeh, ``Asymptotically optimal multiple-access communication
  via distributed rate splitting,'' \emph{IEEE Transactions on Information
  Theory}, vol.~51, no.~1, p. January, 304-319 2007.

\bibitem{knopp1995information}
R.~Knopp and P.~A. Humblet, ``Information capacity and power control in
  single-cell multiuser communications,'' in \emph{Proc. IEEE International
  Conference on Communications}, Seattle, WA, June 1995, pp. 331--335.

\bibitem{Jafar:IT2008}
V.~R. Cadambe and S.~A. Jafar, ``Interference alignment and degrees of freedom
  of the {$K$}-user interference channel,'' \emph{IEEE Transactions on
  Information Theory}, vol.~54, no.~8, pp. 3425 -- 3441, July 2008.

\bibitem{maddah2008communication}
M.~A. Maddah-Ali, A.~S. Motahari, and A.~K. Khandani, ``Communication over{
  MIMO} {X} channels: Interference alignment, decomposition, and performance
  analysis,'' \emph{IEEE Transactions on Information Theory}, vol.~54, no.~8,
  pp. 3457--3470, August 2008.

\bibitem{han1981new}
T.~Han and K.~Kobayashi, ``A new achievable rate region for the interference
  channel,'' \emph{IEEE Transactions on Information Theory}, vol.~27, no.~1,
  pp. 49--60, January 1981.

\bibitem{chong2008han}
H.-F. Chong, M.~Motani, H.~K. Garg, and H.~El~Gamal, ``On the
  {H}an--{K}obayashi region for the interference channel,'' \emph{IEEE
  Transactions on Information Theory}, vol.~54, no.~7, pp. 3188--3195, June
  2008.

\bibitem{etkin2008gaussian}
R.~H. Etkin, N.~David, and H.~Wang, ``Gaussian interference channel capacity to
  within one bit,'' \emph{IEEE Transactions on information theory}, vol.~54,
  no.~12, pp. 5534--5562, December 2008.

\bibitem{carleial1975case}
A.~Carleial, ``A case where interference does not reduce capacity,'' \emph{IEEE
  Transactions on Information Theory}, vol.~21, no.~5, pp. 569--570, September
  1975.

\bibitem{sato1981capacity}
H.~Sato, ``The capacity of the {G}aussian interference channel under strong
  interference,'' \emph{IEEE Transactions on Information Theory}, vol.~27,
  no.~6, pp. 786--788, November 1981.

\bibitem{benzel1979capacity}
R.~Benzel, ``The capacity region of a class of discrete additive degraded
  interference channels,'' \emph{IEEE Transactions on Information Theory},
  vol.~25, no.~2, pp. 228--231, March 1979.

\bibitem{gamal1982capacity}
A.~El~Gamal and M.~Costa, ``The capacity region of a class of deterministic
  interference channels,'' \emph{IEEE Transactions on Information Theory},
  vol.~28, no.~2, pp. 343--346, March 1982.

\bibitem{cadambe2009capacity}
V.~R. Cadambe, S.~A. Jafar, and S.~Vishwanath, ``The capacity region of a class
  of deterministic {Z}-channels,'' in \emph{Proc. IEEE International Symposium
  on Information Theory}, Seoul, South Korea, June 2009, pp. 2634--2638.

\bibitem{chong2007capacity}
H.-F. Chong, M.~Motani, and H.~K. Garg, ``The capacity region of a class of
  interference channels,'' in \emph{Proc. IEEE International Symposium on
  Information Theory}, Nice, France, June 2007, pp. 2856--2860.

\bibitem{bresler2008two}
G.~Bresler and D.~N.~C. Tse, ``The two-user {G}aussian interference channel:
  {A} deterministic view,'' \emph{European Transactions on Telecommunications},
  vol.~19, no.~4, pp. 333--354, April 2008.

\bibitem{Villacres2018}
G.~Villacr\'es, T.~Koch, A.~Sezgin, and G.~Vazquez-Vilar, ``Robust signaling
  for bursty interference,'' \emph{Entropy}, vol.~20, no.~11, p. 870, November
  2018.

\bibitem{YiSun20}
X.~{Yi} and H.~{Sun}, ``Opportunistic treating interference as noise,''
  \emph{IEEE Transactions on Information Theory}, vol.~66, no.~1, pp. 520--533,
  January 2020.

\bibitem{YiSum20TCOM}
------, ``Opportunistic topological interference management,'' \emph{IEEE
  Transactions on Communications}, vol.~68, no.~1, pp. 521--535, January 2020.

\bibitem{wang2014sliding}
L.~Wang, E.~Sasoglu, and Y.-H. Kim, ``Sliding-window superposition coding for
  interference networks,'' in \emph{Proc. IEEE International Symposium on
  Information Theory}, Honolulu, HI, June 2014, pp. 2749--2753.

\bibitem{bandemer2015optimal}
B.~Bandemer, A.~El~Gamal, and Y.-H. Kim, ``Optimal achievable rates for
  interference networks with random codes,'' \emph{IEEE Transactions on
  Information Theory}, vol.~61, no.~12, pp. 6536--6549, December 2015.

\bibitem{tuan2017superposition}
H.~D. Tuan, H.~H.~M. Tam, H.~H. Nguyen, T.~Q. Duong, and H.~V. Poor,
  ``Superposition signaling in broadcast interference networks,'' \emph{IEEE
  Transactions on Communications}, vol.~65, no.~11, pp. 4646 -- 4656, November
  2017.

\bibitem{yagi2011multi}
H.~Yagi and H.~V. Poor, ``Multi-level rate-splitting for synchronous and
  asynchronous interference channels,'' in \emph{Proc. IEEE International
  Symposium on Information Theory}, St. Petersburg, Russia, July 2011, pp.
  2080--2084.

\bibitem{zhao2012maximum}
Y.~Zhao, C.~W. Tan, A.~S. Avestimehr, S.~N. Diggavi, and G.~J. Pottie, ``On the
  maximum achievable sum-rate with successive decoding in interference
  channels,'' \emph{IEEE Transactions on Information Theory}, vol.~58, no.~6,
  pp. 3798--3820, June 2012.

\bibitem{geng2015optimalityJ}
C.~Geng, N.~Naderializadeh, A.~S. Avestimehr, and S.~A. Jafar, ``On the
  optimality of treating interference as noise,'' \emph{IEEE Transactions on
  Information Theory}, vol.~61, no.~4, pp. 1753--1767, April 2015.

\bibitem{Tajer:IT2016}
M.~Ashraphijuo, A.~Tajer, C.~Gong, and X.~Wang, ``A receiver-centric approach
  to interference management: Fairness and outage optimization,'' \emph{IEEE
  Transactions on Information Theory}, vol.~62, no.~10, pp. 5619--5642, October
  2016.

\bibitem{huang2012degrees}
C.~Huang, S.~A. Jafar, S.~Shamai~(Shitz), and S.~Vishwanath, ``On degrees of
  freedom region of {MIMO} networks without channel state information at
  transmitters,'' \emph{IEEE Transactions on Information Theory}, vol.~58,
  no.~2, pp. 849--857, February 2012.

\bibitem{zhu2011degrees}
Y.~Zhu and D.~Guo, ``The degrees of freedom of isotropic {MIMO} interference
  channels without state information at the transmitters,'' \emph{IEEE
  Transactions on Information Theory}, vol.~58, no.~1, pp. 341--352, January
  2012.

\bibitem{vaze2012degree}
C.~S. Vaze and M.~K. Varanasi, ``The degree-of-freedom regions of {MIMO}
  broadcast, interference, and cognitive radio channels with no {CSIT},''
  \emph{IEEE Transactions on Information Theory}, vol.~58, no.~8, pp.
  5354--5374, August 2012.

\bibitem{gou2011degrees}
T.~Gou, S.~A. Jafar, and C.~Wang, ``On the degrees of freedom of finite state
  compound wireless networks,'' \emph{IEEE Transactions on Information Theory},
  vol.~57, no.~6, pp. 3286--3308, June 2011.

\bibitem{shin2016mimo}
W.~Shin, B.~Lee, B.~Shim, and J.~Lee, ``A {MIMO} relay with delayed feedback
  can improve {DoF} in $k$-user {MISO} interference channel with no {CSIT},''
  \emph{IEEE Transactions on Vehicular Technology}, vol.~65, no.~12, pp.
  10\,188--10\,192, December 2016.

\bibitem{jeon2017degrees}
Y.-S. Jeon, N.~Lee, and R.~Tandon, ``Degrees of freedom and achievable rate of
  wide-band multi-cell multiple access channels with no {CSIT},'' \emph{IEEE
  Transactions on Communications}, vol.~66, no.~4, pp. 1772--1786, April 2017.

\bibitem{morales2019degrees}
M.~Morales-C{\'e}spedes, L.~Vandendorpe, and A.~G. Armada, ``Degrees of freedom
  of 2-tier networks without channel state information at the transmitter,''
  \emph{IEEE Signal Processing Letters}, vol.~26, no.~2, pp. 382--386, February
  2019.

\bibitem{jafar2012blind}
S.~A. Jafar, ``Blind interference alignment,'' \emph{IEEE Journal of Selected
  Topics in Signal Processing}, vol.~6, no.~3, pp. 216--227, June 2012.

\bibitem{lu2013blind}
Y.~Lu and W.~Zhang, ``Blind interference alignment in the {$K$}-user {MISO}
  interference channel,'' in \emph{Proc. IEEE Global Communications
  Conference}, Atlanta, GA, June 2013, pp. 3464--3469.

\bibitem{lu2014blind}
Y.~Lu, W.~Zhang, and K.~B. Letaief, ``Blind interference alignment with
  diversity in $ k $-user interference channels,'' \emph{IEEE Transactions on
  Communications}, vol.~62, no.~8, pp. 2850--2859, August 2014.

\bibitem{jafar2010exploiting}
S.~A. Jafar, ``Exploiting channel correlations-simple interference alignment
  schemes with no {CSIT},'' in \emph{Proc. IEEE Global Communications
  Conference}, Miami, FL, December 2010.

\bibitem{gou2011aiming}
T.~Gou, C.~Wang, and S.~A. Jafar, ``Aiming perfectly in the dark-blind
  interference alignment through staggered antenna switching,'' \emph{IEEE
  Transactions on Signal Processing}, vol.~59, no.~6, pp. 2734--2744, June
  2011.

\bibitem{wang2011improved}
C.~Wang, H.~C. Papadopoulos, S.~A. Ramprashad, and G.~Caire, ``Improved blind
  interference alignment in a cellular environment using power allocation and
  cell-based clusters,'' in \emph{Proc. IEEE International Conference on
  Communications}, Kyoto, Japan, June 2011.

\bibitem{akoum2012data}
S.~Akoum, C.~S. Chen, M.~Debbah, and R.~W. Heath, ``Data sharing coordination
  and blind interference alignment for cellular networks,'' in \emph{Proc. IEEE
  Global Communications Conference}, Anaheim, CA, December 2012, pp.
  4273--4277.

\bibitem{wang2014degrees}
C.~Wang, ``Degrees of freedom characterization: The 3-user {SISO} interference
  channel with blind interference alignment,'' \emph{IEEE Communications
  Letters}, vol.~18, no.~5, pp. 757--760, May 2014.

\bibitem{castanheira2017retrospective}
D.~Castanheira, A.~Silva, and A.~Gameiro, ``Retrospective interference
  alignment: Degrees of freedom scaling with distributed transmitters,''
  \emph{IEEE Transactions on Information Theory}, vol.~63, no.~3, pp.
  1721--1730, March 2017.

\bibitem{chen2017blind}
X.~Chen, Z.~Zhang, L.~Zheng, L.~Wu, J.~Dang, P.-S. Lu, and C.~Sun, ``Blind
  interference alignment in two-cell {Z} interference {MIMO} channel,''
  \emph{IEEE Access}, vol.~5, pp. 10\,526--10\,532, June 2017.

\bibitem{jafar2013topological}
S.~A. Jafar, ``Topological interference management through index coding,''
  \emph{IEEE Transactions on Information Theory}, vol.~60, no.~1, pp. 529--568,
  January 2013.

\bibitem{naderializadeh2014interference}
N.~Naderializadeh and A.~S. Avestimehr, ``Interference networks with no {CSIT}:
  Impact of topology,'' \emph{IEEE Transactions on Information Theory},
  vol.~61, no.~2, pp. 917--938, February 2014.

\bibitem{morales2014blind}
M.~Morales-C{\'e}spedes, J.~Plata-Chaves, D.~Toumpakaris, S.~A. Jafar, and
  A.~Garc{\i}, ``Blind interference alignment for cellular networks,''
  \emph{IEEE Transactions on Signal Processing}, vol.~63, no.~1, pp. 41--56,
  January 2014.

\bibitem{yang2015degrees}
H.~Yang, W.~Shin, and J.~Lee, ``Degrees of freedom for {$K$}-user {SISO}
  interference channels with blind interference alignment,'' in \emph{Proc.
  Asilomar Conference on Signals, Systems, and Computers}, Pacific Grove, CA,
  November 2015, pp. 1097--1101.

\bibitem{Akhlaghi:CL2011}
S.~Akhlaghi and M.~Baghani, ``On the average achievable rate of block fading
  decentralized interference channel,'' \emph{IEEE Communications Letters},
  vol.~15, no.~9, pp. 992--994, September 2011.

\bibitem{vahid2017binary}
A.~Vahid, M.~A. Maddah-Ali, A.~S. Avestimehr, and Y.~Zhu, ``Binary fading
  interference channel with no {CSIT},'' \emph{IEEE Transactions on Information
  Theory}, vol.~63, no.~6, pp. 3565--3578, June 2017.

\bibitem{zhu2016layered}
Y.~Zhu and C.~Shen, ``On layered erasure interference channels without {CSI} at
  transmitters,'' in \emph{Proc. IEEE International Symposium on Information
  Theory}, Barcelona, Spain, July 2016, pp. 710--714.

\bibitem{raja2009two}
A.~Raja, V.~M. Prabhakaran, and P.~Viswanath, ``The two-user compound
  interference channel,'' \emph{IEEE Transactions on Information Theory},
  vol.~55, no.~11, pp. 5100--5120, November 2009.

\bibitem{zhu2011ergodic}
Y.~Zhu and D.~Guo, ``Ergodic fading {Z}-interference channels without state
  information at transmitters,'' \emph{IEEE Transactions on Information
  Theory}, vol.~57, no.~5, pp. 2627--2647, May 2011.

\bibitem{lin2016ergodic}
P.-H. Lin, E.~A. Jorswieck, and R.~F. Schaefer, ``On ergodic fading {G}aussian
  interference channels with statistical {CSIT},'' in \emph{Proc. IEEE
  Information Theory Workshop}, Cambridge, UK, September 2016, pp. 454--458.

\bibitem{lin2019stochastic}
P.-H. Lin, E.~A. Jorswieck, C.~R. Janda, M.~Mittelbach, and R.~F. Schaefer,
  ``On stochastic orders and fading {G}aussian multi-user channels with
  statistical {CSIT},'' in \emph{Proc. IEEE International Symposium on
  Information Theory}, Paris, France, June 2019, pp. 1497--1501.

\bibitem{sebastian2015rate}
J.~Sebastian, C.~Karakus, S.~Diggavi, and I.-H. Wang, ``Rate splitting is
  approximately optimal for fading {G}aussian interference channels,'' in
  \emph{Proc. Annual Allerton Conference on Communication, Control and
  Computing}, Monticello, IL, September 2015, pp. 315--321.

\bibitem{sebastian2018approximate}
J.~Sebastian, C.~Karakus, and S.~Diggavi, ``Approximate capacity of fast fading
  interference channels with no instantaneous {CSIT},'' \emph{IEEE Transactions
  on Communications}, vol.~66, no.~12, pp. 6015 -- 6027, December 2018.

\bibitem{carleial1978interference}
A.~Carleial, ``Interference channels,'' \emph{IEEE Transactions on Information
  Theory}, vol.~24, no.~1, pp. 60--70, January 1978.

\bibitem{sason2004achievableJ}
I.~Sason, ``On achievable rate regions for the {G}aussian interference
  channel,'' \emph{IEEE Transactions on Information Theory}, vol.~50, no.~6,
  pp. 1345--1356, June 2004.

\bibitem{zohdytajersh:2020}
M.~Zohdy, A.~Tajer, and S.~Shamai~(Shitz), ``Distributed interference
  management: {A} broadcast approach,'' \emph{IEEE Transactions on
  Communications}, vol.~69, no.~1, pp. 149--163, January 2021.

\bibitem{Gong:TCOM2012}
C.~Gong, O.~Abu-Ella, A.~Tajer, and X.~Wang, ``Constrained group decoder for
  interference channels,'' \emph{Journal of Communications, Special Issue on
  Future Directions in Computing and Networking}, vol.~7, no.~5, pp. 382--390,
  May 2012.

\bibitem{Katz05}
M.~{Katz} and S.~Shamai~(Shitz), ``Transmitting to colocated users in wireless
  ad hoc and sensor networks,'' \emph{IEEE Transactions on Information Theory},
  vol.~51, no.~10, pp. 3540--3563, October 2005.

\bibitem{Katz06}
------, ``Relaying protocols for two colocated users,'' \emph{IEEE Transactions
  on Information Theory}, vol.~52, no.~6, pp. 2329--2344, June 2006.

\bibitem{Katz07}
------, ``On the outage probability of a multiple-input single-output
  communication link,'' \emph{IEEE Transactions on Wireless Communications},
  vol.~6, no.~11, pp. 4120--4128, June 2007.

\bibitem{Katz09}
------, ``Cooperative schemes for a source and an occasional nearby relay in
  wireless networks,'' \emph{IEEE Transactions on Information Theory}, vol.~55,
  no.~11, pp. 5138--5160, November 2009.

\bibitem{AsSh06TwoHop}
A.~{Steiner} and S.~Shamai~(Shitz), ``Single-user broadcasting protocols over a
  two-hop relay fading channel,'' \emph{IEEE Transactions on Information
  Theory}, vol.~52, no.~11, pp. 4821--4838, November 2006.

\bibitem{AsAsSh07}
A.~{Steiner}, A.~{Sanderovich}, and S.~Shamai~(Shitz), ``Broadcast cooperation
  strategies for two colocated users,'' \emph{IEEE Transactions on Information
  Theory}, vol.~53, no.~10, pp. 3394--3412, October 2007.

\bibitem{BraginskiyAsSh12}
E.~{Braginskiy}, A.~{Steiner}, and S.~Shamai~(Shitz), ``Oblivious sequential
  decode and forward cooperative strategies for the wireless relay channel,''
  \emph{IEEE Transactions on Communications}, vol.~60, no.~11, pp. 3228--3238,
  November 2012.

\bibitem{Zamani14}
M.~{Zamani} and A.~K. {Khandani}, ``Broadcast approaches to the diamond
  channel,'' \emph{IEEE Transactions on Information Theory}, vol.~60, no.~1,
  pp. 623--642, January 2014.

\bibitem{SimoneSh09}
O.~Simeone, O.~Somekh, E.~Erkip, H.~V. Poor, and S.~Shamai~(Shitz), ``A
  broadcast approach to robust communications over unreliable multi-relay
  networks,'' in \emph{Proc. IEEE Information Theory and Applications
  Workshop}, San Diego, CA, February 2009, pp. 334--340.

\bibitem{Baghani16}
M.~{Baghani}, S.~{Akhlaghi}, and V.~{Golzadeh}, ``Average achievable rate of
  broadcast strategy in relay-assisted block fading channels,'' \emph{IET
  Communications}, vol.~10, no.~3, pp. 346--355, March 2016.

\bibitem{Akhlaghi19}
S.~Akhlaghi and S.~A. Khodam~Hoseini, ``Power allocation strategies in
  block-fading two-way relay networks,'' \emph{Journal of Communication
  Engineering}, vol.~8, no.~2, pp. 313--324, 2019.

\bibitem{Attia14}
M.~A. {Attia}, M.~{Shaqfeh}, K.~{Seddik}, and H.~{Alnuweiri}, ``Power
  optimization for layered transmission over decode-and-forward relay
  channels,'' in \emph{Proc. IEEE International Wireless Communications and
  Mobile Computing Conference}, Nicosia, Cyprus, August 2014, pp. 594--599.

\bibitem{JNL04}
J.~{N. Laneman}, D.~N.~C. Tse, and G.~Wornell, ``Cooperative diversity in
  wireless networks: Efficient protocols and outage behavior,'' \emph{IEEE
  Transactions on Information Theory}, vol.~50, no.~12, pp. 3062--3080,
  December 2004.

\bibitem{DENIZ05}
D.~Gunduz and E.~Erkip, ``Opportunistic cooperation and power control
  strategies for delay-limited capacity,'' in \emph{Proc. Conference on
  Information Sciences and Systems}, Baltimore, MD, March 2005.

\bibitem{YUKSEL04}
M.~{Yuksel} and E.~{Erkip}, ``Diversity gains and clustering in wireless
  relaying,'' in \emph{Proc. IEEE International Symposium on Information
  Theory}, Chicago, IL, July 2004.

\bibitem{BOYER04}
J.~{Boyer}, D.~D. {Falconer}, and H.~{Yanikomeroglu}, ``On the aggregate {SNR}
  of amplified relaying channels,'' in \emph{Proc. IEEE Global Communications
  Conference}, Dallas, TX, December 2004, pp. 3394--3398.

\bibitem{LIU05}
Z.~Liu, V.~Stankovic, and Z.~Xiong, ``Practical compress-and-forward code
  design for the half-duplex relay channel,'' in \emph{Proc. IEEE Conference on
  Information Sciences and Systems}, Baltimore, MD, March 2005.

\bibitem{WYNER76}
A.~{Wyner} and J.~{Ziv}, ``The rate-distortion function for source coding with
  side information at the decoder,'' \emph{IEEE Transactions on Information
  Theory}, vol.~22, no.~1, pp. 1--10, 1976.

\bibitem{BERGER71}
T.~Berger, \emph{Rate Distortion Theory, A Mathematical Basis for Data
  Compression}.\hskip 1em plus 0.5em minus 0.4em\relax Englewood Cliffs, New
  Jersey: Prentice-Hall, 1971.

\bibitem{Abramowitz65}
M.~Abramowitz and I.~{Stegun (Eds.)}, \emph{Handbook of Mathematical
  Functions}.\hskip 1em plus 0.5em minus 0.4em\relax National Bureau of
  Standards, 1964; re-issued by Dover Publications, New York, 1965.

\bibitem{SM04}
Y.~{Steinberg} and N.~{Merhav}, ``{On successive refinement for the Wyner-Ziv
  problem},'' \emph{IEEE Transactions on Information Theory}, vol.~50, no.~8,
  pp. 1636--1654, August 2004.

\bibitem{ishwar}
P.~Ishwar, R.~Puri, K.~Ramchandran, and S.~S. Pradhan, ``On rate-constrained
  distributed estimation in unreliable sensor networks,'' \emph{IEEE Journal on
  Selected Areas in Communications}, vol.~34, no.~4, pp. 765--775, April 2005.

\bibitem{chenberger}
J.~Chen and T.~Berger, ``Robust distributed source coding,'' \emph{IEEE
  Transactions on Information Theory}, vol.~54, no.~8, pp. 3385--3398, August
  2008.

\bibitem{AsLupuKatzSh12}
A.~{Steiner}, V.~{Lupu}, U.~{Katz}, and S.~Shamai~(Shitz), ``The spectral
  efficiency of successive cancellation with linear multiuser detection for
  randomly spread {CDMA},'' \emph{IEEE Transactions on Information Theory},
  vol.~58, no.~5, pp. 2850--2873, May 2012.

\bibitem{KimPark20}
J.~{Kim} and S.~{Park}, ``Broadcast coding and successive refinement for
  layered {UE} cooperation in multi-user downlink,'' \emph{IEEE Wireless
  Communications Letters}, vol.~9, no.~6, pp. 893--896, June 2020.

\bibitem{AsSh20_Bottleneck}
A.~{Steiner} and S.~Shamai~(Shitz), ``Broadcast approach for the information
  bottleneck channel,'' in \emph{Proc. IEEE International Conference on
  Microwaves, Antennas, Communications and Electronic Systems}, Tel Aviv,
  Israel, November 2019.

\bibitem{SteinerShamai2020}
------, ``Broadcast approach under information bottleneck capacity
  uncertainty,'' in \emph{Proc. IEEE Information Theory and Applications
  Workshop}, San Diego, CA, February 2020.

\bibitem{steinerSh2020bottleneck}
A.~Steiner and S.~Shamai~(Shitz), ``Broadcast approach for the information
  bottleneck channel,'' \emph{IEEE Transactions on Communications}, 2021.

\bibitem{Liang14}
Y.~{Liang}, L.~{Lai}, H.~V. {Poor}, and S.~Shamai~(Shitz), ``A broadcast
  approach for fading wiretap channels,'' \emph{IEEE Transactions on
  Information Theory}, vol.~60, no.~2, pp. 842--858, February 2014.

\bibitem{Liang12}
------, ``An improved broadcast approach for fading wiretap channels,'' in
  \emph{Proc. IEEE International Symposium on Information Theory}, Cambridge,
  MA, July 2012.

\bibitem{Tulino09}
A.~{Tulino}, G.~{Caire}, and S.~Shamai~(Shitz), ``{Broadcast approach for the
  sparse-input random-sampled MIMO Gaussian channel},'' in \emph{Proc. IEEE
  International Symposium on Information Theory}, Honolulu, HI, July 2014, pp.
  621--625.

\bibitem{Simone09}
O.~{Simeone}, O.~{Somekh}, E.~{Erkip}, H.~V. {Poor}, and S.~Shamai~(Shitz),
  ``Multirelay channel with non-ergodic link failures,'' in \emph{Proc. IEEE
  Information Theory Workshop on Networking and Information Theory}, Volos,
  Greece, June 2009, pp. 52--56.

\bibitem{park2013multilayer}
S.-H. Park, O.~Simeone, O.~Sahin, and S.~Shamai~(Shitz), ``Multi-layer
  transmission and hybrid relaying for relay channels with multiple out-of-band
  relays,'' Tech. Rep., 2013.

\bibitem{ParkSimeoneSahinShamai2014}
S.~H. Park, O.~Simeone, O.~Sahin, and S.~Shamai~(Shitz), ``Robust layered
  transmission and compression for distributed uplink reception in cloud radio
  access networks,'' \emph{IEEE Transactions on Vehicular Technology}, vol.~63,
  no.~1, pp. 204--216, January 2014.

\bibitem{Park_2019}
S.-H. Park, O.~Simeone, and S.~Shamai~(Shitz), ``Robust baseband compression
  against congestion in packet-based fronthaul networks using multiple
  description coding,'' \emph{Entropy}, vol.~21, no.~4, p. 433, April 2019.

\bibitem{SimoneSomek11}
O.~{Simeone}, O.~{Somekh}, E.~{Erkip}, H.~V. {Poor}, and S.~Shamai~(Shitz),
  ``Robust communication via decentralized processing with unreliable backhaul
  links,'' \emph{IEEE Transactions on Information Theory}, vol.~57, no.~7, pp.
  4187--4201, July 2011.

\bibitem{ZouLiang15}
S.~{Zou}, Y.~{Liang}, L.~{Lai}, H.~V. {Poor}, and S.~Shamai~(Shitz),
  ``{Broadcast Networks With Layered Decoding and Layered Secrecy: Theory and
  Applications},'' \emph{Proceedings of the IEEE}, vol. 103, no.~10, pp.
  1841--1856, October 2015.

\bibitem{ZouLiang18}
------, ``Degraded broadcast channel with secrecy outside a bounded range,''
  \emph{IEEE Transactions on Information Theory}, vol.~64, no.~3, pp.
  2104--2120, March 2018.

\bibitem{Karaksik13}
R.~{Karasik}, O.~{Simeone}, and S.~Shamai~(Shitz), ``Robust uplink
  communications over fading channels with variable backhaul connectivity,'' in
  \emph{Proc. IEEE International Symposium on Information Theory}, July 2013,
  pp. 1172--1176.

\bibitem{ROY_6613623}
------, ``Robust uplink communications over fading channels with variable
  backhaul connectivity,'' \emph{IEEE Transactions on Wireless Communications},
  vol.~12, no.~11, pp. 5788--5799, November 2013.

\bibitem{Huleihel15}
W.~{Huleihel}, N.~{Merhav}, and S.~Shamai~(Shitz), ``On compressive sensing in
  coding problems: A rigorous approach,'' \emph{IEEE Transactions on
  Information Theory}, vol.~61, no.~10, pp. 5727--5744, October 2015.

\bibitem{SimoneSh10}
O.~{Simeone}, E.~{Erkip}, and S.~Shamai~(Shitz), ``Robust transmission and
  interference management for femtocells with unreliable network access,''
  \emph{IEEE Journal on Selected Areas in Communications}, vol.~28, no.~9, pp.
  1469--1478, December 2010.

\bibitem{TSEHANLY99}
D.~N.~C. Tse and S.~Hanly, ``Linear multiuser receivers: Effective
  interference, effective bandwidth and user capacity,'' \emph{IEEE
  Transactions on Information Theory}, vol.~45, no.~2, pp. 641--657, March
  1999.

\bibitem{VerduShamai99}
S.~Verd\'u and S.~Shamai~(Shitz), ``Spectral efficiency of {CDMA} with random
  spreading,'' \emph{IEEE Transactions on Information Theory}, vol.~45, no.~2,
  pp. 622--640, March 1999.

\bibitem{ShamaiVerdu01}
S.~Shamai~(Shitz) and S.~Verd\'u, ``The impact of frequency-flat fading on the
  spectral efficiency of {CDMA},'' \emph{IEEE Transactions on Information
  Theory}, vol.~47, no.~4, pp. 1302--1327, May 2001.

\bibitem{shamaiZaidel02}
S.~Shamai~(Shitz), B.~Zaidel, and S.~Verd\'u, ``{Strongest-users-only detectors
  for randomly spread CDMA},'' in \emph{Proc. IEEE International Symposium on
  Information Theory}, Sorrento, Italy, June 2002.

\bibitem{GOLDSMITH04}
M.~Medard, J.~Huang, A.~J. Goldsmith, S.~P. Meyn, and T.~P. Coleman, ``Capacity
  of time-slotted {ALOHA} packetized multiple-access systems over the
  \textsc{AWGN} channel,'' \emph{IEEE Transactions on Wireless Communications
  on Wireless communications}, vol.~3, no.~2, pp. 486--499, March 2004.

\bibitem{CAIRE04_VARIABLERATE}
G.~Caire, D.~Tuninetti, and S.~Verd\'u, ``Variable-rate coding for slowly
  fading {G}aussian multiple-access channels,'' \emph{IEEE Transactions on
  Information Theory}, vol.~50, no.~10, pp. 2271--2292, October 2004.

\bibitem{Koshelev:80}
V.~N. Koshelev, ``Hierarchical coding of discrete sources,'' \emph{Problemy
  Peredachi Informatsii}, vol.~16, no.~3, pp. 31--49, 1980.

\bibitem{Sesia:05}
S.~{Sesia}, G.~{Caire}, and G.~{Vivier}, ``Lossy transmission over slow-fading
  {AWGN} channels: a comparison of progressive and superposition and hybrid
  approaches,'' in \emph{Proc. IEEE International Symposium on Information
  Theory}, Adelaide, Australia, September 2005, pp. 224--228.

\bibitem{EtemadiJafarkhani:06}
F.~{Etemadi} and H.~{Jafarkhani}, ``Optimal layered transmission over
  quasi-static fading channels,'' in \emph{Proc. IEEE International Symposium
  on Information Theory}, Seattle, WA, July 2006, pp. 1051--1055.

\bibitem{CaireNarayanan:05}
G.~{Caire} and K.~{Narayanan}, ``On the distortion {SNR} exponent of hybrid
  digital-analog space-time coding,'' \emph{IEEE Transactions on Information
  Theory}, vol.~53, no.~8, pp. 2867--2878, August 2007.

\bibitem{Gunduz:06}
D.~{Gunduz} and E.~{Erkip}, ``Source and channel coding for quasi-static fading
  channels,'' in \emph{Proc. Asilomar Conference on Signals, Systems, and
  Computers}, Pacific Grove, CA, November 2005, pp. 18--22.

\bibitem{Tishby2004}
G.~Chechik, A.~Globerson, N.~Tishby, and Y.~Weiss, ``Information bottleneck for
  {G}aussian variables,'' in \emph{Proc. Advances in Neural Information
  Processing Systems}, S.~Thrun, L.~K. Saul, and B.~Sch\"{o}lkopf, Eds.,
  Vancouver, Canada, December 2004, pp. 1213--1220.

\bibitem{tishby99information}
N.~Tishby, F.~C. Pereira, and W.~Bialek, ``The information bottleneck method,''
  in \emph{Proc. Annual Allerton Conference on Communication, Control and
  Computing}, Monticello, IL, September 1999, pp. 368--377.

\bibitem{CaireShamai18}
G.~{Caire}, S.~Shamai~(Shitz), A.~{Tulino}, S.~{Verd\"u}, and C.~{Yapar},
  ``Information bottleneck for an oblivious relay with channel state
  information: The scalar case,'' in \emph{Proc. IEEE International Conference
  on the Science of Electrical Engineering}, Eilat, Israel, November 2018.

\bibitem{zohdytajerharvesting}
M.~Zohdy and A.~Tajer, ``Broadcast approach for the single-user energy
  harvesting channel,'' \emph{IEEE Transactions on Communications}, vol.~67,
  no.~5, pp. 3192 -- 3204, May 2019.

\bibitem{lu2015wireless2}
X.~Lu, P.~Wang, D.~Niyato, D.~I. Kim, and Z.~Han, ``Wireless networks with {RF}
  energy harvesting: A contemporary survey,'' \emph{IEEE Communications Surveys
  \& Tutorials}, vol.~17, no.~2, pp. 757--789, November 2015.

\bibitem{lu2015wireless}
X.~Lu, D.~Niyato, P.~Wang, D.~I. Kim, and Z.~Han, ``Wireless charger networking
  for mobile devices: Fundamentals, standards, and applications,'' \emph{IEEE
  Wireless Communications}, vol.~22, no.~2, pp. 126--135, April 2015.

\bibitem{panatik2016energy}
K.~Z. Panatik, K.~Kamardin, S.~A. Shariff, S.~S. Yuhaniz, N.~A. Ahmad, O.~M.
  Yusop, and S.~Ismail, ``Energy harvesting in wireless sensor networks: A
  survey,'' in \emph{Proc. IEEE International Symposium on Telecommunication
  Technologies}, Kuala Lumpur, Malaysia, May 2016, pp. 53--58.

\bibitem{jabbar2010rf}
H.~Jabbar, Y.~S. Song, and T.~T. Jeong, ``{RF} energy harvesting system and
  circuits for charging of mobile devices,'' \emph{IEEE Transactions on
  Consumer Electronics}, vol.~56, no.~1, pp. 247--253, March 2010.

\bibitem{huang2013spatial}
K.~Huang, ``Spatial throughput of mobile ad hoc networks powered by energy
  harvesting,'' \emph{IEEE Transactions on Information Theory}, vol.~59,
  no.~11, pp. 7597--7612, March 2013.

\bibitem{zhang2010energy}
X.~Zhang, H.~Jiang, L.~Zhang, C.~Zhang, Z.~Wang, and X.~Chen, ``An
  energy-efficient {ASIC} for wireless body sensor networks in medical
  applications,'' \emph{IEEE Transactions on Biomedical Circuits and Systems},
  vol.~4, no.~1, pp. 11--18, November 2010.

\bibitem{nishimoto2010prototype}
H.~Nishimoto, Y.~Kawahara, and T.~Asami, ``Prototype implementation of ambient
  {RF} energy harvesting wireless sensor networks,'' in \emph{Proc. IEEE
  Sensors}, Kona, HI, January 2010, pp. 1282--1287.

\bibitem{sudevalayam2011energy}
S.~Sudevalayam and P.~Kulkarni, ``Energy harvesting sensor nodes: Survey and
  implications,'' \emph{IEEE Communications Surveys \& Tutorials}, vol.~13,
  no.~3, pp. 443--461, July 2011.

\bibitem{tajersequentialopt}
A.~Tajer, M.~Zohdy, and K.~Alnajjar, ``Resource allocation under sequential
  resource access,'' \emph{IEEE Transactions on Communications}, vol.~66,
  no.~11, pp. 5608 -- 5620, November 2018.

\bibitem{romero2020rate}
H.~Romero and M.~K. Varanasi, ``Rate splitting, superposition coding and
  binning for groupcasting over the broadcast channel: A general framework,''
  2020.

\bibitem{GohariNair2020}
A.~{Gohari} and C.~{Nair}, ``New outer bounds for the two-receiver broadcast
  channel,'' in \emph{Proc. IEEE International Symposium on Information
  Theory}, Los Angeles, CA, June 2020, pp. 1492--1497.

\bibitem{Weingarten09}
H.~{Weingarten}, T.~{Liu}, S.~Shamai~(Shitz), Y.~{Steinberg}, and
  P.~{Viswanath}, ``The capacity region of the degraded multiple-input
  multiple-output compound broadcast channel,'' \emph{IEEE Transactions on
  Information Theory}, vol.~55, no.~11, pp. 5011--5023, November 2009.

\bibitem{WuXiao2018}
Y.~{Wu}, C.~{Xiao}, Z.~{Ding}, X.~{Gao}, and S.~{Jin}, ``{A Survey on MIMO
  Transmission With Finite Input Signals: Technical Challenges, Advances, and
  Future Trends},'' \emph{Proceedings of the IEEE}, vol. 106, no.~10, pp.
  1779--1833, October 2018.

\bibitem{GengNair2013}
Y.~{Geng}, C.~{Nair}, S.~Shamai~(Shitz), and Z.~V. {Wang}, ``On broadcast
  channels with binary inputs and symmetric outputs,'' \emph{IEEE Transactions
  on Information Theory}, vol.~59, no.~11, pp. 6980--6989, November 2013.

\bibitem{Marton1979}
K.~Marton, ``A coding theorem for the discrete memoryless broadcast channel,''
  \emph{IEEE Transactions on Information Theory}, vol.~25, no.~3, pp. 306--311,
  May 1979.

\bibitem{GelfandPinsker80}
S.~I. Gelfand and M.~S. Pinsker, ``Coding for channel with random parameters,''
  \emph{Problems of Control Theory}, vol.~9, no.~1, pp. 19--31, 1980.

\bibitem{Somekh}
A.~Somekh-Baruch, ``On achievable rates and error exponents for channels with
  mismatched decoding,'' \emph{IEEE Transactions on Information Theory},
  vol.~61, no.~2, pp. 727--740, February 2015.

\bibitem{KornerMarton1975}
J.~{K\"orner} and K.~{Marton}, ``A source network problem involving the
  comparison of two channels,'' \emph{Transactions on Colloquium Information
  Theory}, August 1975.

\bibitem{ElGamal1979}
A.~E. Gamal, ``The capacity of a class of broadcast channels,'' \emph{IEEE
  Transactions on Information Theory}, vol.~25, no.~2, pp. 166--169, May 1979.

\bibitem{Hyadi}
A.~Hyadi, Z.~Rezki, and M.-S. Alouini, ``An overview of physical layer security
  in wireless communication systems with {CSIT} uncertainty,'' \emph{IEEE
  Access}, vol.~4, pp. 6121 -- 6132, September 2016.

\bibitem{Guo13}
D.~{Guo}, S.~Shamai~(Shitz), and S.~Verd\'u, ``The interplay between
  information and estimation measures,'' \emph{Foundations and Trends in Signal
  Processing}, vol.~6, no.~4, pp. 243--429, November 2013.

\bibitem{SAND08}
A.~Sanderovich, S.~Shamai~(Shitz), Y.~Steinberg, and G.~Kramer, ``Communication
  via decentralized processing,'' \emph{IEEE Transactions on Information
  Theory}, vol.~54, no.~7, pp. 3008--3023, July 2008.

\bibitem{Zaidi20}
A.~Zaidi and I.~Estella-Aguerri, ``On the information bottleneck problems:
  Models, connections, applications and information theoretic views,''
  \emph{Entropy}, vol.~22, p. 151, 01 2020.

\bibitem{Aguerri18IZS}
I.~E. Aguerri and A.~Zaidi, ``Distributed information bottleneck method for
  discrete and {G}aussian sources,'' in \emph{Proc. IEEE International Zurich
  Seminar on Information and Communication}, Zurich, Switzerland, February
  2018.

\bibitem{Ugur2018}
Y.~{Ugur}, I.~E. {Aguerri}, and A.~{Zaidi}, ``Vector {G}aussian {CEO} problem
  under logarithmic loss,'' in \emph{Proc. IEEE Information Theory Workshop},
  Guangzhou, China, November 2018.

\bibitem{sholomoDiamond19}
I.~E. {Aguerri}, A.~{Zaidi}, G.~{Caire}, and S.~Shamai~(Shitz), ``On the
  capacity of cloud radio access networks with oblivious relaying,'' in
  \emph{Proc. IEEE International Symposium on Information Theory}, June 2017,
  pp. 2068--2072.

\bibitem{Michaelsholomo19}
X.~{Wu}, A.~{Ozgur}, M.~{Peleg}, and S.~Shamai~(Shitz), ``New upper bounds on
  the capacity of primitive diamond relay channels,'' in \emph{Proc. IEEE
  Information Theory Workshop}, Gotland, Sweden, August 2019.

\bibitem{Urbanke19}
M.~Mondelli, S.~H. Hassani, and R.~Urbanke, ``A new coding paradigm for the
  primitive relay channel,'' \emph{Algorithms}, vol.~12, no.~10, p. 218,
  October 2019.

\bibitem{ZouhairBataineh20}
Z.~Al-qudah, M.~Al~Bataineh, and A.~Musa, ``A novel multiple access diamond
  channel model,'' \emph{International Journal of Communication Systems},
  vol.~33, no.~17, November 2020.

\bibitem{6875354}
A.~{Winkelbauer}, S.~{Farthofer}, and G.~{Matz}, ``{The rate-information
  trade-off for Gaussian vector channels},'' in \emph{Proc. IEEE International
  Symposium on Information Theory}, Honolulu, HI, June 2014, pp. 2849--2853.

\bibitem{Gong:TCOM2011}
C.~Gong, A.~Tajer, and X.~Wang, ``Interference channels with partial group
  decoding,'' \emph{IEEE Transactions on Communications}, vol.~59, no.~11, pp.
  3059 -- 3071, November 2011.

\bibitem{Barak2008}
O.~{Barak}, U.~{Erez}, and D.~{Burshtein}, ``{Bounds on rates of LDPC codes for
  BEC with varying erasure rate},'' in \emph{Proc. IEEE International Symposium
  on Information Theory}, Toronto, Canada, July 2008, pp. 1133--1137.

\bibitem{Golea15}
N.~{Goela}, E.~{Abbe}, and M.~{Gastpar}, ``Polar codes for broadcast
  channels,'' \emph{IEEE Transactions on Information Theory}, vol.~61, no.~2,
  pp. 758--782, February 2015.

\bibitem{Mondelli14}
M.~{Mondelli}, S.~H. {Hassani}, R.~{Urbanke}, and I.~{Sason}, ``Achieving
  marton's region for broadcast channels using polar codes,'' in \emph{Proc.
  IEEE International Symposium on Information Theory}, Honolulu, HI, July 2014,
  pp. 306--310.

\bibitem{Mondelli17}
M.~{Mondelli}, S.~H. {Hassani}, I.~{Maric}, D.~{Hui}, and S.~{Hong},
  ``Capacity-achieving rate-compatible polar codes for general channels,'' in
  \emph{Proc. IEEE Wireless Communications and Networking Conference
  Workshops}, San Francisco, CA, March 2017.

\bibitem{Bhatt17}
A.~{Bhatt}, N.~{Ghaddar}, and L.~{Wang}, ``Polar coding for multiple
  descriptions using monotone chain rules,'' in \emph{Proc. Annual Allerton
  Conference on Communication, Control and Computing}, Monticello, IL,
  September 2017, pp. 565--571.

\bibitem{Li2016}
B.~{Li}, D.~N.~C. Tse, K.~{Chen}, and H.~{Shen}, ``Capacity-achieving rateless
  polar codes,'' in \emph{Proc. IEEE International Symposium on Information
  Theory}, Barcelona, Spain, July 2016.

\bibitem{Boyle}
B.~D. Boyle, J.~M. Walsh, and S.~Weber, ``Channel dependent adaptive modulation
  and coding without channel state information at the transmitter,'' in
  \emph{Proc. IEEE International Conference on Acoustics, Speech and Signal
  ProcessingProcessing}, Vancouver, Canada, May 2013.

\bibitem{Mary}
P.~Mary, J.-M. Gorce, A.~Unsal, and H.~V. Poor, ``Finite blocklength
  information theory: What is the practical impact on wireless
  communications?'' in \emph{Proc. IEEE Global Communications Conference
  (Workshops)}, Washington, DC, December 2016.

\bibitem{ahlswede1989}
R.~Ahlswede and G.~Dueck, ``Identification via channels,'' \emph{IEEE
  Transactions on Information Theory}, vol.~35, no.~1, pp. 15--29, January
  1989.

\bibitem{pereg}
U.~Pereg, H.~Boche, and C.~Deppe, ``Deterministic identification over fading
  channels,'' \emph{arXiv}, 2020.

\bibitem{Qureshi}
J.~Qureshi, C.~Heng~Foh, and J.~Cai, ``Primer and recent developments on
  fountain codes,'' \emph{Recent Advances in Communications and Networking
  Technology}, vol.~2, no.~1, pp. 2--11, July 2013.

\bibitem{Kaspi94}
A.~H. {Kaspi}, ``Rate-distortion function when side-information may be present
  at the decoder,'' \emph{IEEE Transactions on Information Theory}, vol.~40,
  no.~6, pp. 2031--2034, June 1994.

\bibitem{ParkSimeone16}
S.~{Park}, O.~{Simeone}, and S.~{Shamai}, ``Joint optimization of cloud and
  edge processing for fog radio access networks,'' in \emph{Proc. IEEE
  International Symposium on Information Theory}, Barcelona, Spain, July 2016.

\bibitem{KarasikSimeone18}
R.~{Karasik}, O.~{Simeone}, and S.~{Shamai}, ``Fundamental latency limits for
  {D2D}-aided content delivery in fog wireless networks,'' in \emph{Proc. IEEE
  International Symposium on Information Theory}, Vail, CO, June 2018.

\bibitem{Pradhan:FnT2021}
S.~S. Pradhan, A.~Padakandla, and F.~Shirani, ``{An Algebraic and Probabilistic
  Framework for Network Information Theory},'' \emph{Foundations and Trend in
  Communications and Information Theory}, vol.~18, no.~2, pp. 173--379, 2020.

\bibitem{Giordani}
M.~Giordani, M.~Polese, M.~Mezzavilla, S.~Rangan, and M.~Zorzi, ``Toward 6{G}
  networks: Use cases and technologies,'' \emph{IEEE Communications Magazine},
  vol.~58, no.~3, pp. 55--61, March 2020.

\bibitem{Lee2010}
H.-N. Lee, S.-Y. Chung, C.~Fragouli, and Z.-H. Mao, ``Network coding for
  wireless networks,'' \emph{EURASIP Journal on Wireless Communications and
  Networking}, vol. 2010.

\end{thebibliography}


\end{document}